\documentclass[twocolumn,superscriptaddress,floatfix,preprintnumbers,letterpaper]{revtex4-1}
\usepackage{gensymb}
\usepackage{float}
\usepackage{array}
\usepackage{dcolumn}
\usepackage{bm}
\usepackage[colorlinks, pdfborder={0 0 0}, plainpages=false]{hyperref}
\usepackage{graphicx}
\usepackage{xspace}
\usepackage[usenames,dvipsnames]{color}
\usepackage{amssymb}
\usepackage[normalem]{ulem} 
\usepackage{lineno}
\usepackage{tabularx}
\usepackage{letltxmacro}
\usepackage{amsfonts}
\usepackage{amsmath}
\usepackage{amssymb}
\usepackage{bm}
\usepackage{dcolumn}
\usepackage{epsfig}
\usepackage{grffile}
\usepackage[latin1]{inputenc}
\usepackage{latexsym}
\usepackage{rotating}
\usepackage{hyperref}
\usepackage[usenames]{color}
\usepackage{float}
\usepackage{xspace} 
\usepackage{mathrsfs}
\usepackage{subfigure}
\usepackage{multirow}
\usepackage{booktabs}
\usepackage{soul}
\usepackage{xcolor}
\usepackage{colortbl}
\usepackage{etoolbox}

\def\prd{{\it Phys. Rev.} D~}

\def\apjl{{\it Astrophys. J.} Lett~}
\def\apj{{\it Astrophys. J.}}

\def\PRD{{\it Phys. Rev.} D~}

\def\mnras{{\it MNRAS~}}

\newcolumntype{C}[1]{>{\centering\arraybackslash}p{#1}}

\def\lesssim{\mathrel{\hbox{\rlap{\hbox{\lower4pt\hbox{$\sim$}}}\hbox{$<$}}}}
\def\gtrsim{\mathrel{\hbox{\rlap{\hbox{\lower4pt\hbox{$\sim$}}}\hbox{$>$}}}}
\def\alt{\mathrel{\hbox{\rlap{\hbox{\lower4pt\hbox{$\sim$}}}\hbox{$<$}}}}
\def\agt{\mathrel{\hbox{\rlap{\hbox{\lower4pt\hbox{$\sim$}}}\hbox{$>$}}}}

\usepackage{mathtools}

\newenvironment{cititemize2}
{\begin{list}{$\bullet$}
        {\setlength{\topsep}{0pt}
         \setlength{\itemsep}{0pt}
         \setlength{\parsep}{0.25\parsep}
         \settowidth{\labelwidth}{$\bullet$}
         \setlength{\leftmargin}{1em}
}
}
{\end{list}}

\def\gta{\ifmmode {\mathbin{\lower 3pt\hbox   
    {$\,\rlap{\raise 5pt\hbox{$\char'076$}}\mathchar"7218\,$}}}
    \else {${\mathbin{\lower 3pt\hbox
    {$\rlap{\raise 5pt\hbox{$\char'076$}}\mathchar"7218\,$}}}
    $}\fi}
\def\lta{\ifmmode {\,\mathbin{\lower 3pt\hbox   
    {$\,\rlap{\raise 5pt\hbox{$\char'074$}}\mathchar"7218\,$}}}
    \else {${\mathbin{\lower 3pt\hbox
    {$\rlap{\raise 5pt\hbox{$\char'074$}}\mathchar"7218\,$}}}
    $}\fi}

\newcommand{\msun}{{\rm M}_{\odot}}
\newcommand{\beq}{\begin{equation}}
\newcommand{\eeq}{\end{equation}}
\newcommand{\bea}{\begin{eqnarray}}
\newcommand{\eea}{\end{eqnarray}}

\definecolor{darkperiwinkle}{RGB}{102, 102, 128}

\newcommand{\NCSA}{\affiliation{National Center for Supercomputing Applications, University of Illinois at Urbana-Champaign, Urbana, Illinois 61801, USA}}
\newcommand{\ANCSA}{\affiliation{Department of Astronomy, University of Illinois at Urbana-Champaign, Urbana, Illinois 61801, USA}}
\newcommand{\PNCSA}{\affiliation{Department of Physics, University of Illinois at Urbana-Champaign, Urbana, Illinois 61801, USA}}

\newcommand{\BM}{\affiliation{University of Birmingham, Birmingham B15 2TT, United Kingdom}}
\newcommand{\COR}{\affiliation{Cornell Center for Astrophysics and Planetary Science, Cornell University, Ithaca, New York 14853, USA}}

\newcommand{\ICTS}{\affiliation{International Centre for Theoretical Sciences, Tata Institute of Fundamental Research, Bangalore 560012, India}}

\definecolor{light-gray}{gray}{0.9}

\newcommand{\comment}[1]{}

\begin{document}

\title{Observation of eccentric binary black hole mergers with second and third generation gravitational wave detector networks}

\author{Zhuo Chen}\NCSA\PNCSA
\author{E. A. Huerta}\NCSA\PNCSA\ANCSA
\author{Joseph Adamo}\NCSA\PNCSA
\author{Roland Haas}\NCSA
\author{Eamonn O'Shea}\COR
\author{Prayush Kumar}\ICTS\COR
\author{Chris Moore}\BM
\date{\today}

\begin{abstract}
\noindent We introduce an improved version of the Eccentric, Non-spinning, Inspiral-Gaussian-process Merger Approximant (\texttt{ENIGMA}) waveform model that utilizes a more stable and robust numerical method to smoothly connect the analytical relativity-based inspiral evolution with the numerical relativity-based merger phase. We find that this ready-to-use model can: (i) produce physically consistent signals, without reporting any failures, when sampling over 1M samples that were randomly chosen over the \(m_{\{1,\,2\}}\in[5\msun,\,50\msun]\) parameter space, and the entire range of binary inclination angles; (ii) produce waveforms within 0.04 seconds, averaged over 1000 iterations, from an initial gravitational wave frequency \(f_{\textrm{GW}} =15\,\textrm{Hz}\) and at a sample rate of \(8192\,\textrm{Hz}\); and (iii) reproduce the physics of quasi-circular mergers, since its overlap with 
\texttt{SEOBNRv4} waveforms is \({\cal{O}}\geq0.99\) assuming advanced LIGO zero detuned high power noise configuration, and signals generated from \(f_{\textrm{GW}}= 15\,\textrm{Hz}\). We utilize \texttt{ENIGMA} to compute the expected signal-to-noise ratio (SNR) distributions of eccentric binary black hole mergers assuming the existence of second and third generation gravitational wave detector networks. For second generation detectors, we assume the geographical location and detection sensitivities of the twin LIGO detectors, Virgo, KAGRA, LIGO-India, and a  LIGO-type detector in Australia. For third generation detectors, we assume the geographical location of second generation detectors, and use the proposed sensitivities of the Cosmic Explorer and the Einstein Telescope. In the context of advanced LIGO-type detectors, we find that the SNR of eccentric mergers is always larger than quasi-circular mergers for systems with \(e_0\leq0.4\) at \(f_{\textrm{GW}} =10\,\textrm{Hz}\), even if the timespan of eccentric signals is just a third of quasi-circular systems with identical total mass and mass-ratio. For Cosmic Explorer-type detector networks, we find that eccentric mergers have similar SNRs than quasi-circular systems for \(e_0\leq0.3\) at \(f_{\textrm{GW}} =10\,\textrm{Hz}\). Systems with \(e_0\sim0.5\)  at \(f_{\textrm{GW}} =10\,\textrm{Hz}\) have SNRs that range between 50\%-90\% of the SNR produced by quasi-circular mergers, even if these eccentric signals are just between a third to a tenth the length of quasi-circular systems. For Einstein Telescope-type detectors, we find that eccentric mergers have similar SNRs than quasi-circular systems for \(e_0\leq0.4\) at \(f_{\textrm{GW}} =5\,\textrm{Hz}\). The most eccentric events in our sample,  \(e_0\sim0.6\) at \(f_{\textrm{GW}} =5\,\textrm{Hz}\), merge at least five times faster than quasi-circular systems, and may be detectable with SNRs between 50\%-85\% the SNR of quasi-circular mergers with identical binary components.
\end{abstract}

\pacs{Valid PACS appear here}
\maketitle


\section{Introduction}
\label{intro}

The gravitational wave (GW) detection of binary black hole (BBH) mergers 
with the advanced LIGO~\cite{LSC:2015,DII:2016} and advanced Virgo~\cite{Virgo:2015} detectors is now a common occurrence~\cite{LIGOScientific:2018mvr,Abbott:2020uma,LIGOScientific:2020stg}. As these detectors gradually reach their target sensitivity, and more detectors join the existing GW detector network, it is expected that an ever increasing number 
of GW observations will enable statistical analyses that may shed new and detailed 
information about the astrophysical origin of compact binary sources~\cite{Belczynski:2014iua,bel:2016Na,2017NatCo814906S,deMink:2016MNRAS}. In this paper we are particularly 
interested in BBHs formed in dense stellar environments, such as core-collapsed globular clusters or galactic nuclei, and are expected to enter 
the frequency band of ground-based GW detectors with non-negligible eccentricity~\cite{galcen:2018,Sippel:2013,Strader:2012,ssm:2018,sam:2017ApJ,Samsing:2014,sam:2018PhRvD3014S,Leigh:2018MNRAS,ssm:2017,ssm:2018,samsing:2018ApJ140S,lisa:2018b,Huerta:2009,sam:2017ApJ84636S,sam:2018MNRAS1548S,Huerta:2015a,sam:2019MNRAS30S,sam:2018PhRvD3014S,sam:2018MNRAS5436S,Huerta:2014,Anton:2014,samdor:2018MNRAS5445S,samdorII:2018MNRAS4775D,samdor:2018arXiv64S,samjoh:2018Z,rocarl:2018PhRvDR,kremerjoh:2018aK,lopez:2018L,hoang:2017APJ,gon:2017,hpoang:2017,lisa:2018a,MikKoc:2012,Naoz:2013,gondkoc:2018G,antonras:2016ApJ7A,Huerta:2013a,arcakoc:2018A,takkoc:2018ApJT,gondkoc:2018ApJ5G,antoni:2018A,Anto:2015arXiv}. Recent studies suggest that the expected detection rate of eccentric BBH mergers 
for second generation detectors is about
 100 mergers per cubic Gpc per year~\cite{Salemi:2019owp}, and about 1700 mergers per cubic Gpc 
 per year for binary neutron star mergers~\cite{Nitz_2020}. For third generation detectors, it may be possible to 
 observe from several hundred to a few thousand eccentric mergers per cubic Gpc per year~\cite{Huerta:2014,PhysRevD.98.083028}.

The study and modeling of eccentric BBH systems has gained traction in recent years. 
Waveform models that describe the GW emission of these 
sources has rapidly evolved from inspiral-only GW models~\cite{Huerta:2013a,Huerta:2014,Hinder:2010,Tiwari:2019jtz,Ebersold:2019kdc,Boetzel:2019nfw,Ireland:2019tao,Moore:2019xkm,Loutrel:2018ydu,lou:2016arXiv}, to semi-analytical and machine-learning based 
models that describe the inspiral-merger-ringdown evolution of these GW sources~\cite{huerta:2018PhRvD,Huerta:2017a,hinder:2017a,cao:2017,Hinderer:2017,Liu:2019jpg,Chiaramello:2020ehz}. 
Numerical relativity has also been used to obtain insights into the non-linear dynamics of 
these systems throughout merger and ringdown~\cite{huerta_nr_catalog,ian:2017,Habib:2019cui,johnson:2017,ramos_buades}, and to study the impact of higher-order waveform 
modes for the detection of eccentric BBH mergers~\cite{Adam:2018prd}. Recent studies have also shed light on how to 
extract signatures of dynamical formation in GW sources detected by advanced LIGO and Virgo~\cite{Romero_Shaw_2019,Salemi:2019owp,Tiwari:2016,Ramos-Buades:2020eju,wu_ecc_mnras}, estimate 
the eccentricity of GW signals~\cite{PhysRevD.98.083028}, including the recent detection  GW190425~\cite{Romero_Shaw_2020}.

In this article we introduce an improved version of the inspiral-merger-ringdown \texttt{ENIGMA} waveform model~\cite{huerta:2018PhRvD}, 
which describes the GW emission of non-spinning BHs that evolve on moderately eccentric 
orbits, to compare the signal-to-noise (SNR) distribution of quasi-circular and moderately 
eccentric BBH mergers.  This study is motivated by a number of observations, e.g., it has been 
documented in the literature that GWs emitted by BBHs that evolve on moderately eccentric orbits 
exhibit the following properties~\cite{huerta_nr_catalog}: 
(i) the waveform amplitude of eccentric signals tends to be larger than that of quasi-circular ones 
during the early inspiral evolution; (ii) GWs produced by eccentric BBH mergers are shorter than those 
of quasi-circular BBH mergers. In view of these observations, one may naturally like to 
explore the astrophysical scenarios in which eccentric and quasi-circular BBH mergers have 
similar SNR distributions, and to quantify the eccentricity threshold at which condition (i) above no 
longer compensates for condition (ii) leading to BBH mergers that produce GWs whose SNRs are lower 
than quasi-circular ones. This article aims to shed light on these points in the context of 
second and third generation, ground-based GW detector networks.

The rest of this article is organized as follows. Section~\ref{sec:enigma} provides a brief 
description of the new features in the \texttt{ENIGMA} waveform model~\cite{huerta:2018PhRvD}, 
and a number of benchmark analyses 
we have conducted to thoroughly test it. Section~\ref{sec:dat} introduces data analysis tools 
that we use throughout this article. In Section~\ref{sec:networks} we describe the second and 
third generation GW detector networks considered in this study, including the power spectral 
densities (PDSs) 
used to represent each GW detector. We present and discuss our results in 
Sections~\ref{sec:second_results} and \ref{sec:third_results}. We summarize 
our findings and future work in Section~\ref{sec:end}.


\section{ENIGMA waveform model}
\label{sec:enigma}

The \texttt{ENIGMA} model has two pieces. The first combines post-Newtonian (PN) results 
that describe the dynamics of moderately eccentric BBH mergers, and encompasses instantaneous, 
tails and tails-of-tails contributions, and a contribution due to non-linear memory~\cite{huerta:2018PhRvD,Arun:2009PRD,ArunBlanchet:2008,Blanchet:2006}. This 
framework is equivalent to the PN approximant TaylorT4 at 3PN order in the zero eccentricity 
limit. \texttt{ENIGMA} also incorporates higher-order PN corrections for the energy flux of 
quasi-circular binaries and gravitational self-force corrections to the binding energy of compact 
binaries up to 6PN order. The equations of motion that describe the orbital dynamics and the 
radiative evolution of the system are expressed in terms of the gauge-invariant quantity 
\(x=\left(M\,\omega\right)^{2/3}\), where \(M\) represents the total mass of the binary and 
\(\omega\) is the mean orbital frequency. We use this parametrization because it provides 
an accurate description of eccentric BBH mergers when directly compared to numerical 
relativity simulations~\cite{Hinder:2010,huerta:2018PhRvD,habhu:2019}. 

The second part of the \texttt{ENIGMA} model is a stand-alone merger waveform. 
Since \texttt{ENIGMA} is constructed under the assumption that moderately eccentric 
compact binaries circularize before merger, this part of the model is designed to 
describe the late inspiral-merger-ringdown of quasi-circular BBHs. The merger part 
is constructed using Gaussian process emulation~\cite{Rasmussen:2005}, i.e., we 
train a machine learning algorithm to do interpolation using a dataset of numerical 
relativity waveforms that describe non-spinning BHs on quasi-circular orbits~\cite{Boyle:2019kee}. 
Once trained, this Gaussian emulator produces merger waveforms within milliseconds. We 
then conduct an optimization procedure to identify the optimal time at which we can 
smoothly connect the inspiral and merger pieces of the waveform model. The 
inclusion of higher-order post-Newtonian and self-force corrections improves the 
modeling of the inspiral phase, facilitating its attachment to the merger waveform during 
the late-time inspiral evolution. We fine-tune this late-time attachment condition by maximizing the 
overlap, computed using advanced LIGO zero detuned high power configuration and 
a minimum filtering frequency of 15Hz, between quasi-circular \texttt{ENIGMA} 
waveforms and \texttt{SEOBNRv4} waveforms~\cite{Bohe:2016gbl}. By construction, 
the higher-order PN and self-force results we use 
to model the inspiral evolution of \texttt{ENIGMA} are reliable for 
eccentricities \(e_0 \leq 0.6\)---see the analytical expressions used 
for this purpose in Appendix A of~\cite{Huerta:2017a}. Furthermore, since we also assume that the binary systems circularize prior 
to attaching the quasi-circular merger waveform, the waveform approximant will only attach a 
merger waveform is the residual eccentricity prior to merger is \(\leq 0.01\). 
If this condition is not met, then the approximant will output an error stating that the 
residual eccentricity is too large to attach a quasi-circular merger waveform.

\noindent \textbf{New features of the  \texttt{ENIGMA} model} The two key improvements 
to \texttt{ENIGMA} are in the merger attachment and the handling of the binary inclination angle. 

For the merger attachment, we sample the range \(M\omega\in[0.02,\,0.1]\) 
in steps of \(1\times10^{-4}\), and for each frequency value, \(M\omega_i\), we construct 
a complete \texttt{ENIGMA} waveform whose merger phase is attached at \(M\omega_i\). 
Each of these waveforms is then compared with an \texttt{SEOBNRv4} waveform that describes 
the same binary system, densely covering the parameter space \(m_{\{1,\,2\}}\in[5\msun,\,50\msun]\). 

For this calibration procedure, we use standard tools, namely, if \(h\) and \(s\) denote 
\texttt{ENIGMA} and  \texttt{SEOBNRv4} waveforms, \(S_n(f)\) the power spectral density 
(PSD) of a given GW detector, 
and \(\tilde{h}(f)\) the Fourier transform of \(h(t)\), the noise-weighted inner product 
between \(h\) and \(s\) is given by

\beq
\left( h | s\right) = 2 \int^{f_1}_{f_0} \frac{\tilde{h}^{*}(f)\tilde{s}(f) + \tilde{h}(f)\tilde{s}^{*}(f) }{S_n(f)}\mathrm{d}f\,.
\label{inn_pro}
\eeq

\begin{figure*}[htbp!]
	\centerline{
\includegraphics{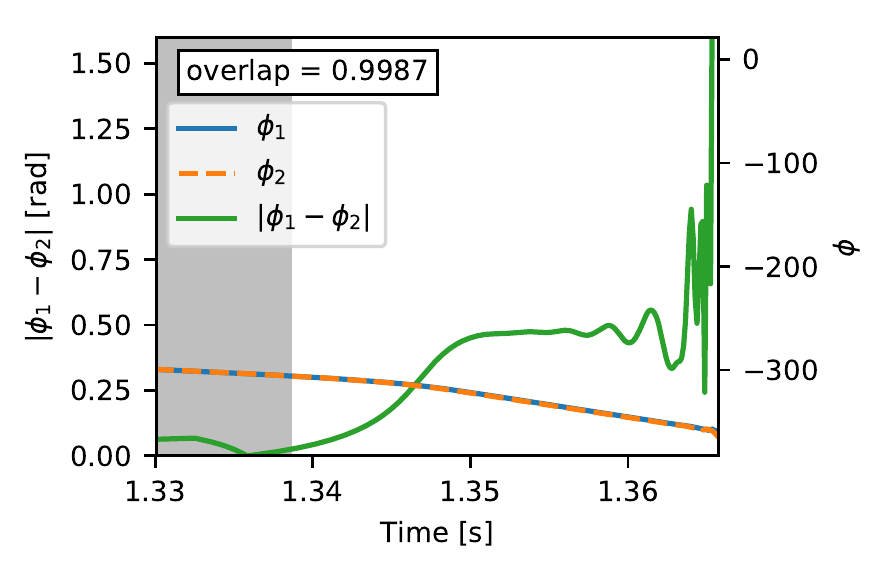}\hfill%
\includegraphics{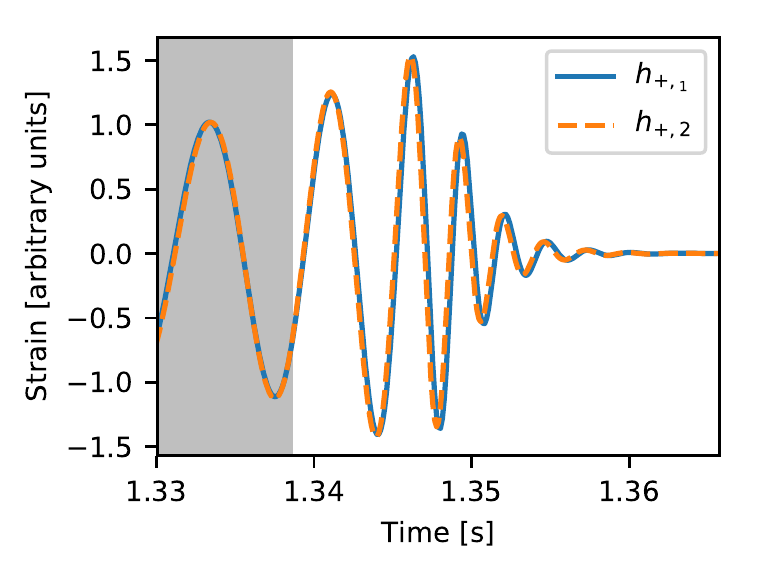}
}
\caption{Comparison between the original \texttt{ENIGMA} model
in~\cite{huerta:2018PhRvD}, and the model introduced in this 
article that has been recalibrated with the updated \texttt{SEOBNRv4} 
model found in LIGO's Algorithm Library~\cite{lalsuite}.
The binary black hole system used for this comparison has total mass 
$M=40\,M_\odot$, mass ratio $q=4$, and initial eccentricity $e_0=0.1$ at a
gravitational wave
frequency $f_\text{GW} = 25\,\mathrm{Hz}$. The left panel
shows the phase difference $\phi_1 - \phi_2$ between the gravitational wave
templates around the point in time where the merger-ringdown part is attached
to the inspiral part. Differences before that point are due to using a
different time integration method and interpolation scheme. Phase
differences beyond that are due to the updated matching parameters obtained
when recalibrating against \texttt{SEOBNRv4}.
The right panel shows the $+$ polarization of the
gravitational wave strain during the same time interval. The shaded area is
the pre-attachment inspiral region and the non-shaded area is the
merger-ringdown part.}
\label{fig:attachment_diff}
\end{figure*}

\noindent For calibration purposes, we set the initial GW frequency to  
\(f_0= f_{\textrm{GW}} =15\,\textrm{Hz}\), and use the zero detuned high power noise curve 
for advanced LIGO~\cite{updated_aligo} with \(f_1=4096\,{\rm Hz}\). 
The waveforms are produced at a sample rate of \(8192\,\textrm{Hz}\). The overlap between 
\(h\) and \(s\) is defined as 

\begin{align}
\label{over}
{\cal{O}}  (h,\,s)&= \underset{ t_c\, \phi_c}{\mathrm{max}}\left(\hat{h}|\hat{s}_{[t_c,\,  \phi_c]}\right)\quad{\rm with}\\
\label{n_overl}
\hat{h}&=h\,\left(h | h\right)^{-1/2}\,,
\end{align}

\noindent where \(\hat{s}_{[t_c,\,  \phi_c]}\)  indicate that the normalized waveform \(\hat{s}\) 
has been time- and phase-shifted. We then determine the attachment frequency value 
\(M\omega^{*}\) that maximizes the overlap between the two aforementioned waveform 
families across the parameter space under consideration, and use it to compute the exact 
attachment time \(t_{\textrm{attach}}\). The previous description summarizes 
the method utilized to calibrate \texttt{ENIGMA} in~\cite{huerta:2018PhRvD}. However, 
one realizes that this procedure only tells us that when 
the orbital frequency during inspiral meets the condition \(\omega_{\textrm{inspiral}}> \omega^{*}\) 
then we know that  \(t_{\textrm{attach}}\) lies within the range $[t - \mathrm{d}t, t]$. Thus, our 
accuracy is dependent on the size of $\mathrm{d}t$, which for values we have used to describe binary 
black hole mergers, i.e., \(\mathrm{d}t=\{1/4096\textrm{s}, 1/8192\textrm{s}\}\),  
$\mathrm{d}t$ may be inaccurate. To address this limitation and better compute \(t_{\textrm{attach}}\), we have implemented 
the bisection root finding method on the range $[t - dt, dt]$ to determine the exact value of 
\(t_{\textrm{attach}}\). After this addition, we have tested and found that \(t_{\textrm{attach}}\) 
so obtained is 
sufficiently accurate up to the limit of floating point precision. Figure~\ref{fig:attachment_diff} 
shows the difference between the two methodologies used to construct complete \texttt{ENIGMA} 
waveforms. 
 
\begin{figure}
\centering
\includegraphics[width=\linewidth]{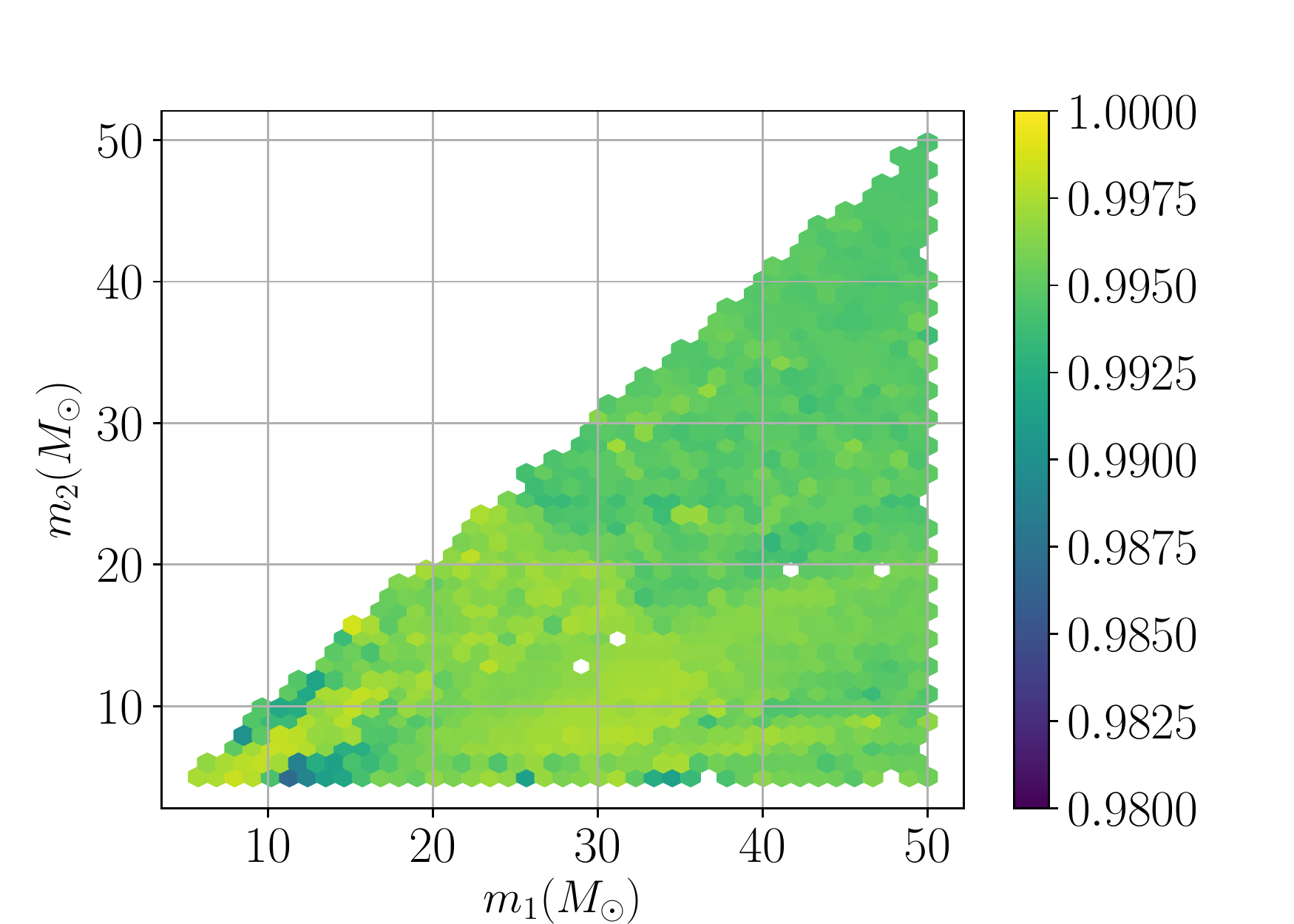}
\caption{Overlap distribution between \texttt{ENIGMA} and \texttt{SEOBNRv4\_ROM} waveforms for component
masses $5\,M_{\odot} \le m_{\{1,\,2\}} \le 50\,M_\odot$. These results were produced assuming advanced LIGO zero detuned high power PSD configuration, and setting a minimum gravitational wave frequency of \(f_{\textrm{GW}}=15\,\textrm{Hz}\) in Equation~\eqref{over}.}
\label{fig:MatchENIGMAvsSEOBNRv4}
\end{figure}

\noindent Using the aforementioned approach to produce complete  \texttt{ENIGMA} waveforms,
Figure~\ref{fig:MatchENIGMAvsSEOBNRv4} presents the overlap between our newly 
recalibrated \texttt{ENIGMA} model in the quasi-circular limit and  \texttt{SEOBNRv4} waveforms. 
It is worth pointing out that when we tested waveform production at scale, we found that we can randomly sample the 
parameter space shown in Figure~\ref{fig:MatchENIGMAvsSEOBNRv4}, and produce 1M waveforms error free. 
In contrast, about 1\% of these waveforms would not be produced with the previous version of \texttt{ENIGMA} due to instabilities in the merger attachment routine.

\begin{figure}
	\centerline{
	\includegraphics[width=0.55\textwidth]{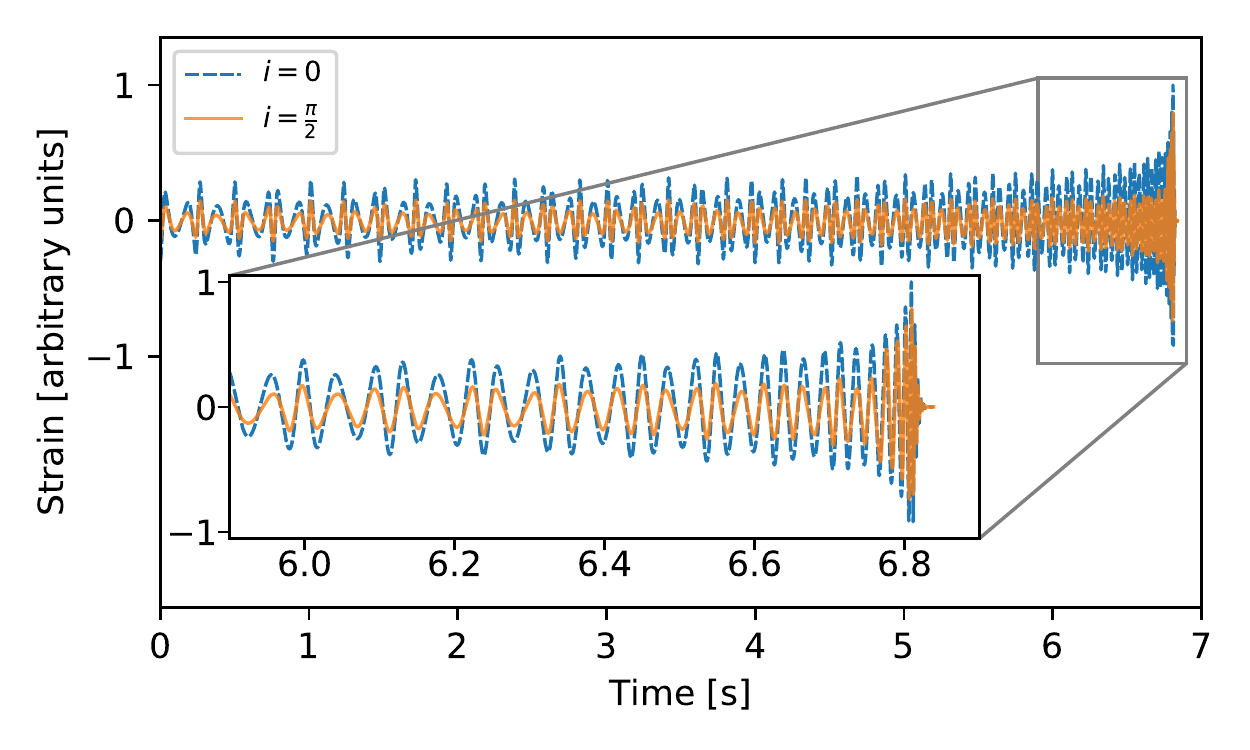}
	}
	\caption{Real part of the waveform strain produced by a binary black hole merger with parameters \(e_0=0.3\) at a gravitational wave frequency of \(f_{\textrm{GW}}=9\textrm{Hz}\), mass-ratio \(q=3\) and total mass \(M=60\msun\). Note the impact of the inclination angle in the amplitude of the waveform signal for two sample cases \(i=\{0,\,\pi/2\}\).} 
	\label{fig:inclination}
\end{figure}

\noindent The second improvement to \texttt{ENIGMA} is the handling of 
the binary inclination angle. In the original version of \texttt{ENIGMA}, 
the calibration was conducted assuming zero inclination. We relax that assumption in 
this study. While for 
quasi-circular mergers the inclination angle enters the amplitude of the plus and 
cross polarizations as a trivial multiplicative factor, the inclination angle modifies the 
amplitude of the plus and cross polarizations in a non-trivial fashion even at 
the leading-order PN correction level, 
as shown in Eqs.~(8) and (9) in~\cite{Huerta:2017a}. In this version of \texttt{ENIGMA} 
we now provide a robust handling of the inclination angle when smoothly connecting the 
inspiral and merger waveforms. To get a glimpse of the importance of the 
inclination angle in the morphology of eccentric mergers, Figure~\ref{fig:inclination} 
presents the real part of the waveform strain that describes a BBH merger with \(e_0=0.3\) at 
\(f_{\textrm{GW}}=9\,\textrm{Hz}\), mass-ratio \(q=3\) and total mass \(M=60\msun\).

Figure~\ref{fig:inclination} shows that in stark contrast with quasi-circular mergers, changing the inclination angle from 
\(i=0\) to \(i=\pi/2\) does not lead to a simple reduction of the waveform amplitude by an overall factor of \(1/2\). 
Rather, we see that the morphology of the two waveforms is rather different, see e.g., the slope of the signals 
near the peaks and valleys. One may understand these changes by looking at 
Equations (8) and (9) in~\cite{Huerta:2017a}. There we see that the inclination angle enters in two 
different pieces. The first one, proportional to \(1+\cos^2 i\), is also present in quasi-circular mergers. However, the second piece,  proportional to \(\sim \sin^2 i\), is the one driving the changes in the morphology of the 
waveforms.

\noindent In Section~\ref{sec:networks} we explore in detail the impact of the binary inclination angle in the SNR distributions of eccentric BBH mergers for second and third generation GW detector networks. We introduce a minimal set of data analysis tools we require for such analyses in the following section. 


\section{Data analysis toolkit}
\label{sec:dat}

The SNR of a detector network is given by the sum of the power of the individual detectors~\cite{Finn:2001}

\begin{equation}
\label{eqn:SNRnet}
\rho^2_N = \sum_{k=1}^{N_D}\rho^2_k\,,
\end{equation}

\noindent where $N_D$ is the number of detectors and where we define the individual SNRs as

\begin{equation}\label{eqn:SNRcomponent}
\rho^2_k = 2\int_{f_0}^{f_1}\frac{|h_k(f)|^2}{S_{h}(f)}{\rm d}f,
\end{equation}

\noindent where $h_k(f)$ is the waveform projected onto the $k$-th detector. Averaging over the 
polarization angle, \(\psi\), we obtain

\begin{equation}\label{eqn:SNRnetavg}
\left<\rho^2_N\right> = 2\sum_k (F_{+,k}^2 + F_{\times, k}^2)\int_{f_0}^{f_1}\frac{|h(f)|^2}{S_{h}(f)}{\rm d}f,
\end{equation}

\noindent where $F_{+,k}$ and $F_{\times, k}$ are the antenna patterns of the individual detectors. 
As noted in~\cite{Schutz:2011}, this integral does not depend on $k$ and may therefore be 
taken outside the sum. Following the conventions described in~\cite{Schutz:2011}, the antenna 
pattern functions are given by

\begin{eqnarray}
F_+ &=& \sin\eta[a\cos(2\psi)+b\sin(2\psi)],\label{eqn:fplusgen}\\
F_\times &=& \sin\eta[b\cos(2\psi)-a\sin(2\psi)]\label{eqn:fcrossgen},
\end{eqnarray}

\noindent where the functions $a$ and $b$ are given by

\begin{align}  
a &=\frac{1}{16}\sin(2\chi)[3-\cos(2\beta)][3-\cos(2\theta)]\cos[2(\phi+\lambda)]+\nonumber\\& \frac{1}{4}\cos(2\chi)\sin(\beta)[3-\cos(2\theta)]\sin[2(\phi+\lambda)]+\nonumber\\&
\frac{1}{4}\sin(2\chi)\sin(2\beta)\sin(2\theta)\cos(\phi+\lambda)+\nonumber\\& \frac{1}{2}\cos(2\chi)\cos(\beta)\sin(2\theta)\sin(\phi+\lambda)+\nonumber\\&\frac{3}{4}\sin(2\chi)\cos^2(\beta)\sin^2(\theta)\,,\label{eqn:a}
\end{align}

\begin{align} 
b &= \cos(2\chi)\sin(\beta)\cos(\theta)\cos[2(\phi+\lambda)]-\nonumber\\&\frac{1}{4}\sin(2\chi)[3-\cos(2\beta)]\cos(\theta)\sin[2(\phi+\lambda)]+\nonumber\\& \cos(2\chi)\cos(\beta)\sin(\theta)\cos(\phi+\lambda)-\nonumber\\&\frac{1}{2}\sin(2\chi)\sin(2\beta)\sin(\theta)\sin(\phi+\lambda)\,.\label{eqn:b}
\end{align}

\noindent The source location is determined by the spherical coordinates on the sky, \((\theta,\,\phi)\); 
and  \((\beta,\,\lambda)\) indicate the latitude and longitude location of GW detectors; the bisector of the 
GW detector's arms points in the \(\chi\) direction which is measured counter-clockwise from East. The 
GW detector arms have an opening angle \(\eta\). As described in~\cite{Schutz:2011}, in this coordinate 
system the celestial coordinates \((\theta,\,\phi)\) are aligned with latitude and longitude in such a way that 
the equators of both systems coincide, and \((\theta=\pi/2,\,\phi=0)\) is in the zenith direction above 
\((\beta=0,\,\lambda=0)\). In what follows, we present results for 

\begin{equation}
\tilde \rho = \langle\rho^2_N\rangle^{1/2}\,.
\label{eq_snr}
\end{equation}

\noindent Furthermore, since we want to compare the SNR distribution of eccentric BBH populations with respect to quasi-circular BBH mergers, we will present our results using the metric

\begin{equation}
\Delta \tilde \rho = 100\times\frac{\tilde \rho (e_0; \theta,\,\phi) -  \tilde \rho (e_0=0; \theta,\,\phi)}{ \tilde \rho_{\textrm{LHV}} (e_0=0; \theta^{*},\,\phi^{*})}\,,
\label{eq:metric}
\end{equation}

\noindent where \(\tilde \rho (e_0; \theta,\,\phi)\) is the detector network SNR of a BBH system with 
initial eccentricity \(e_0\) at a GW frequency \(f_0\);  \(\tilde \rho (e_0=0; \theta,\,\phi)\) is the detector 
network SNR for the same BBH system but now assuming it is quasi-circular; and  
\(\tilde \rho_{\textrm{LHV}} (e_0=0; \theta^{*},\,\phi^{*})\) is the maximum 
SNR of a quasi-circular BBH population across the sky assuming a network consisting 
of GW detectors located at the LIGO Livingston (L), LIGO Hanford (H) and 
Virgo (V) sites---or LHV from now on.

\section{Signal-to-noise ratio distributions for second and third generation gravitational wave detector networks} 
\label{sec:networks}

We consider second and third generation, ground-based GW detector networks, whose geographical locations  are listed in Table~\ref{tab:dets}. We use the following nomenclature in Table~\ref{tab:dets}: LIGO Livingstone:  L; LIGO Hanford: H; Virgo: V; KAGRA: K; LIGO India: LI; and a gravitational wave detector in Australia, AIGO, is labelled as A.

\begin{widetext}
\begin{table*}
\setlength{\tabcolsep}{12pt}
   \centering
   \begin{tabular}{lcccc}
   	\toprule
      Detector & Label & Longitude & Latitude & Orientation\\\hline
	\midrule
      LIGO Livingston & L & 90\degree\ 46' 27.3" W & 30\degree\ 33' 46.4" N & 208.0\degree (WSW) \\
      LIGO Hanford & H & 119\degree\ 24' 27.6" W & 46\degree\ 27' 18.5" N & 279.0\degree (NW) \\
      VIRGO, Italy & V & 10\degree\ 30' 16" E & 43\degree\ 37' 53" N &  333.5\degree(NNW) \\
      KAGRA, Japan & K & 137\degree\ 10'  48" E & 36\degree\ 15' 00" N & 20.0\degree (WNW)  \\
      LIGO India & LI & 74\degree\ 02' 59" E & 19\degree\ 05' 47" N & 270.0\degree (W) \\
      AIGO, Australia & A & 115\degree\ 42' 51" E  & 31\degree\ 21' 29" S & 45.0\degree (NE) \\\hline
      \bottomrule
   \end{tabular}
   \caption{Geographical location of the gravitational wave interferometers considered in this study.}
   \label{tab:dets}
\end{table*}
\end{widetext}

\noindent The PSD of each detector used in these studies was obtained from the open 
source files at~\cite{updated_aligo,ads_other_detectors}, and are shown in Figure~\ref{fig:psds_all}. Note that 
we have assumed that the opening angle of all detectors is \(\eta=\pi/2\). The actual networks 
we consider for second-generation GW detectors are:

 \begin{figure}
	\centerline{
	\includegraphics[width = 0.5\textwidth]{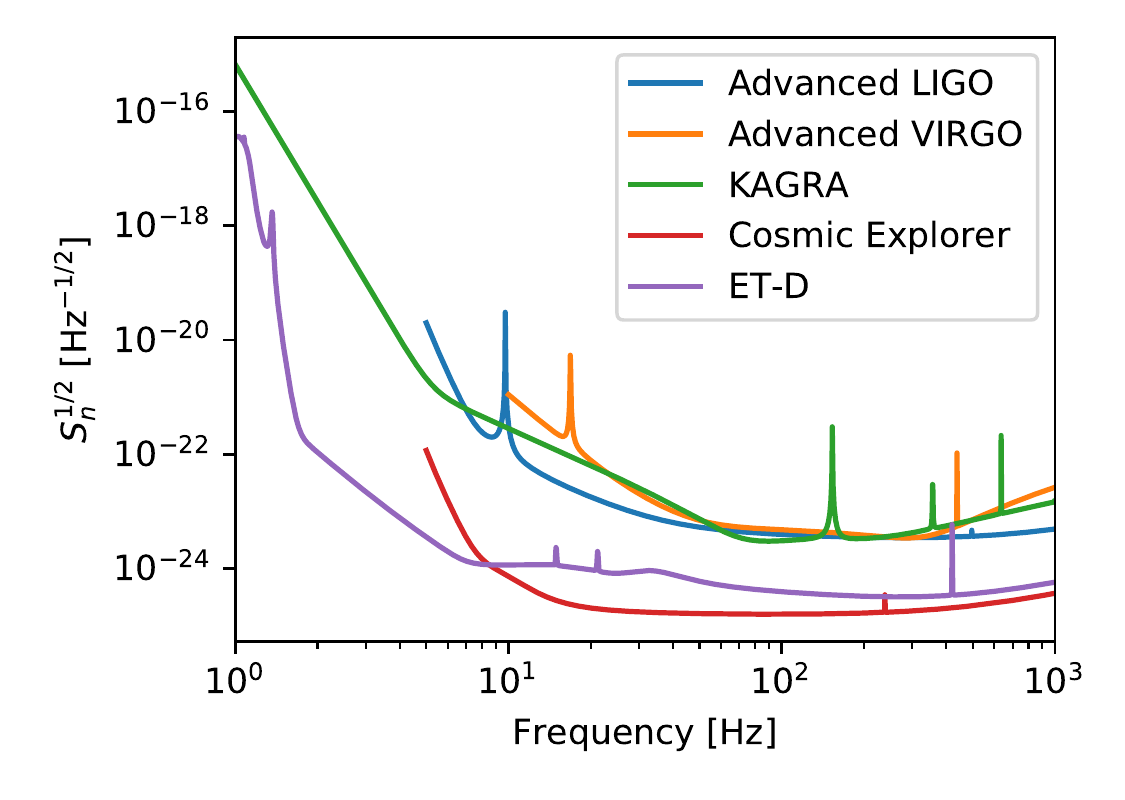}
	}
	\caption{Power Spectral Density configurations of advanced LIGO, advance Virgo, Kagra, Cosmic Explorer, and the Einstein Telescope. We used the data provided in~\cite{updated_aligo,ads_other_detectors} to model the sensitivity of these second and third generation gravitational wave detectors.}
	\label{fig:psds_all}
\end{figure}

\begin{cititemize2}
\item L+H+V. Herein we assume the target PSD for advanced LIGO for both L\&H, and the target PSD for V.
\item L+H+V+K. As the network above, and now also using the target PSD for K.
\item L+H+V+K+LI+A. As the  network above, and setting the PSDs for LI\&A to be the same as for L.
\end{cititemize2}

For third generation detector networks that involve Cosmic Explorer~\cite{reitze2019cosmic} 
or the Einstein Telescope, we will consider the 
same geographical locations for individual detectors as listed in Table~\ref{tab:dets}, and 
replace PSD with the target  
sensitivity of Cosmic Explorer and the Einstein Telescope, configuration D (ET-D)~\cite{Punturo:2010zz,amann2020siteselection}, as described in~\cite{updated_aligo,ads_other_detectors}.

\begin{figure*}[!hbt]
\centerline{
\includegraphics{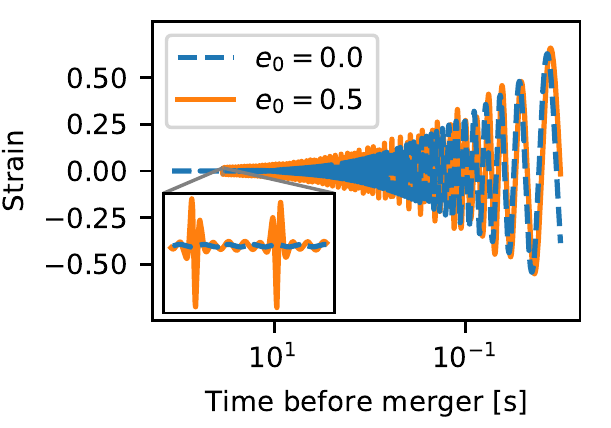}\hfill%
\includegraphics{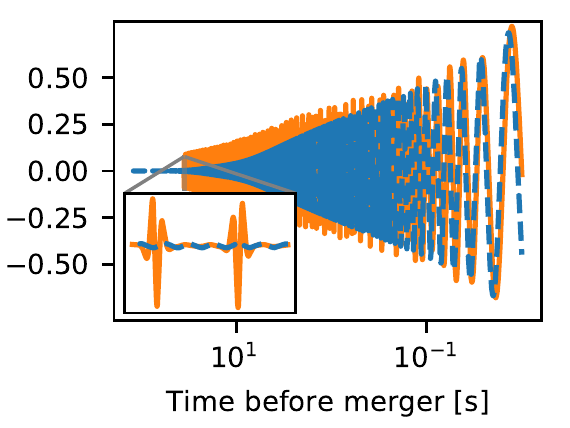}\hfill%
\includegraphics{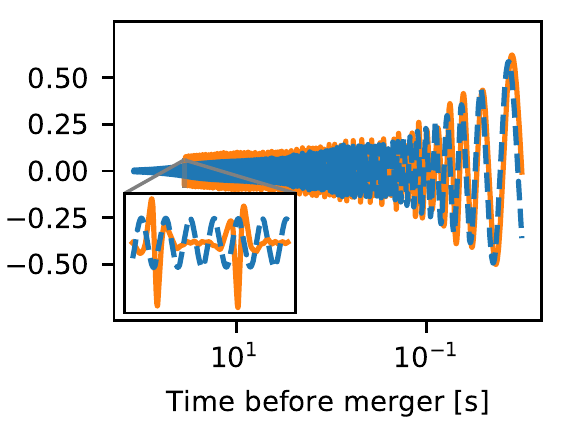}\hfill%
}
\caption{Effect of the power spectral sensitivity of Advanced LIGO (left), Cosmic Explorer
(center) and ET-D (right) on a waveform signal that describes a binary system with component masses
 \((50\msun,\,10\msun)\), and
eccentricity of $e_0 = 0$ (blue, dashed curve) and $e_0 = 0.5$ (orange, solid
curve), both measured at a gravitational wave frequency $f_{\textrm{GW}} = 4\,\textrm{Hz}$. The quasi-circular 
signal is 126.27s long, while the eccentric signal is 37.32s long. Eccentricity-driven amplitude amplifications are 
clearly noticeable throughout the entire evolution (the orange signal is louder throughout, see also insets which
show data in a 1s interval at the beginning of the eccentric signal).
Data are normalized to have a maximum amplitude of unity.}
\label{fig:whitened_wf}
\end{figure*}

To motivate the results we present in the following sections, 
in Figure~\ref{fig:whitened_wf} we present a waveform 
that describes the merger of a BBH with component masses  \((50\msun,\,10\msun)\), 
and \(e_0=0.5\) at $f_{\textrm{GW}} = 4\,\textrm{Hz}$. The signal is whitened with the target 
PSDs for advanced LIGO, Cosmic Explorer 
and the ET-D. We notice that Cosmic Explorer and ET-D magnify amplifications in the 
amplitude of the waveform that 
are driven by orbital eccentricity. The key point is that these amplifications occur at lower 
frequencies, or early times in the waveform signal, in such a way that when we compute 
the SNR of these eccentric mergers, amplitude magnifications compensate for the shrinkage 
in timespan that is also driven by orbital eccentricity. For this particular case, 
\textit{this eccentric waveform is just a fifth the length of its quasi-circular counterpart}. However, as
we discuss below, the SNR of eccentric and quasi-circular systems are comparable. These 
observations will drive the presentation of our results in the following sections.

\subsection{SNR distributions for second generation detector networks}
\label{sec:second_results}

In this section we present our findings for second generation 
detector networks. We have found that we can extract the most 
physics from the observation of eccentric BBH mergers when we take 
into the account two key properties that drive the evolution of 
eccentric mergers, i.e., the reduction in the length of waveform signals, 
and the rate at which SNR is accumulated. 

\begin{figure*}[!hbt]
	\centerline{
	\includegraphics[width = 0.33\textwidth]{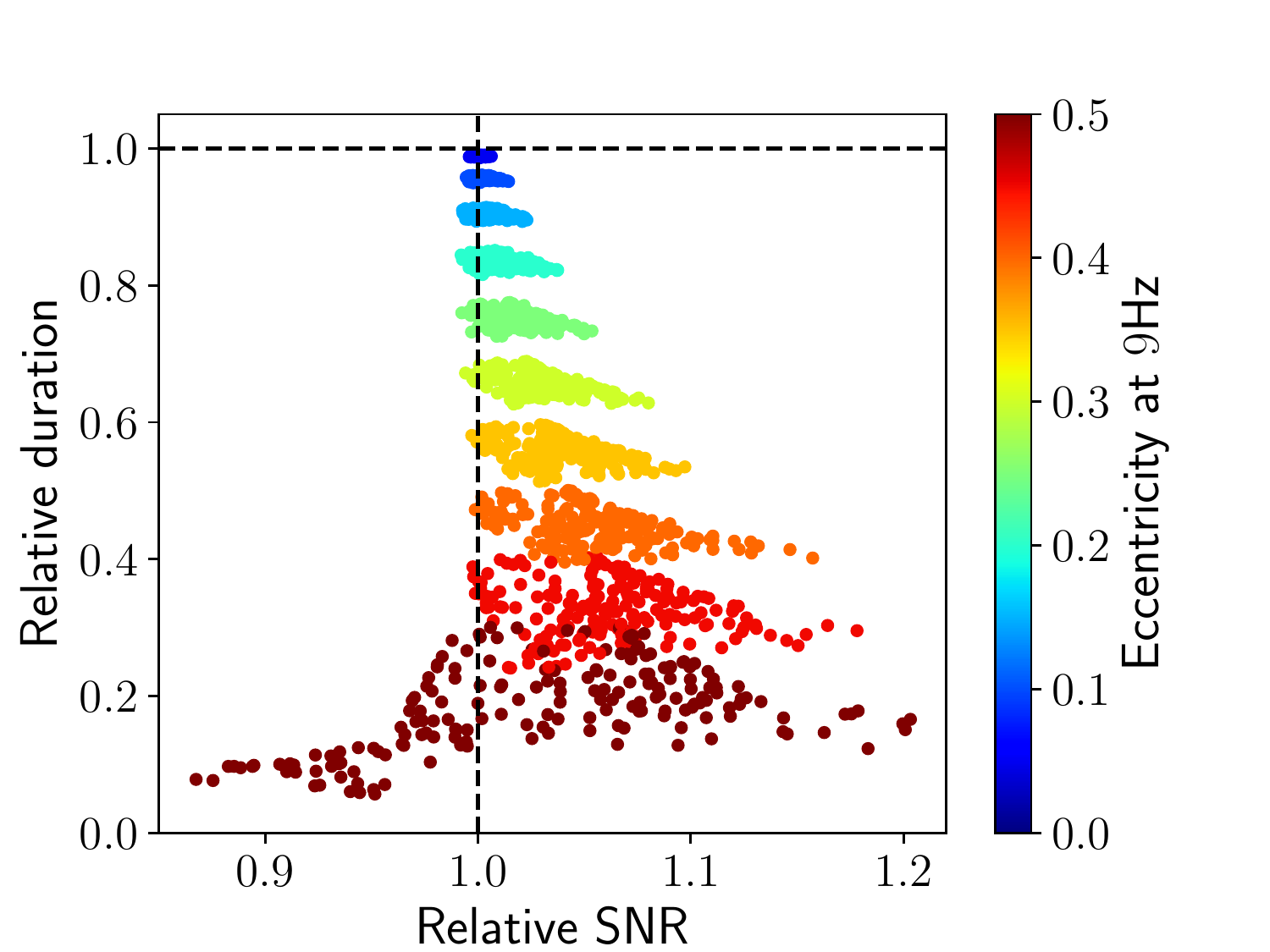}
	\includegraphics[width = 0.33\textwidth]{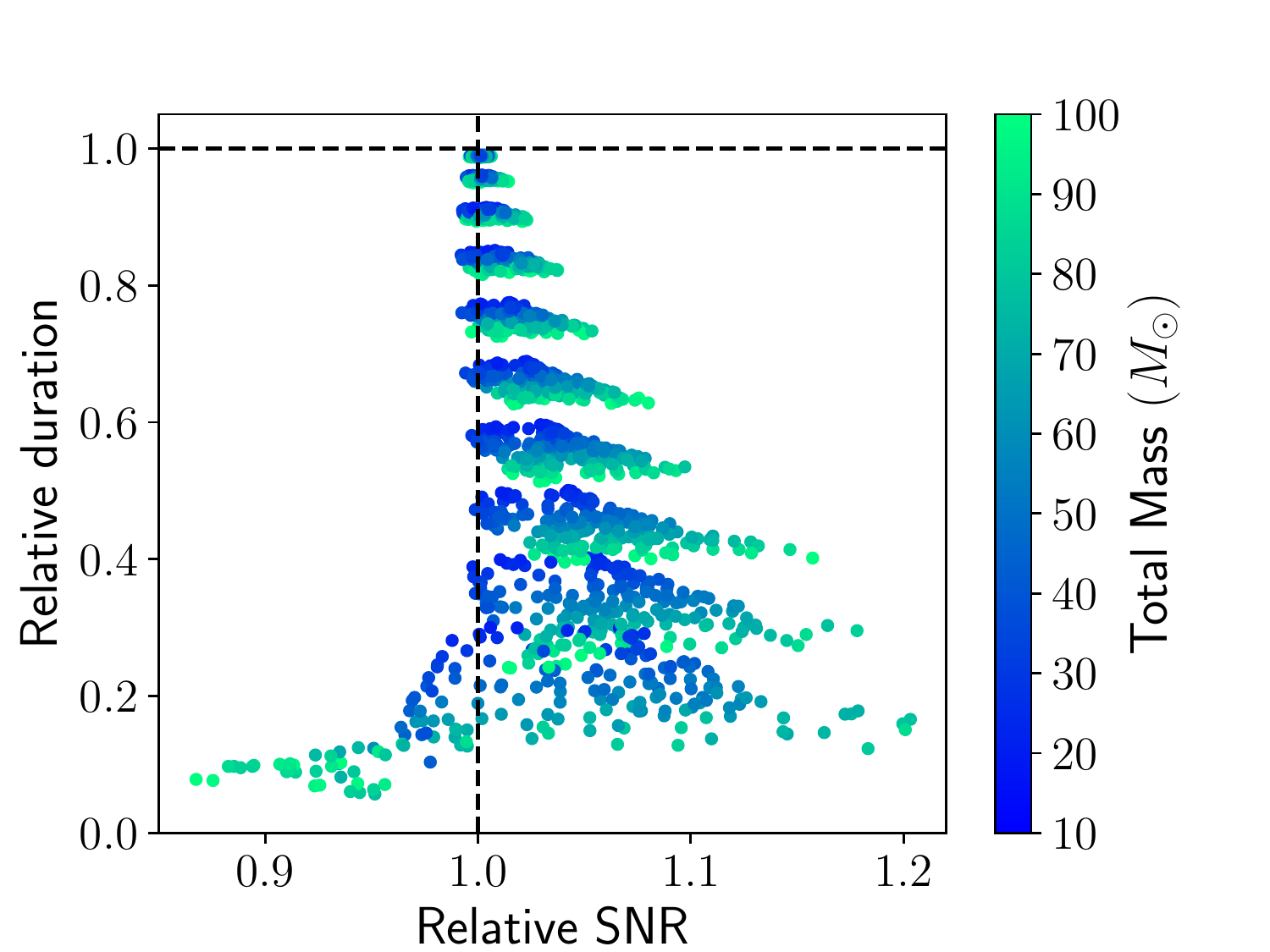}
	\includegraphics[width = 0.33\textwidth]{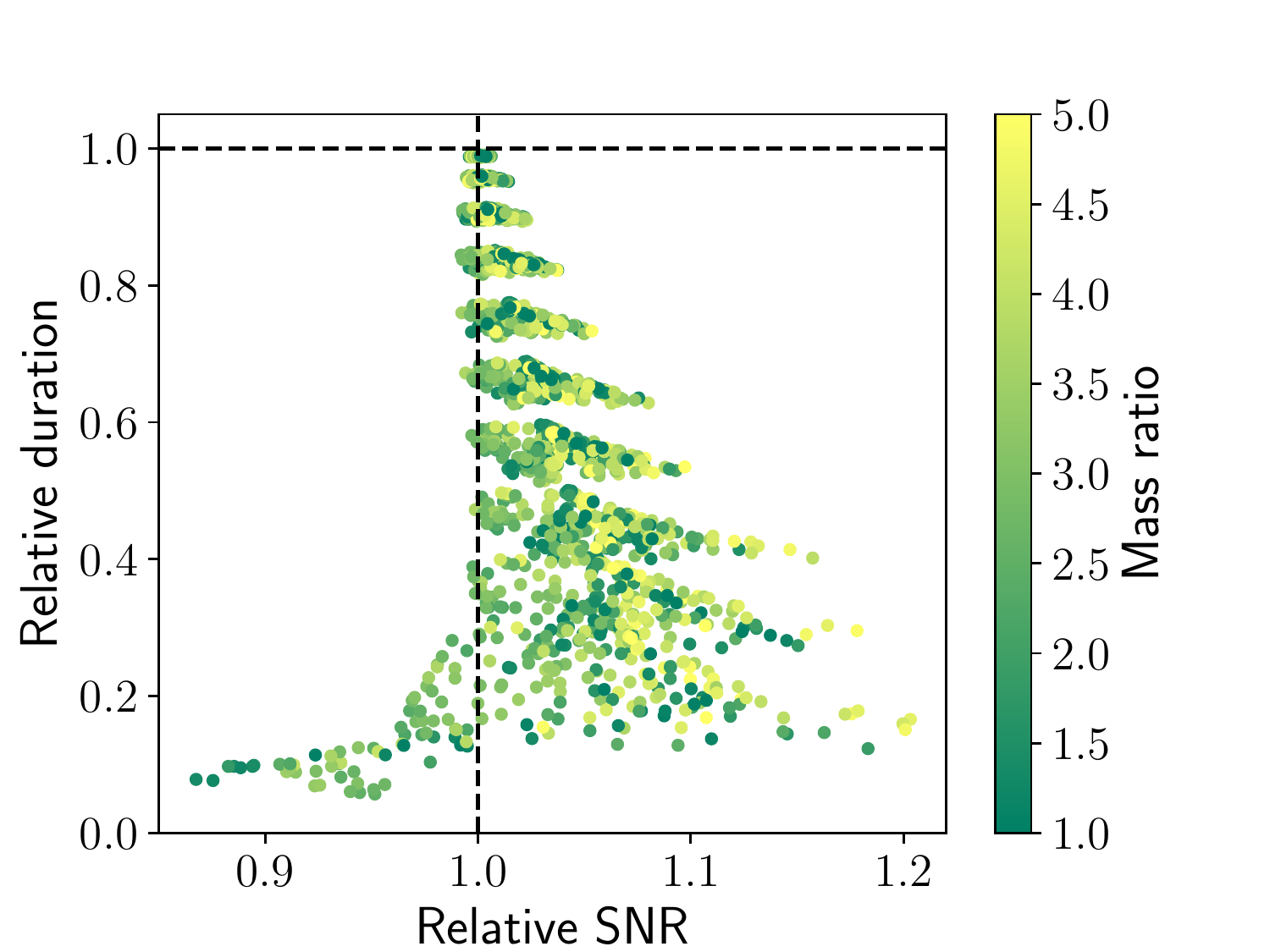}
	}
	\caption{Length and signal-to-noise-ratio of eccentric binary black hole mergers relative to quasi-circular systems. Both panels show systems with total mass \(20\msun\leq M\leq 100\msun\) and mass-ratios \(1\leq q\leq5\). In these panels, \(\texttt{Relative SNR}\rightarrow\texttt{SNR(eccentric)}/\texttt{SNR(quasi-circular)}\), and \(\texttt{Relative duration}\rightarrow\texttt{duration(eccentric)}/\texttt{duration(quasi-circular)}\).}
	\label{fig:snr_time_ligo}
\end{figure*}

To motivate this discussion, Figure~\ref{fig:snr_time_ligo} presents 
SNR distributions of eccentric BBH mergers, normalized with respect to quasi-circular 
BBH systems, assuming eccentricities \(e_0\leq0.5\) measured at 
\(f_{\textrm{GW}}=9\textrm{Hz}\). The SNR distributions are computed by 
setting \(f_0=11\,\textrm{Hz}\) and \(f_1=4096\,\textrm{Hz}\)  in 
Eq.~\eqref{eqn:SNRcomponent}, and assuming a single advanced LIGO detector. 

The panels in Figure~\ref{fig:snr_time_ligo} present results for BBH systems with 
total masses \(20\msun\leq M\leq 100\msun\) and mass-ratios \(1\leq q\leq5\). These panels show 
the interplay between total mass and mass-ratio in the SNR distribution of eccentric mergers.  
These results show that heavier binaries have larger SNRs, and that they tend to have 
shorter duration in band.
These results show that the SNR 
of eccentric mergers will be between \([0.8-1.2]\times \textrm{SNR}\) of quasi-circular mergers 
that have the same mass-ratio and total mass. 
The most dramatic finding in these studies is that for these extreme scenarios in which 
the SNR of \(e_0\sim0.5\) mergers is similar to quasi-circular events, the waveform length of 
eccentric systems is about eight times shorter for equal mass systems, 
and five times shorter for \(q=5\) systems.
In other words, the rate at which SNR is accumulated for eccentric mergers is 
significantly faster than for quasi-circular mergers. We will study this point in further detail below.

\noindent In order to provide a visual representation of the importance of having access to 
a global detector network, we have quantified the SNR sky distribution for 
a few sample cases with component masses 
\(m_{\{1,\,2\}}= \{\left(30\msun, 30\msun\right), \left(50\msun, 10\msun\right)\}\), 
with eccentricities \(e_0\leq0.5\) measured at \(f_{\textrm{GW}}=9\textrm{Hz}\). 
The SNR distributions are computed by setting \(f_0=11\,\textrm{Hz}\) and \(f_1=4096\,\textrm{Hz}\) in 
Eq.~\eqref{eqn:SNRcomponent}. Highlights of these results include: 

\begin{itemize}
\item Figure~\ref{fig:equal_mass_LIGO_type} shows that for systems with 
component masses \((30\msun,\,30\msun)\), a gradual increase in eccentricity drives a 
corresponding boost in SNR. This effect reaches a maximum for \(e_0=0.4\), with a net SNR 
increase of \(\Delta \textrm{SNR} \sim12\%\) with respect to quasi-circular systems. 
Systems with larger eccentricity, \(e_0=0.5\), have similar SNRs than quasi-circular mergers. 
In other words, for \(e_0\geq0.4\) BBH mergers the increase in amplitude 
of the signal at lower frequencies no longer compensates for the shrinkage in the waveform length. 
We can extract this information by looking at the sky maps from top to bottom. 
If we now look at them from left to right, we learn that a larger detector network increases the 
loudness of signals across the sky, while also forming distinct detection hot spots, e.g., 
see the sky map for \(e_0=0.4\) assuming a 6 detector network. 
\item Figure~\ref{fig:five_mass_LIGO_type} shows that asymmetric mass-ratio BBH mergers 
produce louder mergers than comparable mass-ratio systems for larger values of 
eccentricity. In this case, 
\((50\msun,\,10\msun)\) systems report a net increase in SNR, 
\(\Delta \textrm{SNR} \sim30\%\) for \(e_0=0.5\).
\end{itemize}

\newcommand{\wid}{0.34}
\begin{figure*}
	\centerline{
		\includegraphics[width=\textwidth]{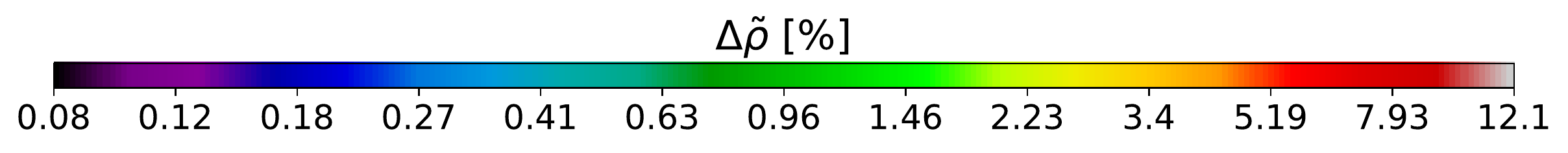}
	}
	\centerline{
		\includegraphics[width=\wid\textwidth]{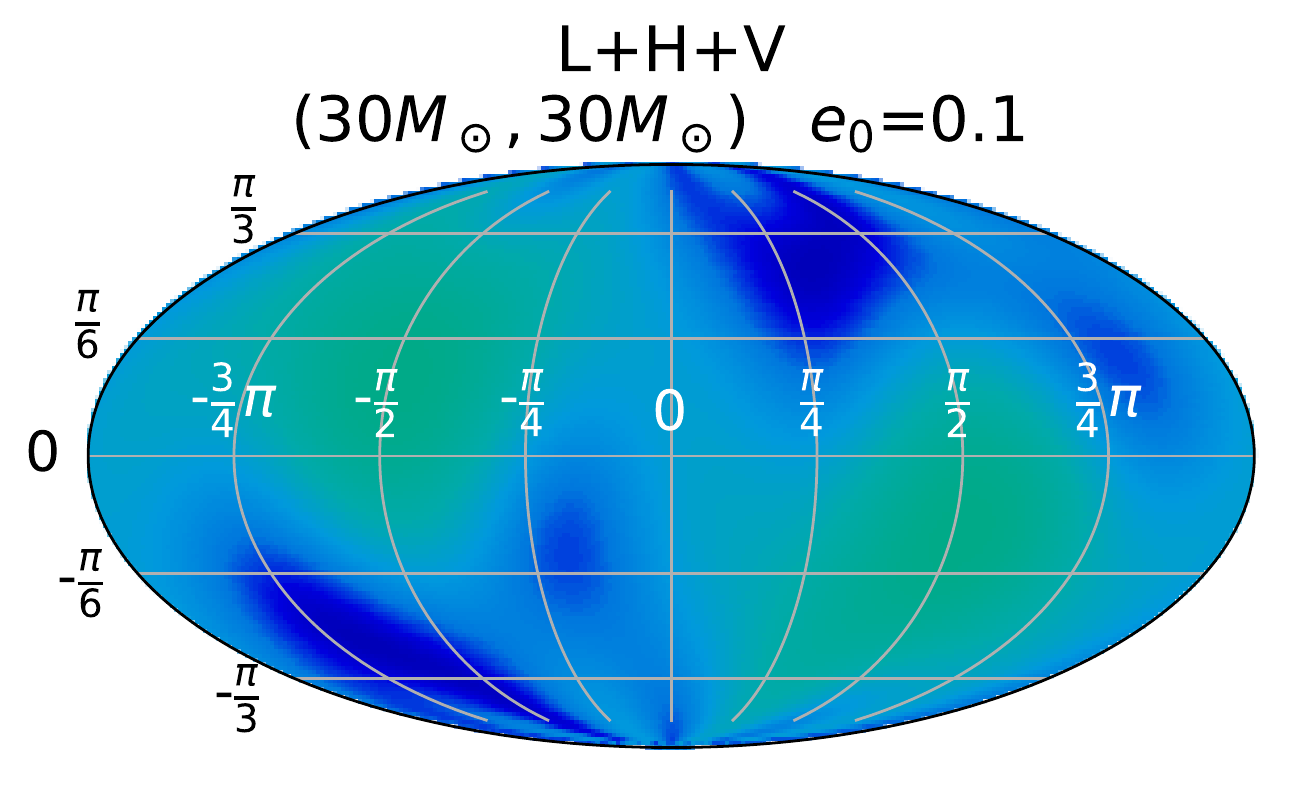}
		\includegraphics[width=\wid\textwidth]{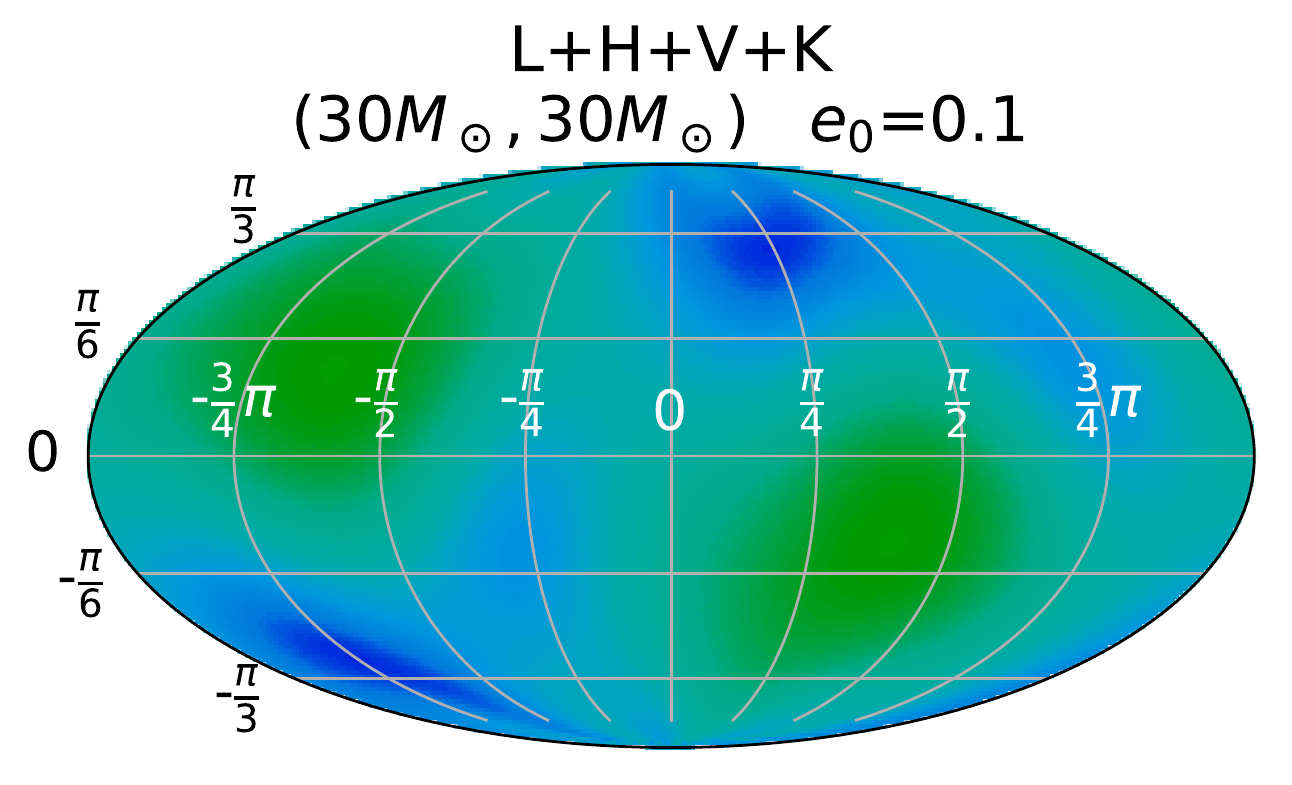}
		\includegraphics[width=\wid\textwidth]{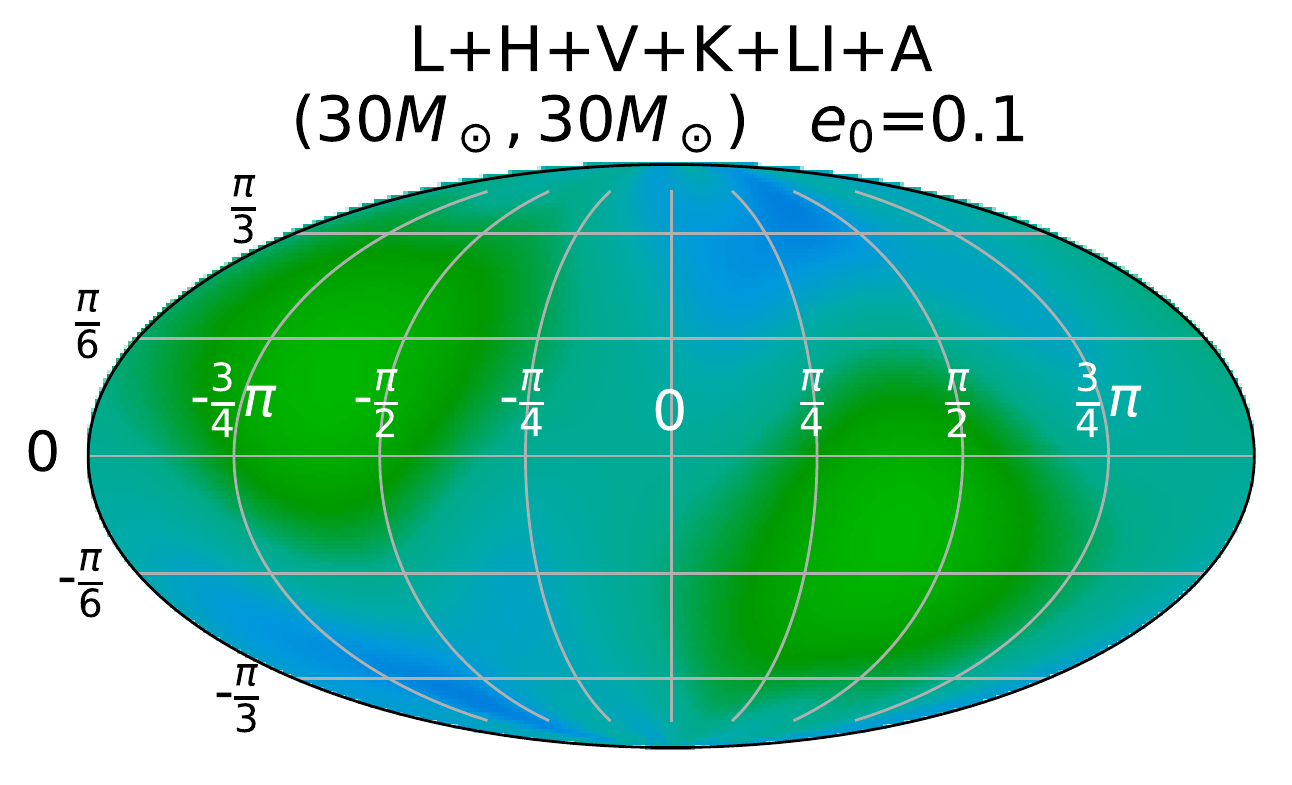}
	}
	\centerline{
		\includegraphics[width=\wid\textwidth]{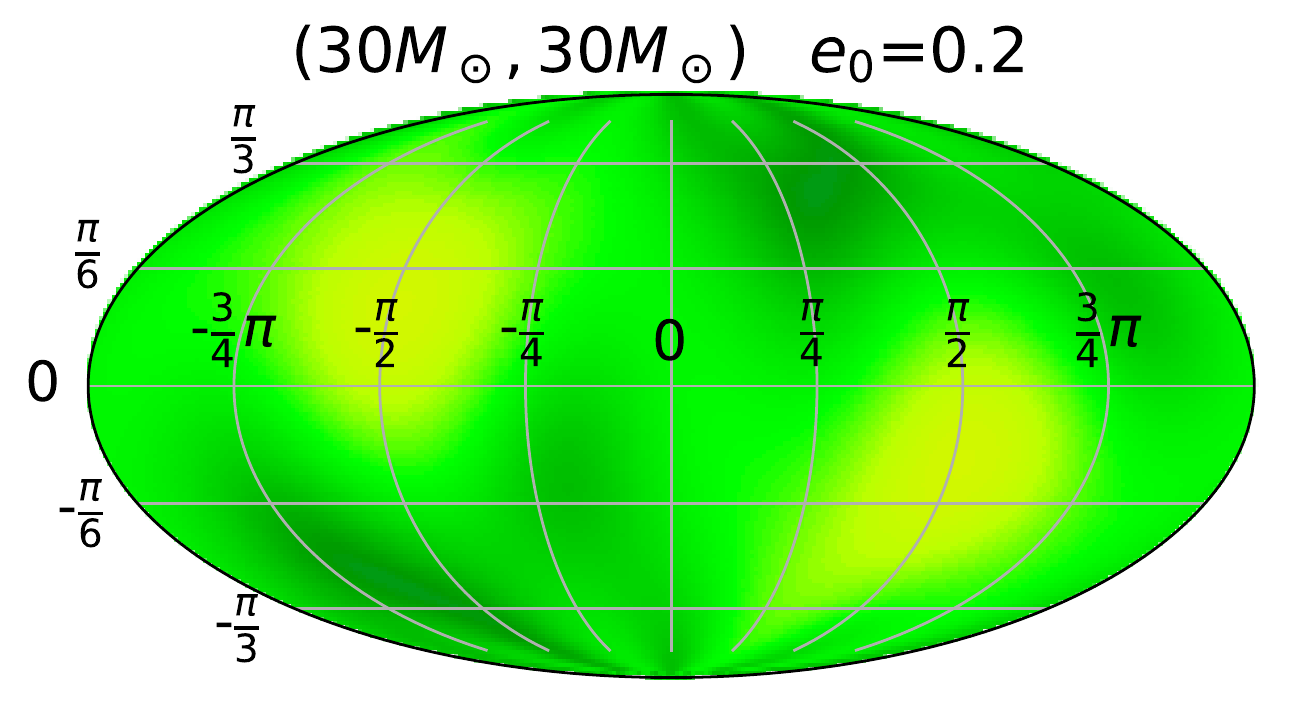}
		\includegraphics[width=\wid\textwidth]{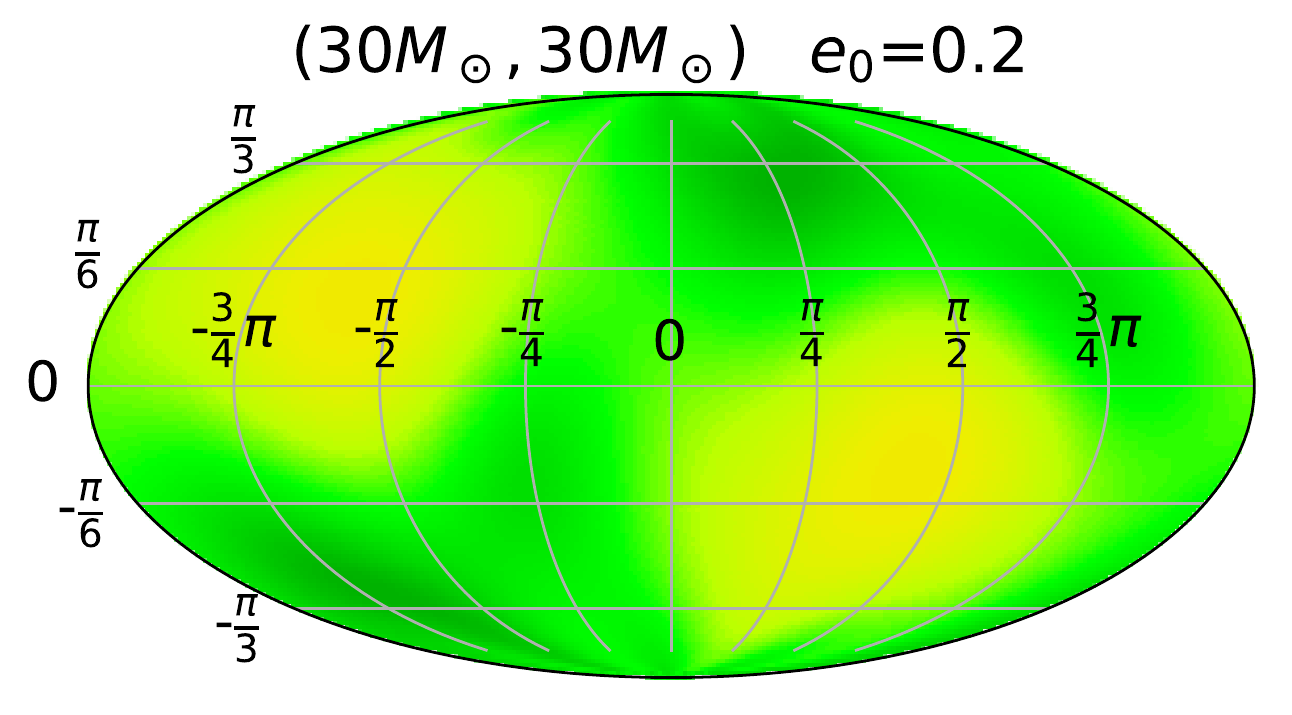}
		\includegraphics[width=\wid\textwidth]{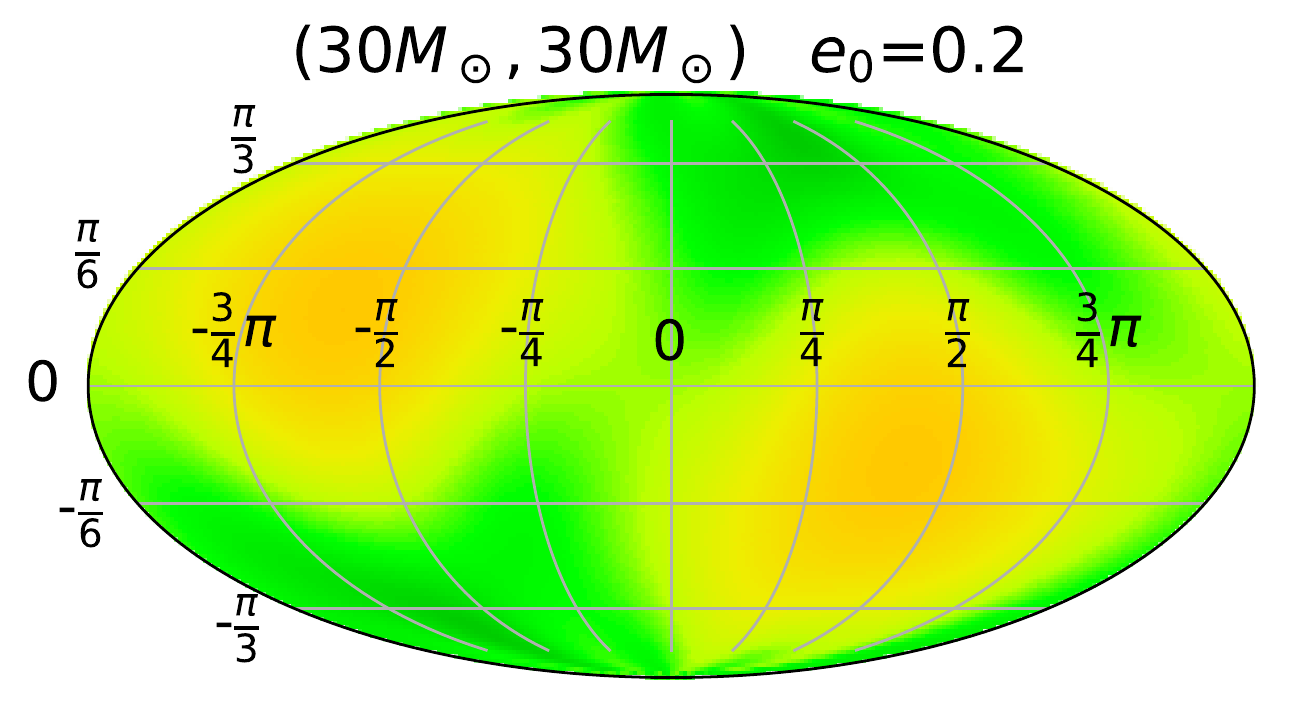}
	}
	\centerline{
		\includegraphics[width=\wid\textwidth]{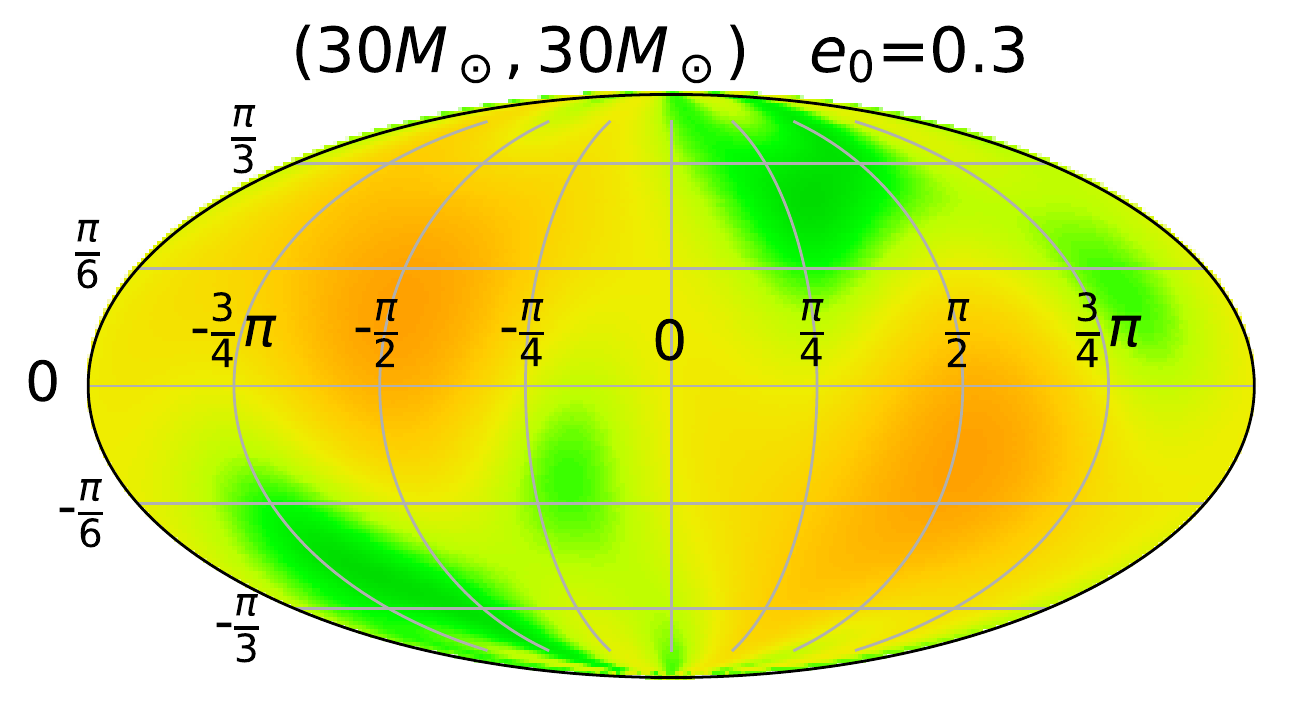}
		\includegraphics[width=\wid\textwidth]{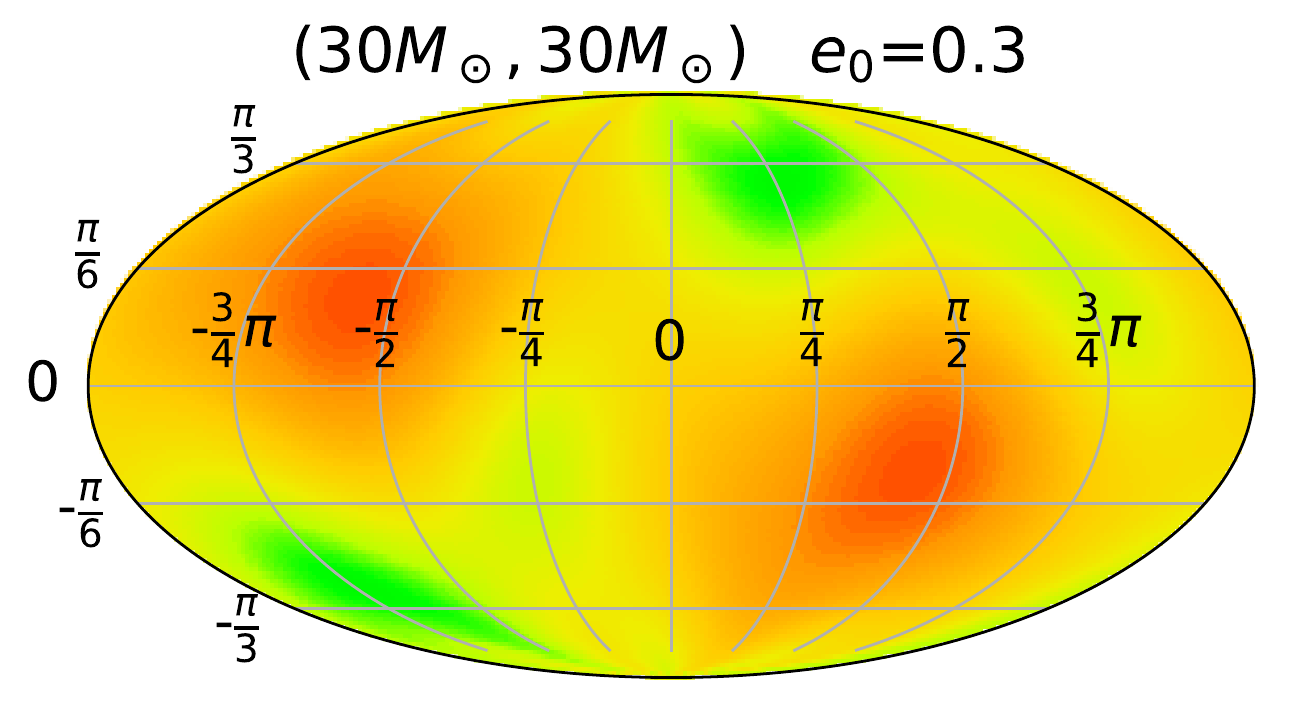}
		\includegraphics[width=\wid\textwidth]{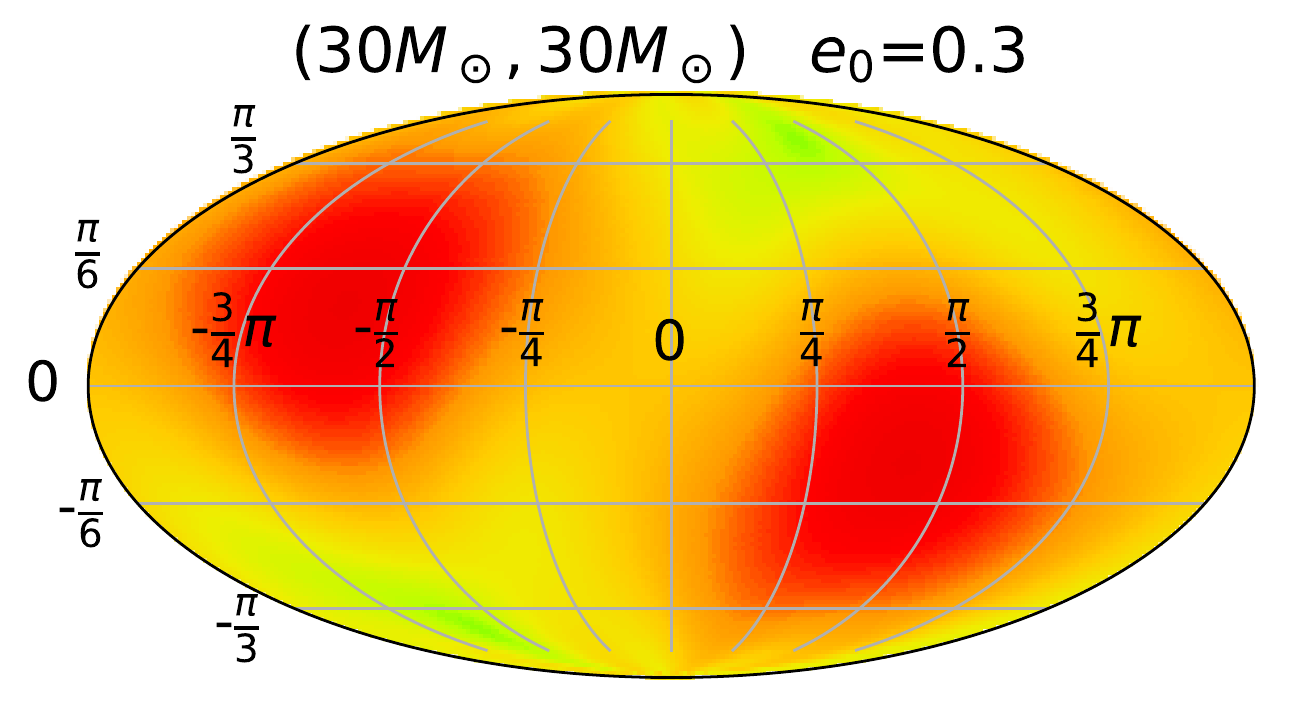}
	}
	\centerline{
		\includegraphics[width=\wid\textwidth]{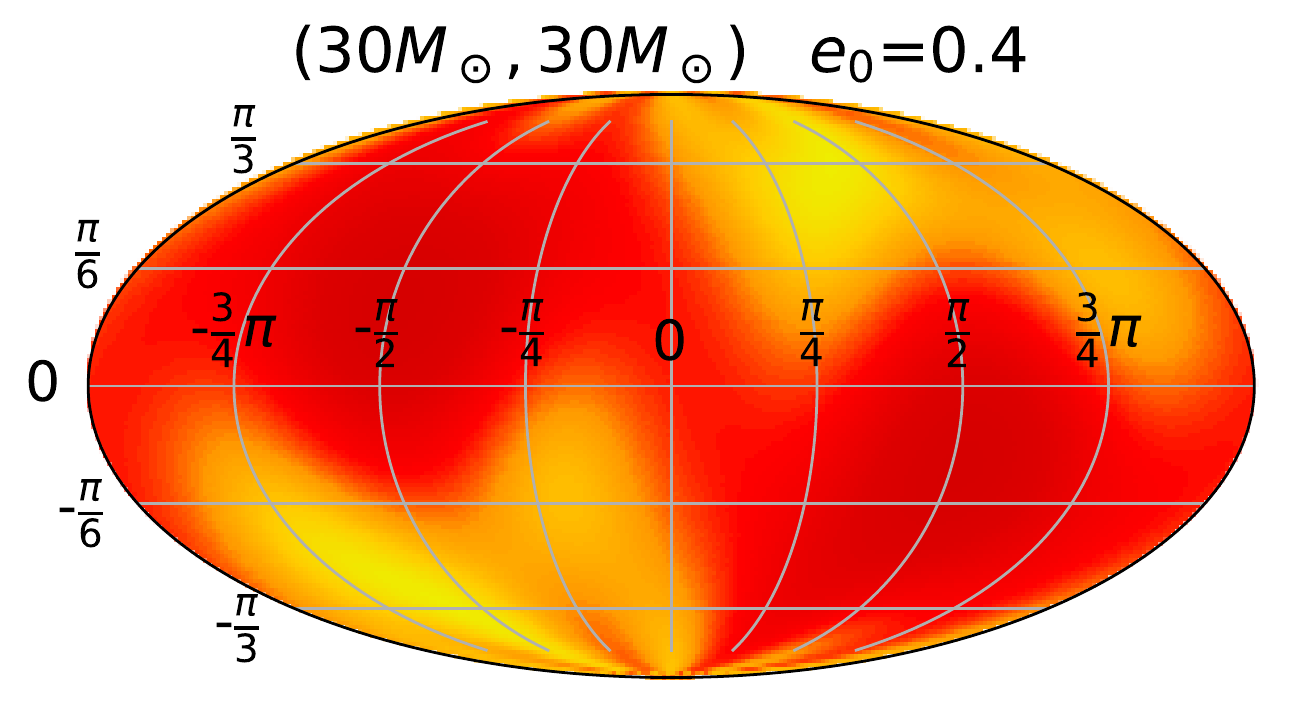}
		\includegraphics[width=\wid\textwidth]{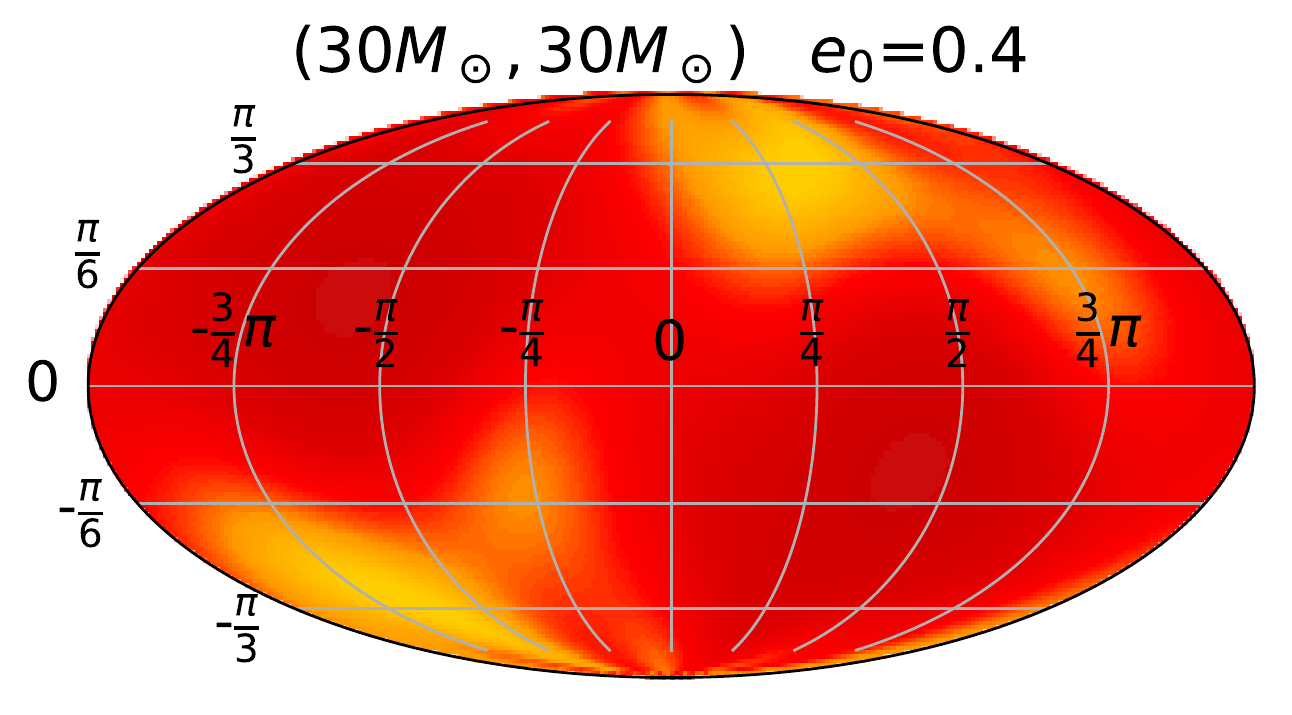}
		\includegraphics[width=\wid\textwidth]{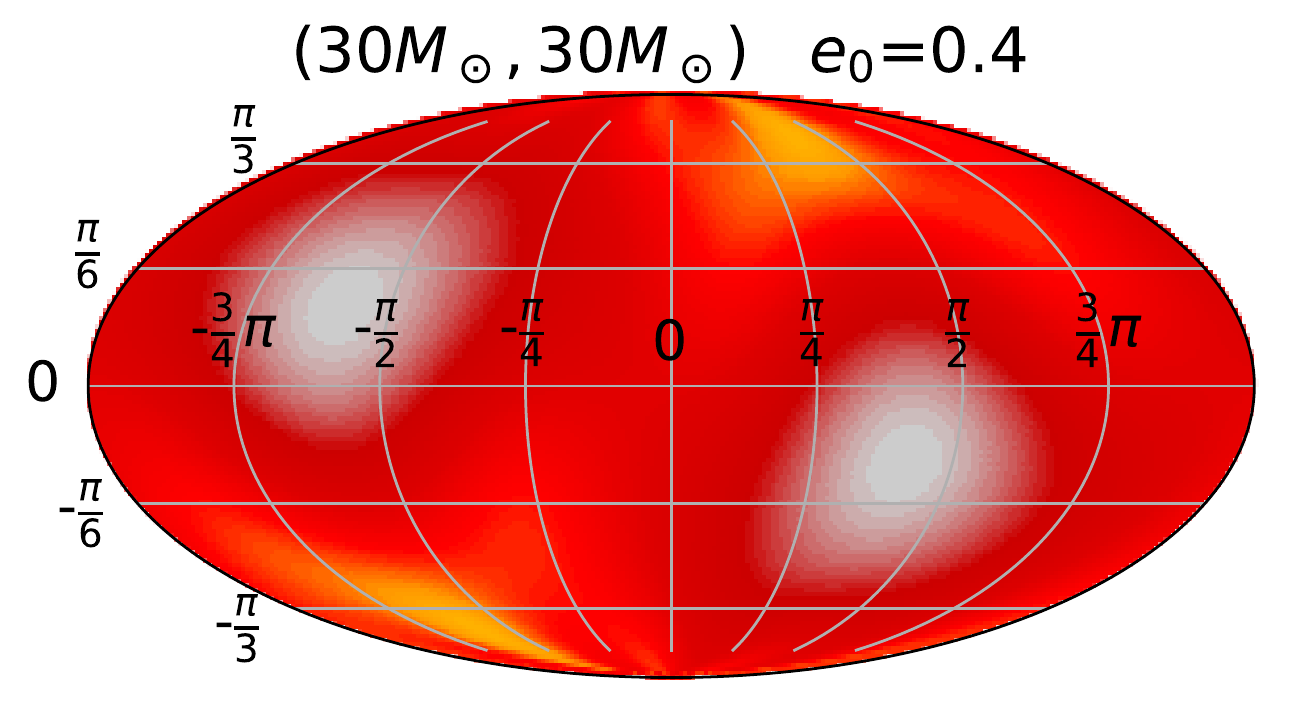}
	}
	\centerline{
		\includegraphics[width=\wid\textwidth]{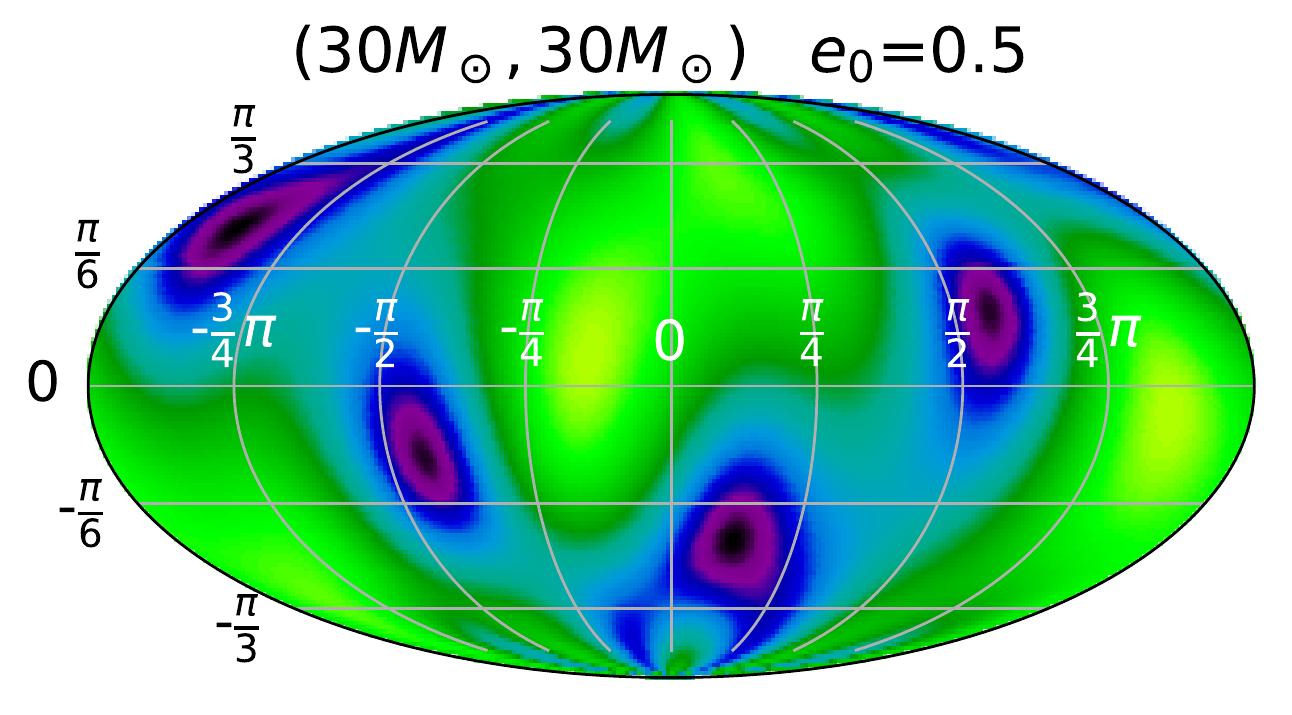}
		\includegraphics[width=\wid\textwidth]{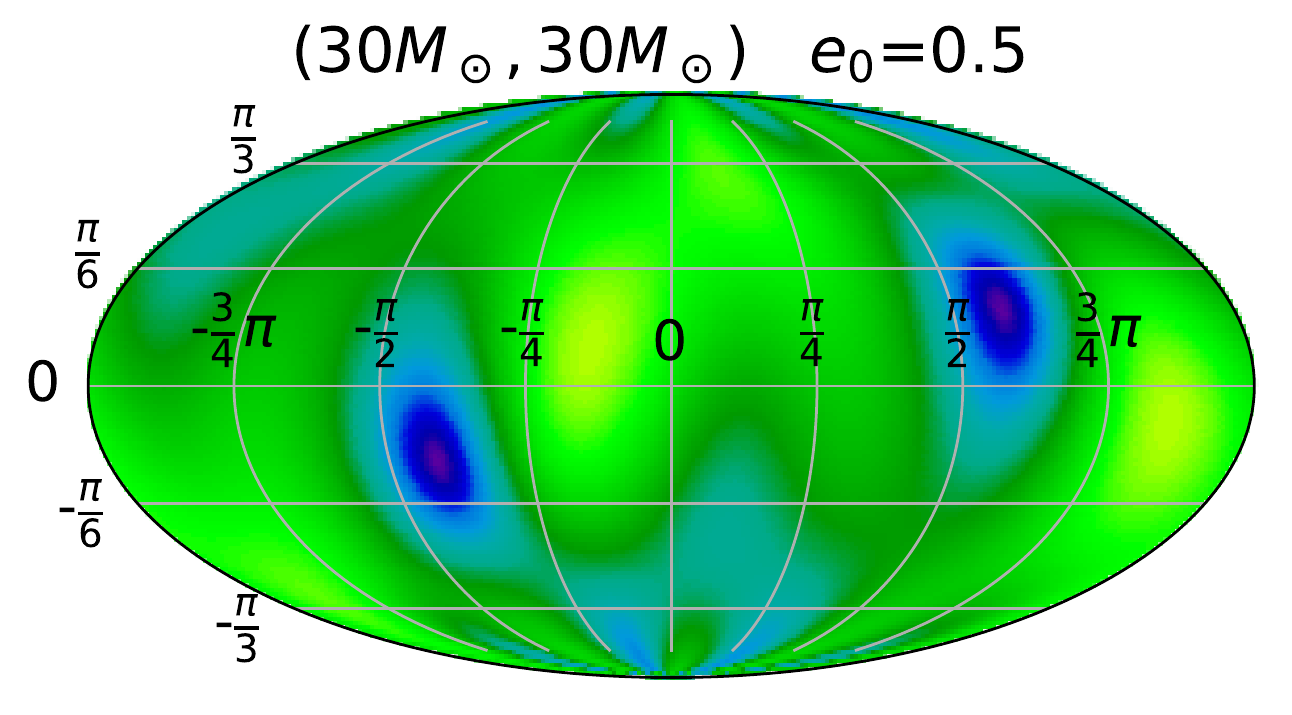}
		\includegraphics[width=\wid\textwidth]{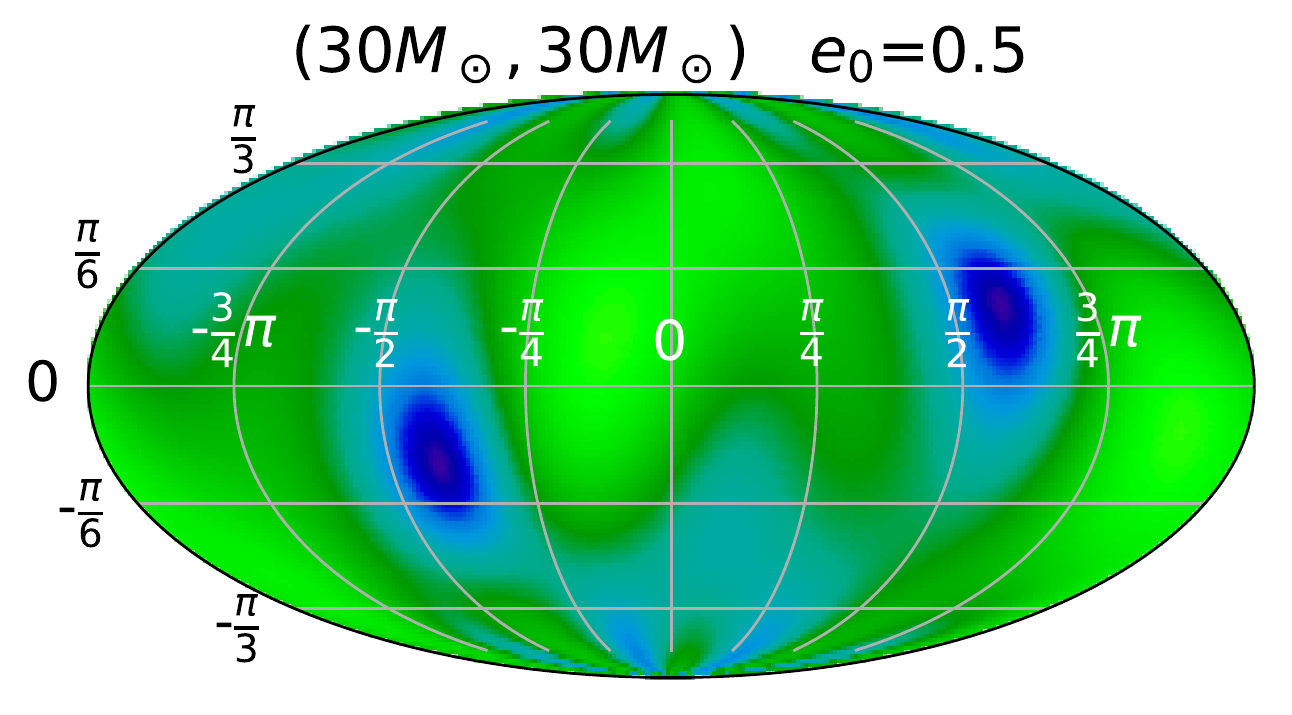}
	}
	\caption{Using the metric provided in Eq.~\ref{eq:metric}, these sky maps show the relative signal-to-noise 
	difference between eccentric and quasi-circular black hole mergers with component masses \((30\msun,\,30\msun)\). We use the Mollweide projection \((\vartheta, \varphi) \rightarrow (\pi/2-\theta, \phi-\pi)\), averaged over polarization angles, and set the binary inclination angle to $i = \pi/4$. The range of eccentricity increases from top  to bottom, while the size of the detector network increases from left to right. Additional results for $i = 0$ are presented in Appendix~\ref{app:ap1}.} 
	\label{fig:equal_mass_LIGO_type}
\end{figure*}

\begin{figure*}
	\centerline{
		\includegraphics[width=\textwidth]{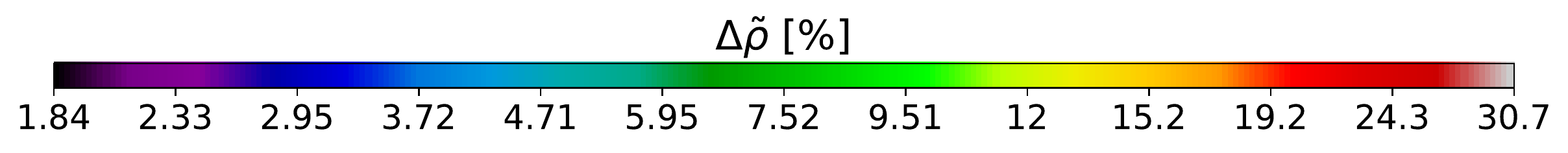}
	}
	\centerline{
		\includegraphics[width=\wid\textwidth]{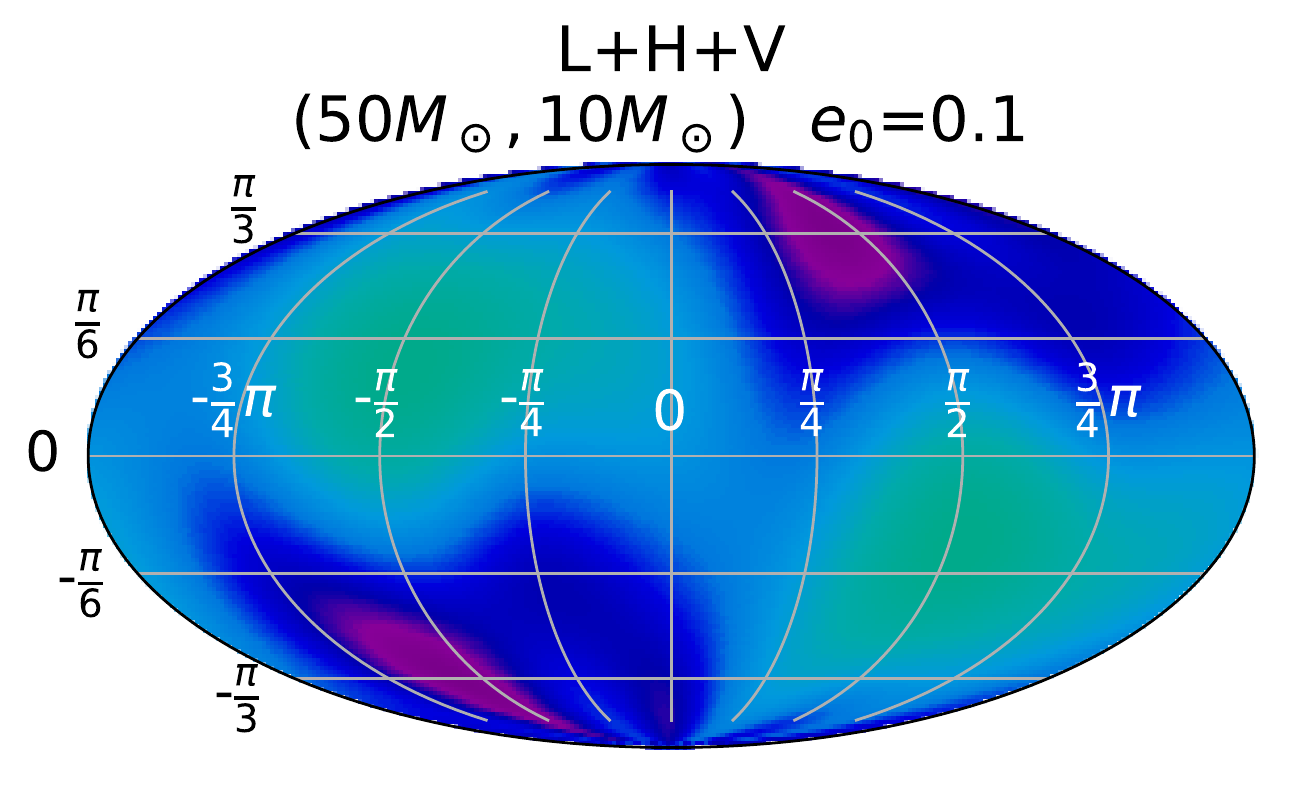}
		\includegraphics[width=\wid\textwidth]{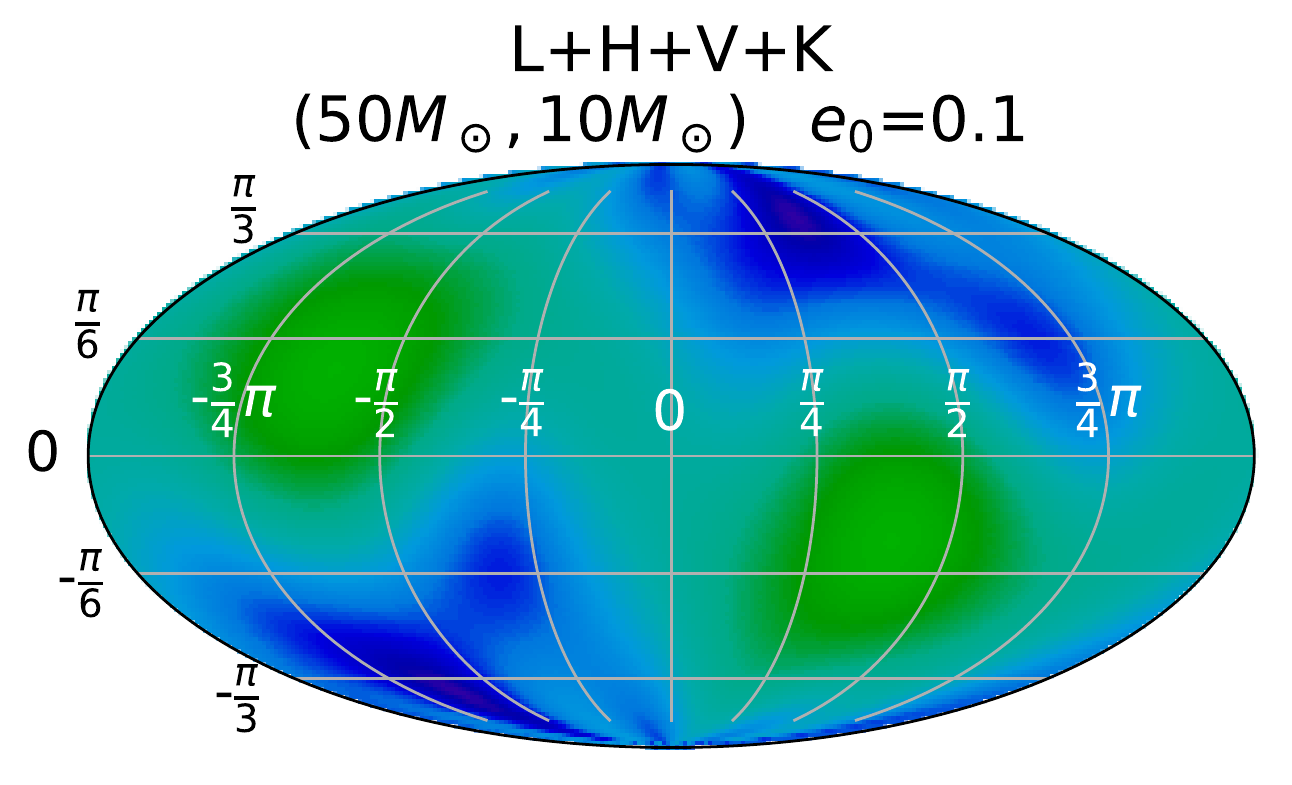}
		\includegraphics[width=\wid\textwidth]{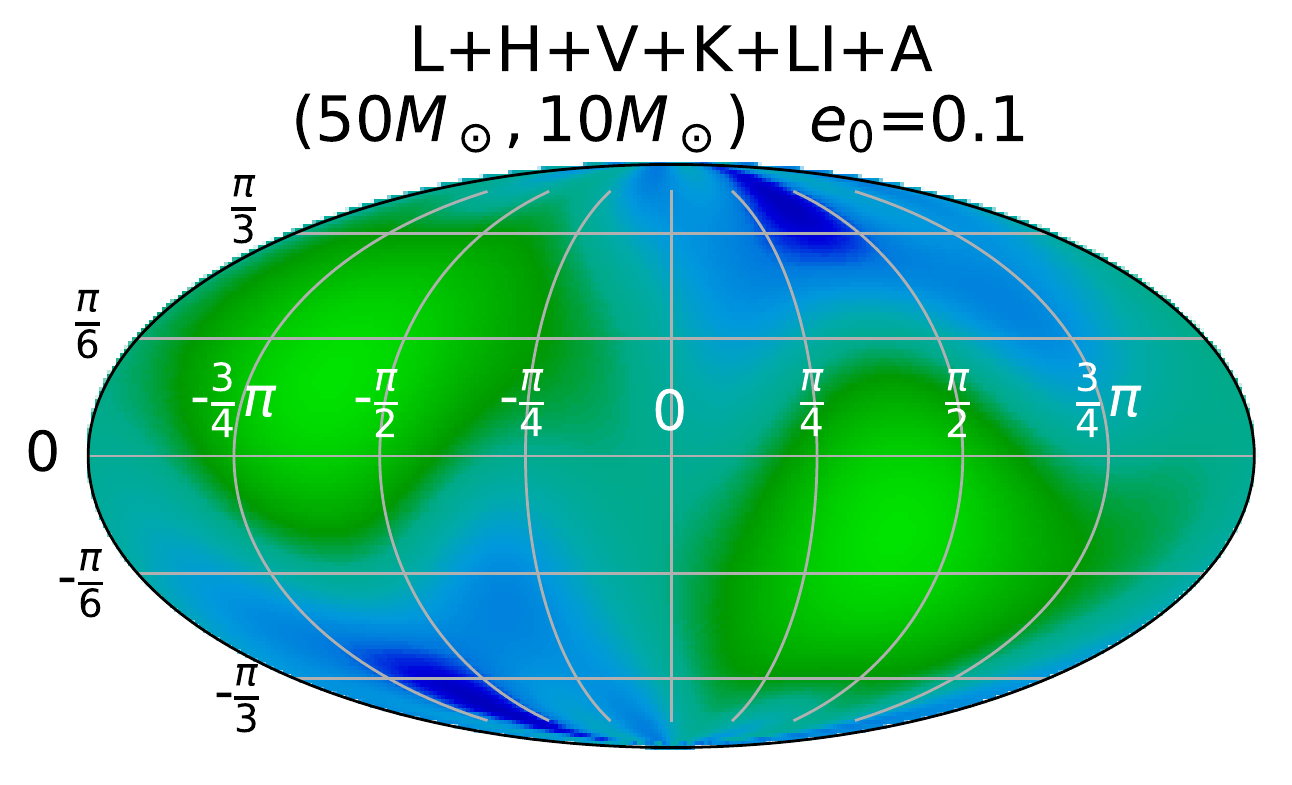}
	}
	\centerline{
		\includegraphics[width=\wid\textwidth]{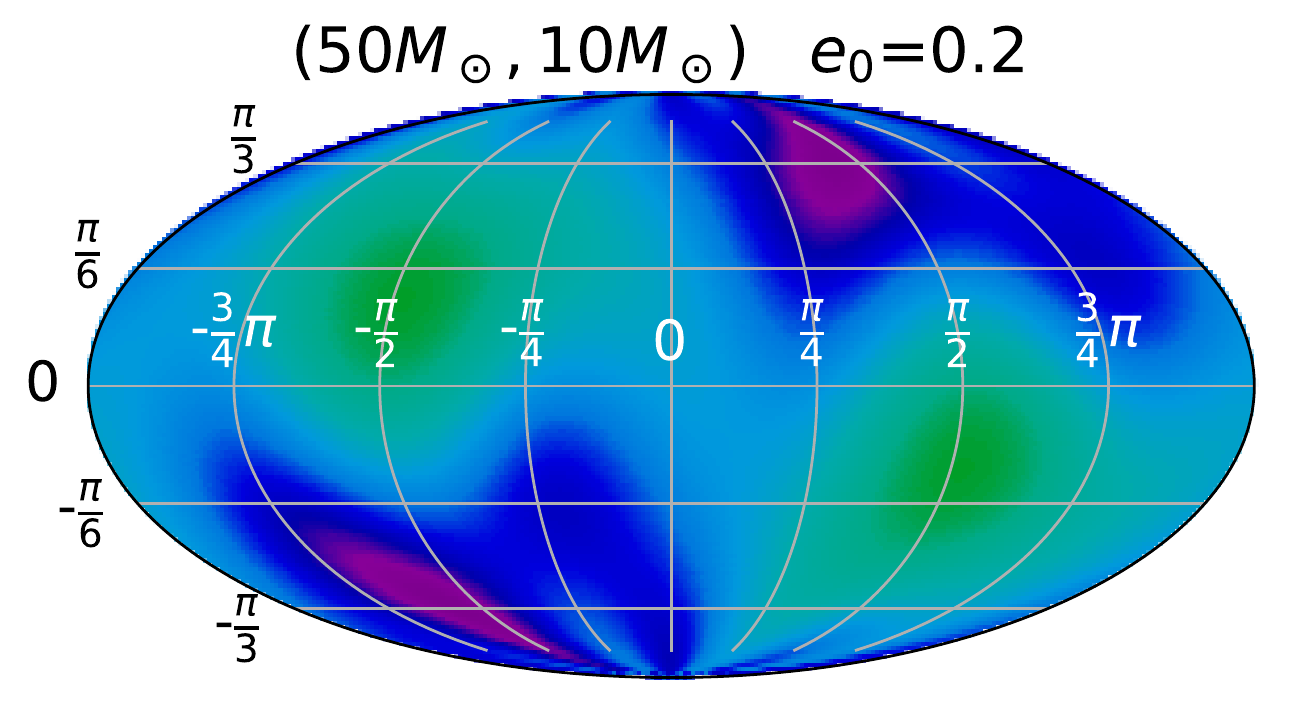}
		\includegraphics[width=\wid\textwidth]{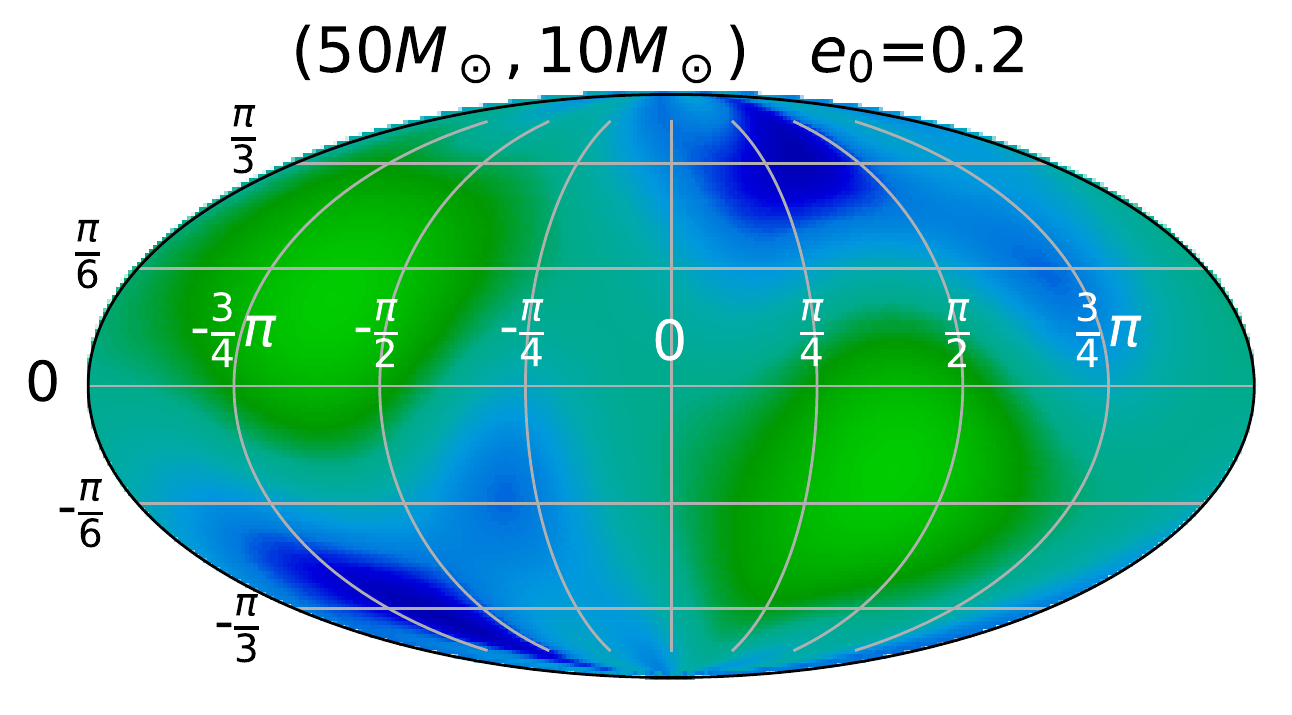}
		\includegraphics[width=\wid\textwidth]{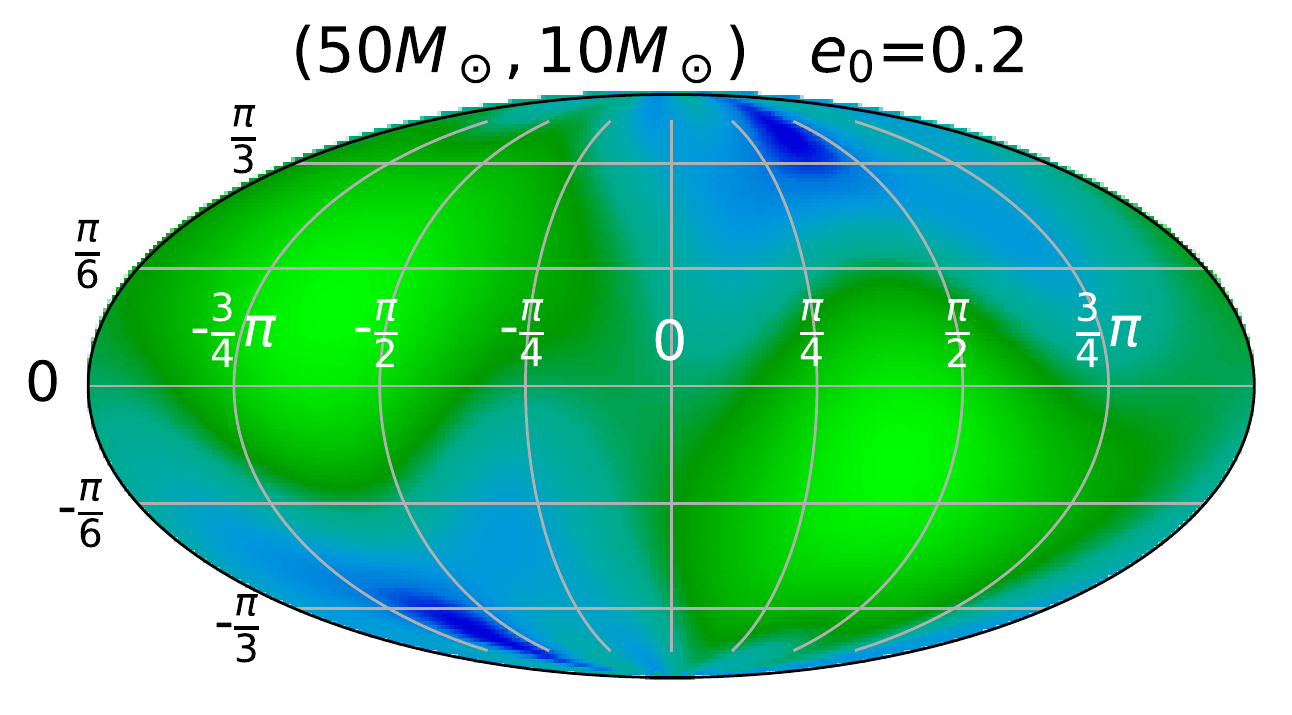}
	}
	\centerline{
		\includegraphics[width=\wid\textwidth]{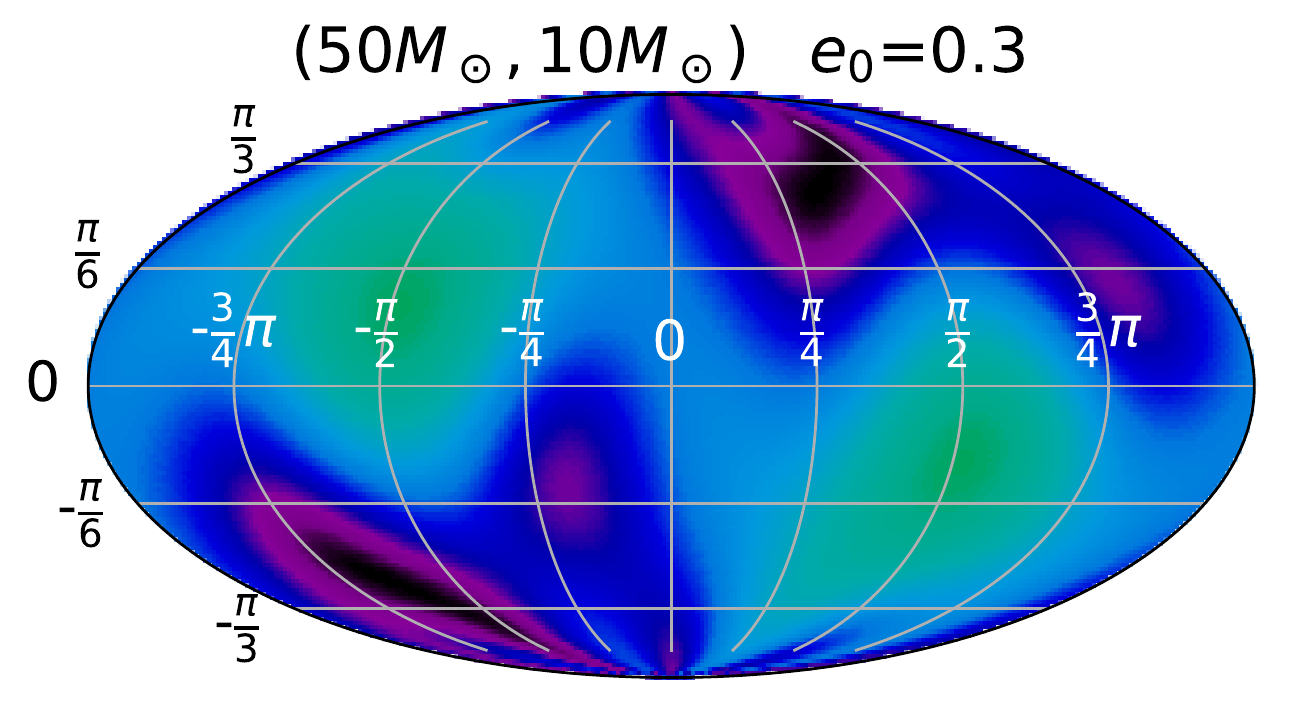}
		\includegraphics[width=\wid\textwidth]{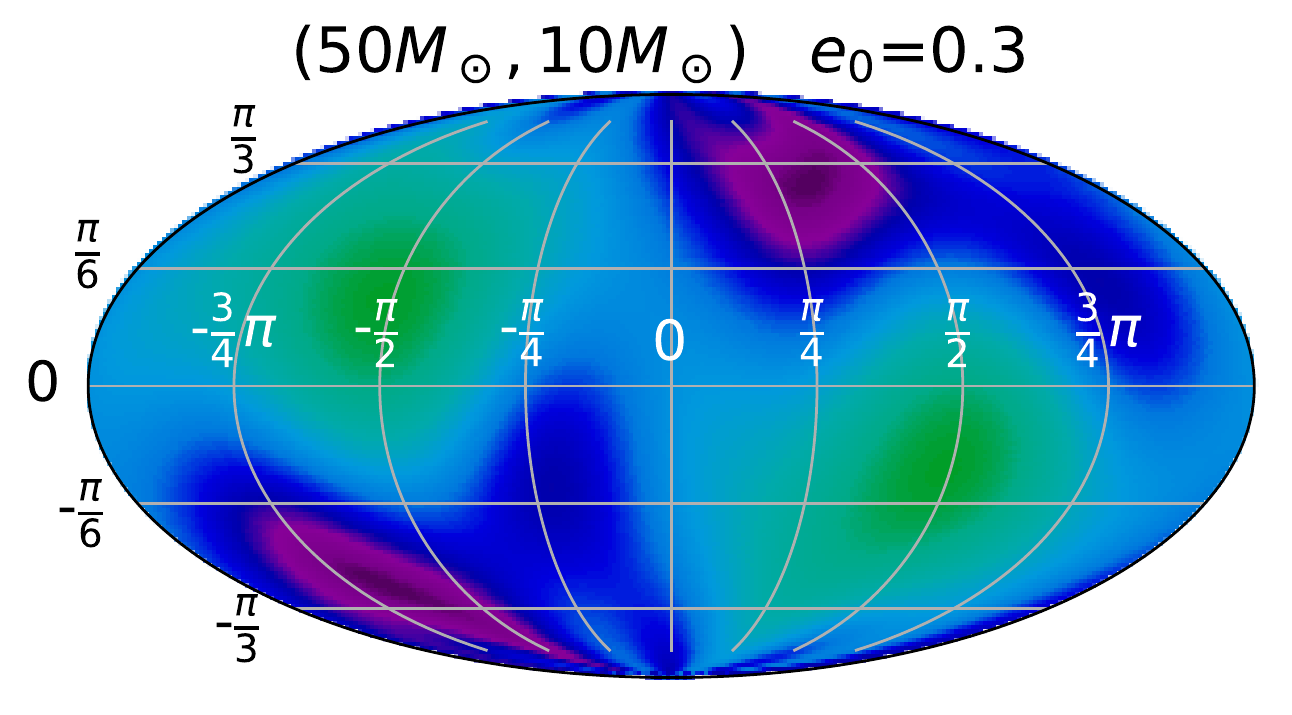}
		\includegraphics[width=\wid\textwidth]{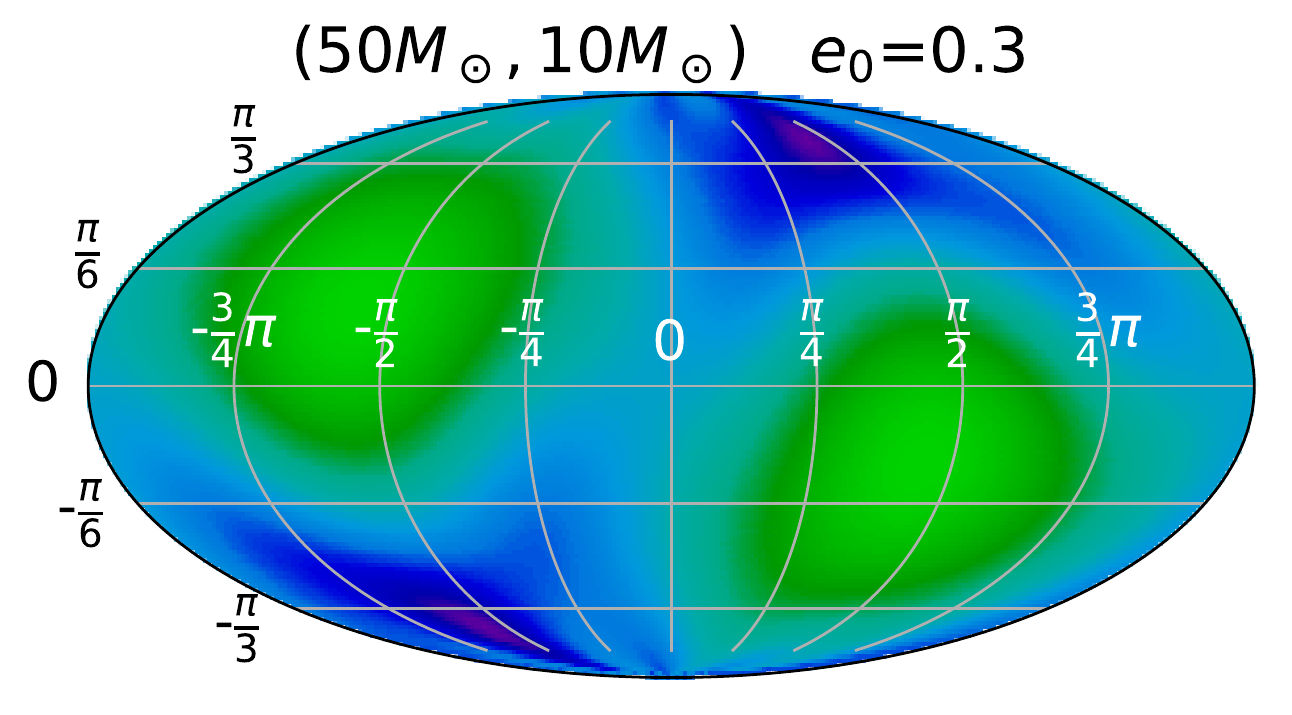}
	}
	\centerline{
		\includegraphics[width=\wid\textwidth]{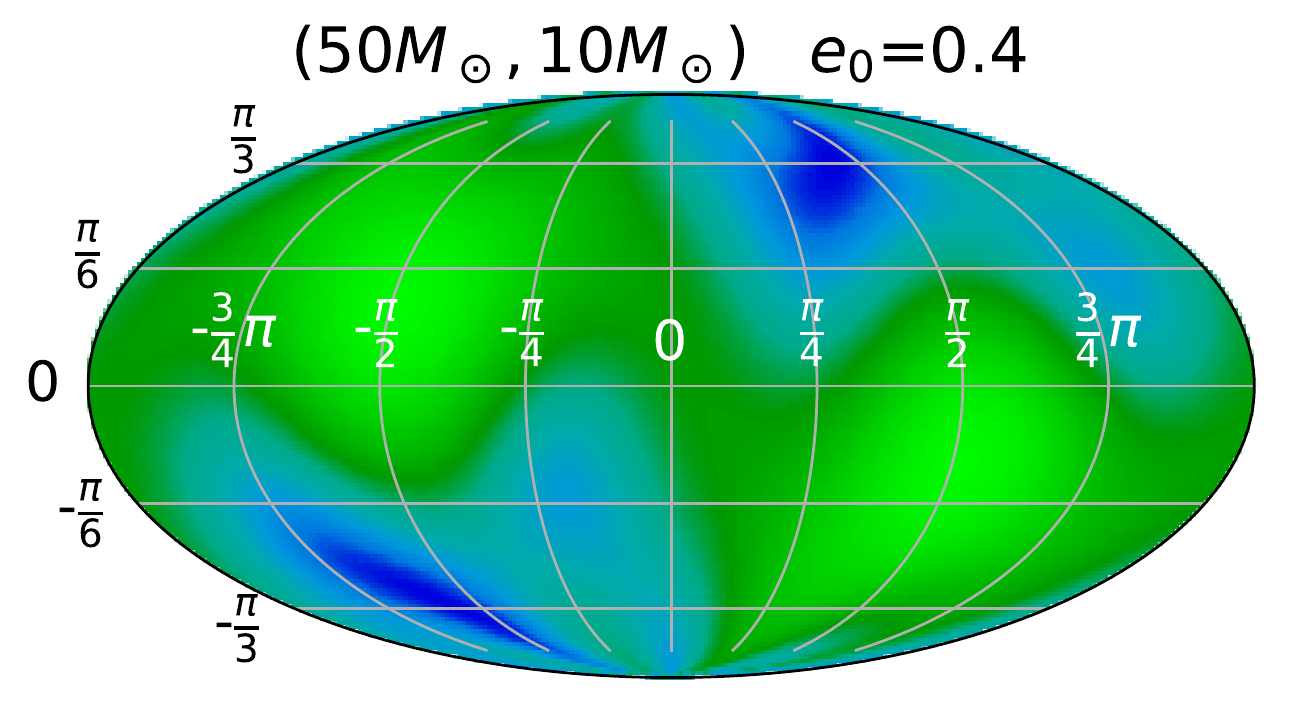}
		\includegraphics[width=\wid\textwidth]{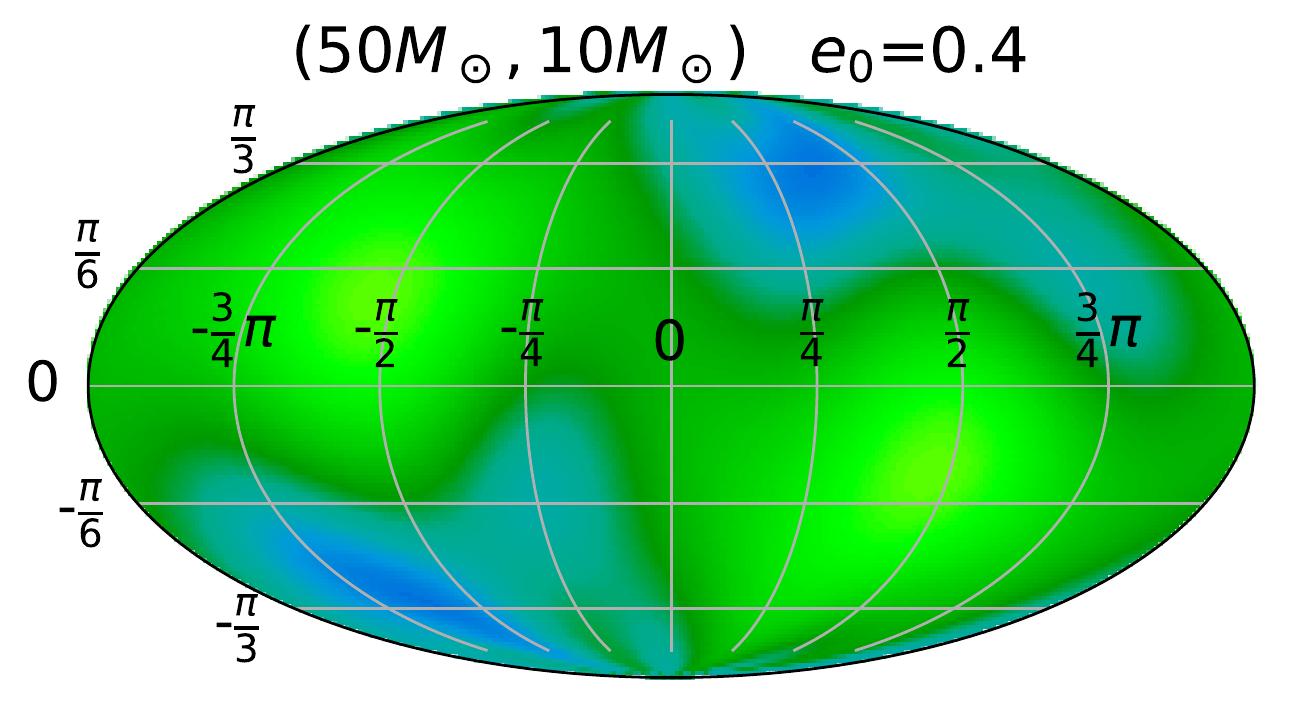}
		\includegraphics[width=\wid\textwidth]{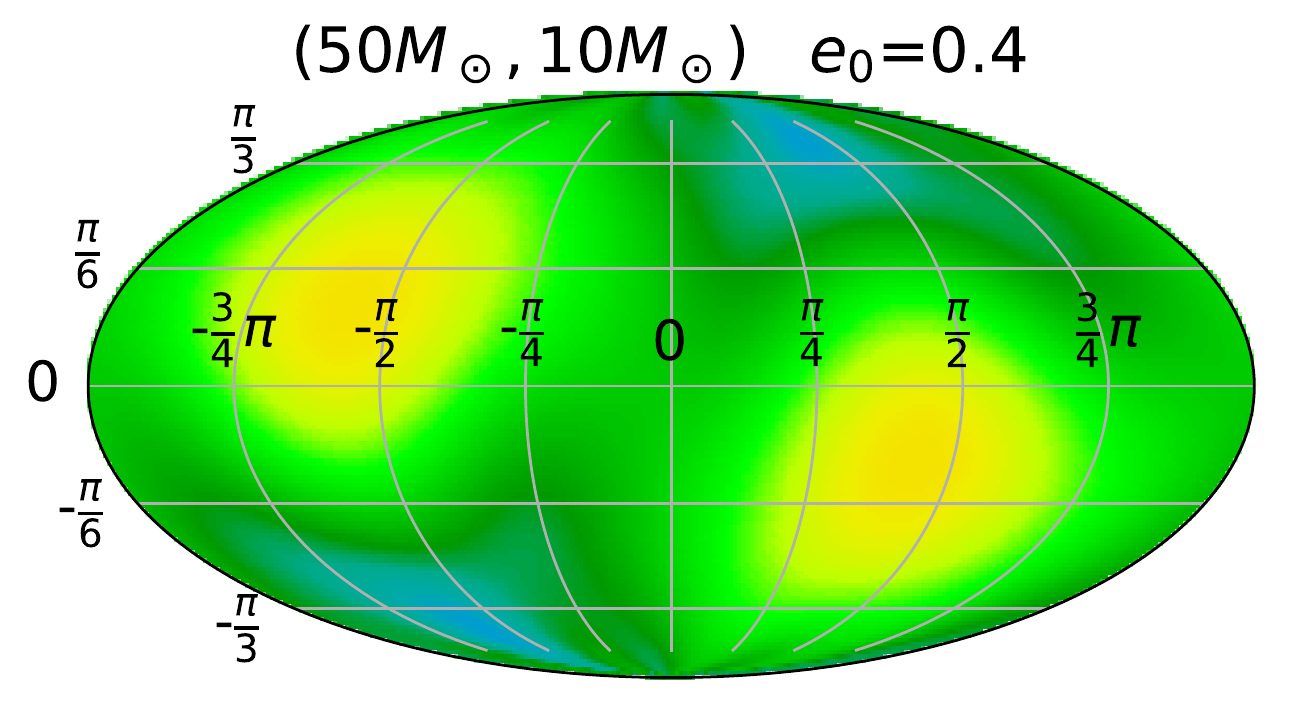}
	}
	\centerline{
		\includegraphics[width=\wid\textwidth]{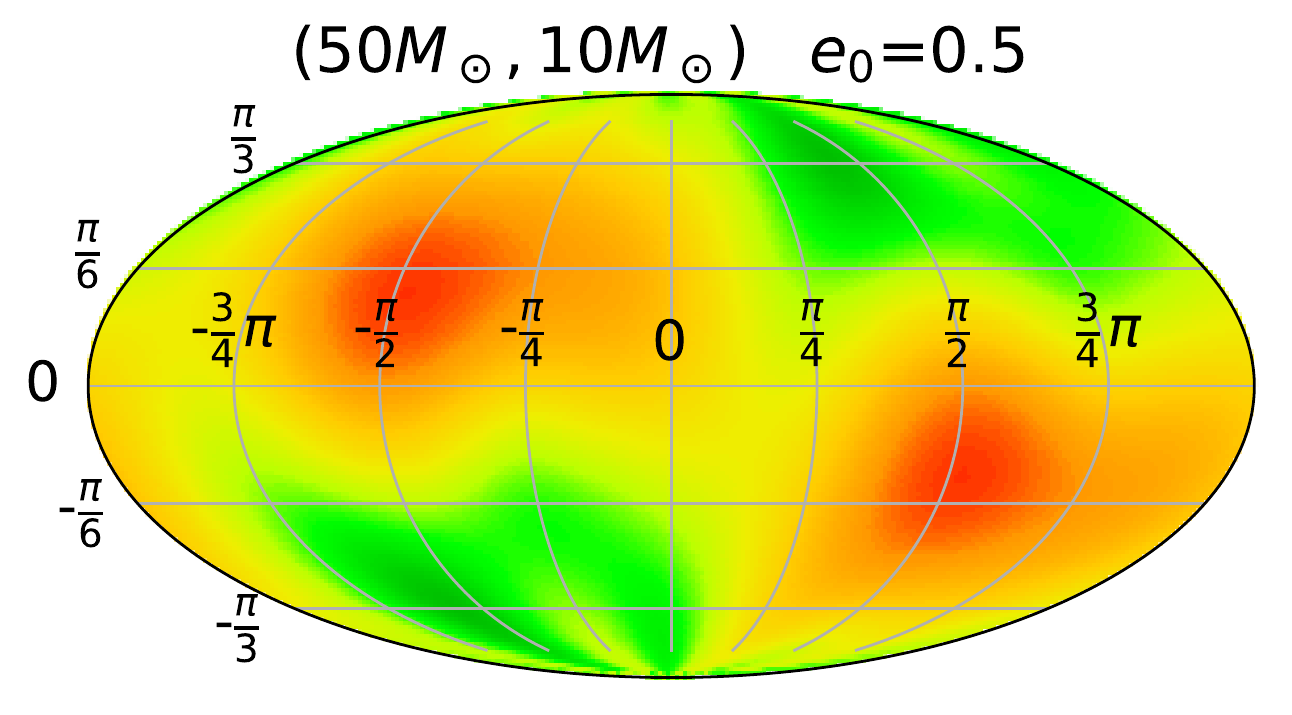}
		\includegraphics[width=\wid\textwidth]{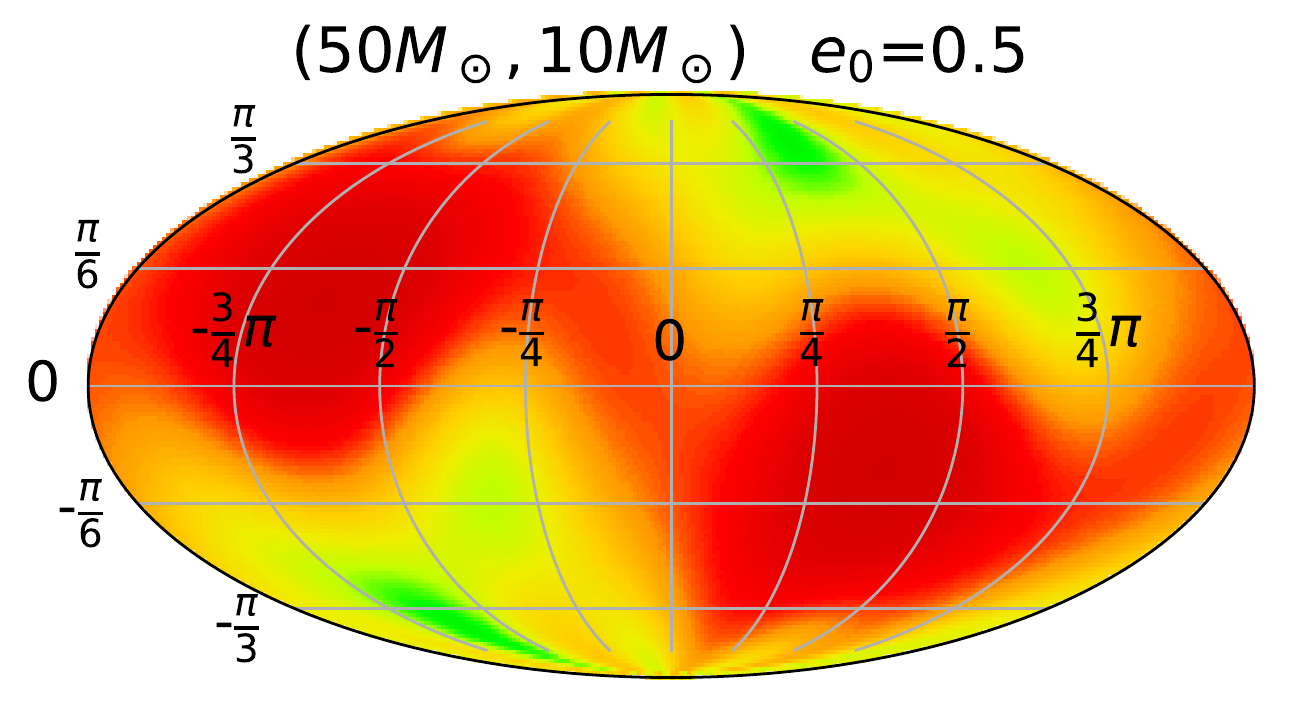}
		\includegraphics[width=\wid\textwidth]{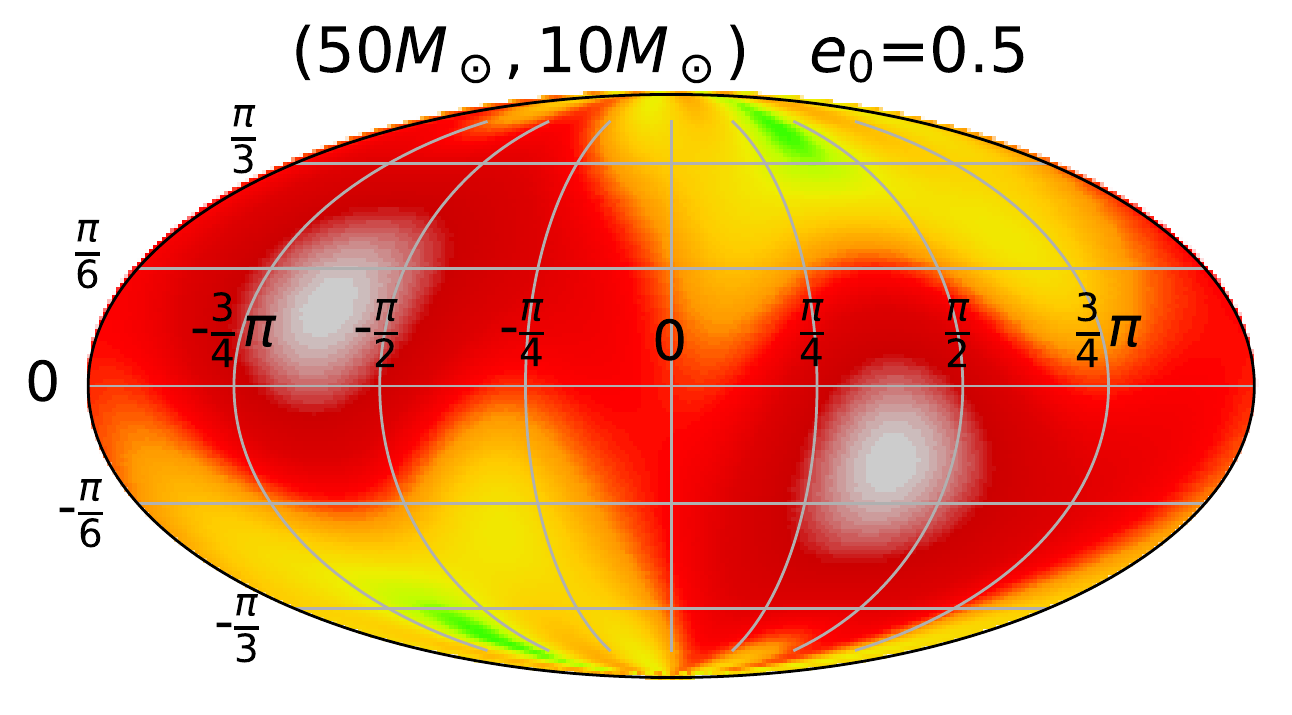}
	}
	\caption{As Figure~\ref{fig:equal_mass_LIGO_type} but now for black hole mergers with component masses \((50\msun,\,10\msun)\).} 
	\label{fig:five_mass_LIGO_type}
\end{figure*}

\clearpage

\subsection{Third generation detector networks}
\label{sec:third_results}

We consider two configurations for third generation detector networks. 
In these two cases the geographical location of the GW detectors is identical to those listed 
in Table~\ref{tab:dets}. The only difference is that we replace the PSD of each of these detectors 
using the target sensitivity of either Cosmic Explorer or ET-D.

\noindent \textbf{Cosmic Explorer} This is a proposed L-shaped detector, based in the US, whose arms 
will be 10 times longer that advanced LIGO's. Details regarding Cosmic Explorer's layout, scale, 
technology and science scope may be found at~\cite{cosmic_exp}. 

\begin{figure*}
	\centerline{
	\includegraphics[width = 0.33\textwidth]{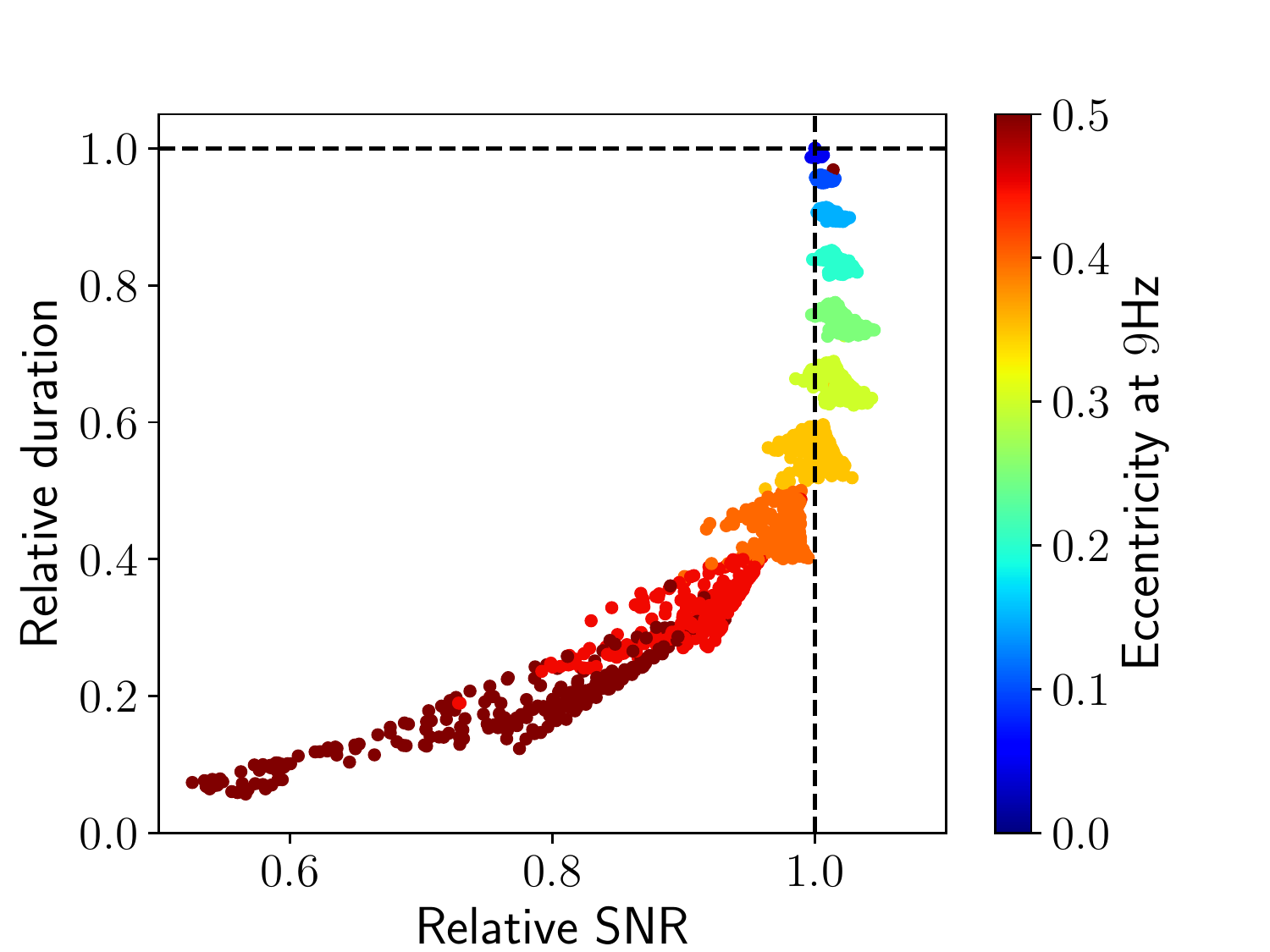}
	\includegraphics[width = 0.33\textwidth]{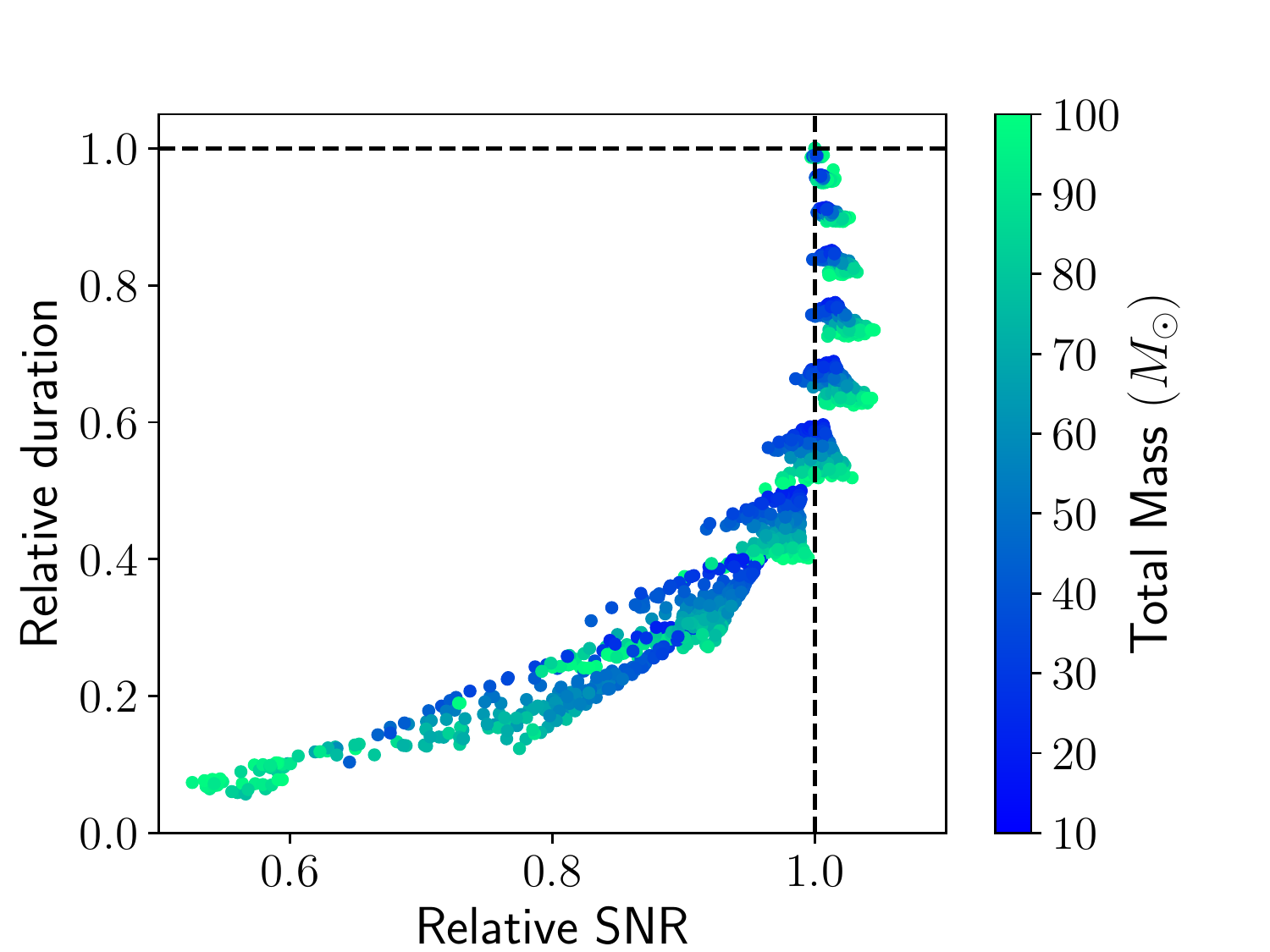}
	\includegraphics[width = 0.33\textwidth]{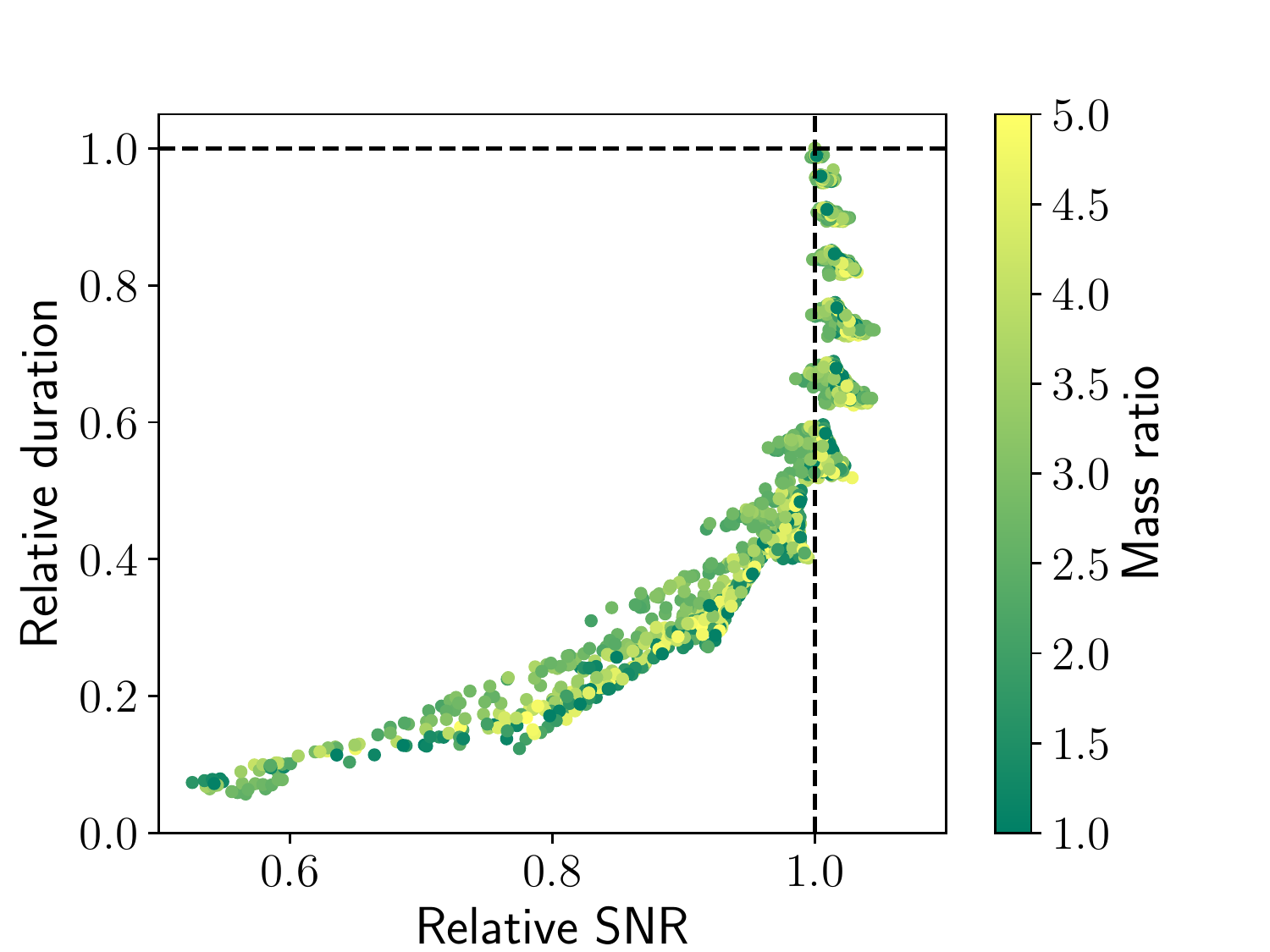}
	}
	\centerline{
	\includegraphics[width = 0.33\textwidth]{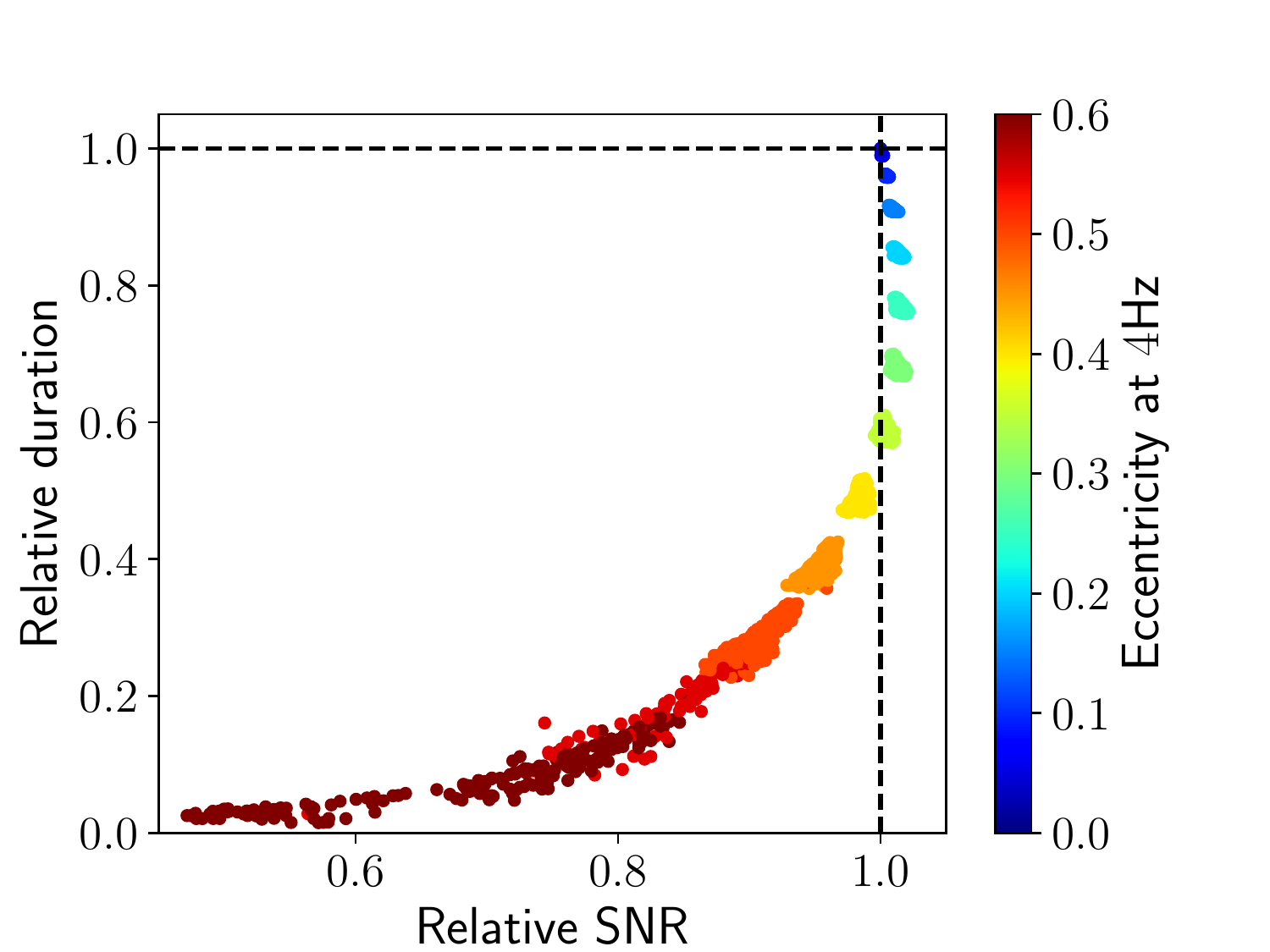}
	\includegraphics[width = 0.33\textwidth]{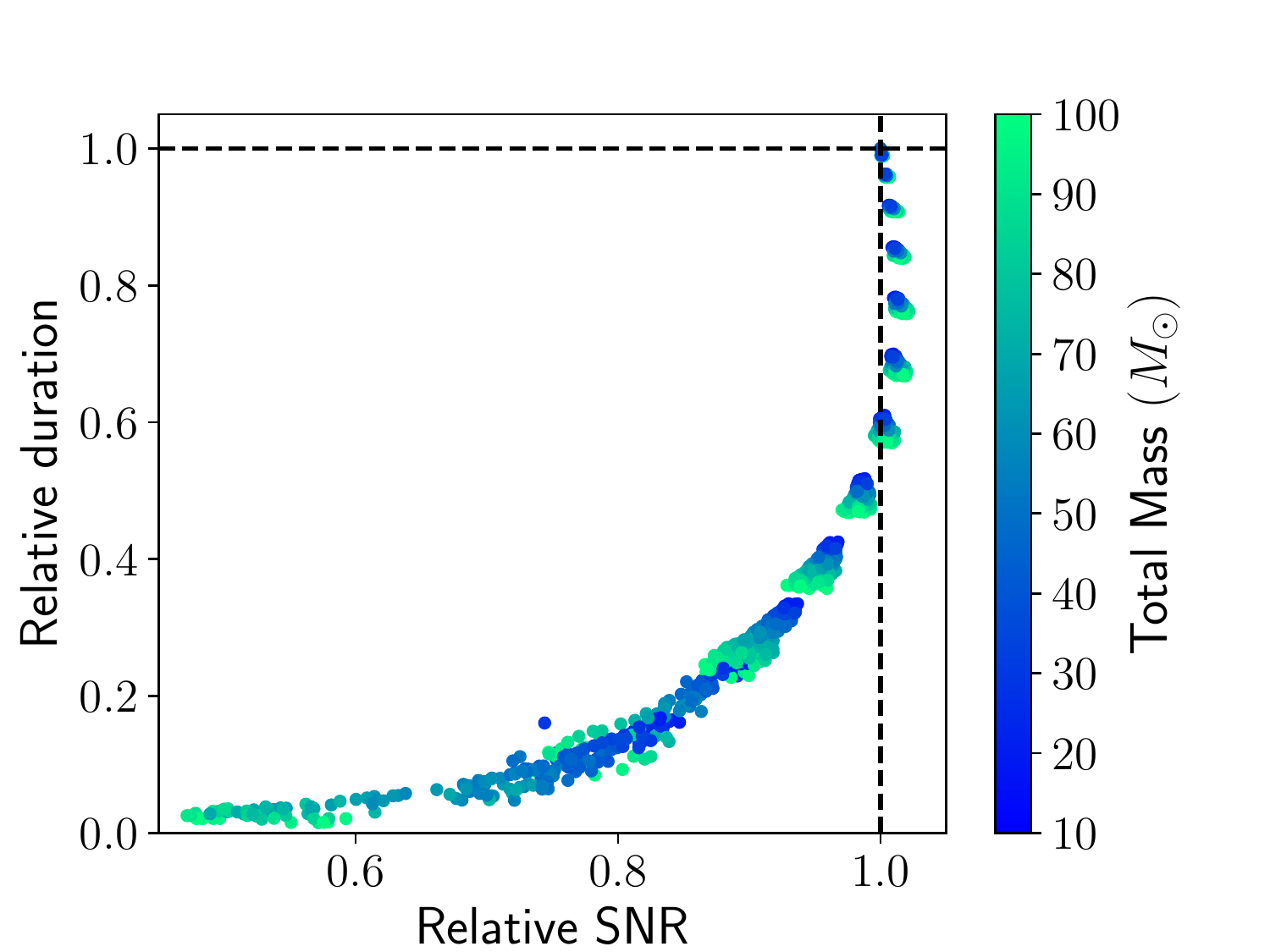}
	\includegraphics[width = 0.33\textwidth]{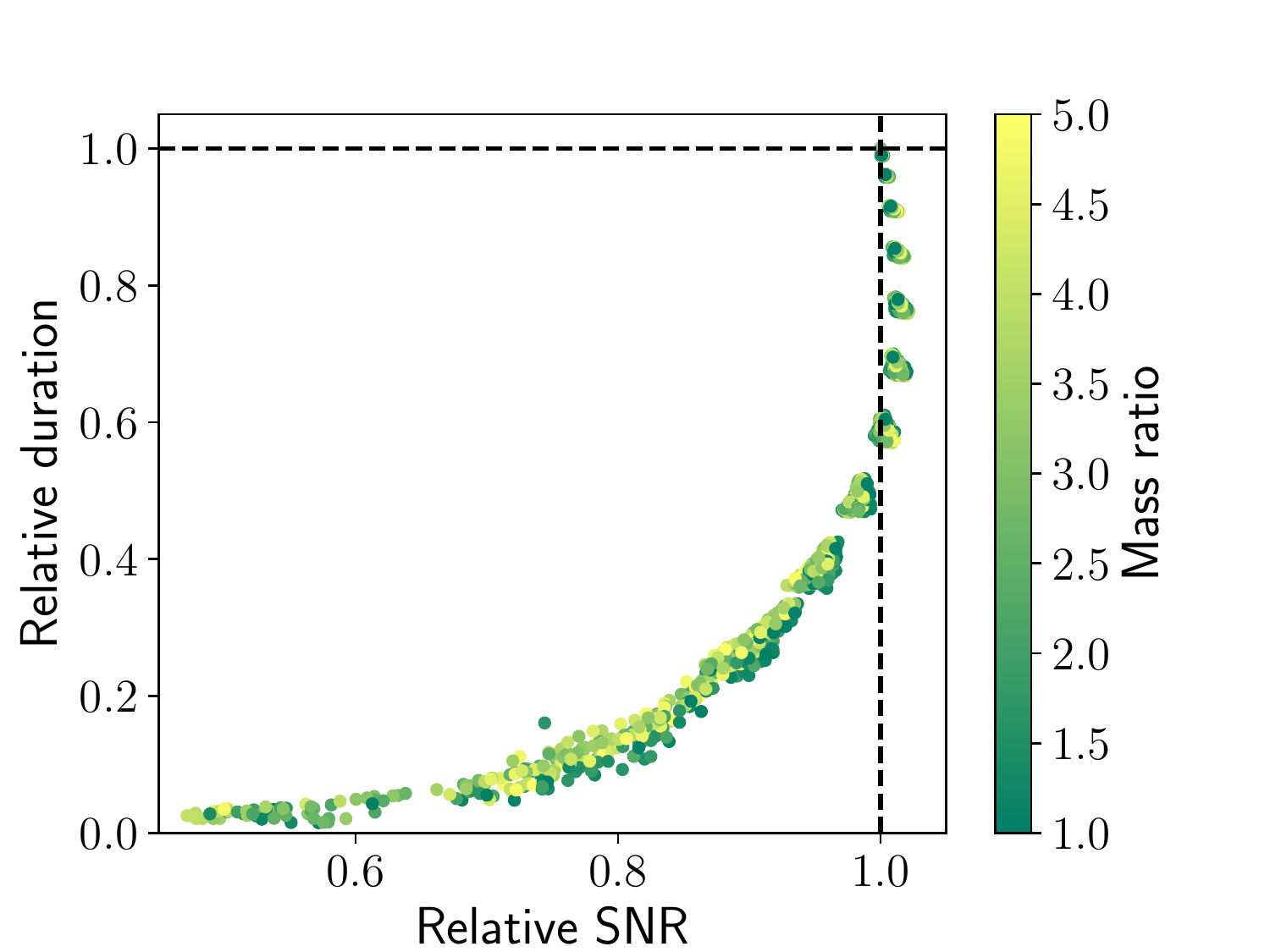}
	}
	\caption{As Figure~\ref{fig:snr_time_ligo}, but now for Cosmic Explorer (top panels) and ET-D (bottom panels). As before,  these results present systems with total mass \(20\msun\leq M\leq 100\msun\) and mass-ratios \(1\leq q\leq5\).}
	\label{fig:snr_time_CE_and_ETD}
\end{figure*}

To provide a high level description of the importance of eccentricity, total mass and 
mass-ratio for the observation of BBH mergers, we consider a single Cosmic Explorer detector, 
and compute the SNR and waveform length of BBH systems with total mass \(20\msun \leq M \leq 100\), 
mass-ratios \(1\leq q \leq 5\), and \(e_0\leq0.5\) 
measured at \(f_{\textrm{GW}} = 9\,\textrm{Hz}\). We compute SNR distributions 
setting \(f_0=11\,\textrm{Hz}\) and \(f_1=4096\,\textrm{Hz}\) in Eq.~\eqref{eqn:SNRcomponent}. 
We have set these parameters for the computation of SNR to provide a direct 
comparison to advanced LIGO-type detectors. The key difference is that Cosmic Explore provides 
enhanced sensitivity at lower frequencies, as shown in Figure~\ref{fig:psds_all}. These 
results, presented in the top panels of Figure~\ref{fig:snr_time_CE_and_ETD}, 
show that the SNR distributions of eccentric and quasi-circular mergers 
are comparable up to eccentricities \(e_0\leq0.4\). However, there is a significant drop in SNR 
for systems with \(e_0\sim0.5\). 

\begin{figure*}[p]
	\centerline{
		\includegraphics[width=\textwidth]{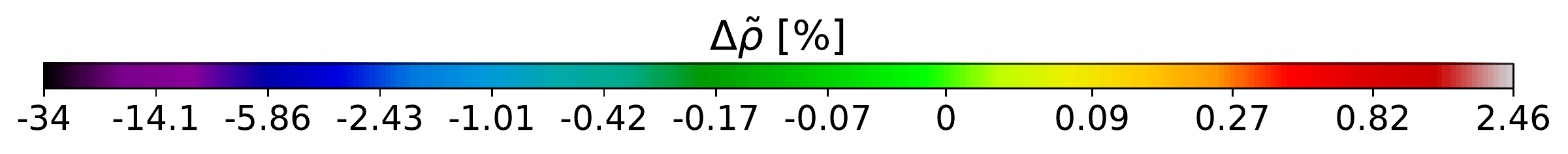}
	}
	\centerline{
		\includegraphics[width=\wid\textwidth]{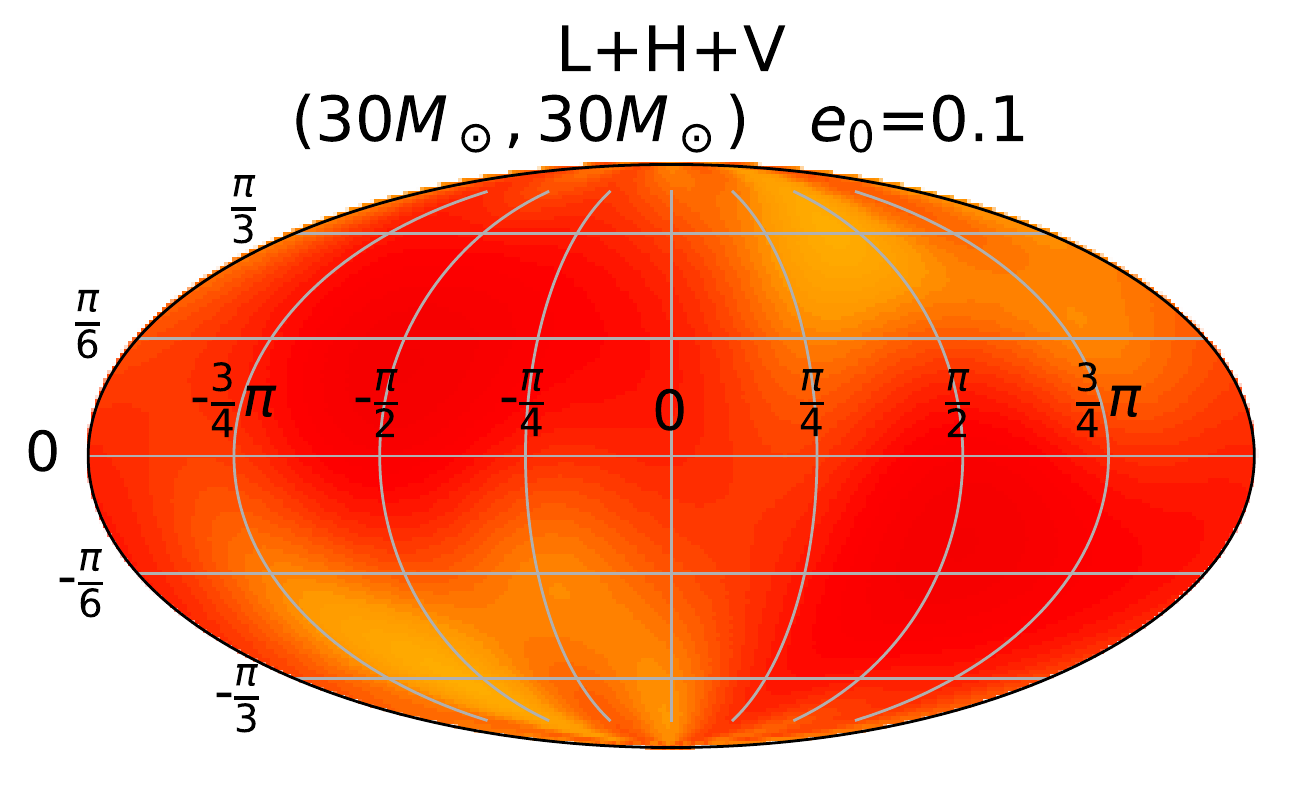}
		\includegraphics[width=\wid\textwidth]{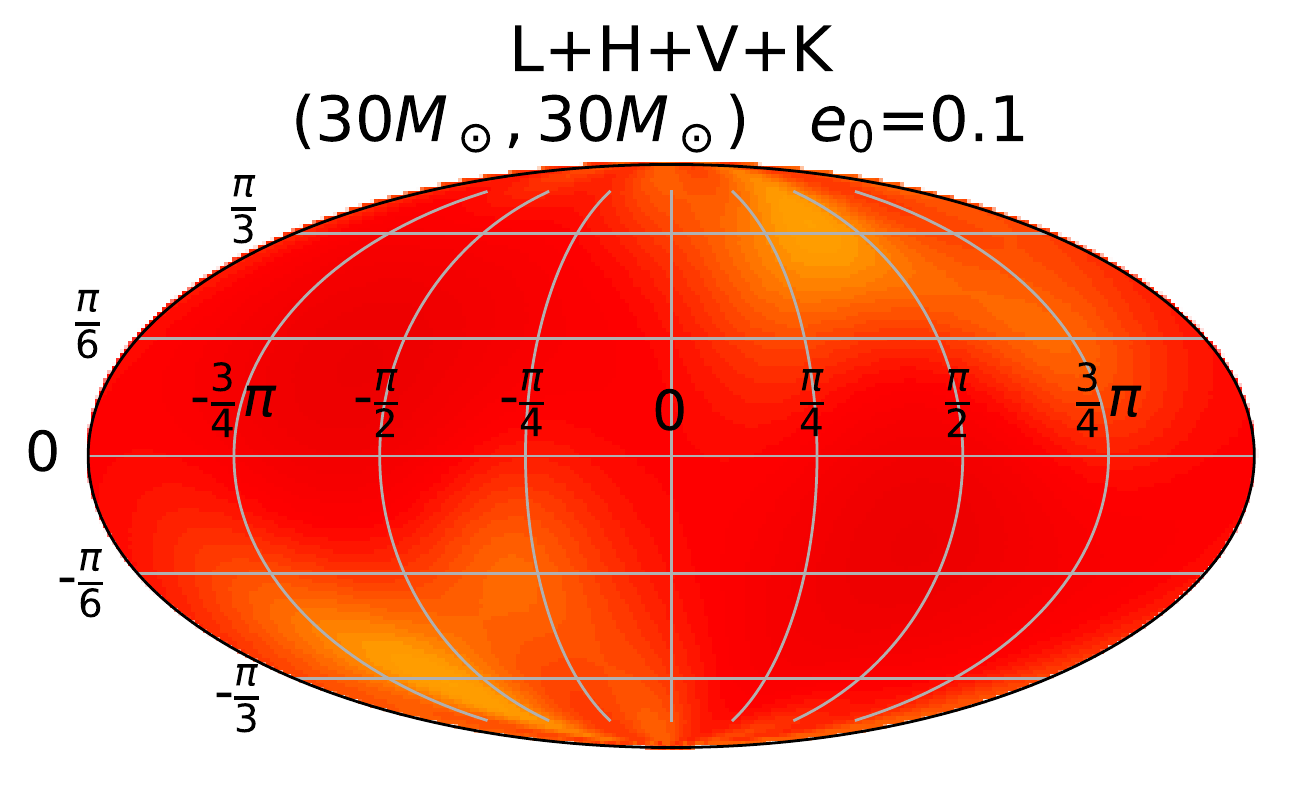}
		\includegraphics[width=\wid\textwidth]{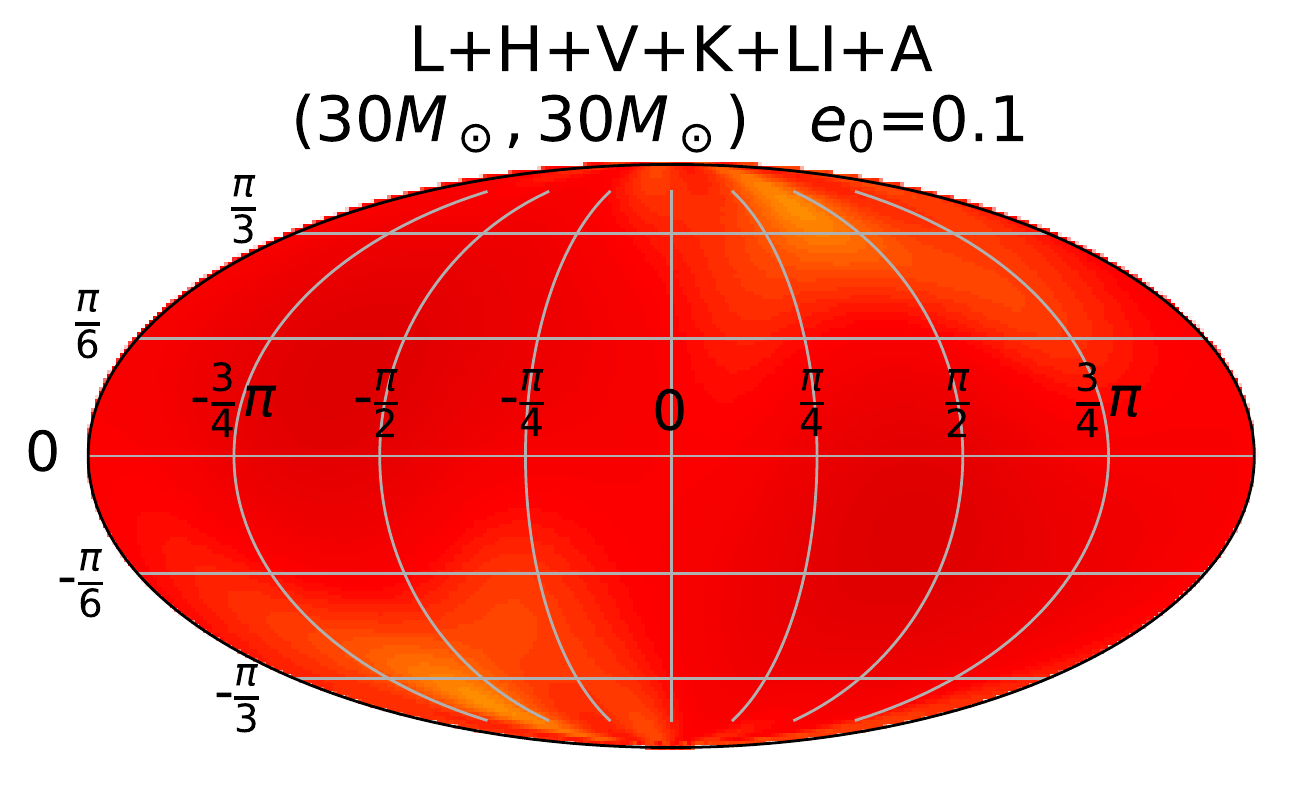}
	}
	\centerline{
		\includegraphics[width=\wid\textwidth]{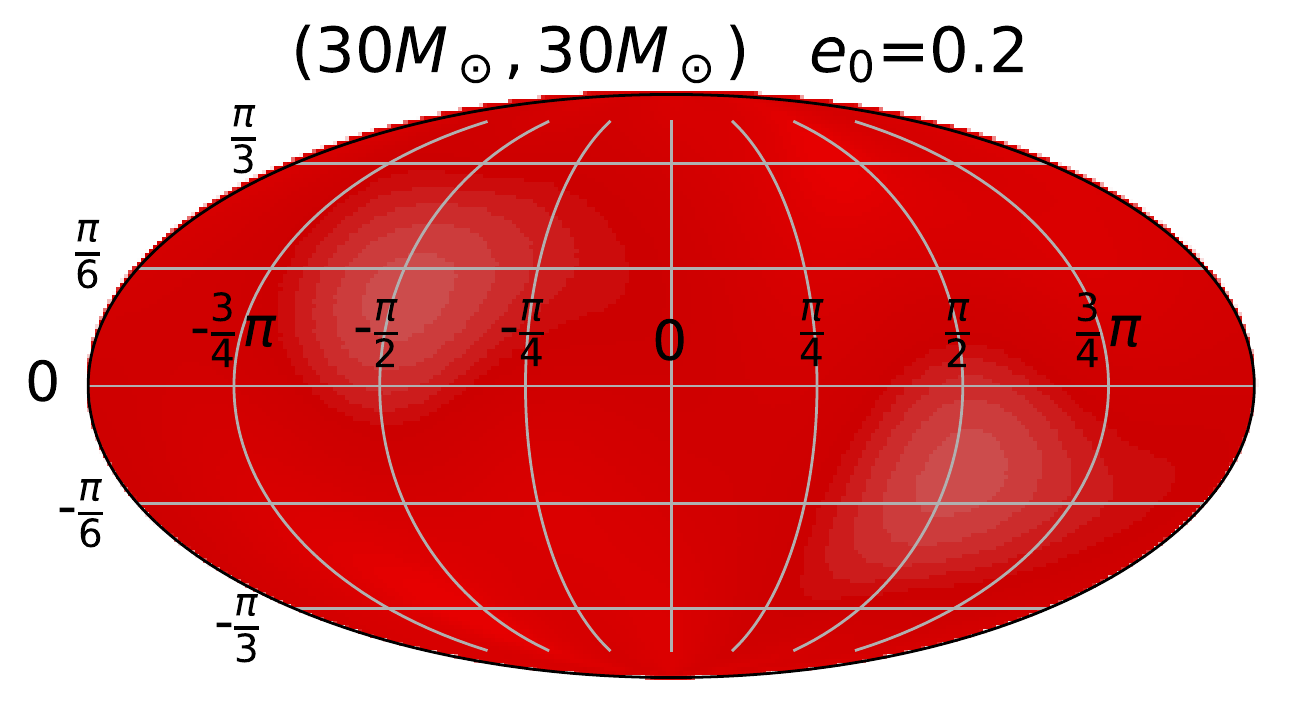}
		\includegraphics[width=\wid\textwidth]{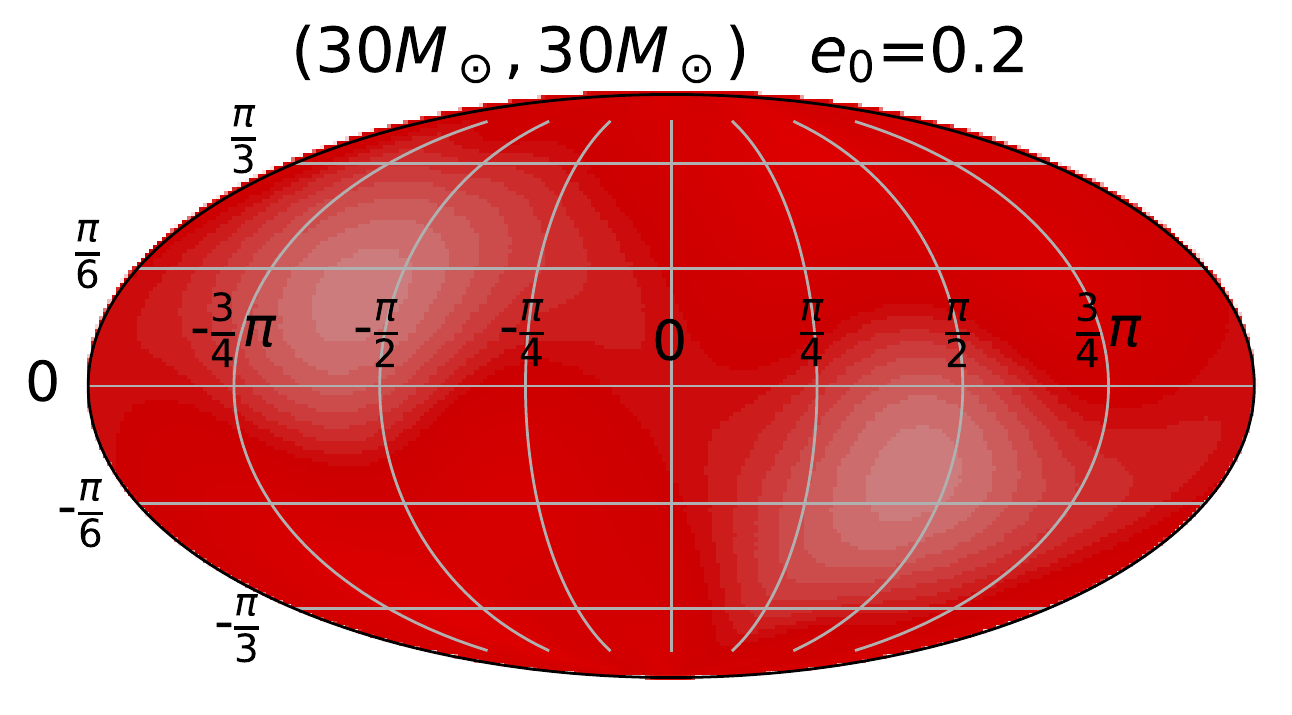}
		\includegraphics[width=\wid\textwidth]{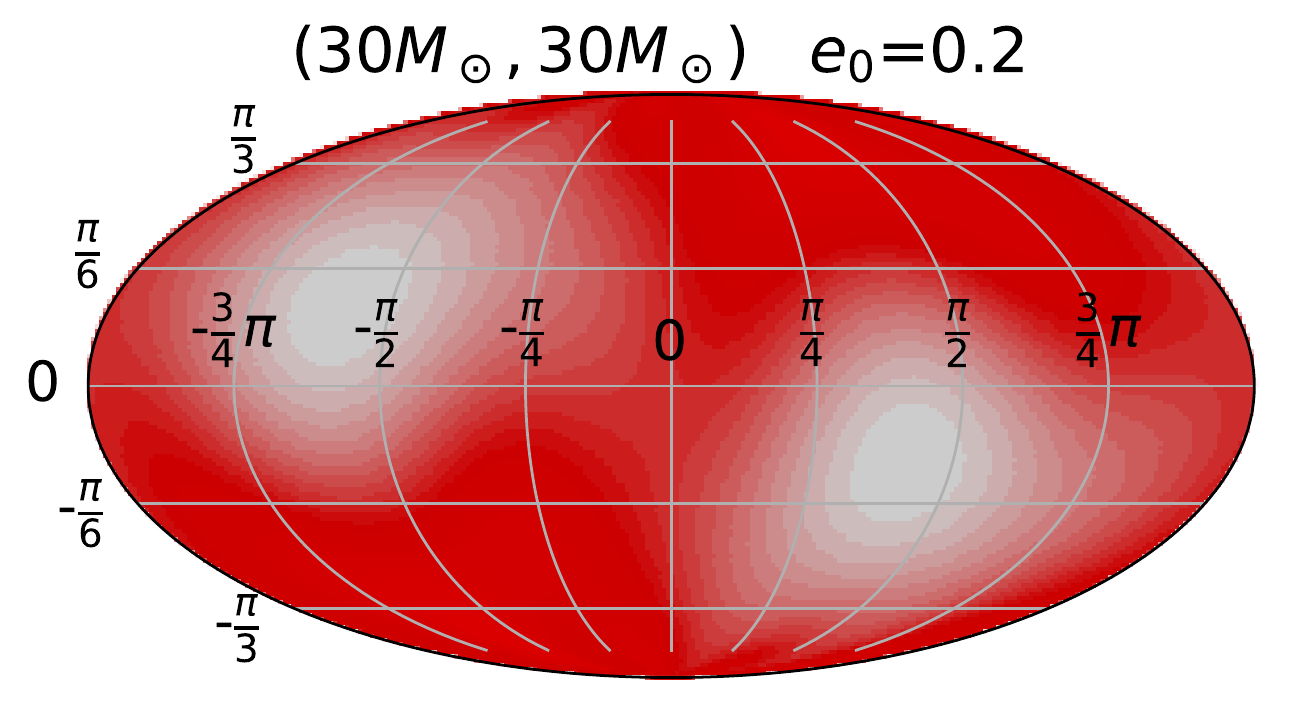}
	}
	\centerline{
		\includegraphics[width=\wid\textwidth]{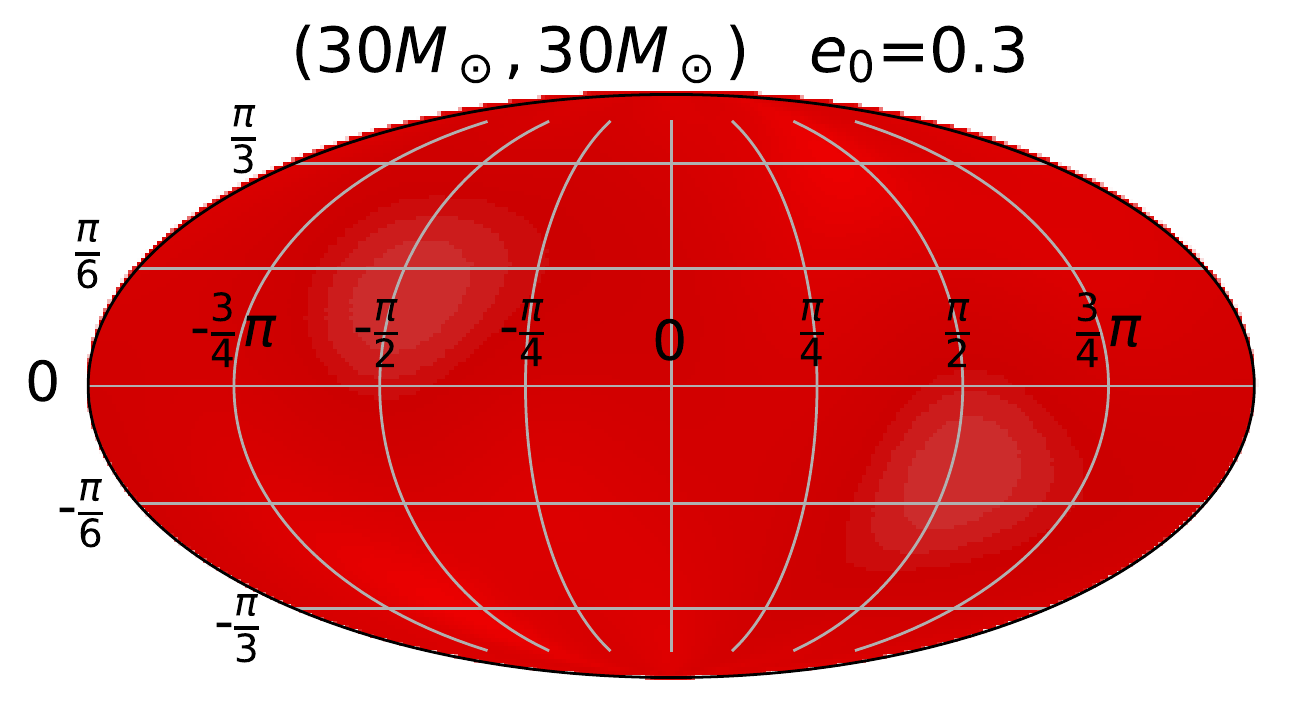}
		\includegraphics[width=\wid\textwidth]{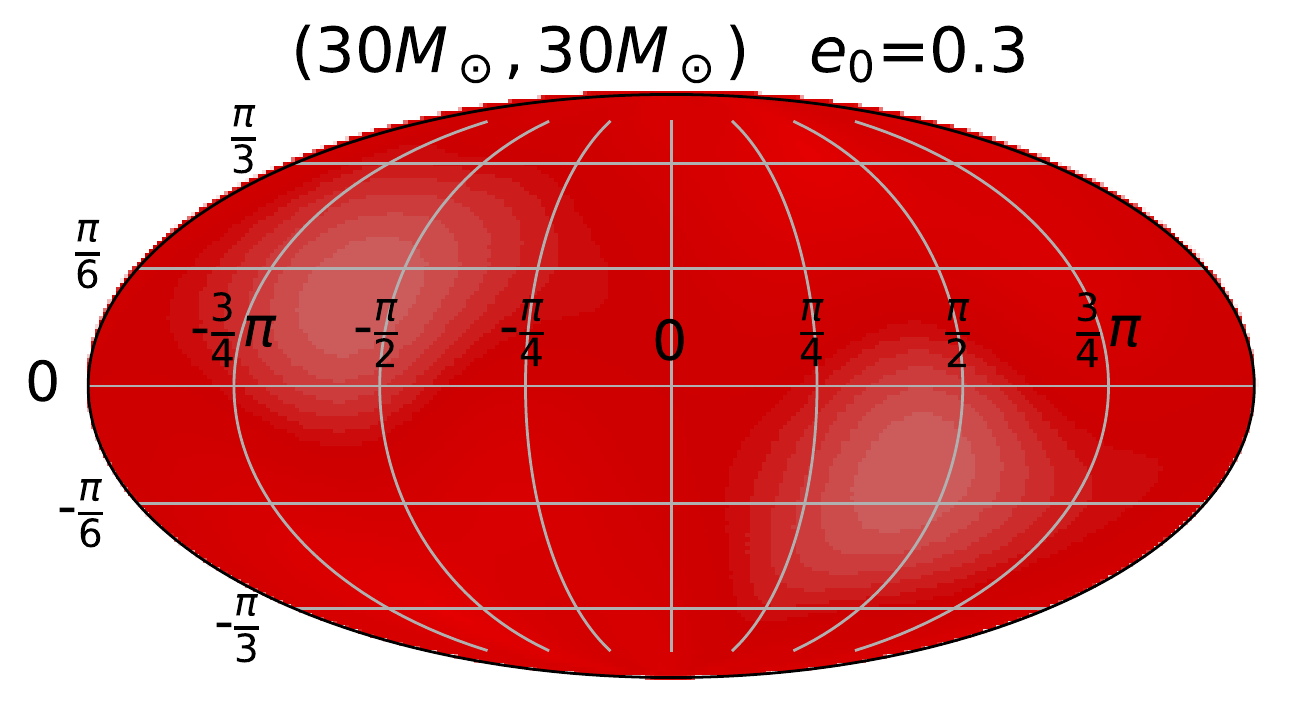}
		\includegraphics[width=\wid\textwidth]{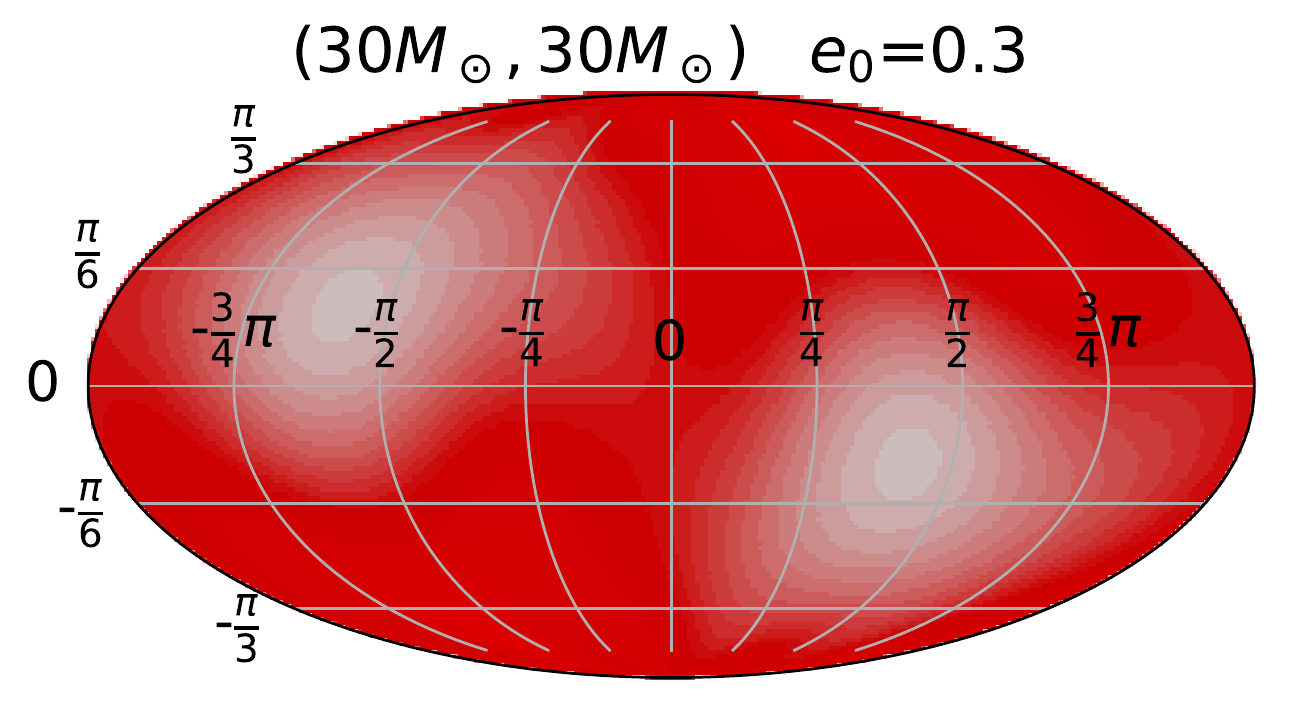}
	}
	\centerline{
		\includegraphics[width=\wid\textwidth]{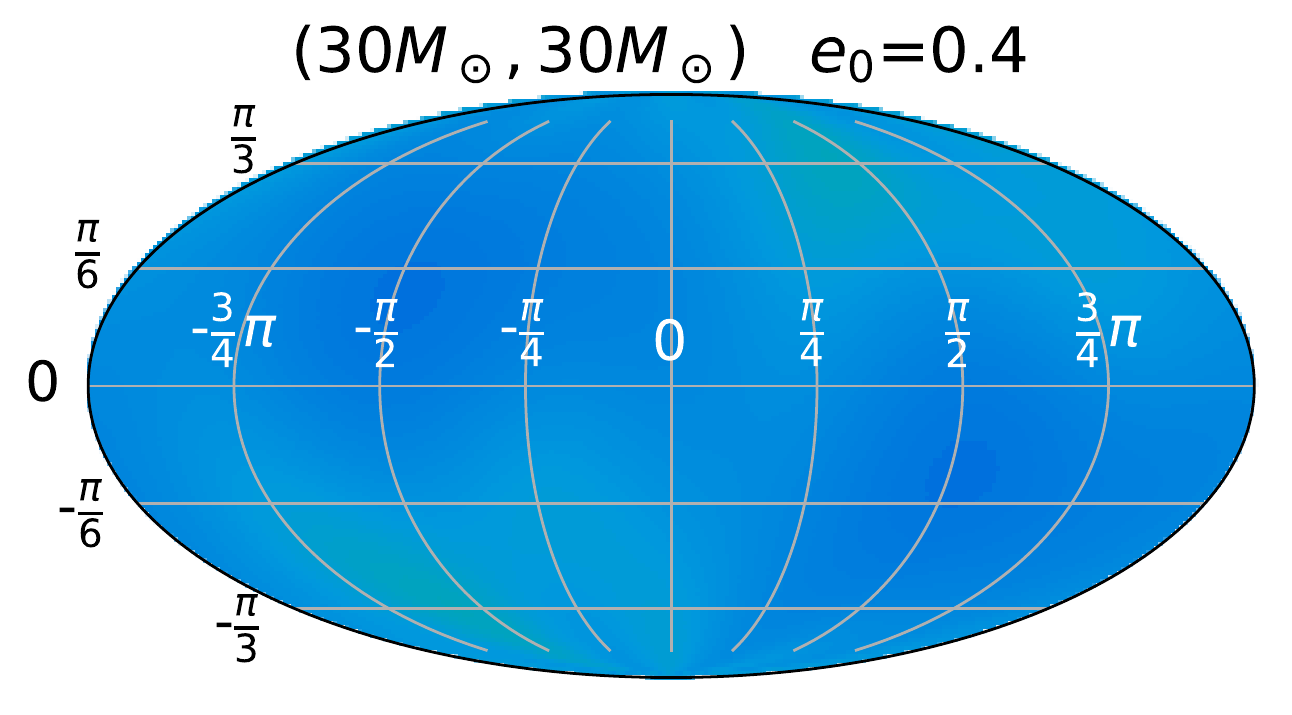}
		\includegraphics[width=\wid\textwidth]{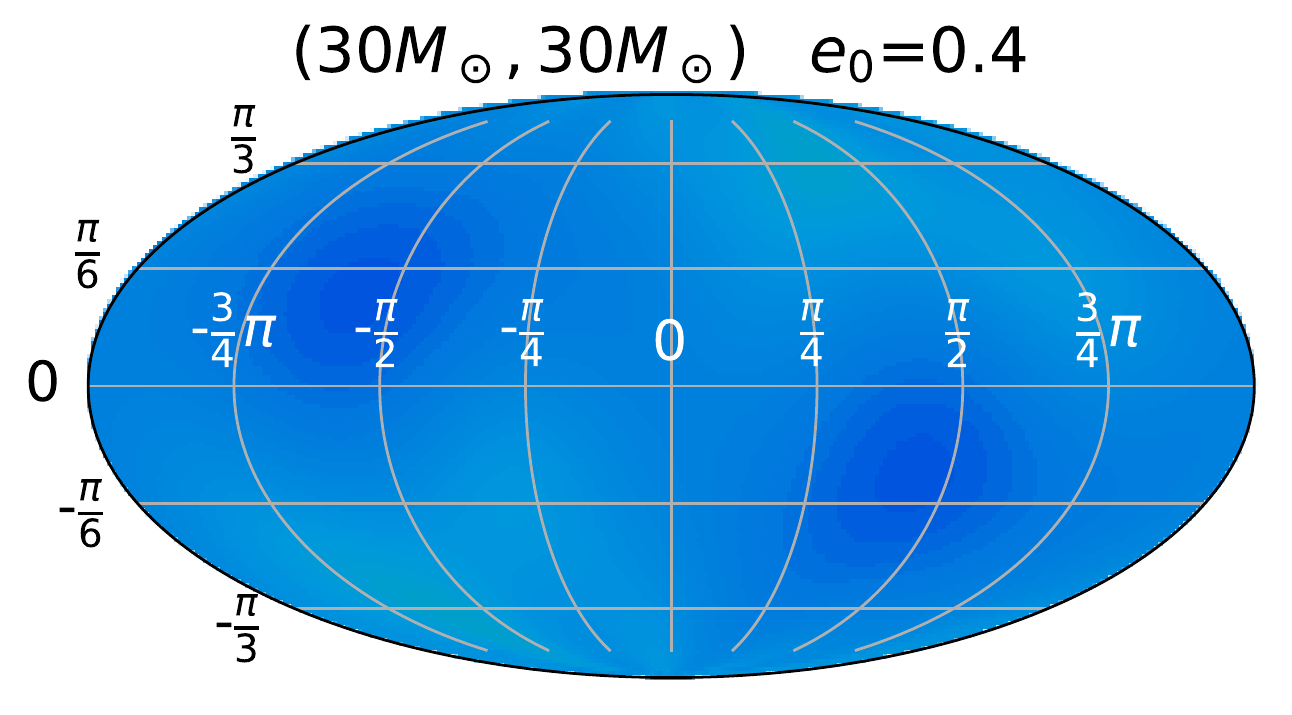}
		\includegraphics[width=\wid\textwidth]{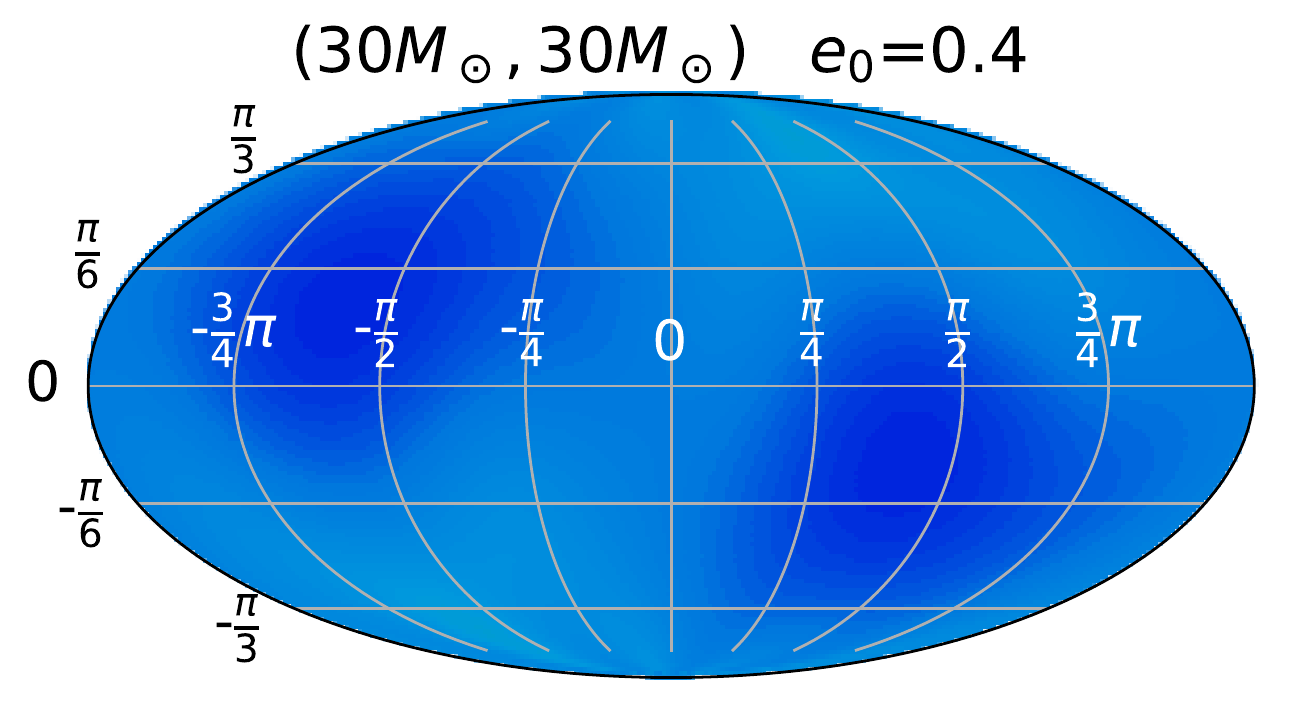}
	}
	\centerline{
		\includegraphics[width=\wid\textwidth]{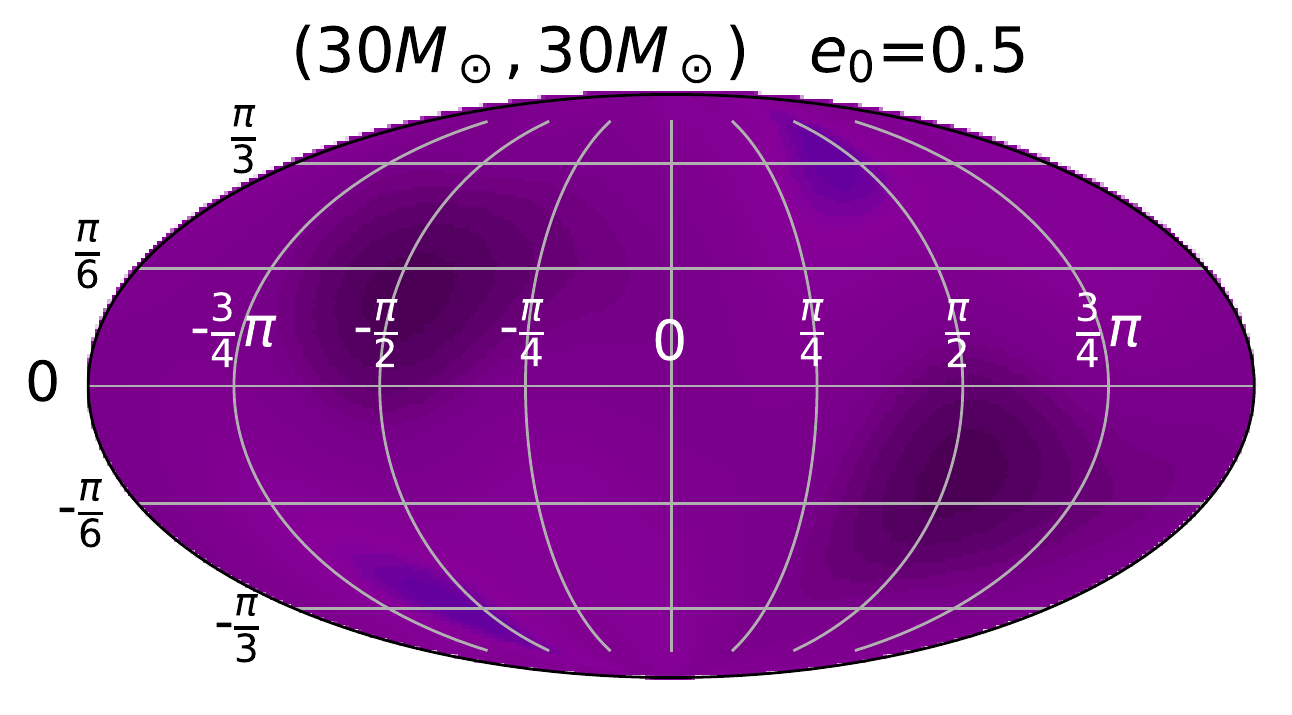}
		\includegraphics[width=\wid\textwidth]{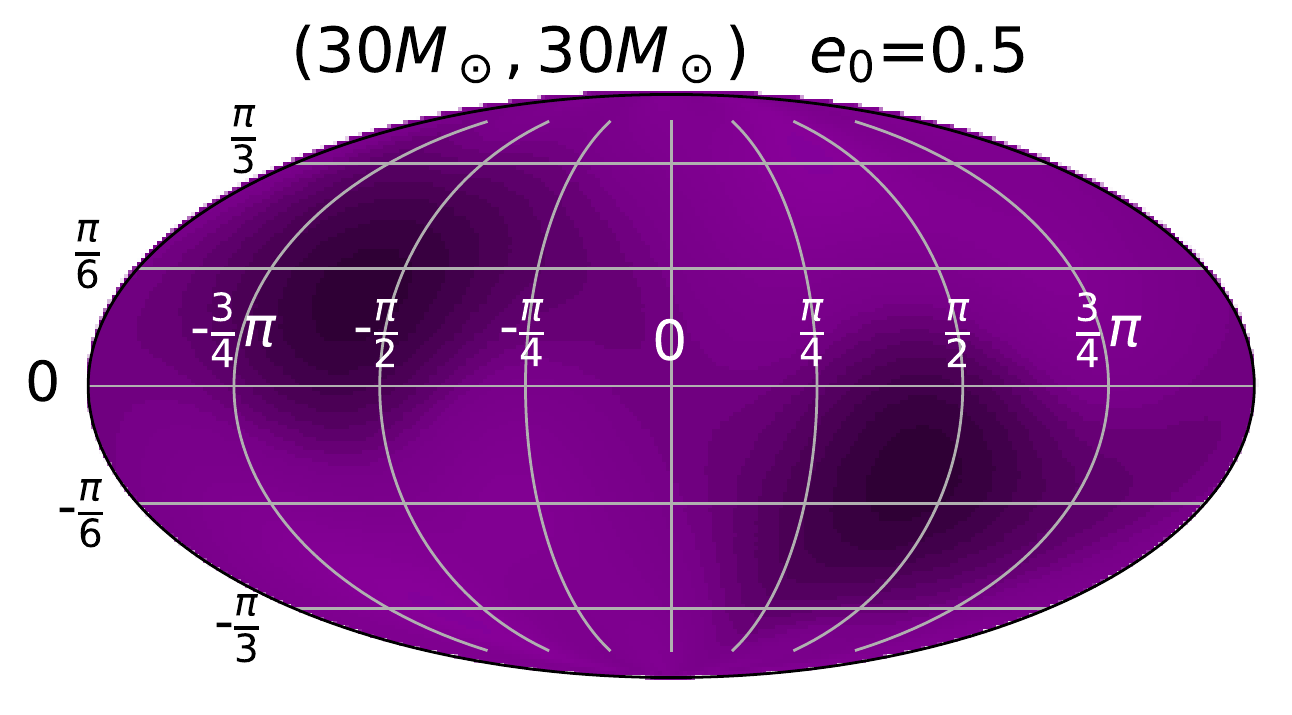}
		\includegraphics[width=\wid\textwidth]{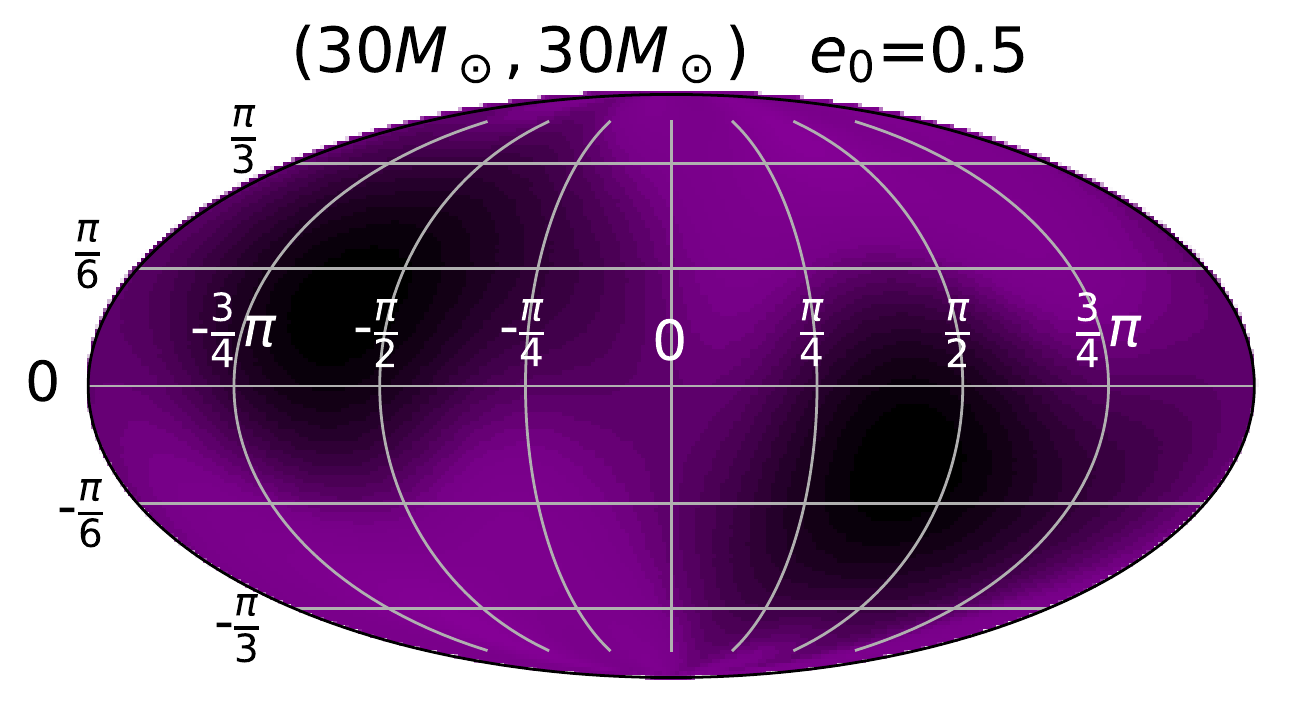}
	}
	\caption{As Figure~\ref{fig:equal_mass_LIGO_type}, but now assuming that the PSD of each detector corresponds to the target sensitivity of Cosmic Explorer. These results are produced assuming binary black hole mergers with component masses \((30\msun,\,30\msun)\). As before, we have used the Mollweide projection, averaged over polarization angles, and set the binary inclination angle to \(i=\pi/4\). See Appendix~\ref{app:ap1} for results assuming  \(i=0\).} 
	\label{fig:ce_equal_mass}
\end{figure*}

\begin{figure*}[p]
	\centerline{
		\includegraphics[width=\textwidth]{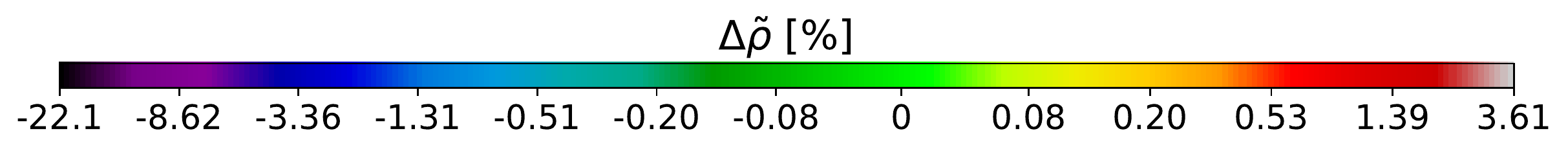}
	}
	\centerline{
		\includegraphics[width=\wid\textwidth]{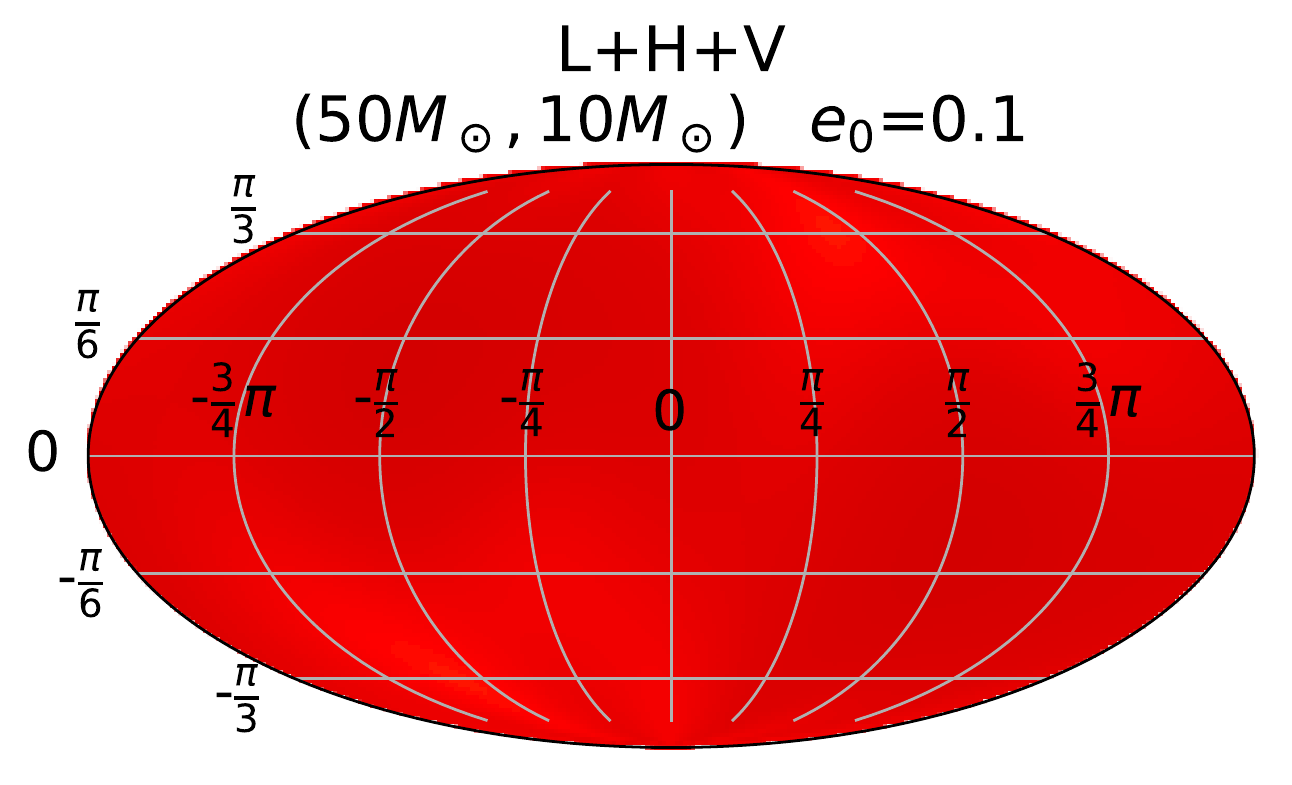}
		\includegraphics[width=\wid\textwidth]{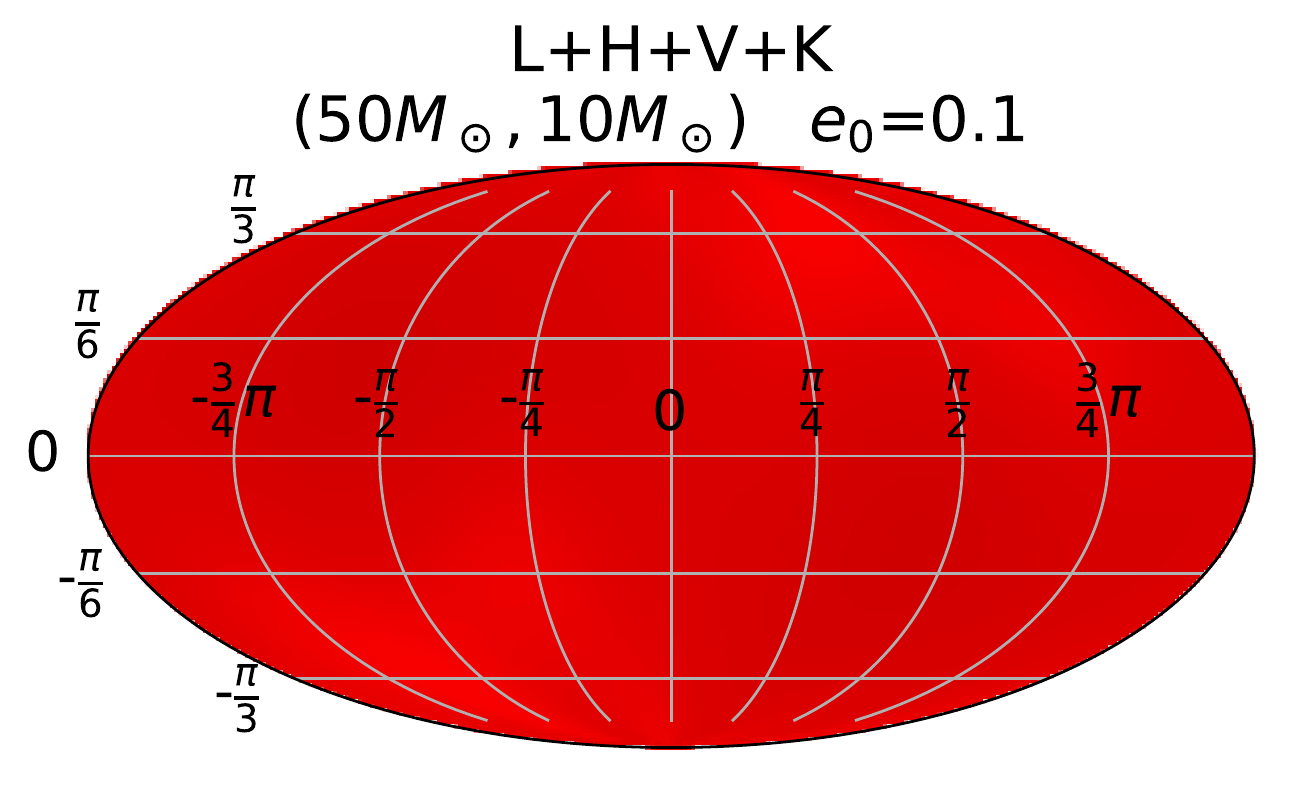}
		\includegraphics[width=\wid\textwidth]{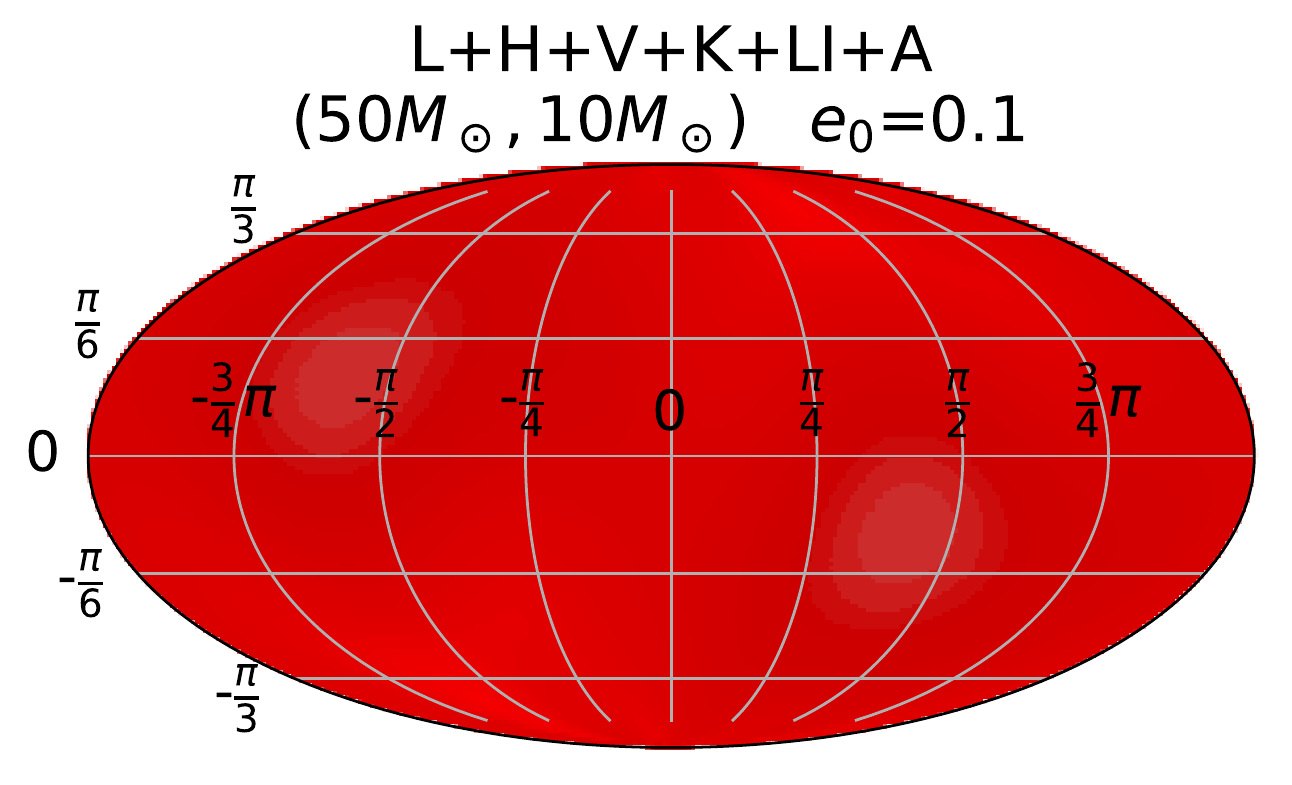}
	}
	\centerline{
		\includegraphics[width=\wid\textwidth]{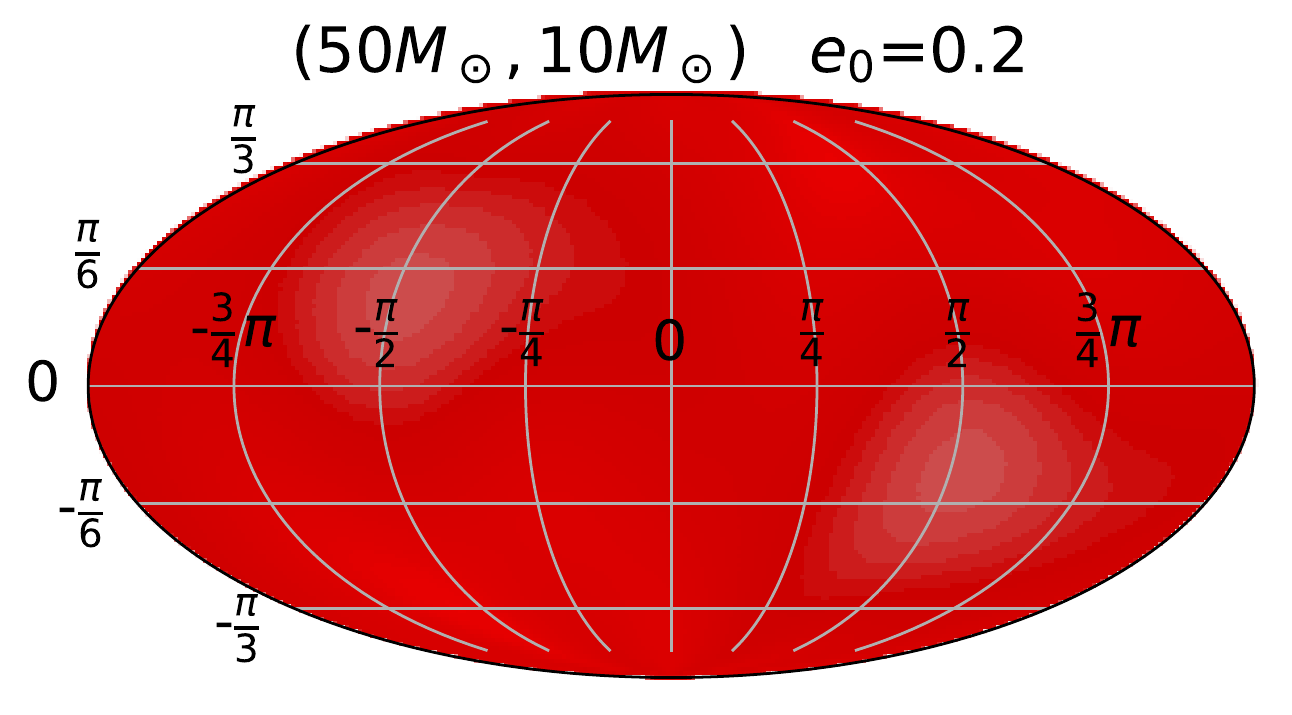}
		\includegraphics[width=\wid\textwidth]{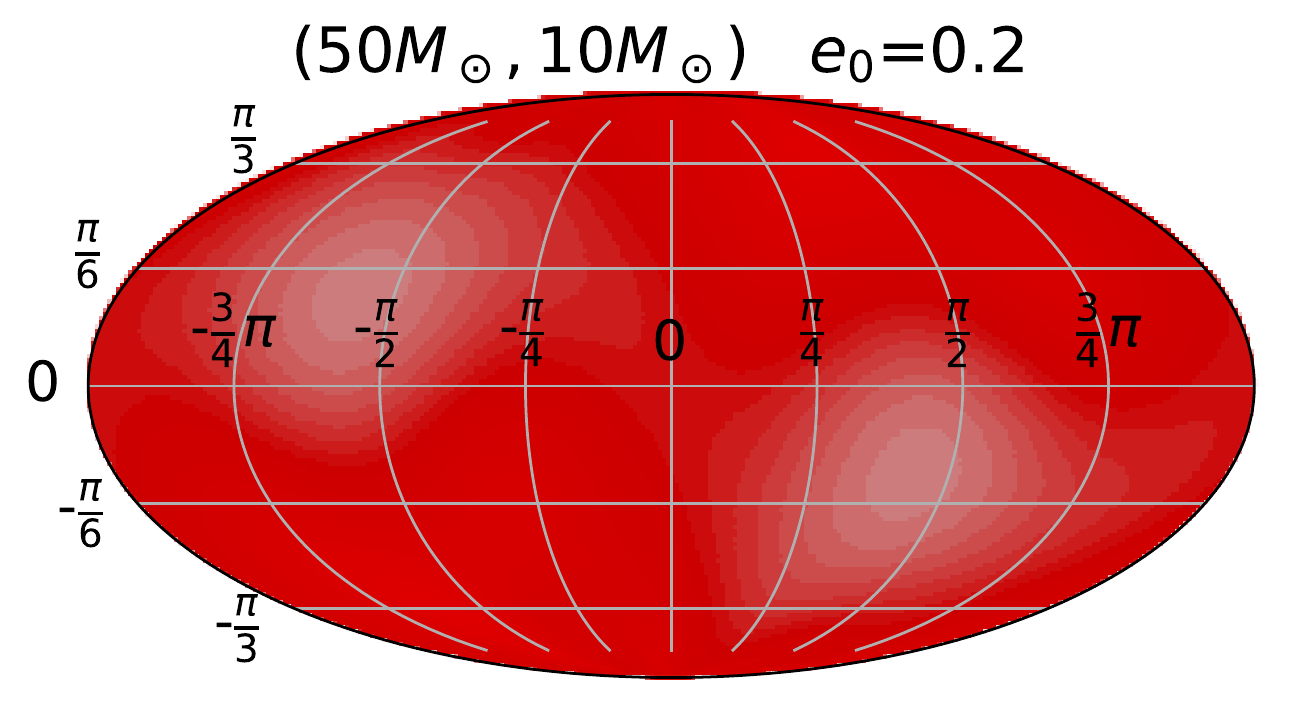}
		\includegraphics[width=\wid\textwidth]{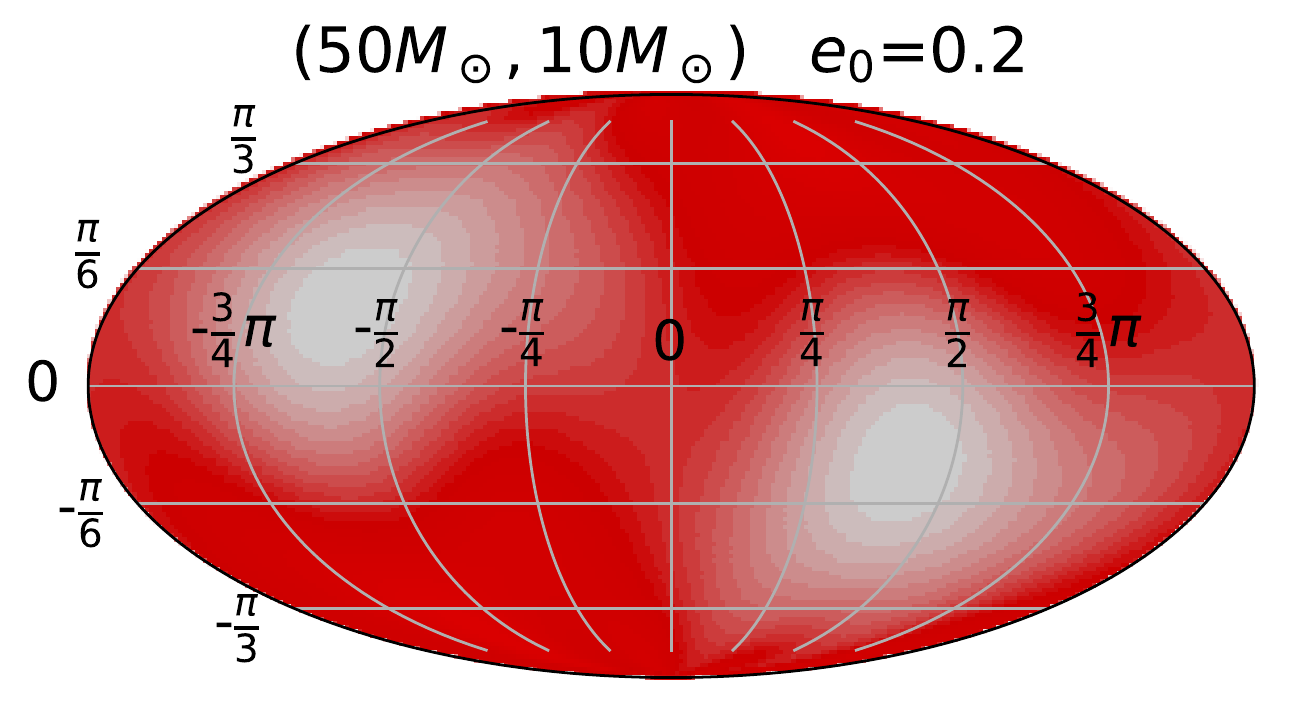}
	}
	\centerline{
		\includegraphics[width=\wid\textwidth]{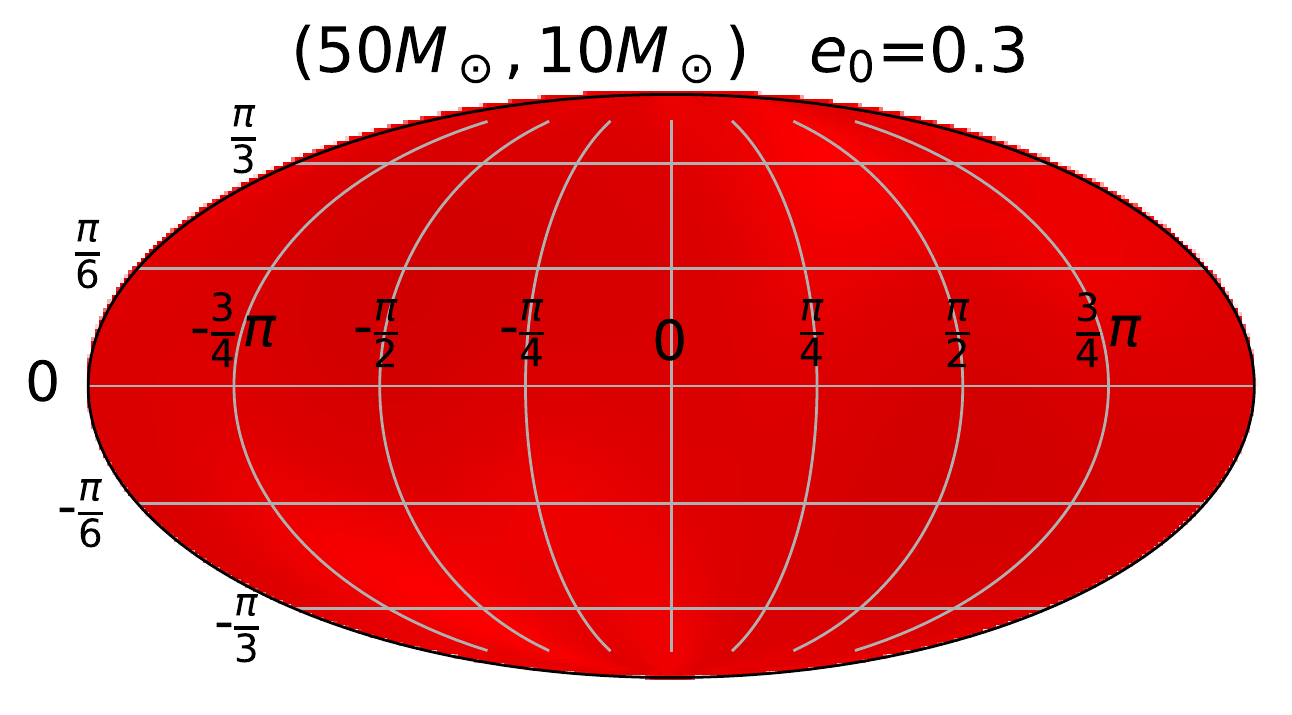}
		\includegraphics[width=\wid\textwidth]{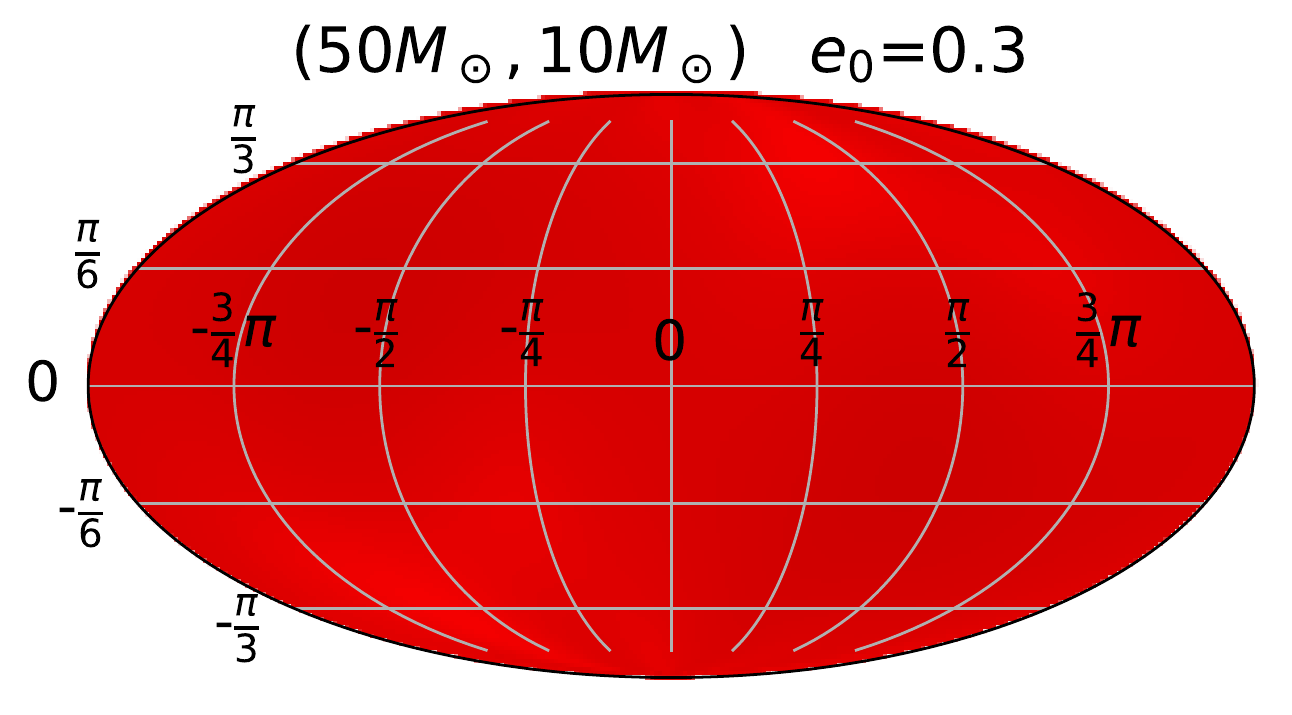}
		\includegraphics[width=\wid\textwidth]{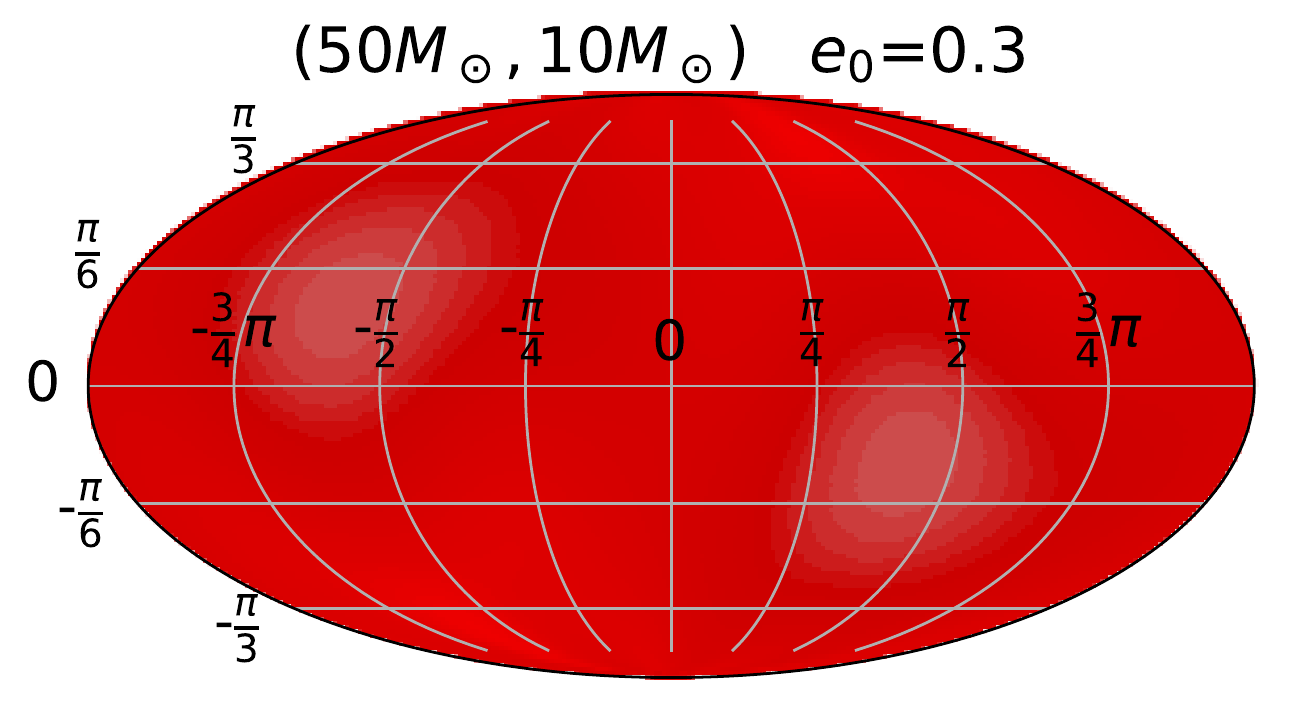}
	}
	\centerline{
		\includegraphics[width=\wid\textwidth]{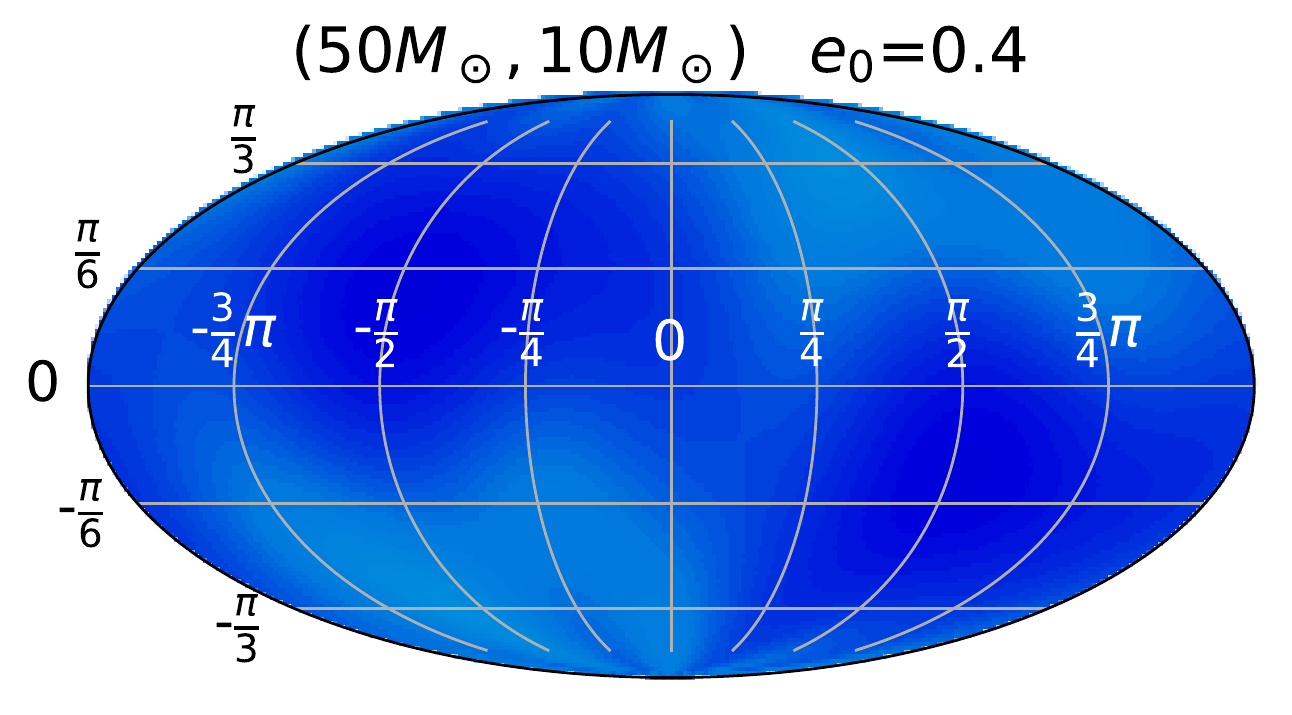}
		\includegraphics[width=\wid\textwidth]{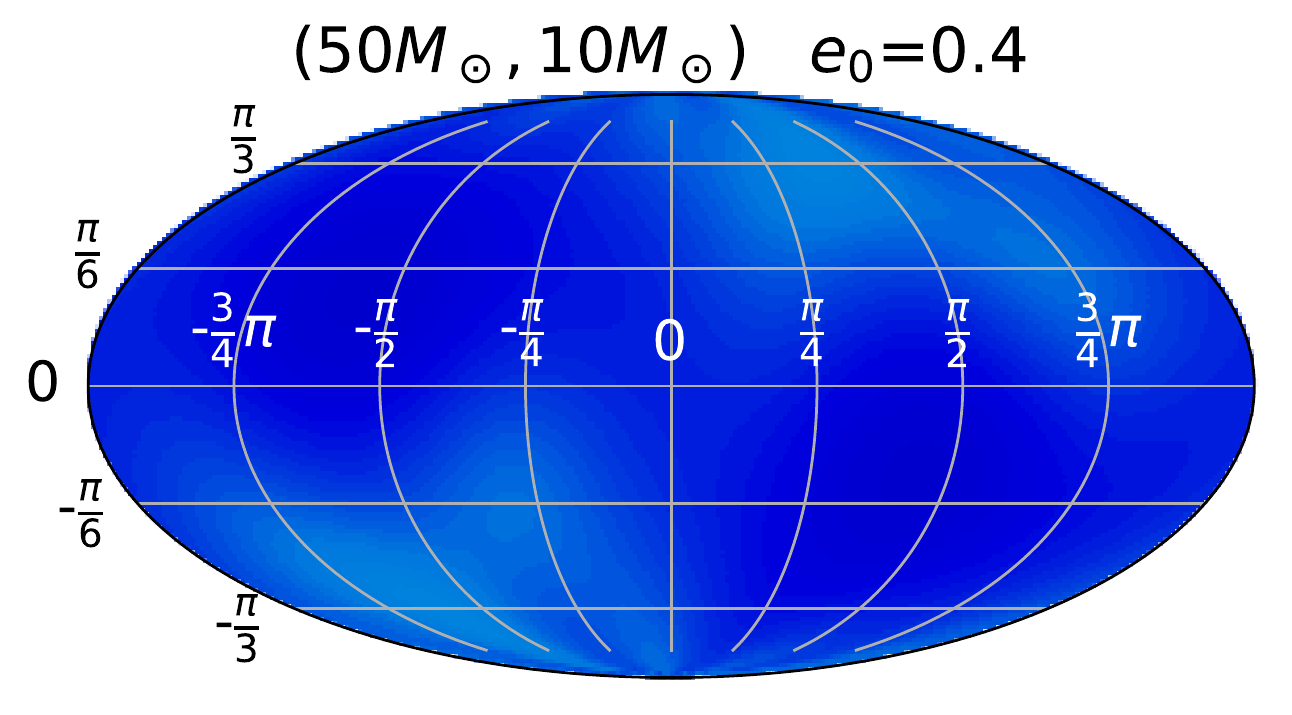}
		\includegraphics[width=\wid\textwidth]{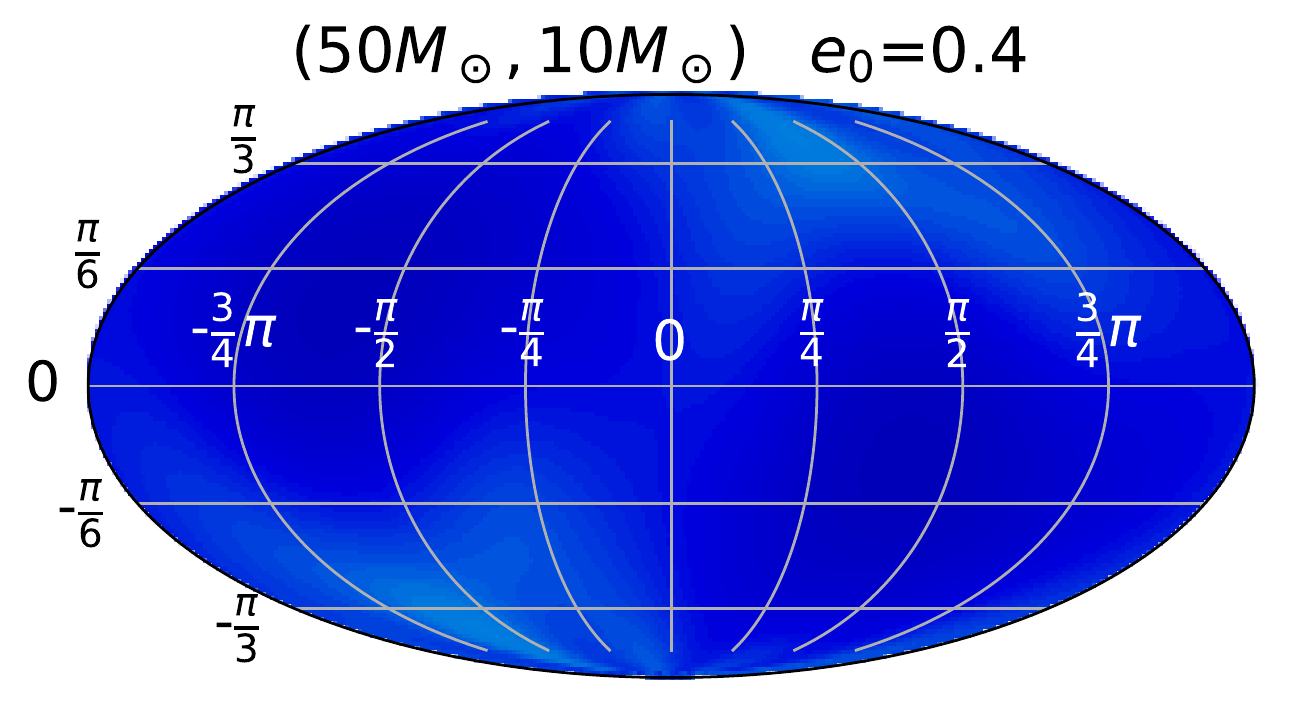}
	}
	\centerline{
		\includegraphics[width=\wid\textwidth]{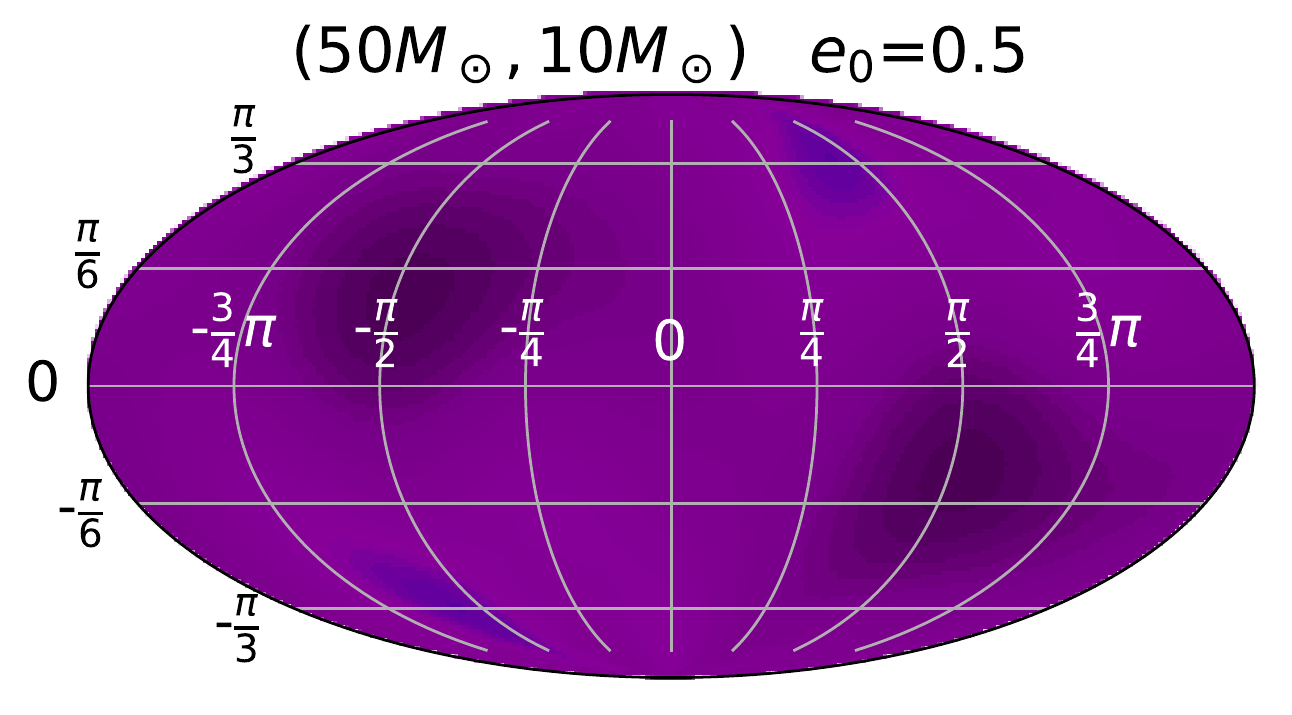}
		\includegraphics[width=\wid\textwidth]{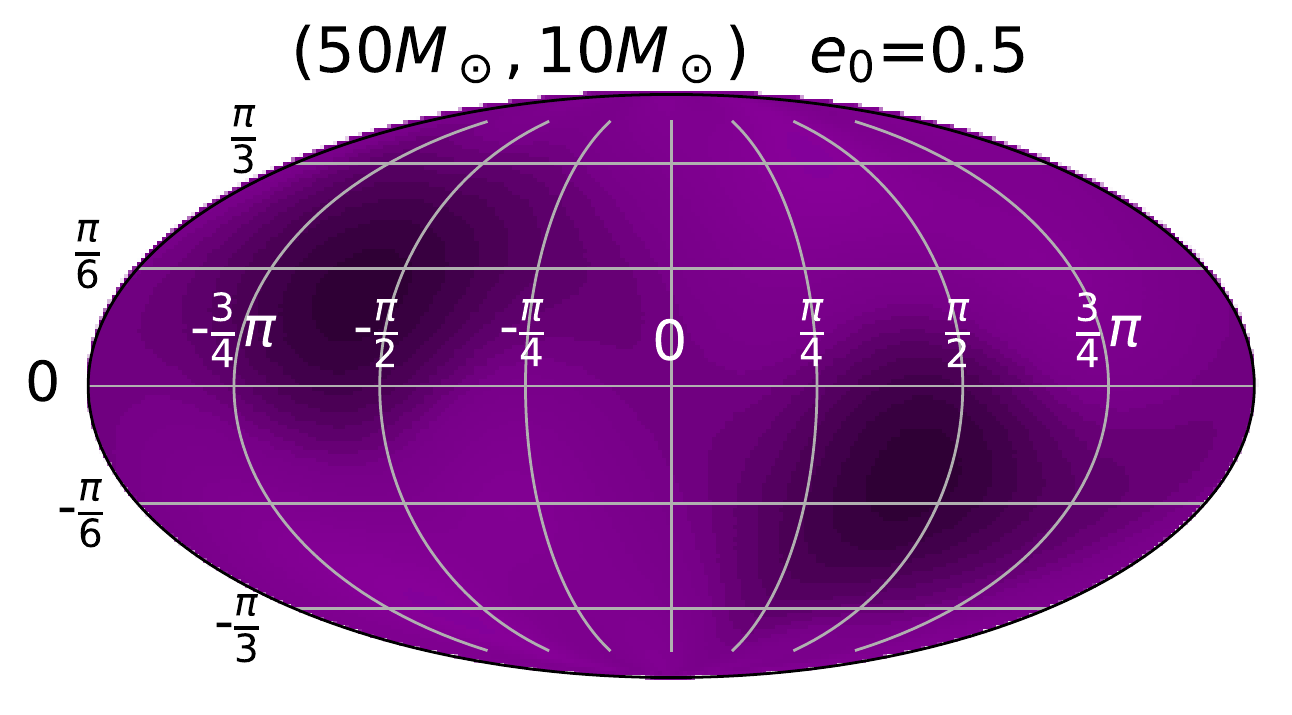}
		\includegraphics[width=\wid\textwidth]{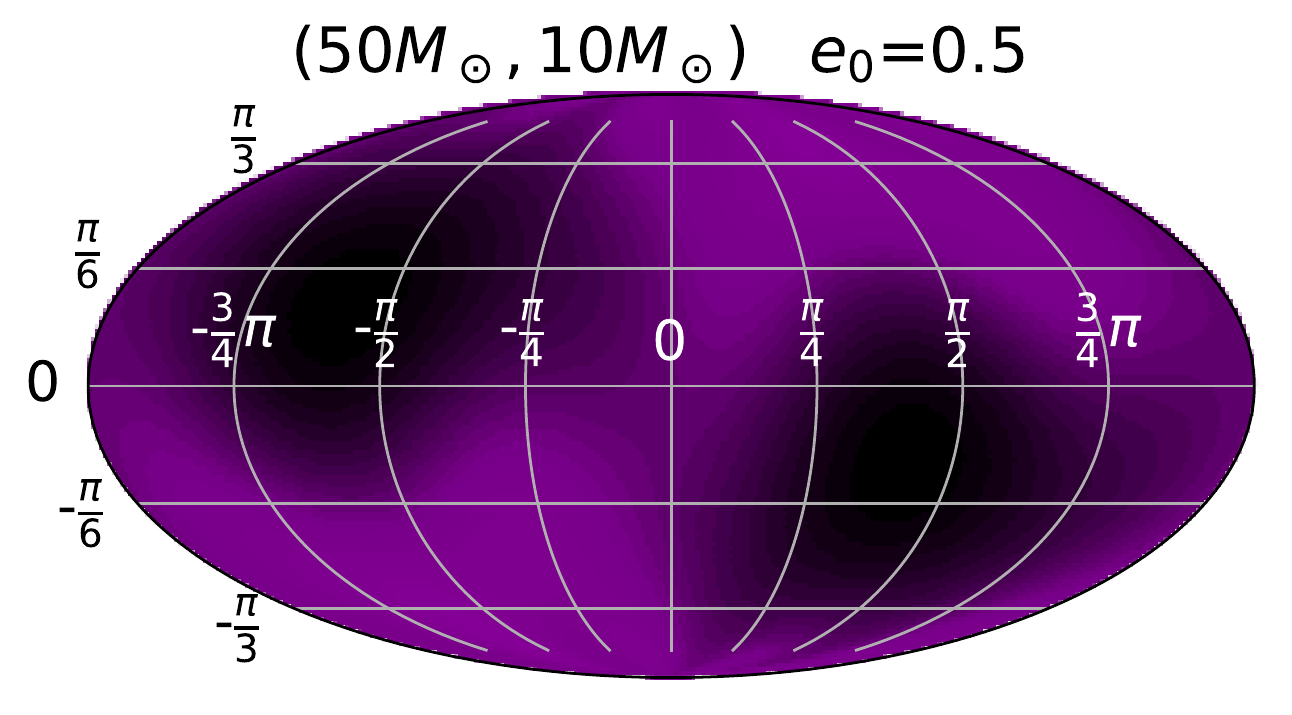}
	}
	\caption{As Figure~\ref{fig:ce_equal_mass} but now for component masses \((50\msun,\,10\msun)\).} 
	\label{fig:ce_five_mass}
\end{figure*}

We also provide results to visualize the SNR sky distribution for a detector network of Cosmic 
Explorers for two cases that describe BBH mergers with 
component masses \(m_{\{1,\,2\}}= \{\left(30\msun, 30\msun\right), \left(50\msun, 10\msun\right)\}\), 
with eccentricities \(e_0\leq0.5\) measured at \(f_{\textrm{GW}}=9\textrm{Hz}\). 
The SNR distributions are computed by setting \(f_0=11\,\textrm{Hz}\) and \(f_1=4096\,\textrm{Hz}\) in 
Eq.~\eqref{eqn:SNRcomponent}. Relevant findings include:

\begin{itemize}
\item Figure~\ref{fig:ce_equal_mass} shows that, for equal mass-ratio BBH mergers, 
eccentric and quasi-circular systems have comparable SNR sky distributions for \(e_0\leq0.4\). 
We notice that in some cold spots, quasi-circular systems are \(\sim34\%\) louder than 
\(e_0=0.5\) BBH systems. Compared to similar systems presented in 
Figure~\ref{fig:equal_mass_LIGO_type} for second generation detector networks, 
we notice that Cosmic Explorer will be able to clearly tell apart these two types of 
populations through their SNR distributions for eccentricities \(e_0\geq0.4\).
\item Figure~\ref{fig:ce_five_mass} presents SNR sky distributions for BBH mergers 
with masses \((50\msun,\,10\msun)\). These results show that larger mass-ratios tend to 
attenuate SNR suppression for the most eccentric systems. Notice that while for 
equal mass-ratio systems with \(e_0\sim0.5\) the SNR drops by \(\sim 36\%\), in this case the 
SNR is \(\sim 24\%\) the SNR of quasi-circular mergers with identical total mass and 
mass-ratio. This finding is also consistent with results presented for second generation 
detector networks.
\end{itemize}

In summary, we find that in the context of the Cosmic Explorer, the SNR of eccentric and quasi-circular 
BBH mergers is comparable for eccentricities \(e_0\geq0.4\). As we discuss in the following section, we can 
disentangle this apparent similarity when we look at the rate at which SNR is accumulated. We also 
find that quasi-circular mergers are significantly louder than eccentric ones for \(e_0\sim0.5\).

\noindent \textbf{Einstein Telescope D} This is a proposed third generation detector, based in 
Europe, that aims to improve the target sensitivity of advanced LIGO-type detectors by an 
order of magnitude. In this study, we assume a geometry in which the detectors arms are 
10km long, L-shaped, and consider the ET-D sensitivity 
curve for all calculations. Information regarding the conceptual design and the 
fundamental physics that may be accomplished with the Einstein Telescope may be found at~\cite{Maggiore_2020JCAP,einstein_telescope}. 

As we did for Cosmic Explorer, we begin this discussion by providing a high level overview 
of the detectability of eccentric BBH mergers with ET-D. It is worth mentioning that given the 
planned sensitivity for ET-D at lower frequencies, we can now produce waveforms from 
\(f_{\textrm{GW}} = 4\,\textrm{Hz}\), which enables us to explore systems with a broader 
range of eccentricities, e.g., \(e_0\leq0.6\). We compute the SNR distributions setting 
\(f_0=5\,\textrm{Hz}\) and \(f_1=4096\,\textrm{Hz}\)  in Eq.~\eqref{eqn:SNRcomponent}. 

The bottom panels in Figure~\ref{fig:snr_time_CE_and_ETD} show that the SNR of eccentric 
and quasi-circular systems is comparable for eccentricities \(e_0\leq0.5\). This is worth 
highlighting in view that eccentric BBH mergers with \(e_0\sim0.5\) had significantly lower 
SNRs than quasi-circular mergers with identical total mass and mass-ratio when observed 
with Cosmic Explorer. In the 
case of ET-D, we find that quasi-circular BBH mergers become louder than eccentric mergers 
for \(e_0\sim0.6\). We may understand this result if we consider that the improved sensitivity of ET-D 
at lower frequencies amplifies the enhancement in the waveform signal, driven by eccentricity corrections, 
thereby compensating for the signal reduction of such highly eccentric systems.

\begin{figure*}[p]
	\centerline{
		\includegraphics[width=\textwidth]{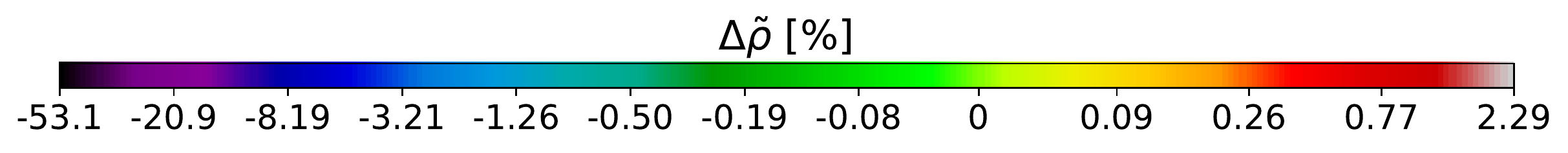}
	}
	\centerline{
		\includegraphics[width=\wid\textwidth]{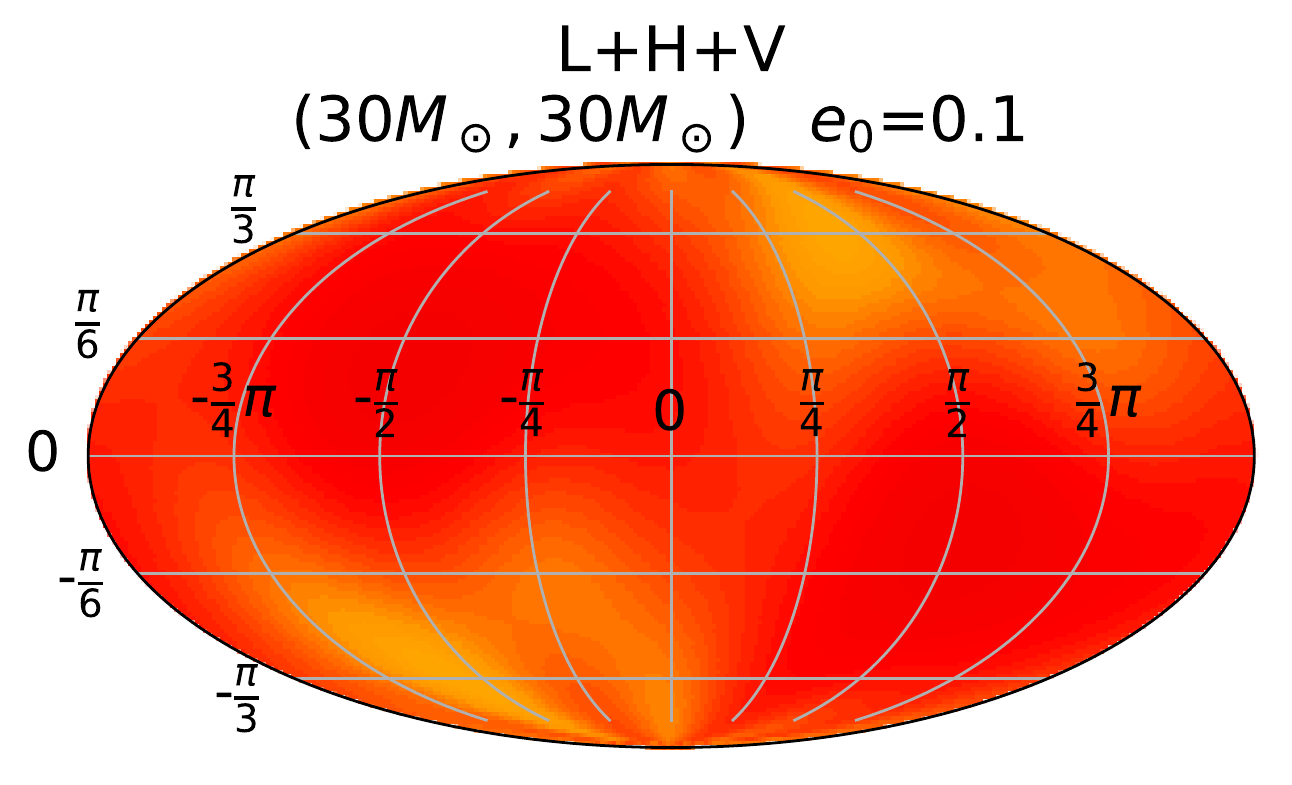}
		\includegraphics[width=\wid\textwidth]{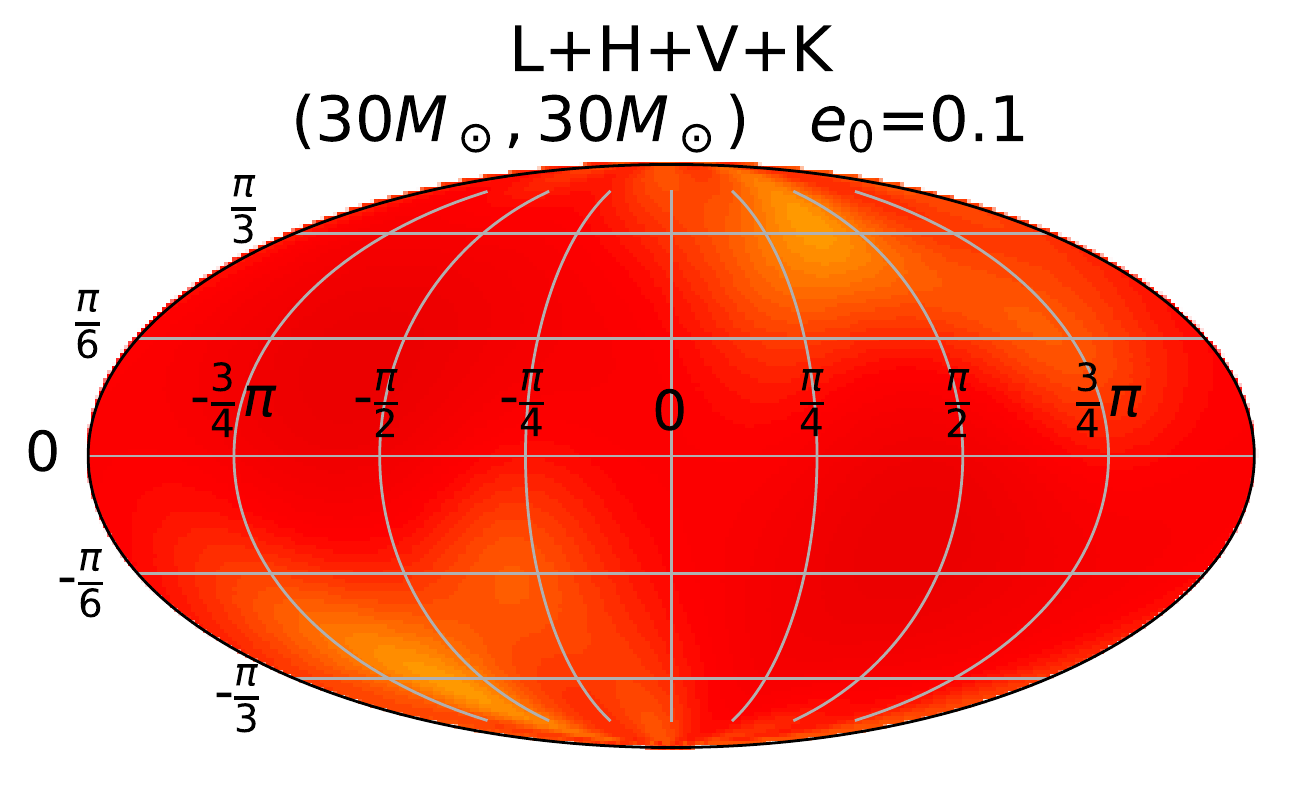}
		\includegraphics[width=\wid\textwidth]{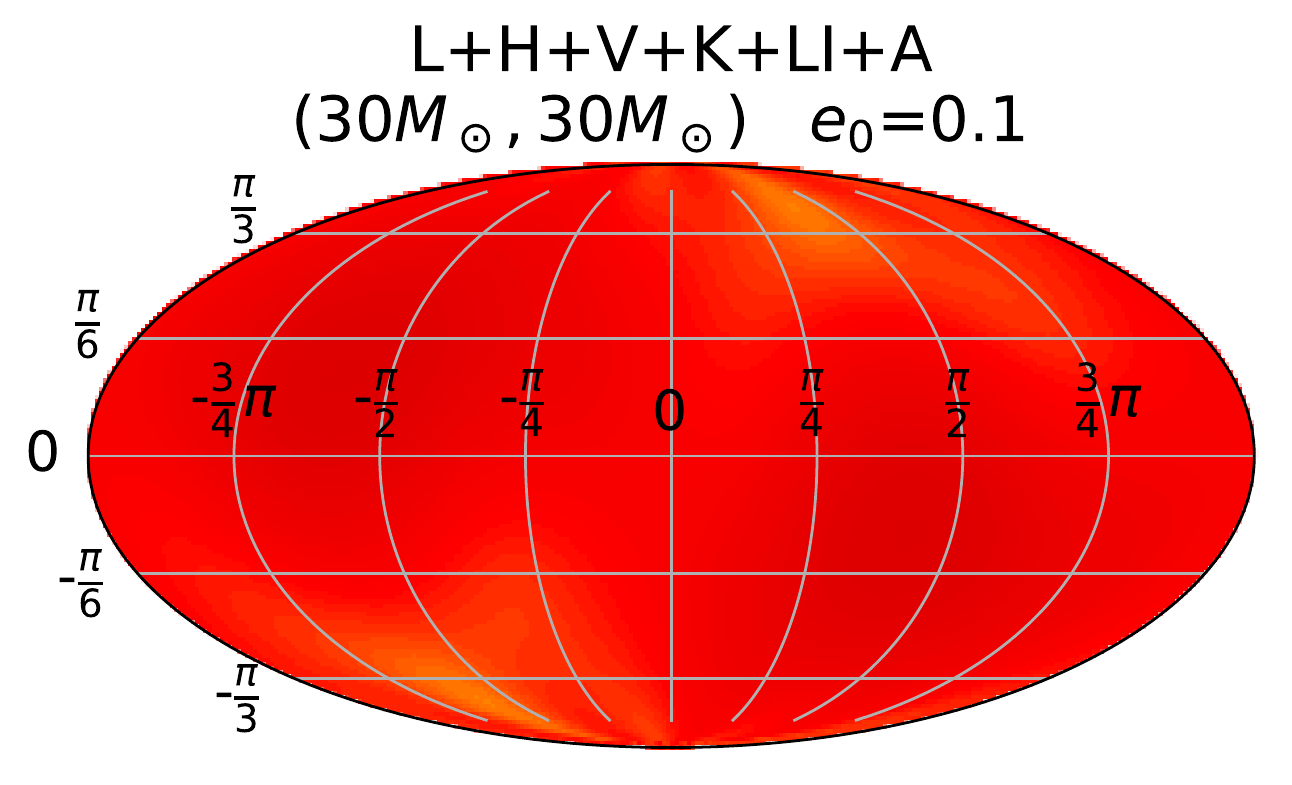}
	}
	\centerline{
		\includegraphics[width=\wid\textwidth]{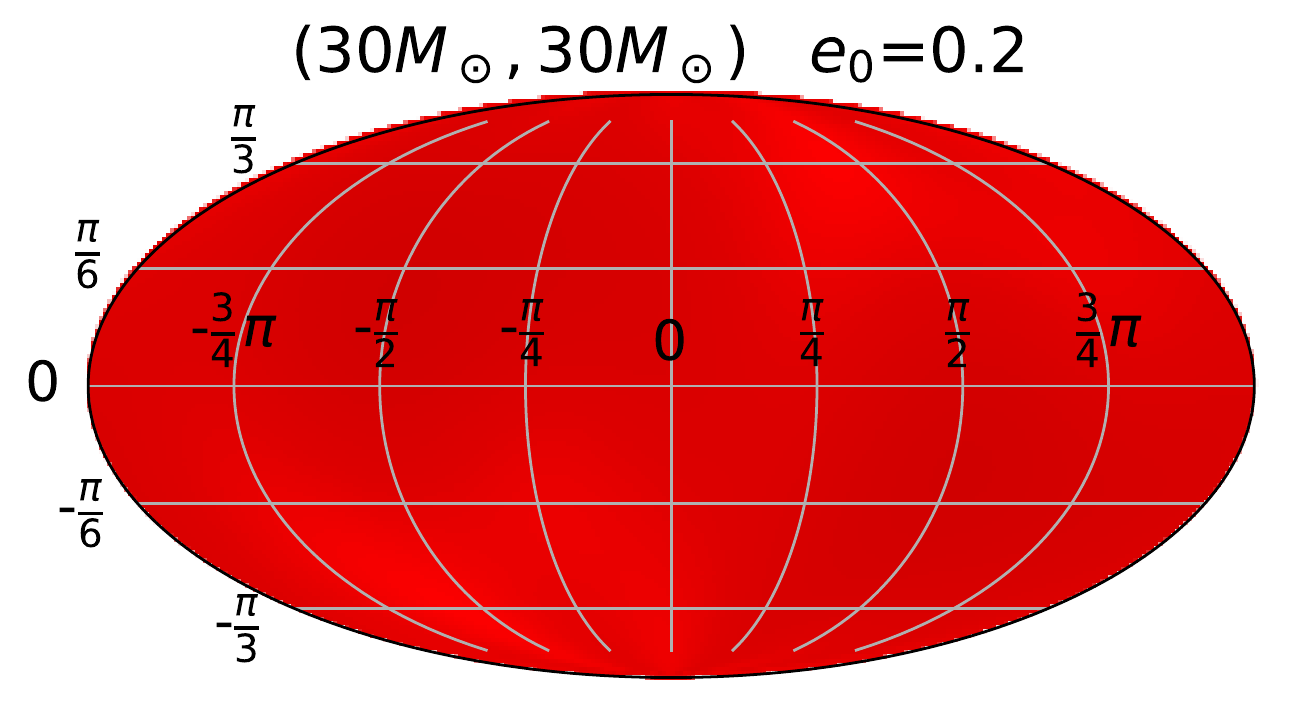}
		\includegraphics[width=\wid\textwidth]{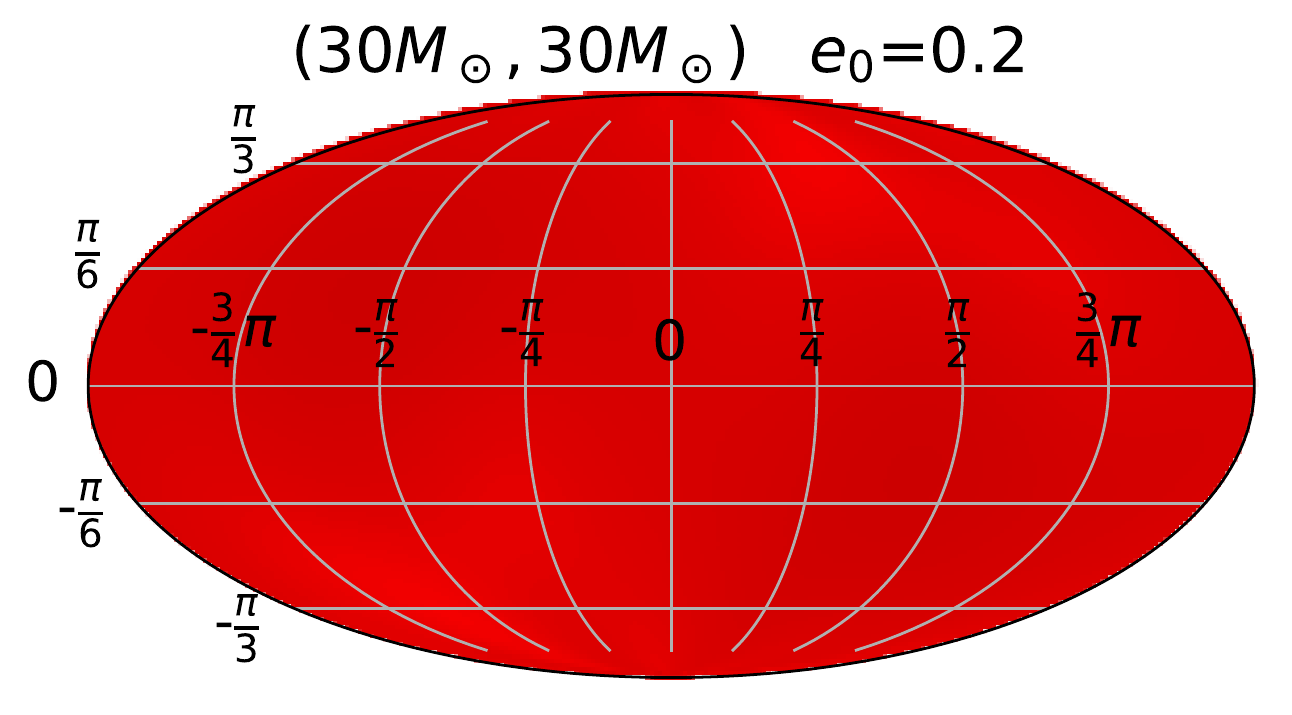}
		\includegraphics[width=\wid\textwidth]{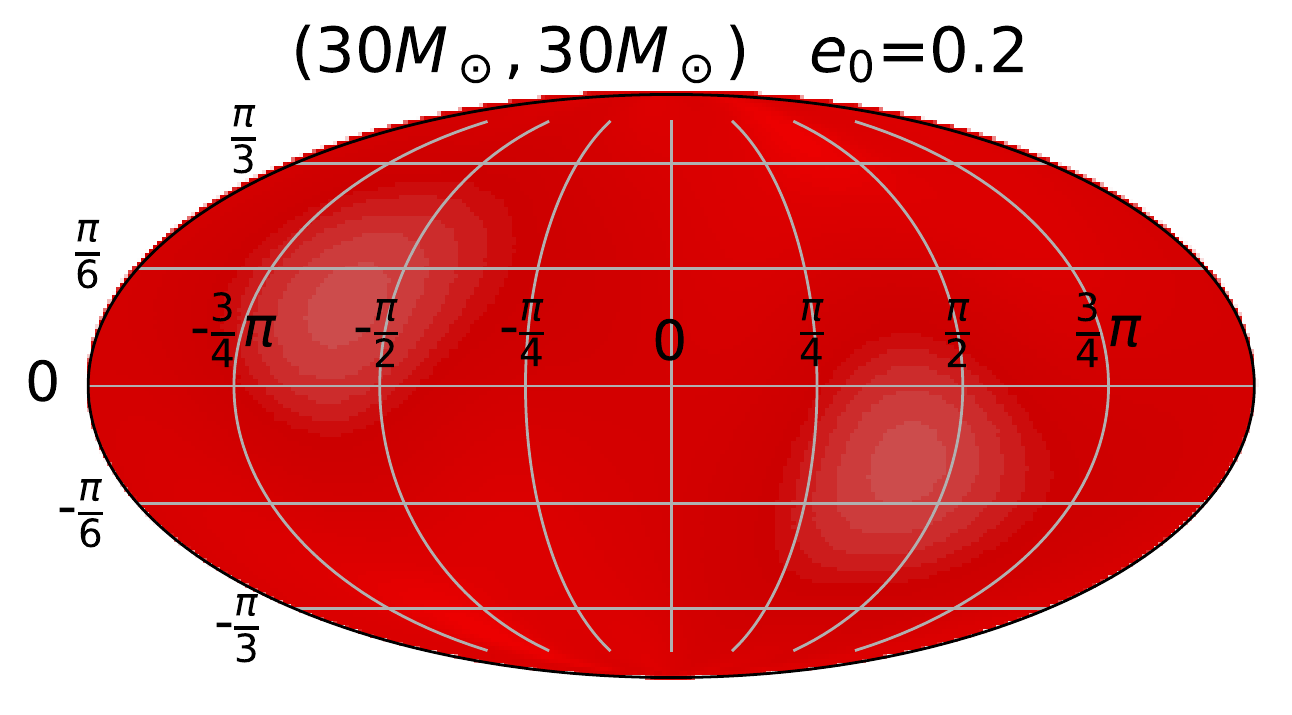}
	}
	\centerline{
		\includegraphics[width=\wid\textwidth]{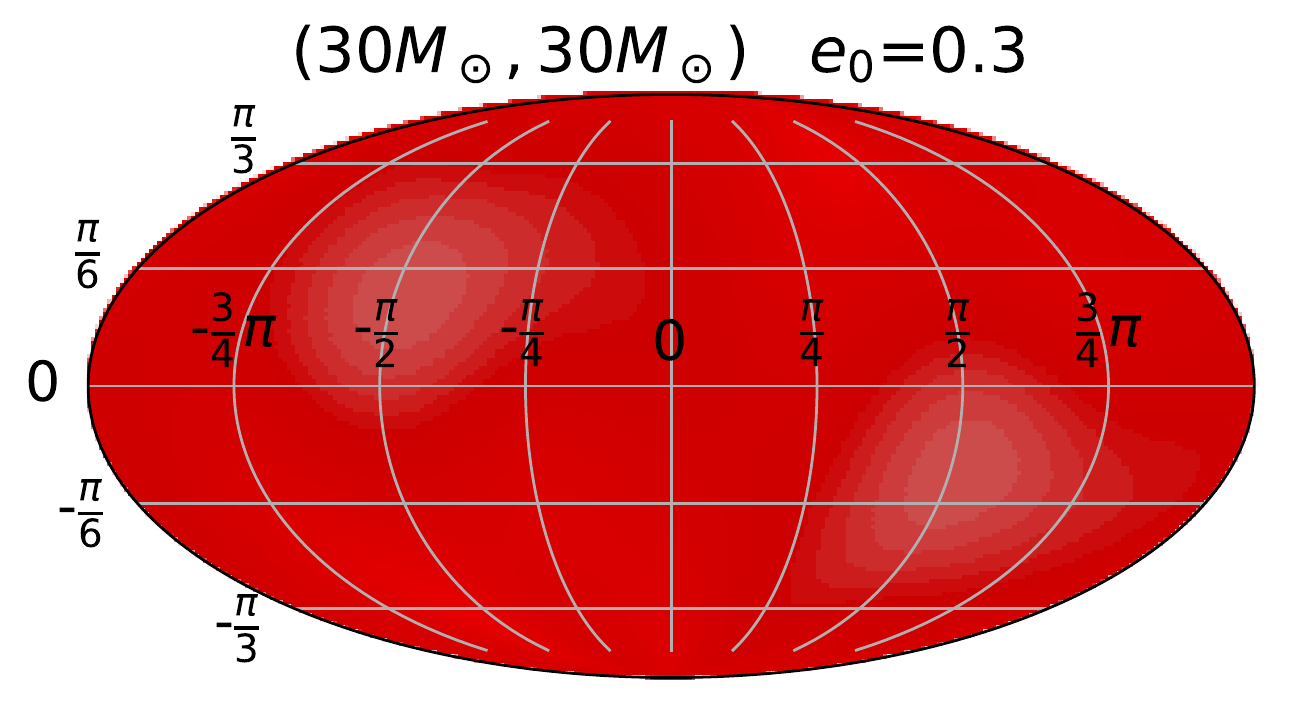}
		\includegraphics[width=\wid\textwidth]{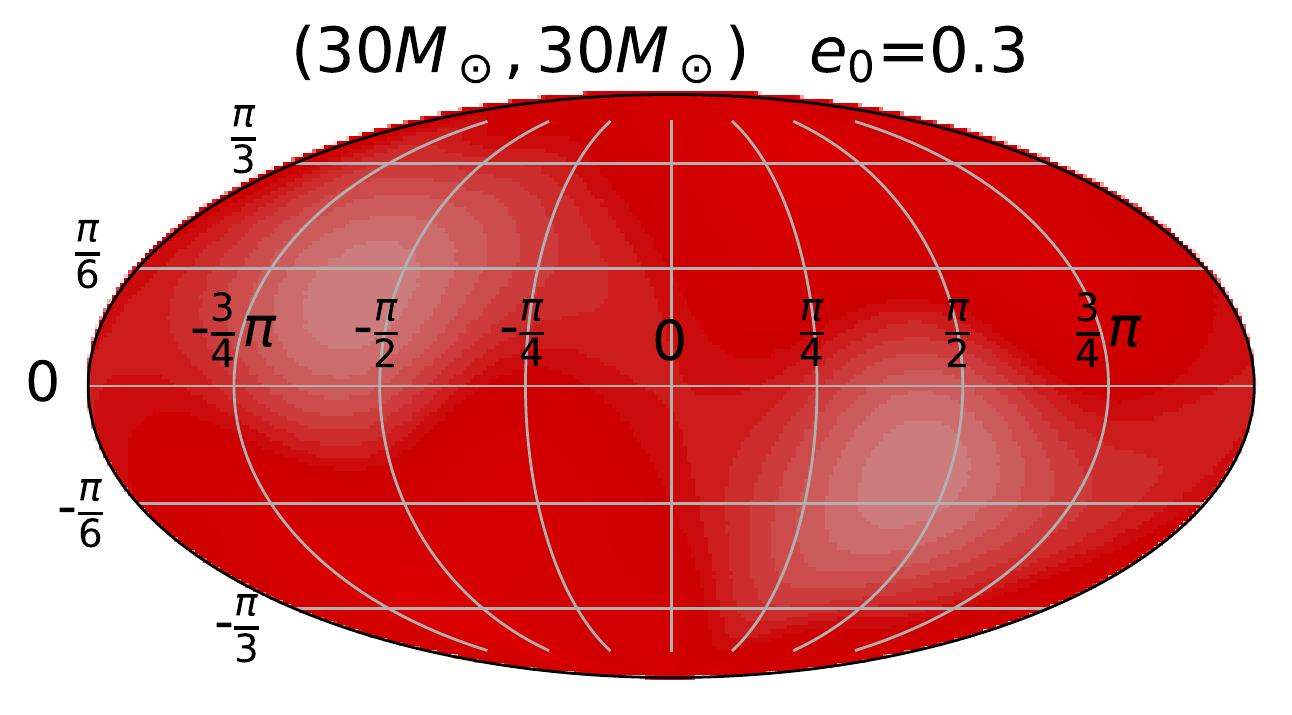}
		\includegraphics[width=\wid\textwidth]{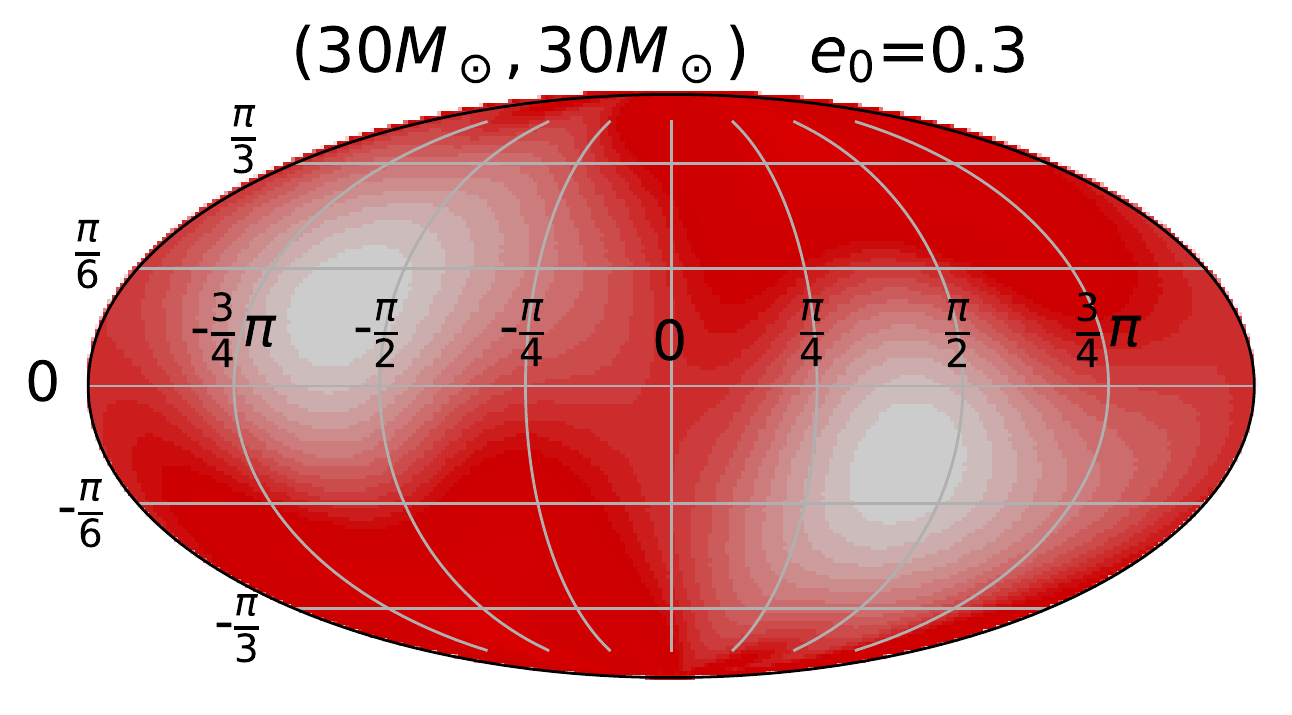}
	}
	\centerline{
		\includegraphics[width=\wid\textwidth]{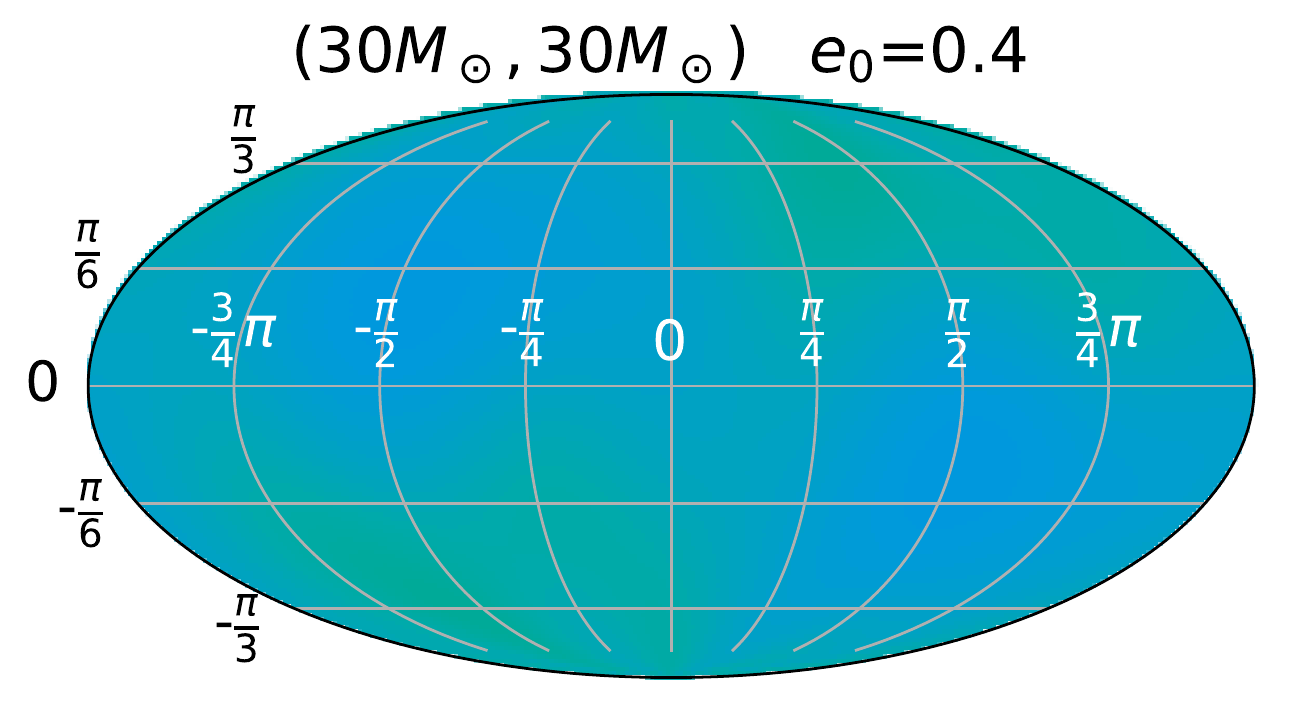}
		\includegraphics[width=\wid\textwidth]{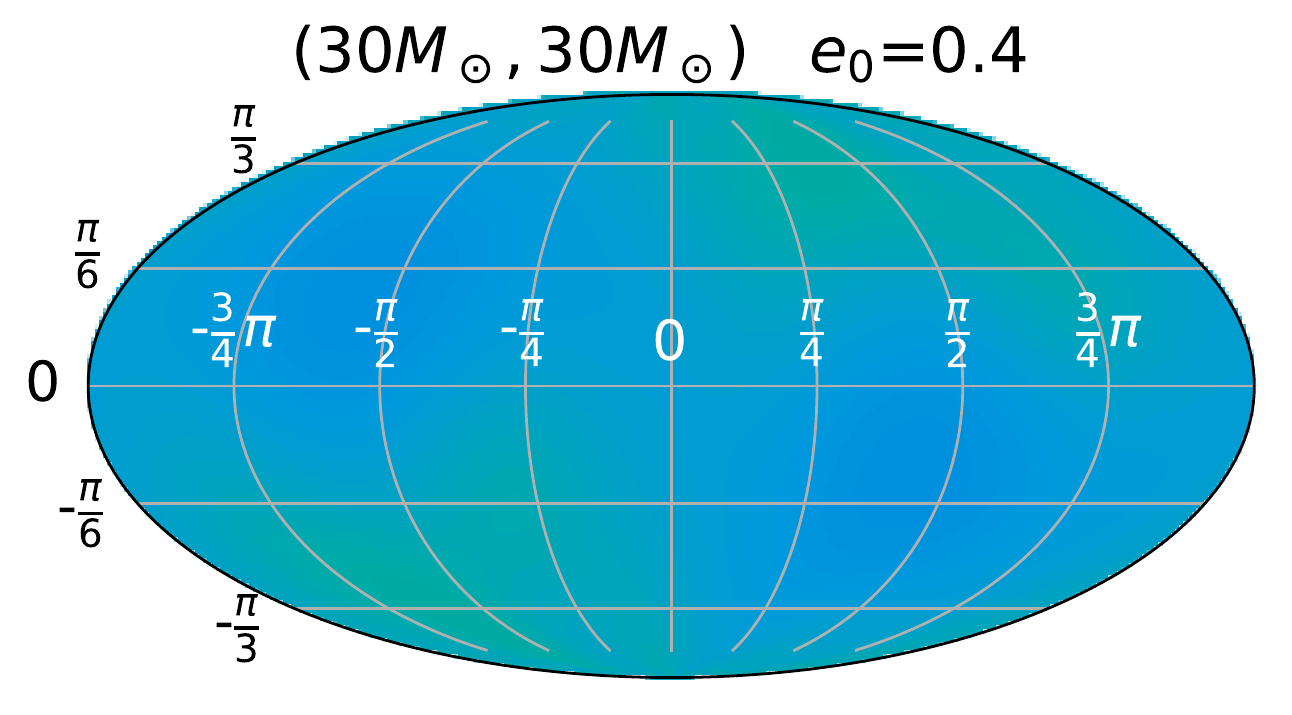}
		\includegraphics[width=\wid\textwidth]{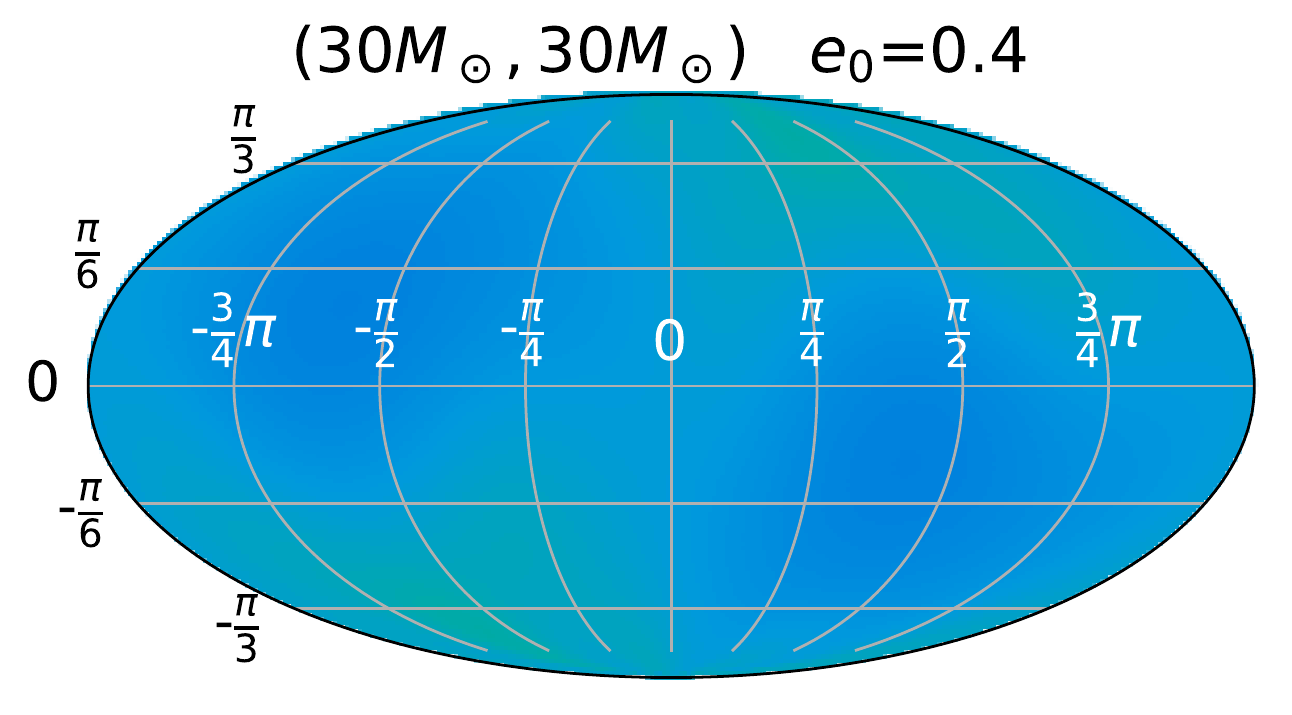}
	}
	\centerline{
		\includegraphics[width=\wid\textwidth]{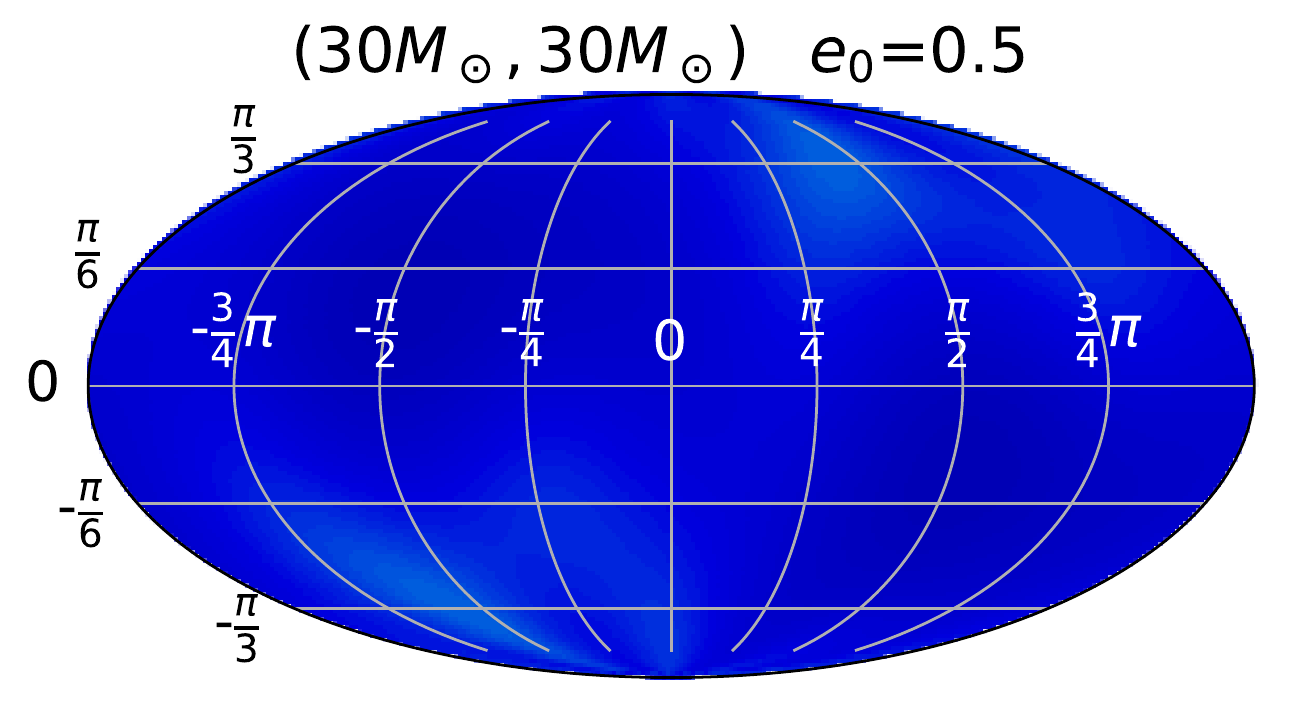}
		\includegraphics[width=\wid\textwidth]{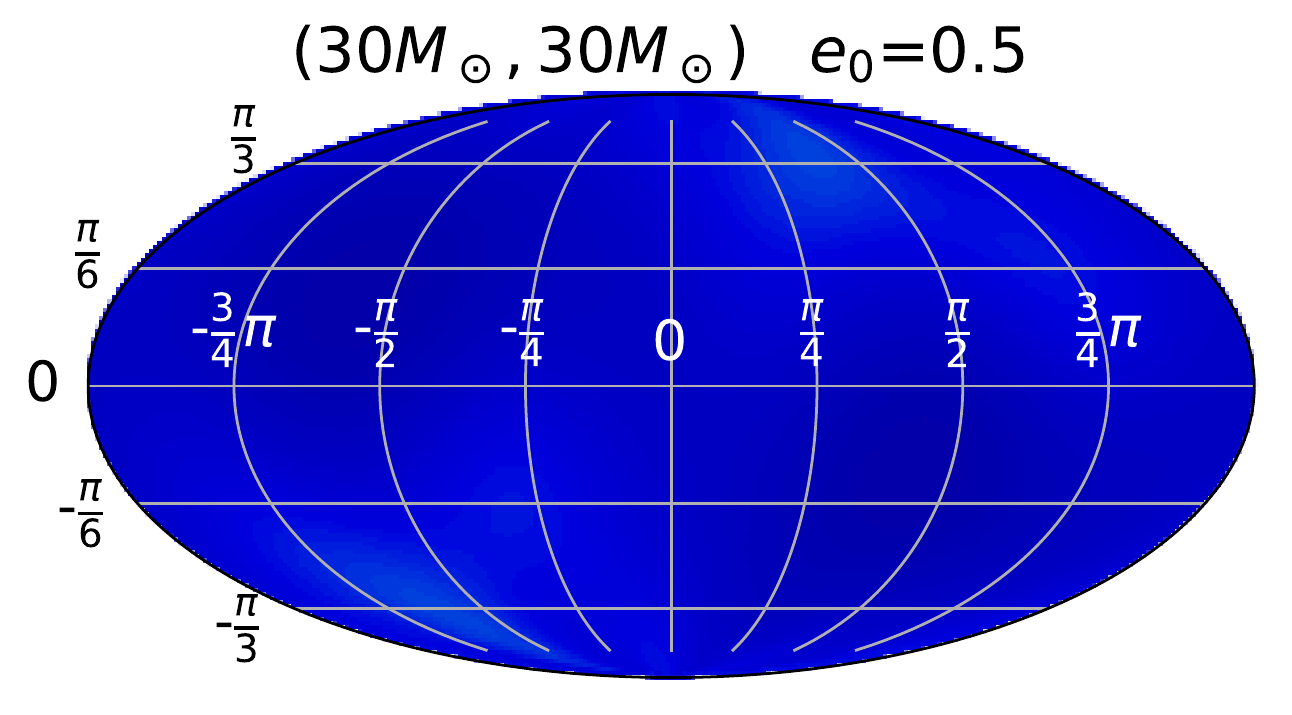}
		\includegraphics[width=\wid\textwidth]{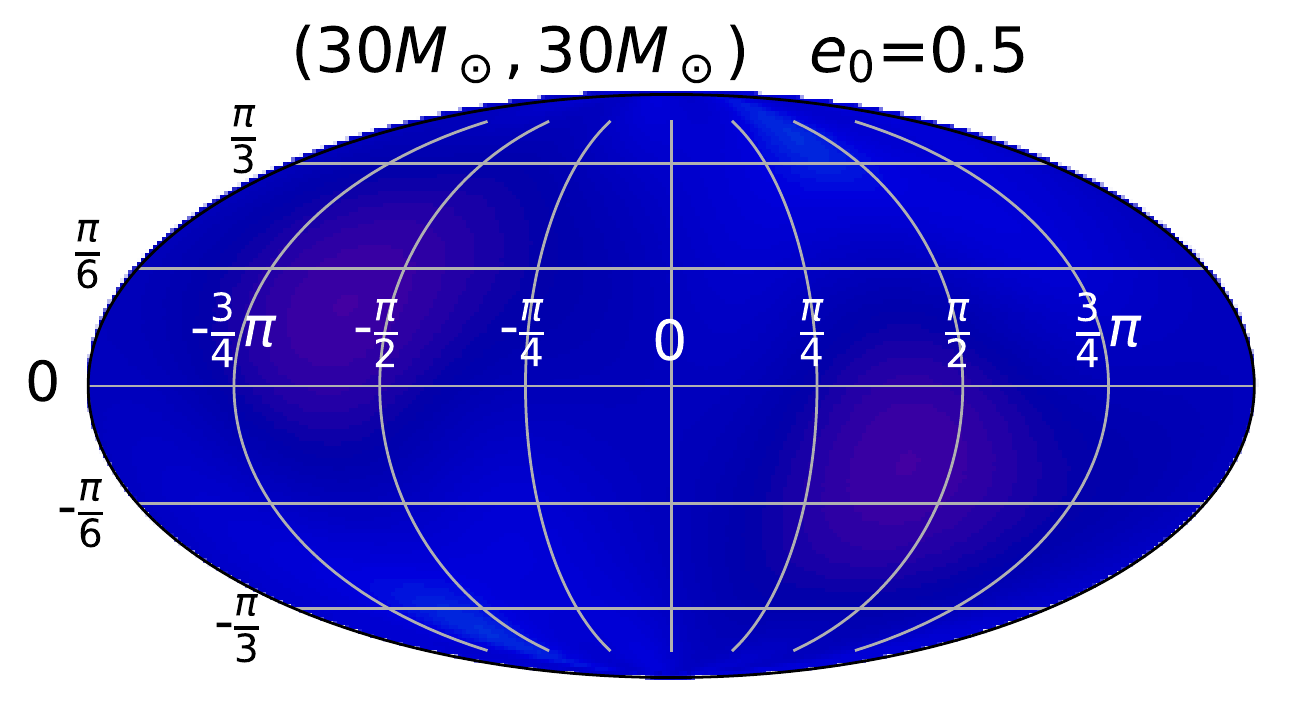}
	}
	\centerline{
		\includegraphics[width=\wid\textwidth]{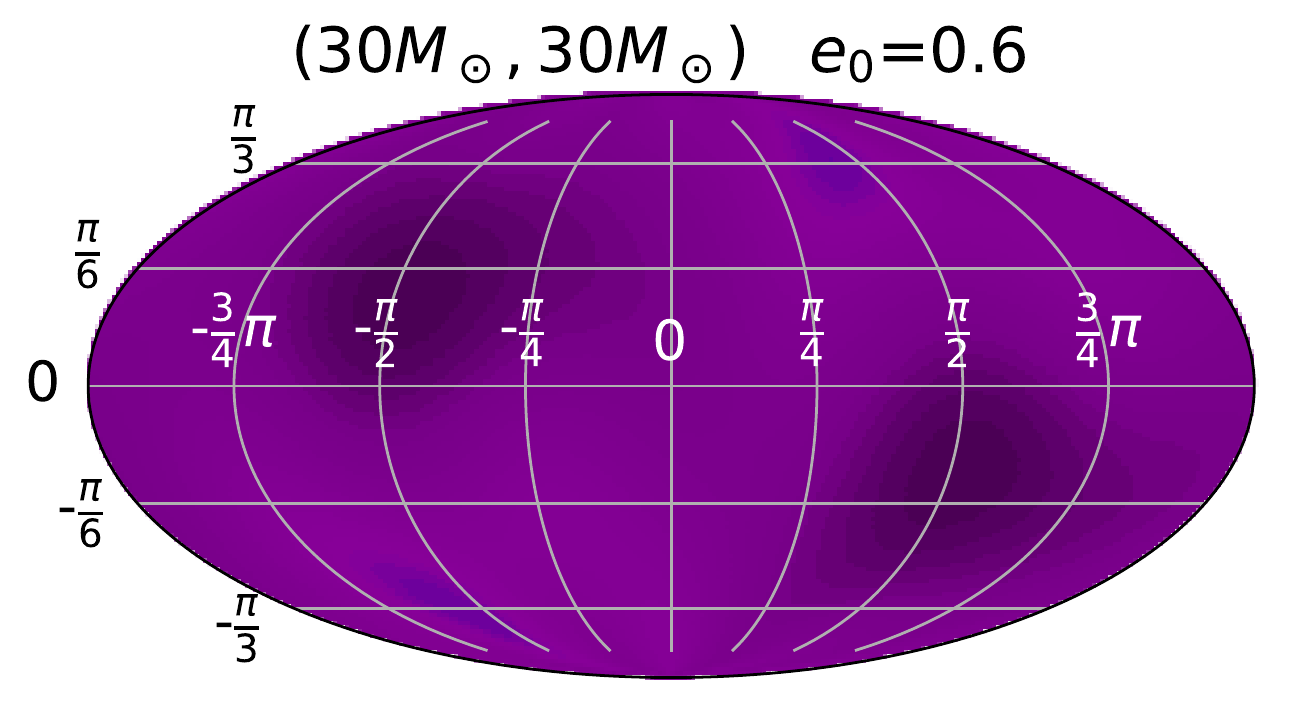}
		\includegraphics[width=\wid\textwidth]{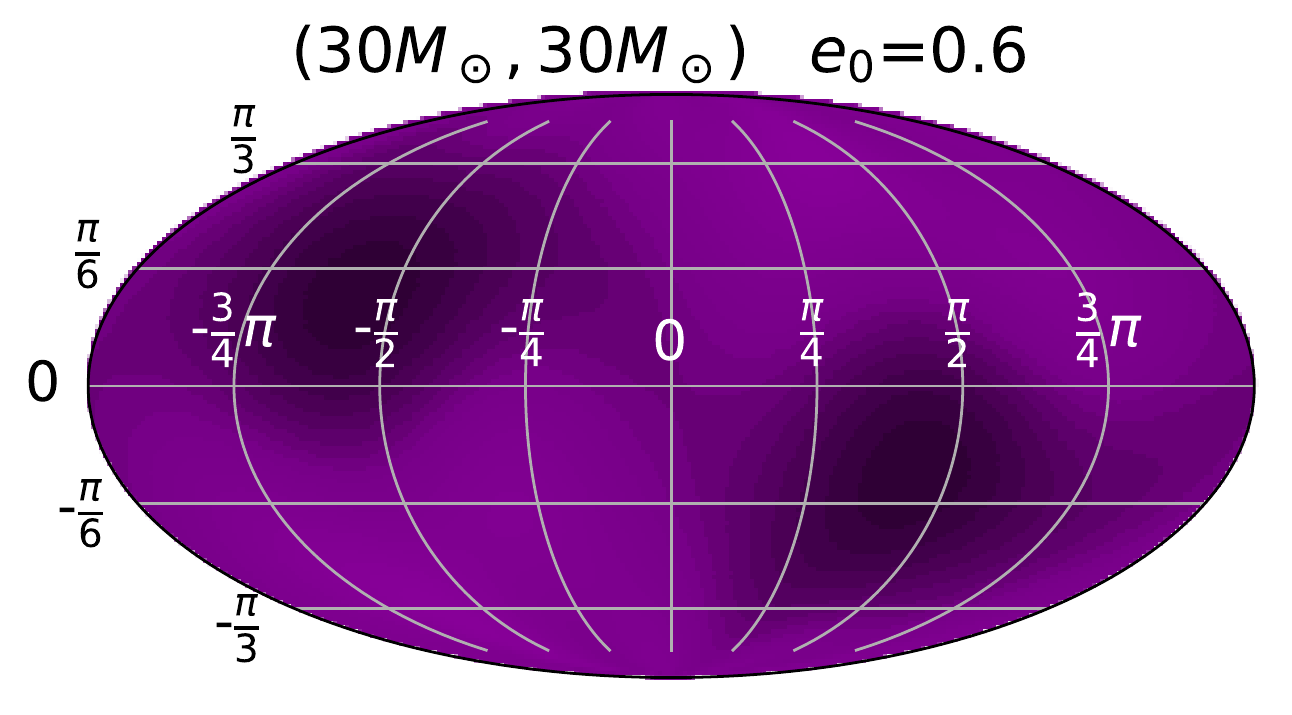}
		\includegraphics[width=\wid\textwidth]{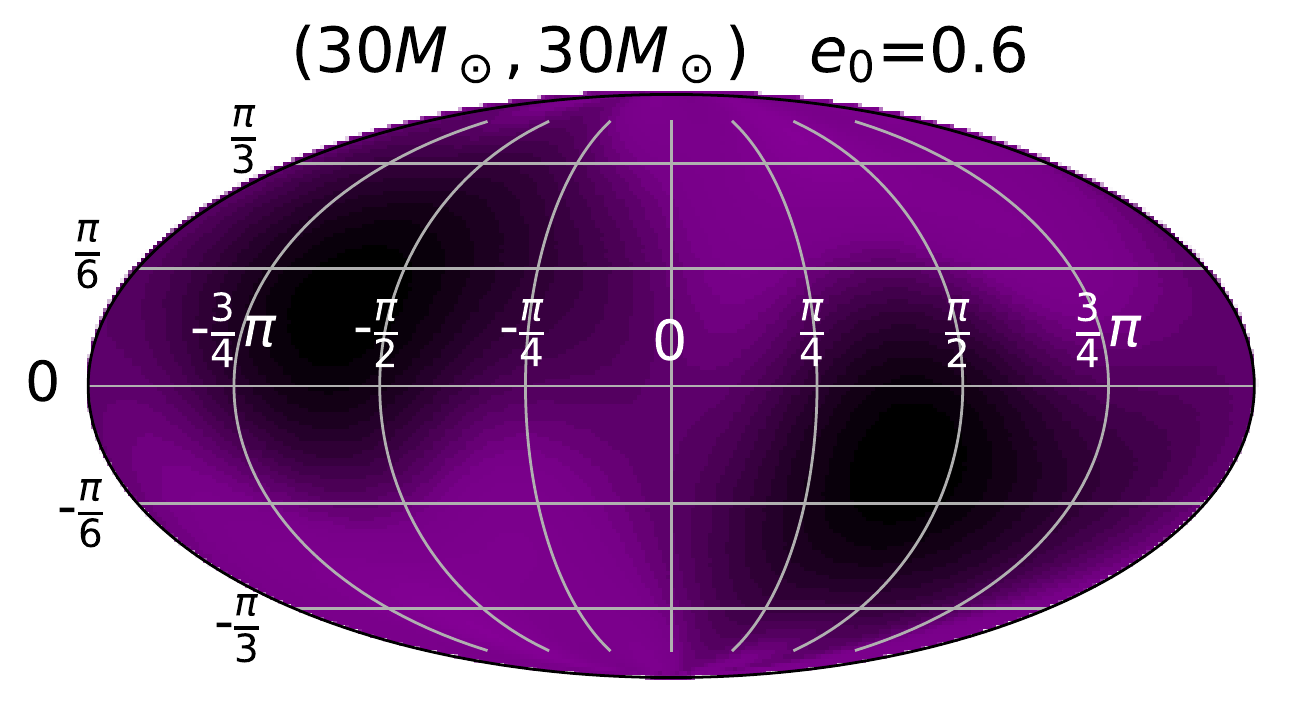}
	}
	\caption{As Figure~\ref{fig:equal_mass_LIGO_type}, but now assuming that the PSD of each detector corresponds to the target sensitivity of ET-D. These results are produced assuming binary black hole mergers with component masses \((30\msun,\,30\msun)\). As before, we have used the Mollweide projection, averaged over polarization angles, and set the binary inclination angle to \(i=\pi/4\). See Appendix~\ref{app:ap1} for results assuming  \(i=0\).} 
	\label{fig:etd_equal_mass}
\end{figure*}

\begin{figure*}[p]
	\centerline{
		\includegraphics[width=\textwidth]{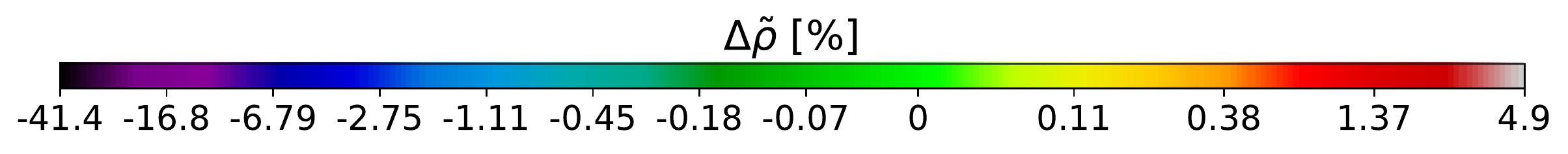}
	}
	\centerline{
		\includegraphics[width=\wid\textwidth]{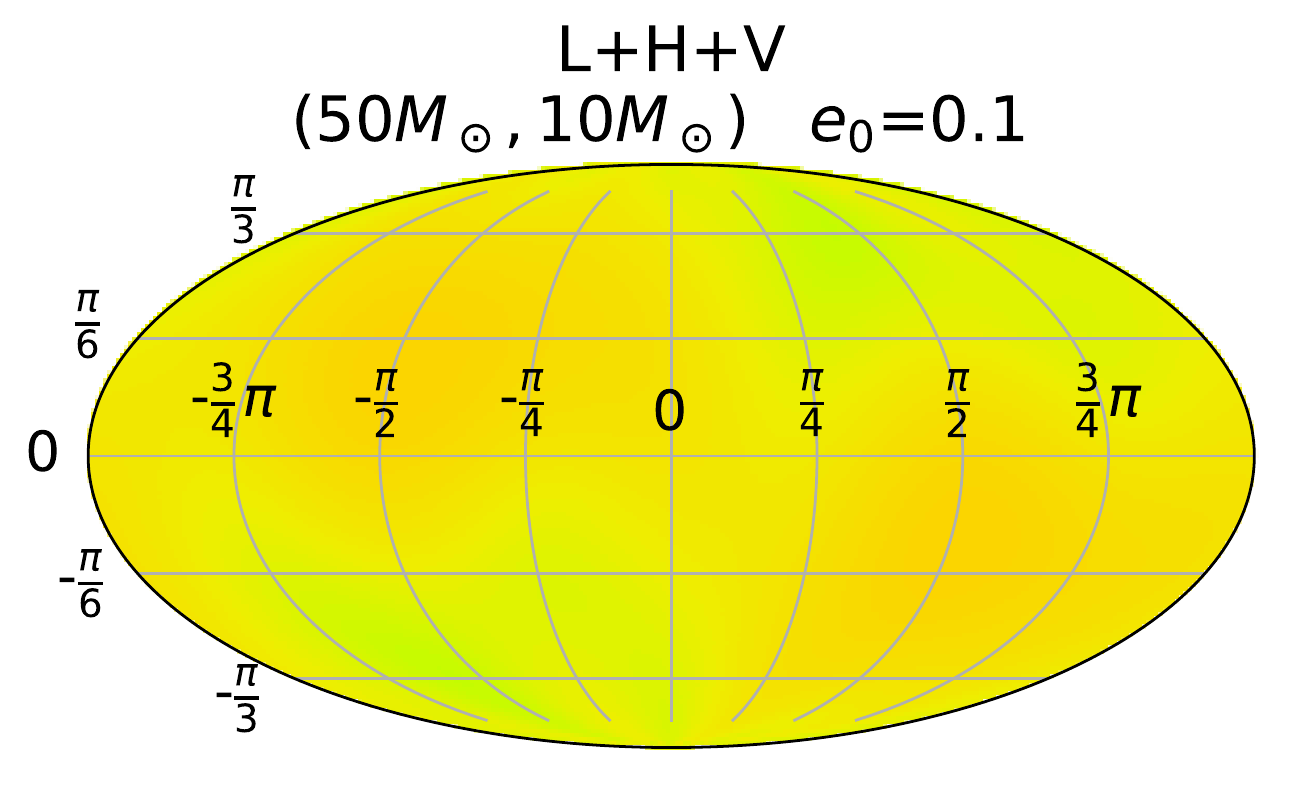}
		\includegraphics[width=\wid\textwidth]{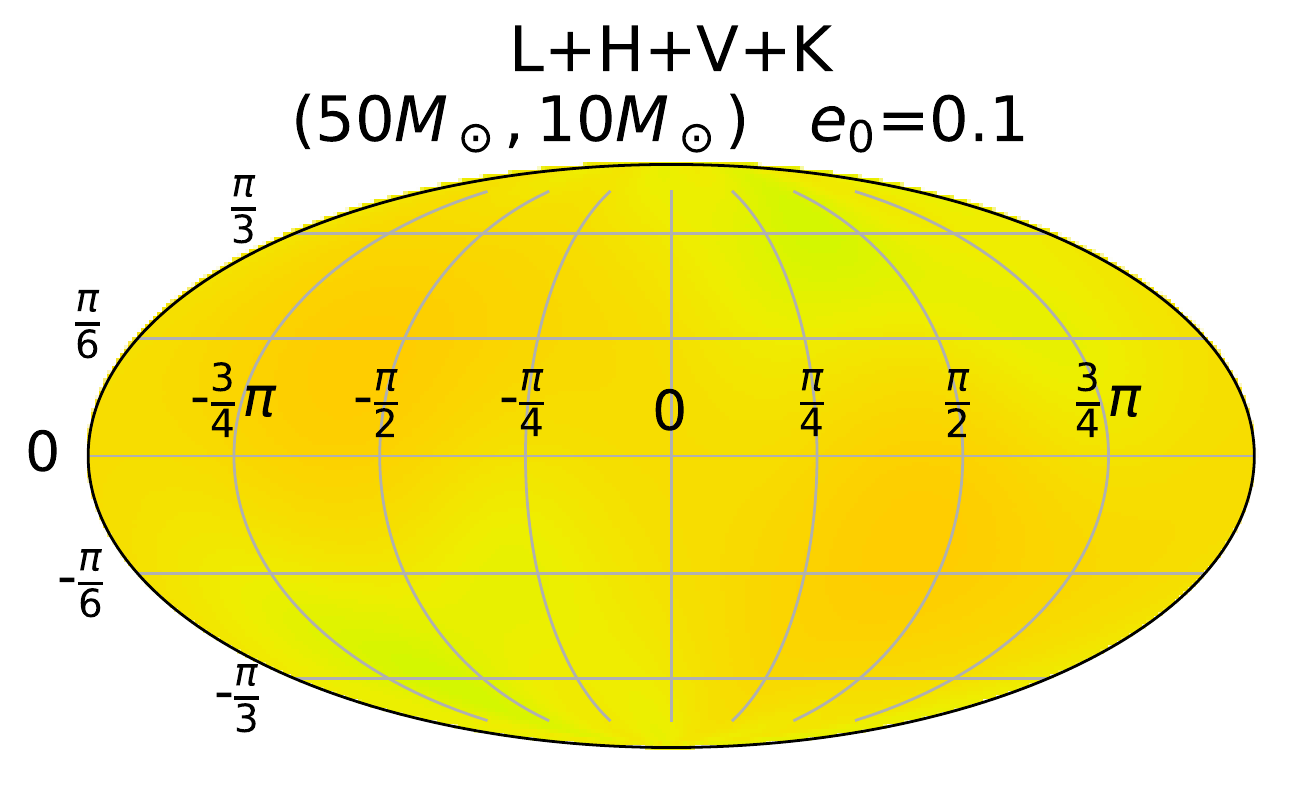}
		\includegraphics[width=\wid\textwidth]{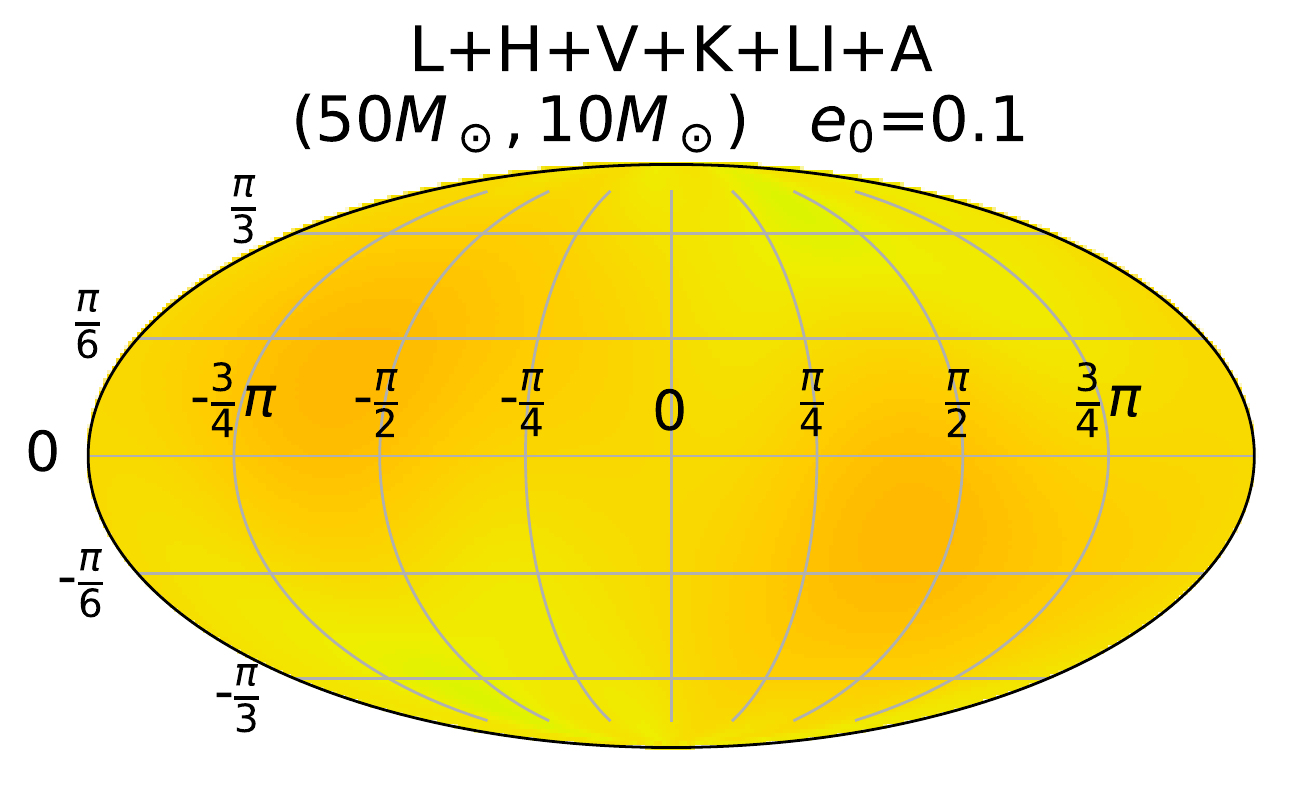}
	}
	\centerline{
		\includegraphics[width=\wid\textwidth]{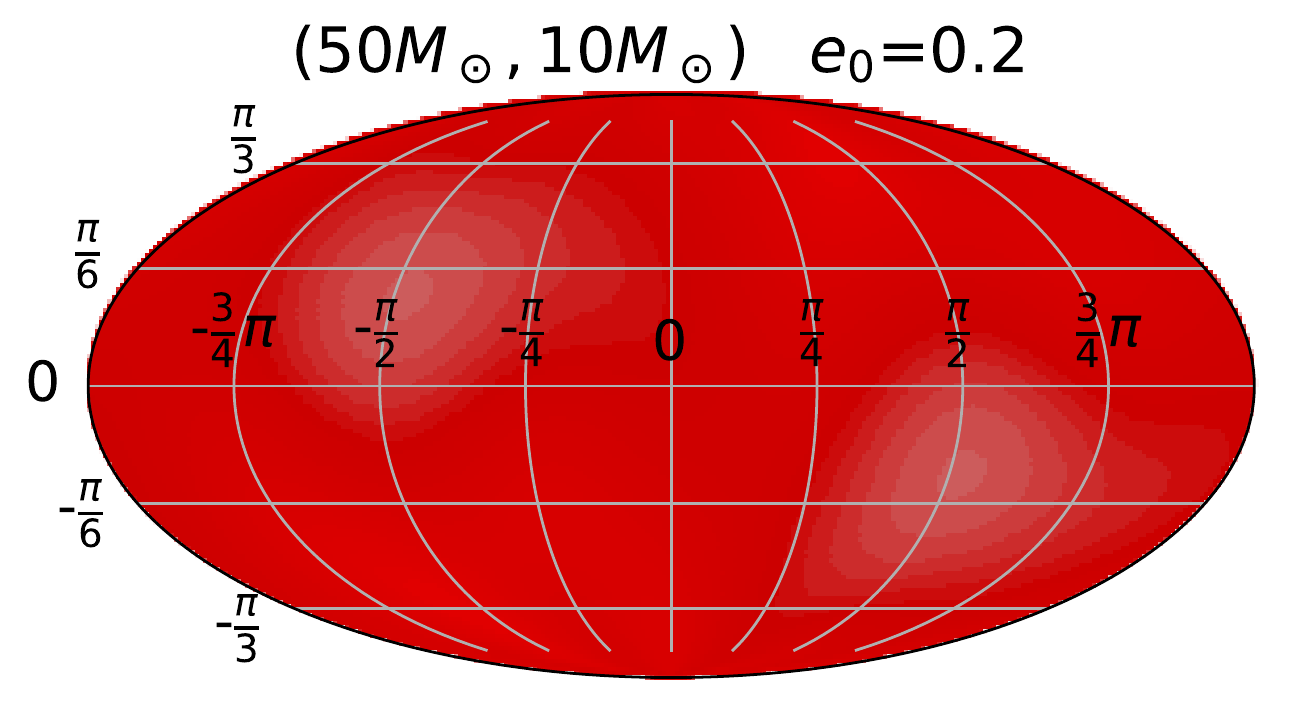}
		\includegraphics[width=\wid\textwidth]{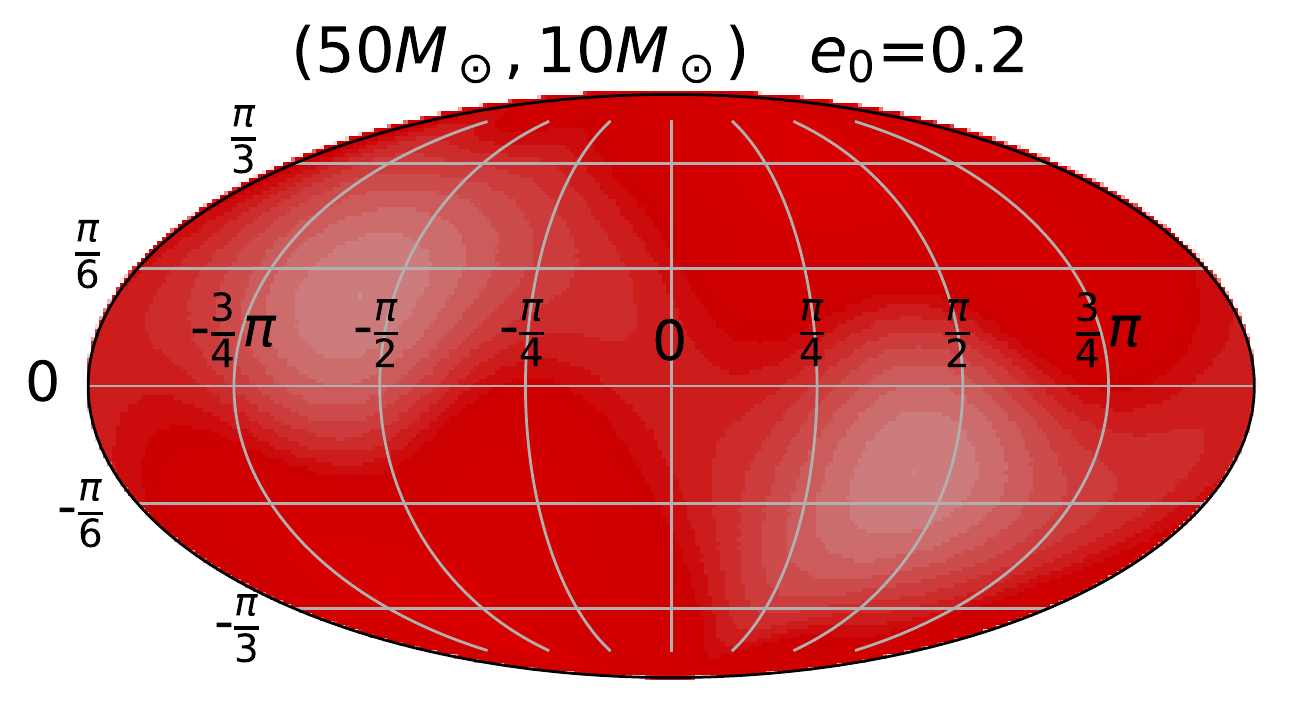}
		\includegraphics[width=\wid\textwidth]{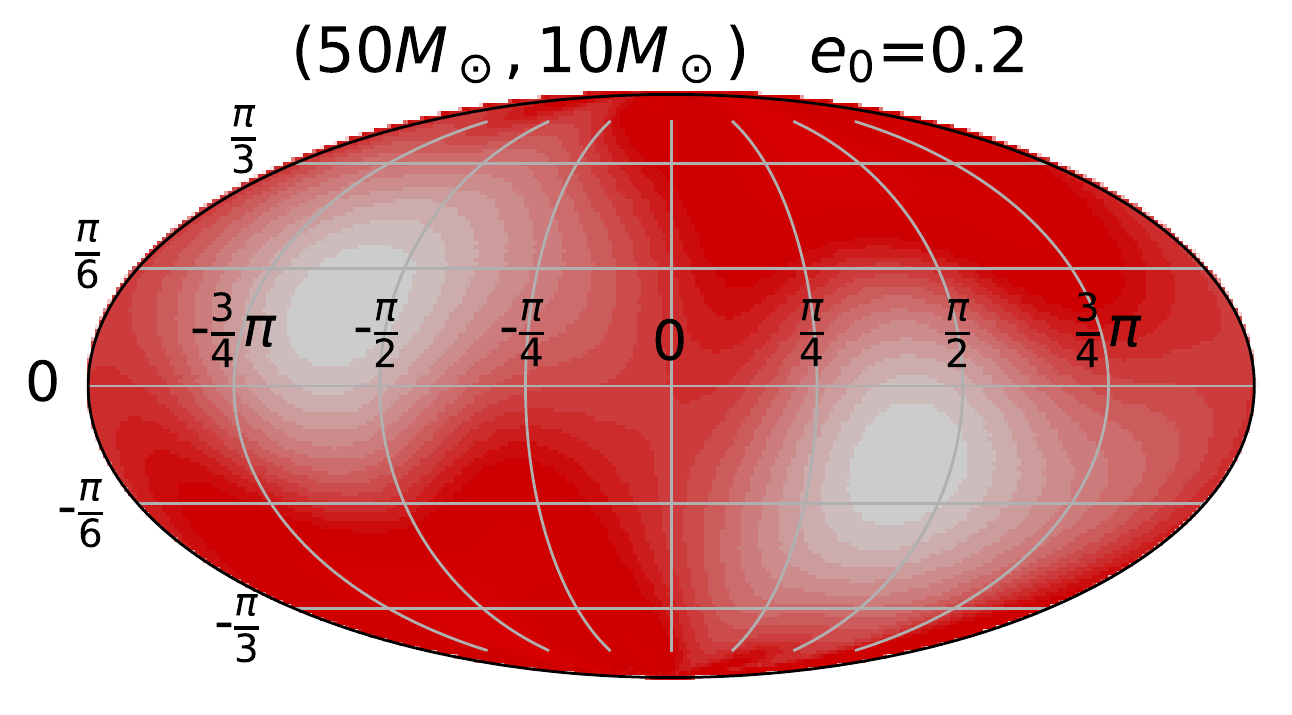}
	}
	\centerline{
		\includegraphics[width=\wid\textwidth]{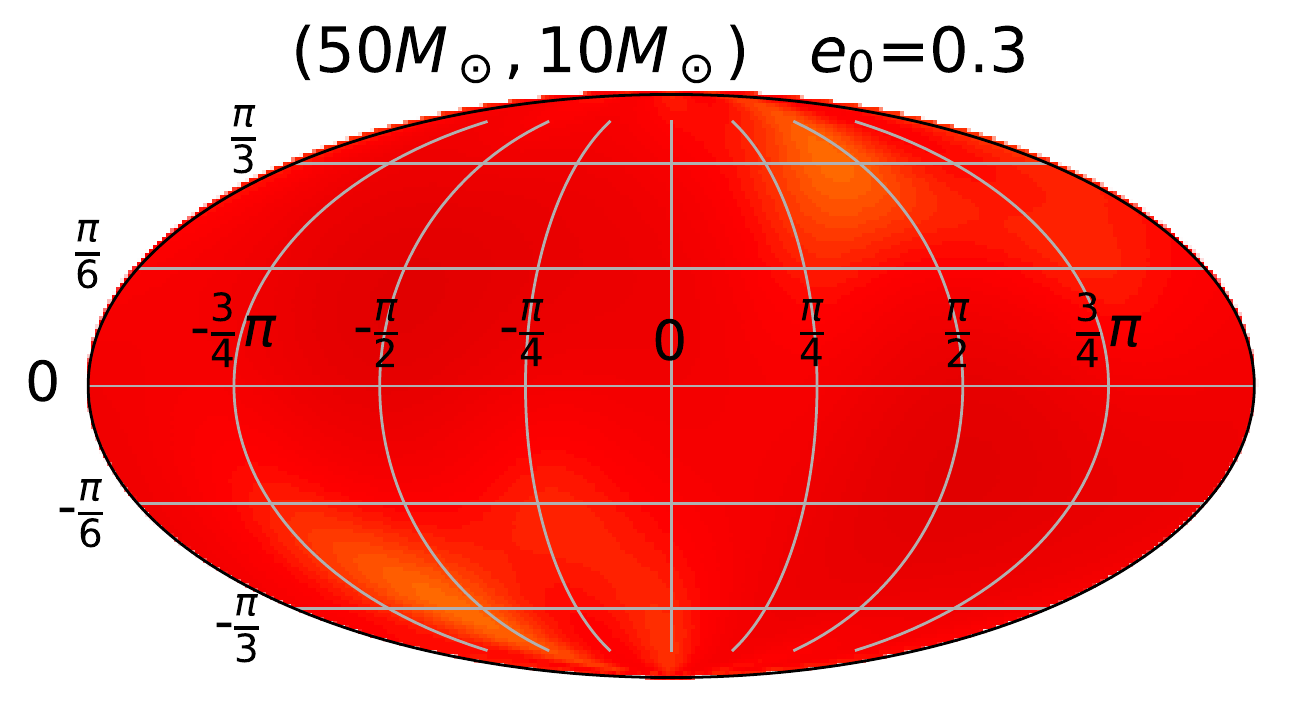}
		\includegraphics[width=\wid\textwidth]{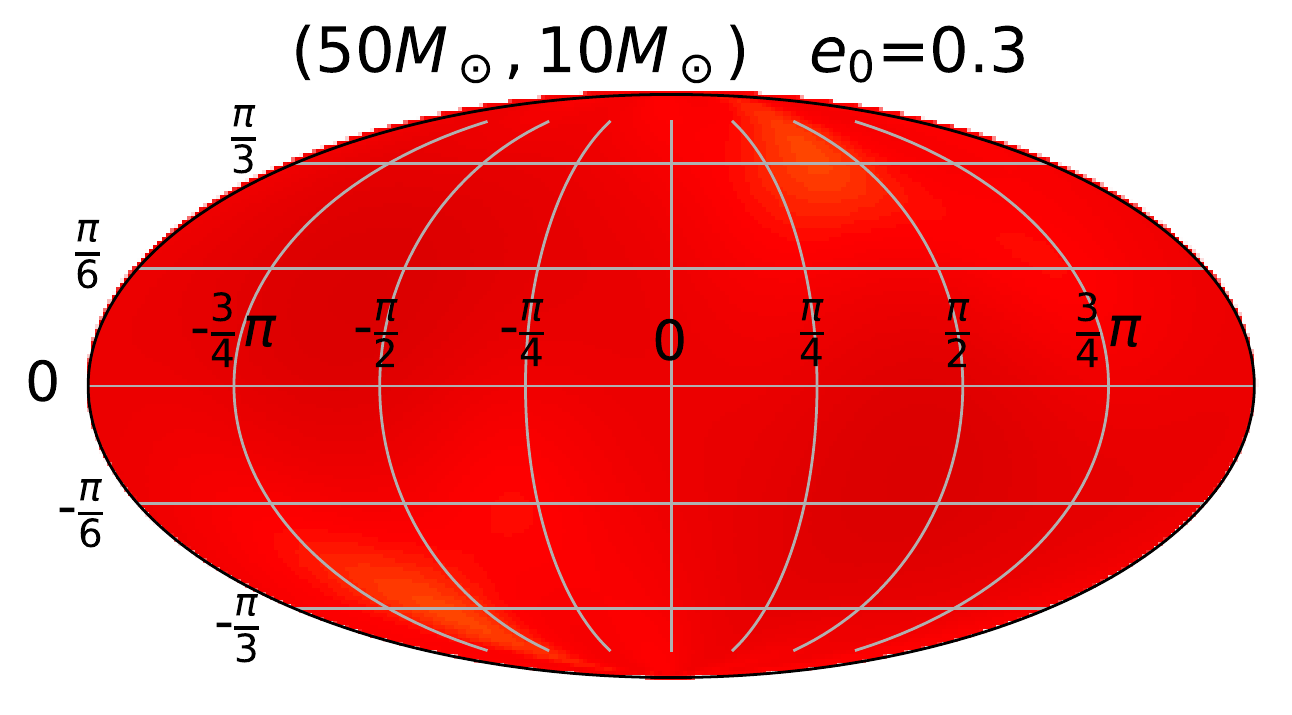}
		\includegraphics[width=\wid\textwidth]{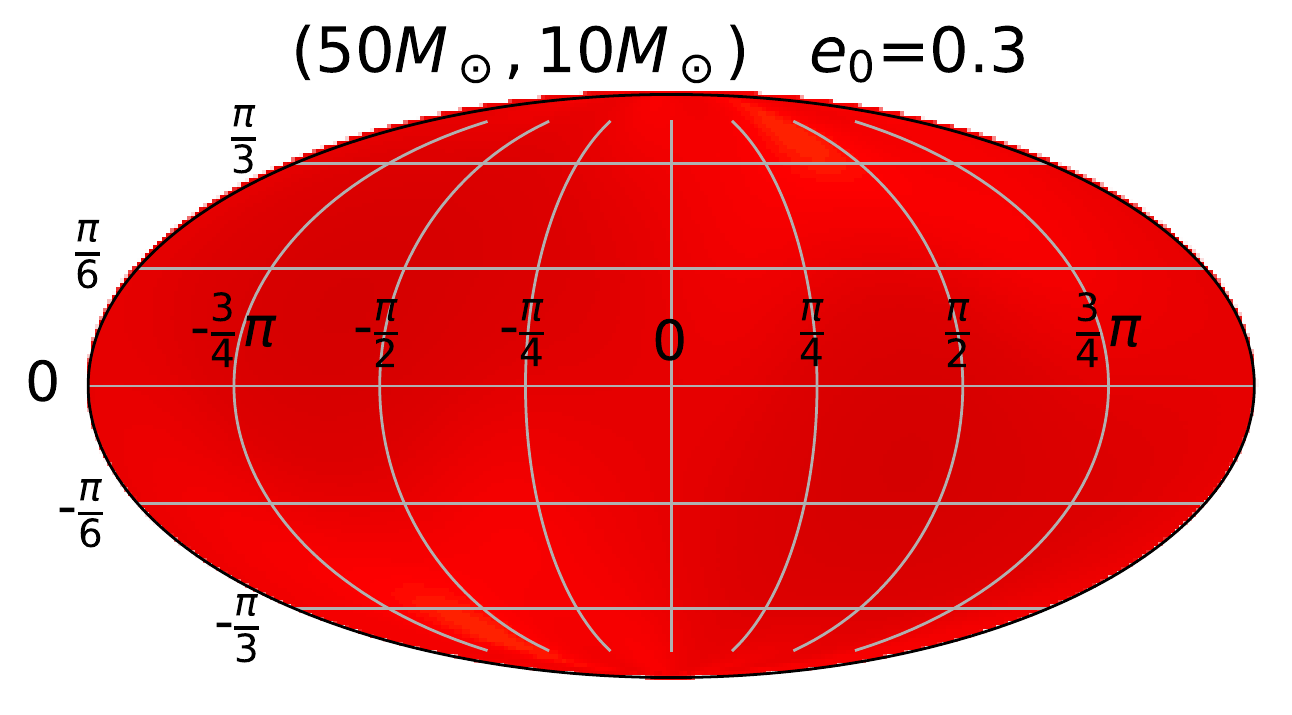}
	}
	\centerline{
		\includegraphics[width=\wid\textwidth]{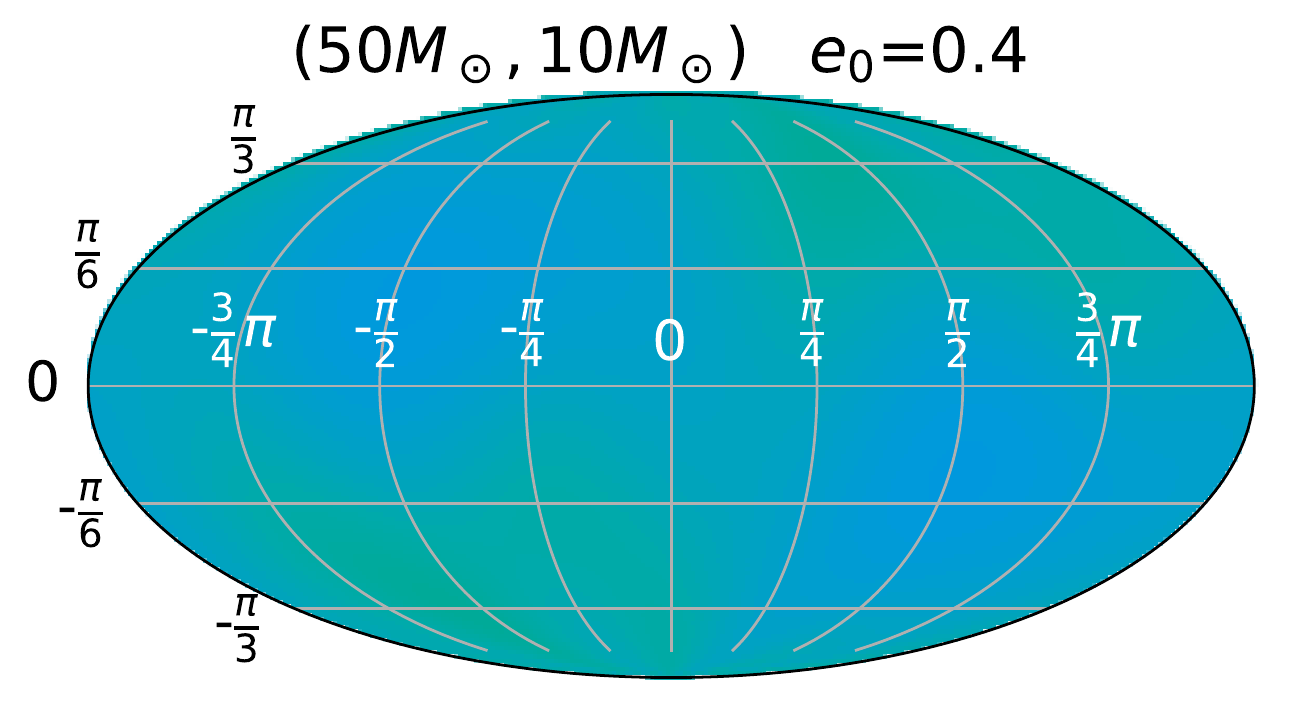}
		\includegraphics[width=\wid\textwidth]{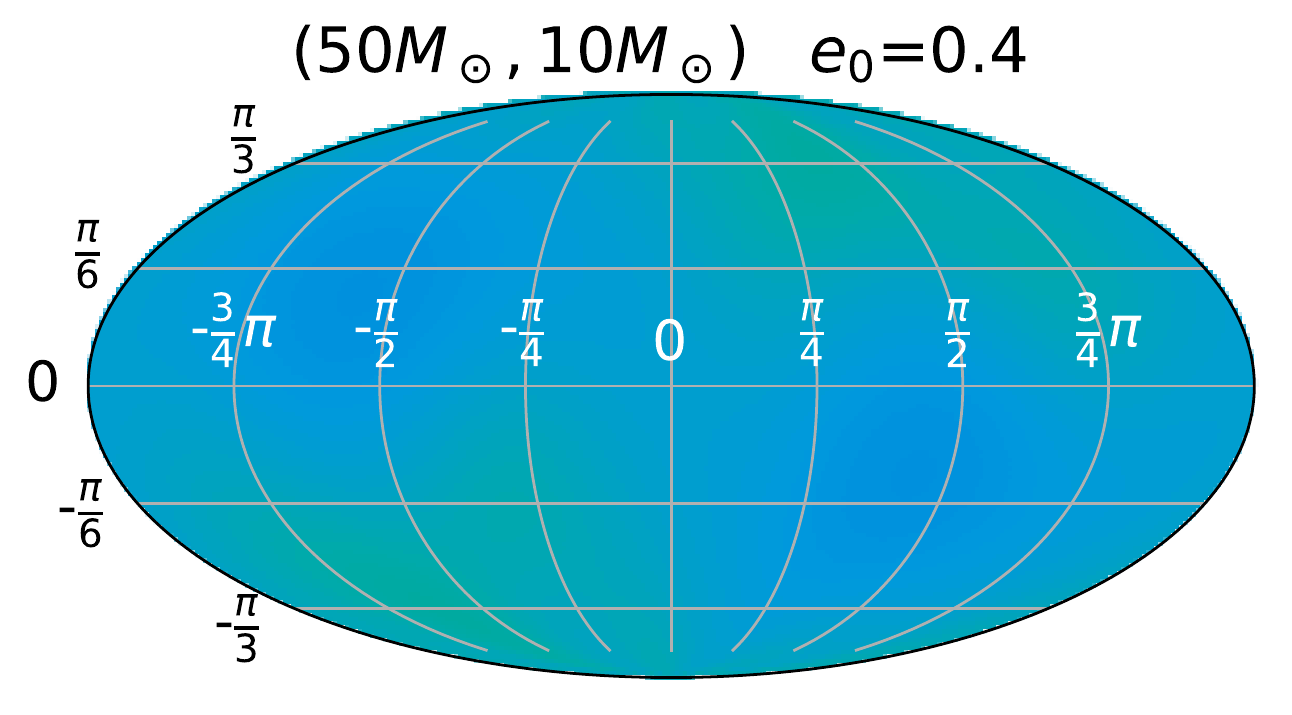}
		\includegraphics[width=\wid\textwidth]{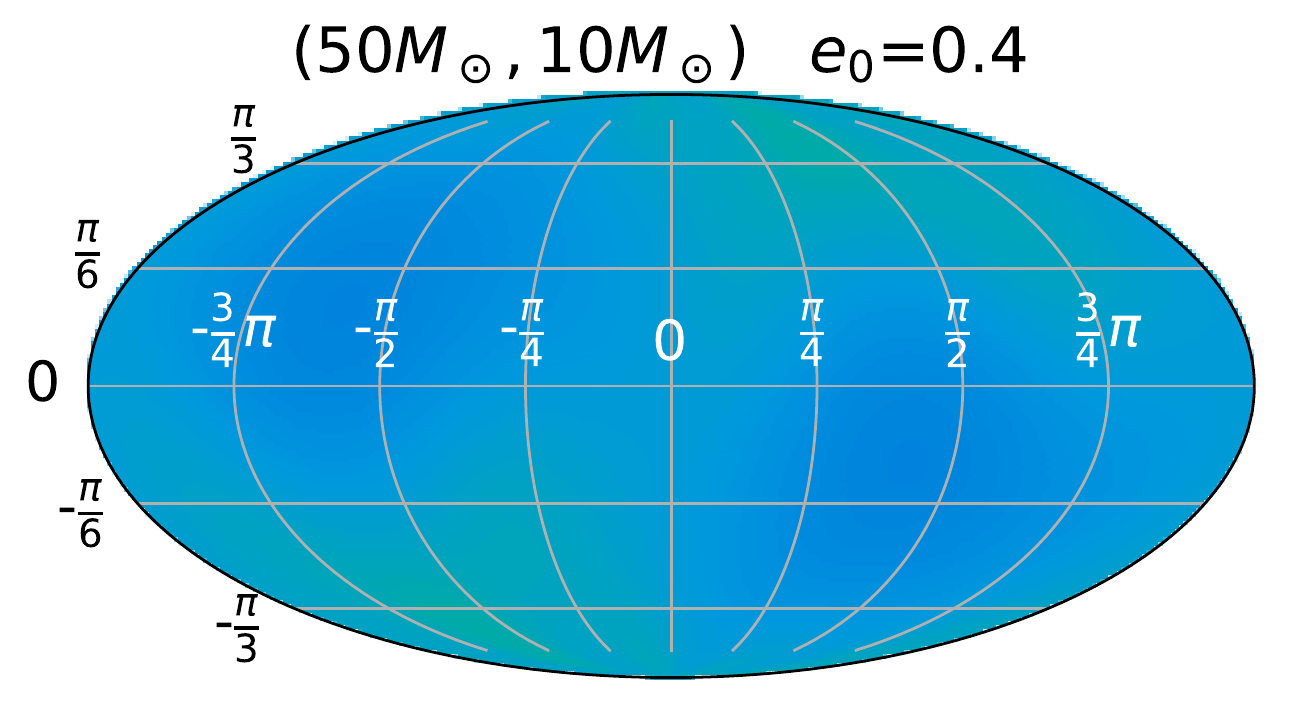}
	}
	\centerline{
		\includegraphics[width=\wid\textwidth]{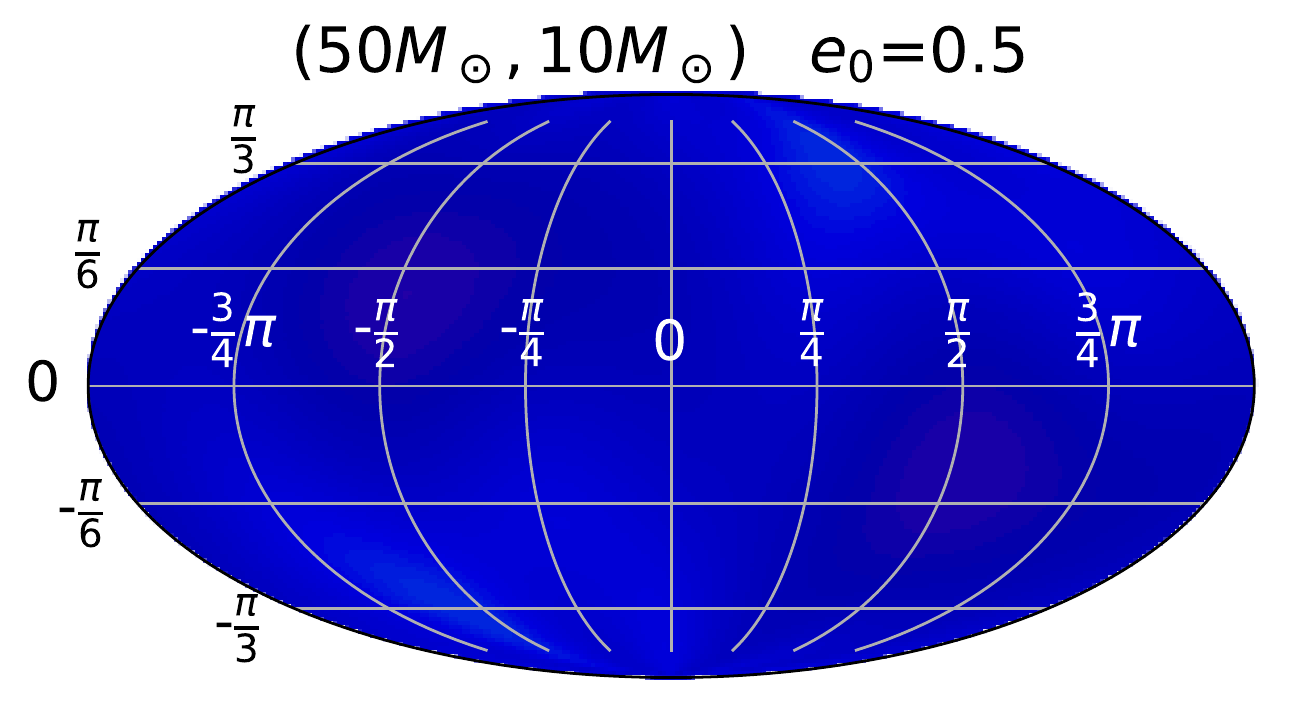}
		\includegraphics[width=\wid\textwidth]{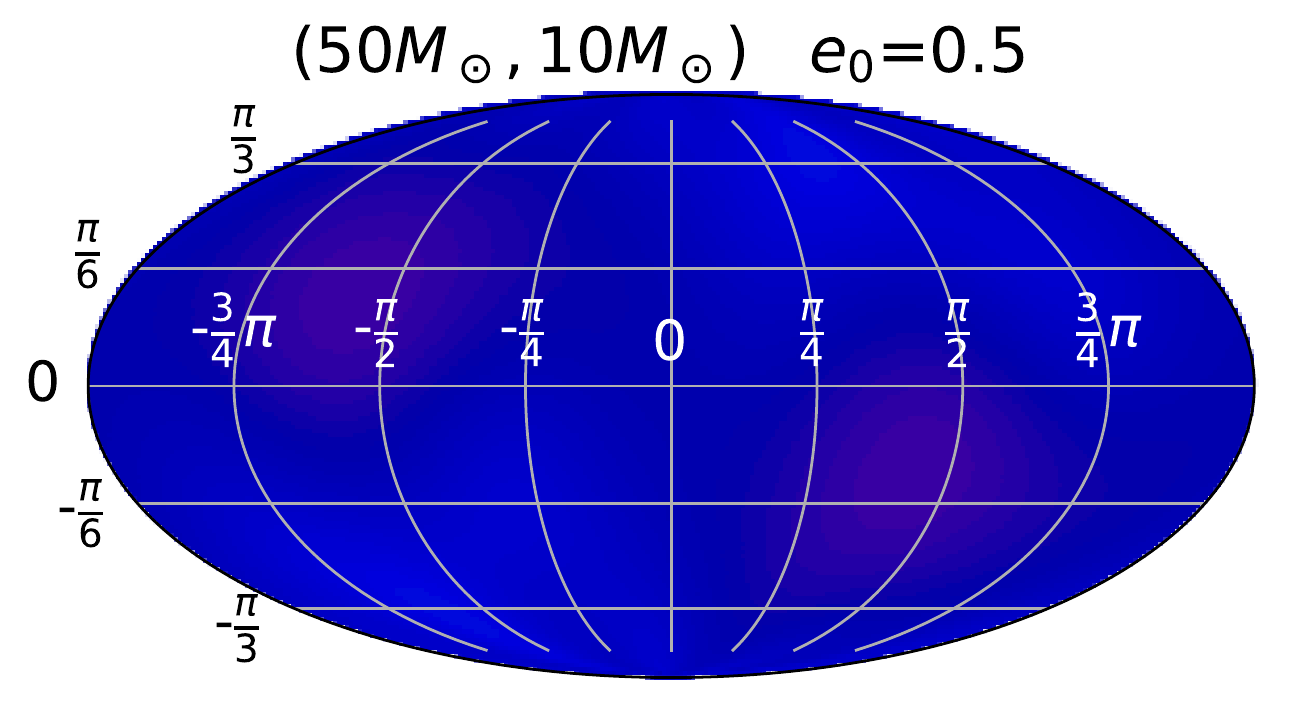}
		\includegraphics[width=\wid\textwidth]{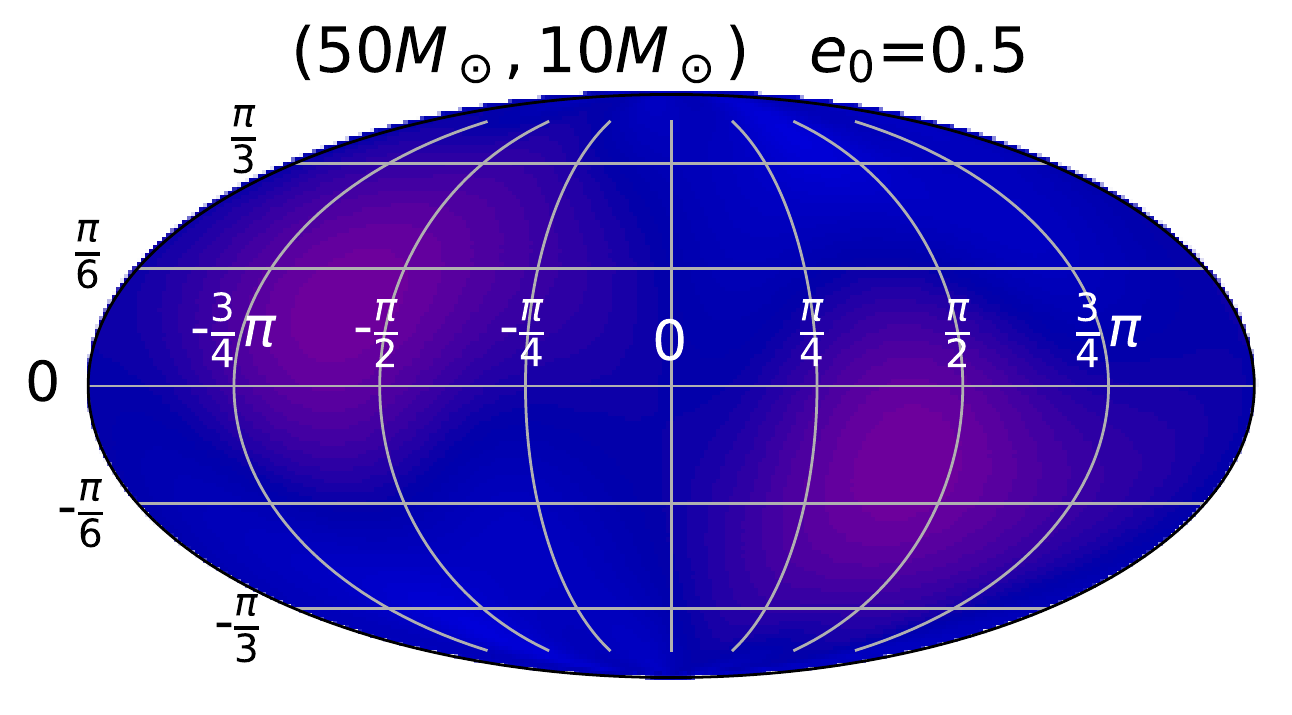}
	}
	\centerline{
		\includegraphics[width=\wid\textwidth]{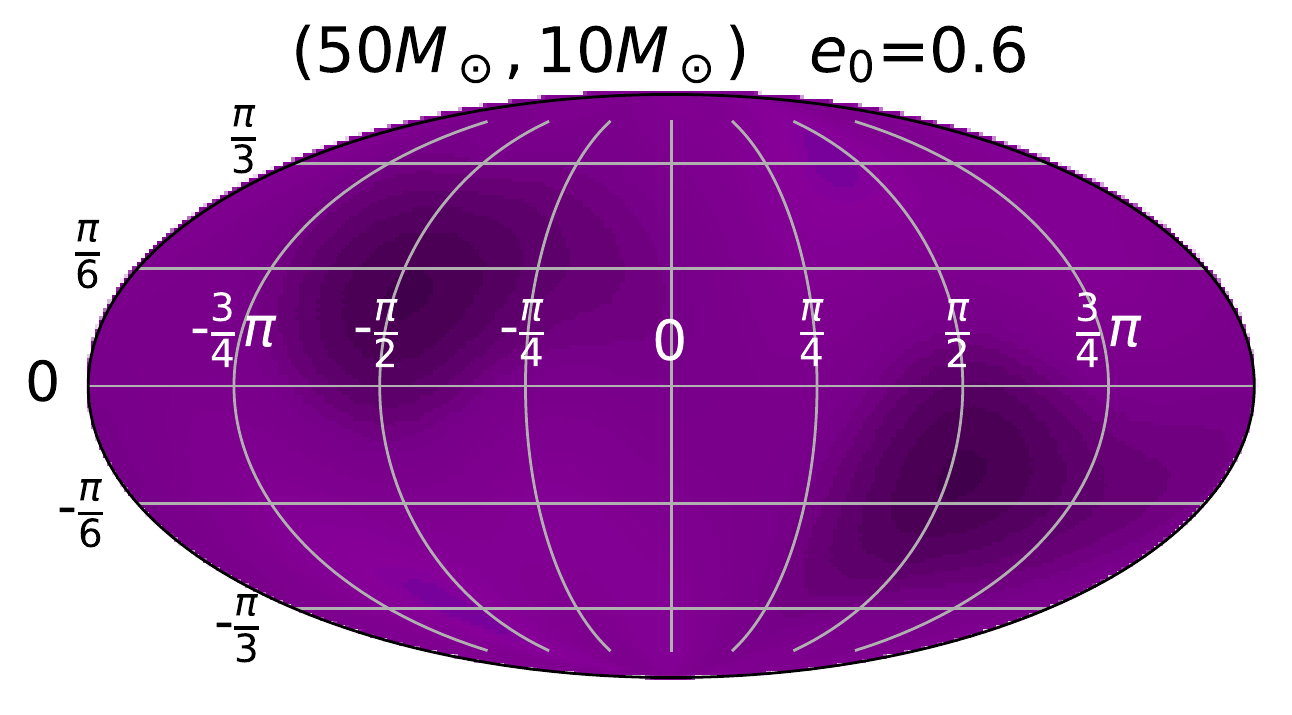}
		\includegraphics[width=\wid\textwidth]{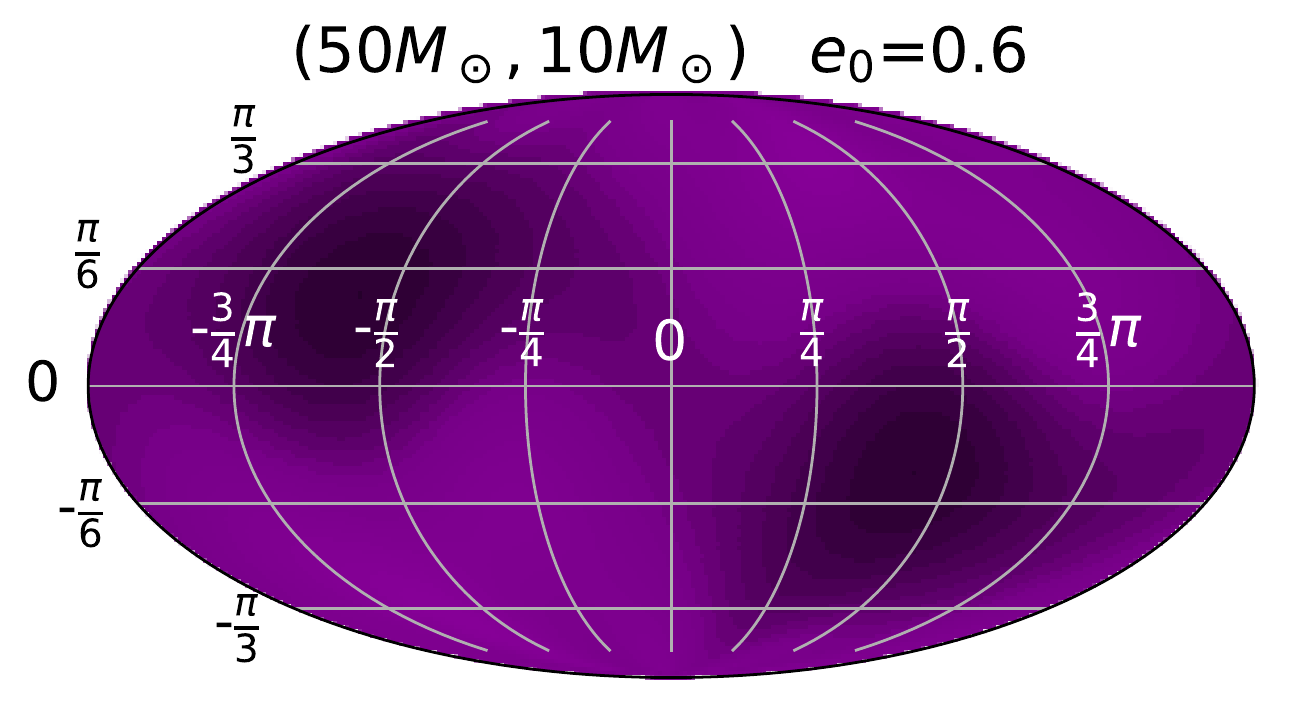}
		\includegraphics[width=\wid\textwidth]{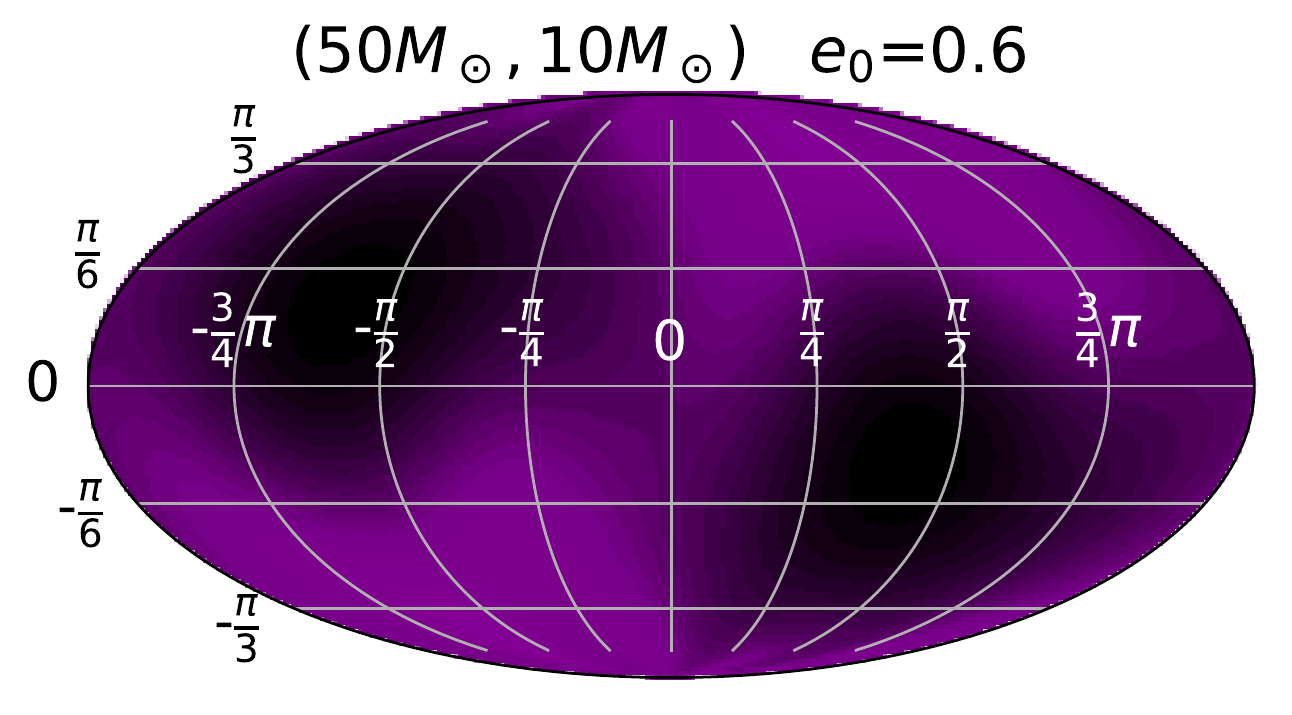}
	}
	\caption{As Figure~\ref{fig:etd_equal_mass} but now for component masses \((50\msun,\,10\msun)\).} 
	\label{fig:etd_five_mass}
\end{figure*}

We have also produce SNR sky distributions for ET-D detector networks. For consistency, we present 
results for systems with 
component masses \(m_{\{1,\,2\}}= \{\left(30\msun, 30\msun\right), \left(50\msun, 10\msun\right)\}\), 
with eccentricities \(e_0\leq0.6\) are measured at \(f_{\textrm{GW}}=4\textrm{Hz}\). 
The SNR distributions are computed by setting \(f_0=5\,\textrm{Hz}\) and \(f_1=4096\,\textrm{Hz}\) in 
Eq.~\eqref{eqn:SNRcomponent}. Some observations we draw from these results include:

\begin{itemize}
\item We notice that for equal mass binaries, Figure~\ref{fig:etd_equal_mass}, 
the SNR sky distributions of quasi-circular and eccentric mergers are comparable 
for eccentricities in the range \(e_0<0.5\). This is truly remarkable if one considers that these 
eccentric systems have less than half the timespan of their quasi-circular counterparts. 
These results underscore the impact that third generation GW detectors will have at 
boosting the amplitude amplifications driven by eccentricity at lower frequencies, 
which will actually compensate for the dramatic reduction in waveform timespan 
for signals with \(e_0\sim0.5\). 
\item As we have discussed before, asymmetric mass-ratio systems tend to attenuate 
the SNR suppression in eccentric mergers. For instance, for the most eccentric systems in our sample, 
see Figure~\ref{fig:etd_five_mass}, the SNR reduction for \(e_0\sim0.6\) mergers 
is about 40\% the SNR of an equivalent system with zero 
eccentricity. In contrast, Figure~\ref{fig:etd_equal_mass} shows that for comparable mass-ratio mergers, 
quasi-circular systems are 50\% louder than \(e_0\sim0.6\) BBH mergers with the same total mass and 
mass-ratio.
\end{itemize}

These results present a number of unique properties of eccentric mergers that may be used 
to differentiate them from quasi-circular BBHs. It is worth mentioning that while spin corrections 
may mimic the physics of moderately eccentric mergers, as shown in~\cite{Huerta:2017a}, 
the effects of eccentricity we have discussed are unique for moderate values of 
eccentricity for second or third generation GW detector networks. No spin 
effects are capable of amplifying the waveform amplitude \textrm{and} shortening the 
waveform timespan in the way shown in Figures~\ref{fig:whitened_wf}, \ref{fig:snr_time_ligo}, and~\ref{fig:snr_time_CE_and_ETD}, or to accelerate the 
accumulation of SNR at the rate we presented in this section.

\section{Conclusions}
\label{sec:end}

We have introduced an improved version of the \texttt{ENIGMA} model that provides a 
numerically stable approach to attach the merger waveform to the inspiral evolution, as well as the inclusion of 
 the binary inclination angle, which is a key feature that not only modifies the 
amplitude evolution of eccentric mergers, but also encodes additional physics even at leading order post-Newtonian 
corrections. We have tested the robustness of our waveform generator, finding that we can randomly 
sample the \(m_{\{1,\,2\}} \in [5\msun,\,50\msun]\) parameter space, varying the 
inclination angle, and produce over 1M physically consistent waveforms without reporting any errors. 
We then established its accuracy in the quasi-circular limit by computing overlaps with 
the \texttt{SEOBNRv4} model, finding overlaps \({\cal{O}}\geq0.99\) over the parameter 
space of applicability for \texttt{ENIGMA}. We benchmarked \texttt{ENIGMA}, finding that, 
averaged over 1000 iterations, it can produce a single waveform from 
\(f_{\textrm{GW}}=15\,\textrm{Hz}\) at a sample rate of \(8192\,\textrm{Hz}\) within 0.04 seconds.

We then explored the properties of SNR sky distributions produced by eccentric and quasi-circular 
BBH mergers in the context of second and third generation GW detectors. Our results indicate that 
SNR sky distributions tend to present eccentric and quasi-circular mergers as similar events, which 
may only be differentiated for rather eccentric BBH populations. We have introduced a complementary 
method to tell apart these two populations that consists of studying the rate at which SNR accumulates 
in eccentric mergers, demonstrating that such approach exhibits tell-tale signatures that may not be 
mimicked by other orbital effects, such as spin corrections.

With these studies we introduce a ready-to-use \texttt{ENIGMA} model that may be readily put 
at work to conduct 
GW searches of eccentric events in similar manner to those presented 
in~\cite{Nitz_2020,lenon2020measuring}, or to constrain the eccentricity of all 
BBH mergers detected by LIGO and Virgo to date. It may be timely and relevant 
to assess the use of  \texttt{ENIGMA} waveforms to recover the astrophysical parameters of 
eccentric BBH mergers using well characterized eccentric numerical relativity 
waveforms~\cite{huerta_nr_catalog,ian:2017,Habib:2019cui,ramos_buades} embedded 
in real advanced LIGO noise to 
quantify the biases that are introduced by noise and intrinsic waveform errors 
in parameter estimation studies. 
These studies will be pursued in the near future.

\section{Acknowledgements}

We gratefully acknowledge National Science Foundation (NSF) awards OAC-1931561, 
OAC-1934757, OAC-1550514, OAC-1931280, OAC-2004879, PHY-1912081, NSF-1659702, 
and the Sherman Fairchild foundation.
Compute resources were provided
by XSEDE using allocation TG-PHY160053. 
PK acknowledges the support of the Department of Atomic Energy, Government of India, under project no. RTI4001.
This research is part of the Blue Waters sustained-petascale computing
project, which is supported by the National Science Foundation (awards
OCI-0725070 and ACI-1238993) the State of Illinois, and as of December, 2019,
the National Geospatial-Intelligence Agency. Blue Waters is a joint effort of
the University of Illinois at Urbana-Champaign and its National Center for
Supercomputing Applications.
We acknowledge support from the NCSA and the 
\href{http://gravity.ncsa.illinois.edu}{NCSA Gravity Group}.
This work made use of the Illinois Campus Cluster, a computing resource that 
is operated by the Illinois Campus Cluster Program (ICCP) in conjunction with 
the NCSA and which is supported by funds from the University of Illinois 
at Urbana-Champaign~\cite{CCUUIUC}.


\appendix 

\clearpage
\begin{widetext}

\section{Additional signal-to-noise sky distributions for eccentric binary black hole mergers}
\label{app:ap1}

In this section Figures~\ref{fig:ligo_equal_opt}-\ref{fig:etd_five_opt} present SNR sky distributions for eccentric BBH mergers with inclination angle \(i=0\). These results mirror those we presented in the main body of the article for second and third generation GW detector networks.

\begin{figure*}
	\centerline{
		\includegraphics[width=\textwidth]{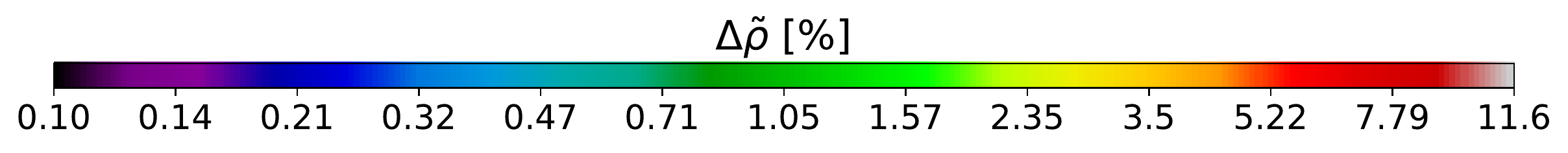}
	}
	\centerline{
		\includegraphics[width=\wid\textwidth]{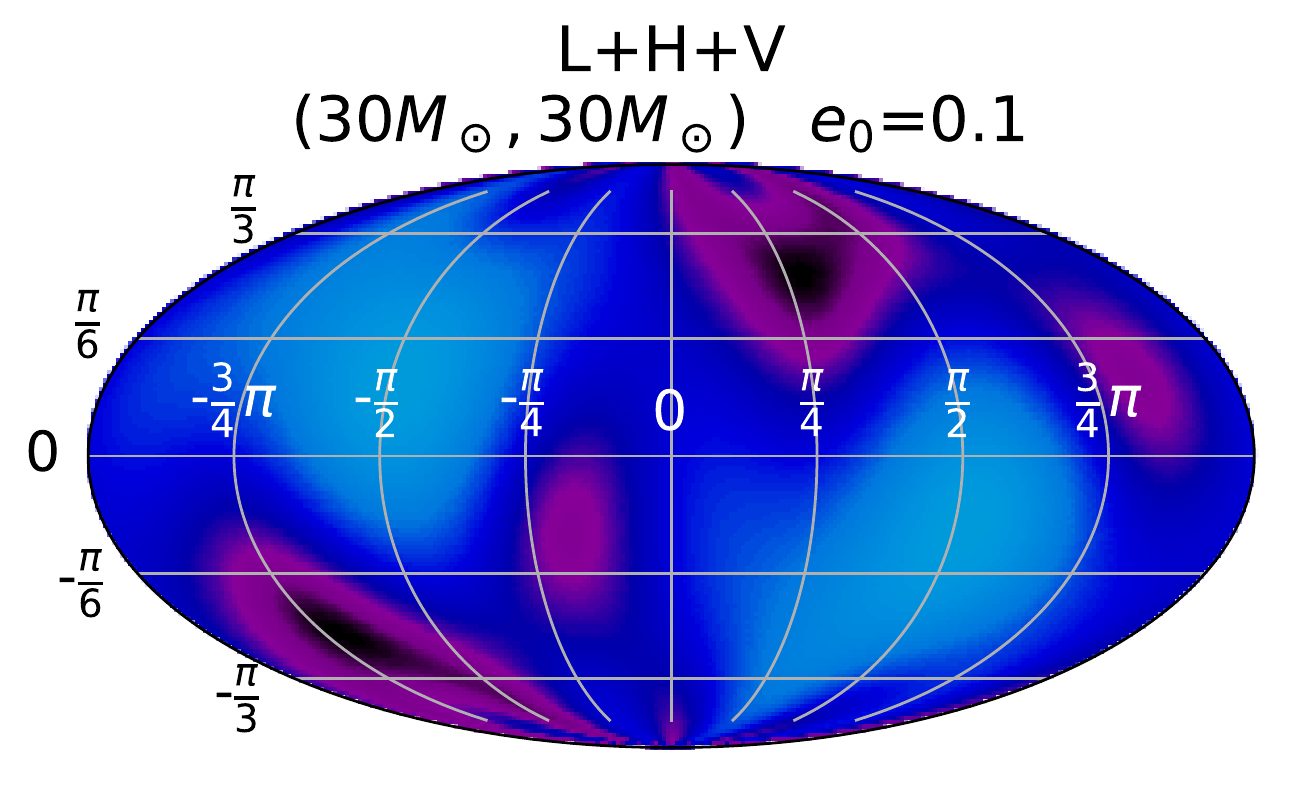}
		\includegraphics[width=\wid\textwidth]{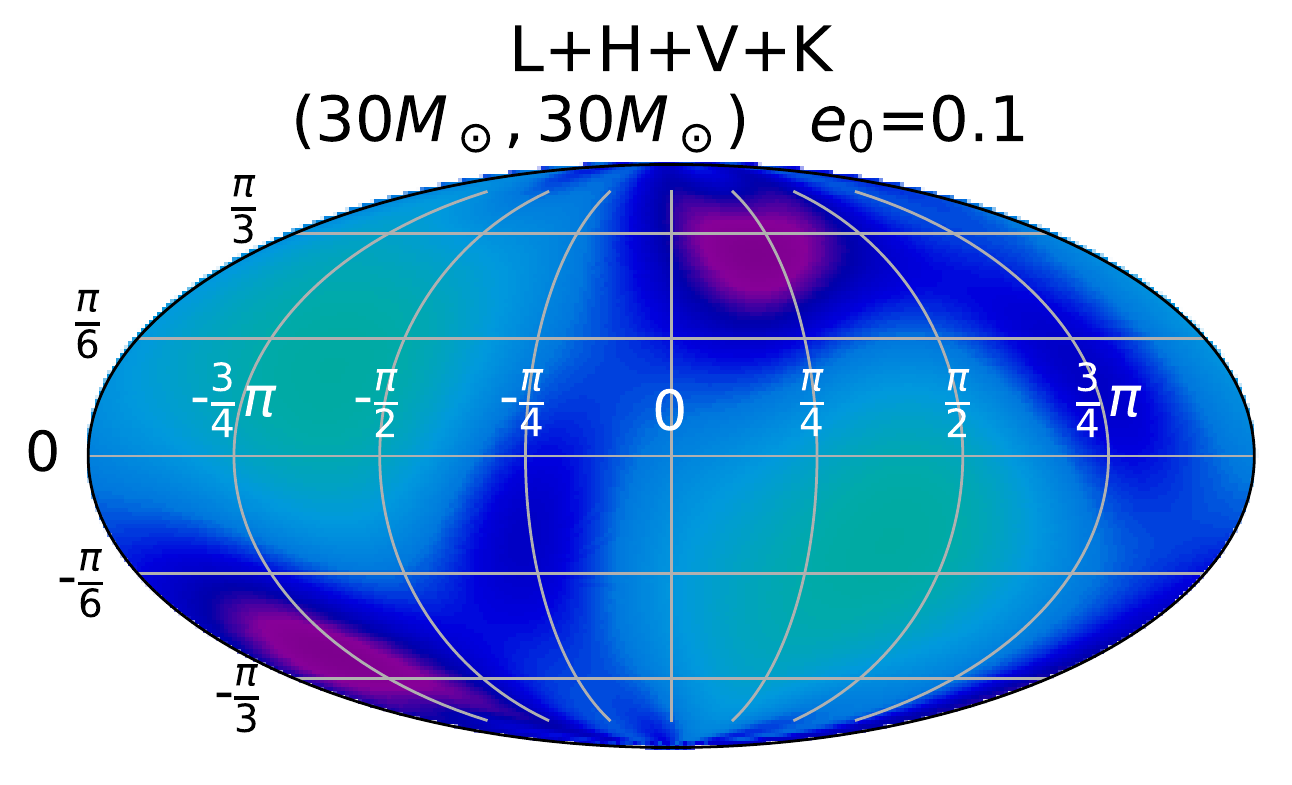}
		\includegraphics[width=\wid\textwidth]{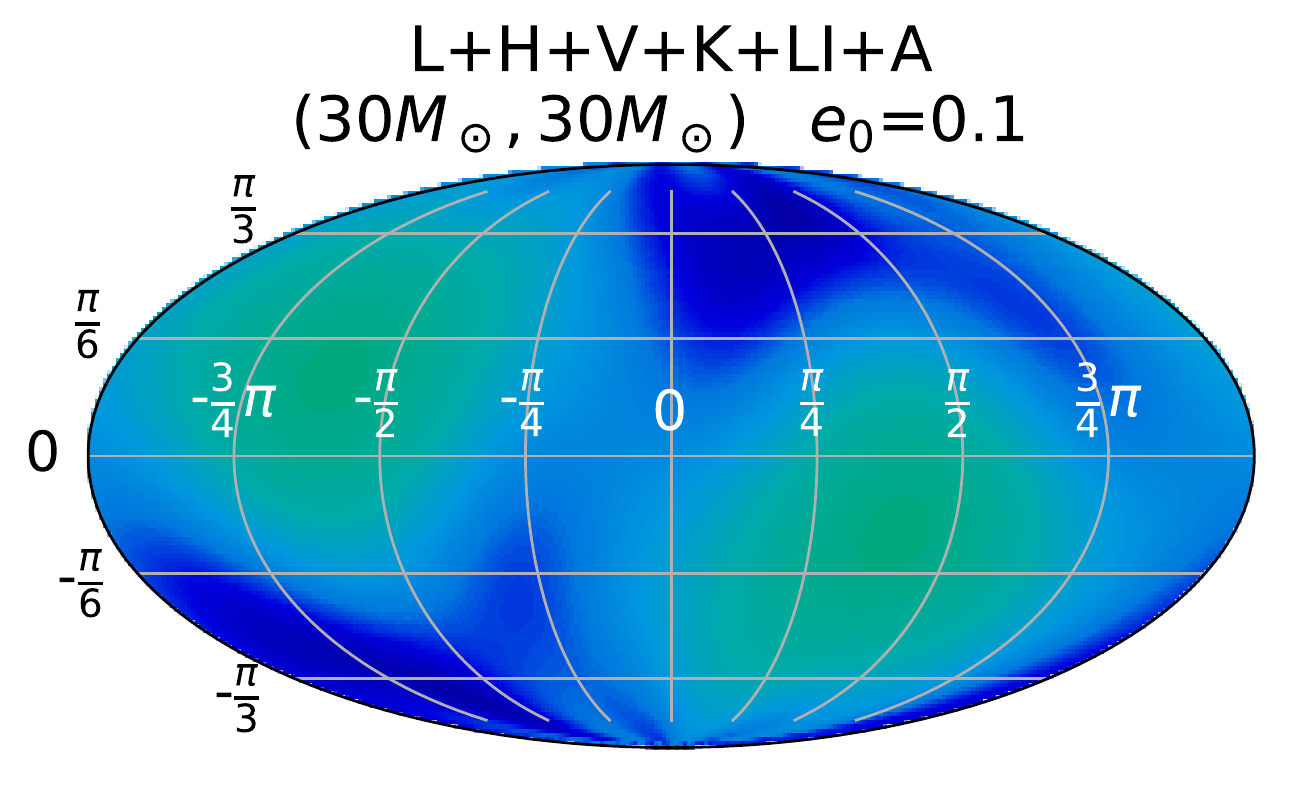}
	}
	\centerline{
		\includegraphics[width=\wid\textwidth]{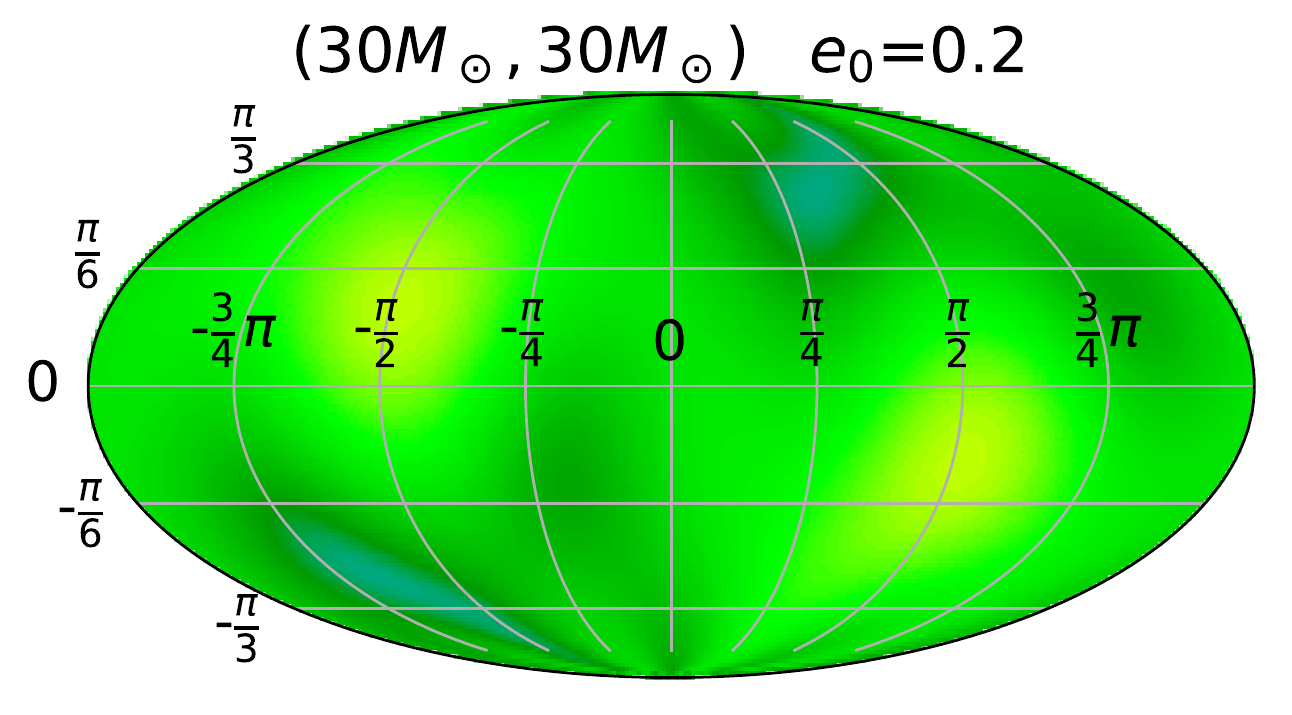}
		\includegraphics[width=\wid\textwidth]{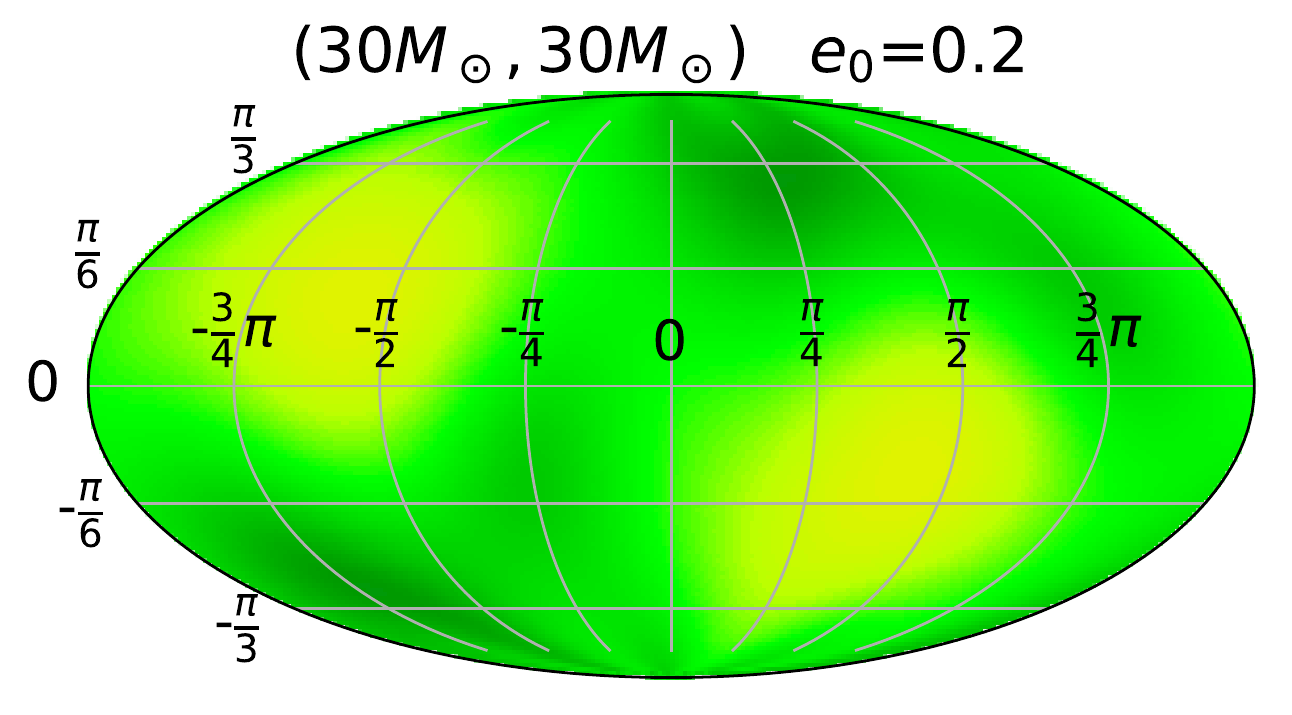}
		\includegraphics[width=\wid\textwidth]{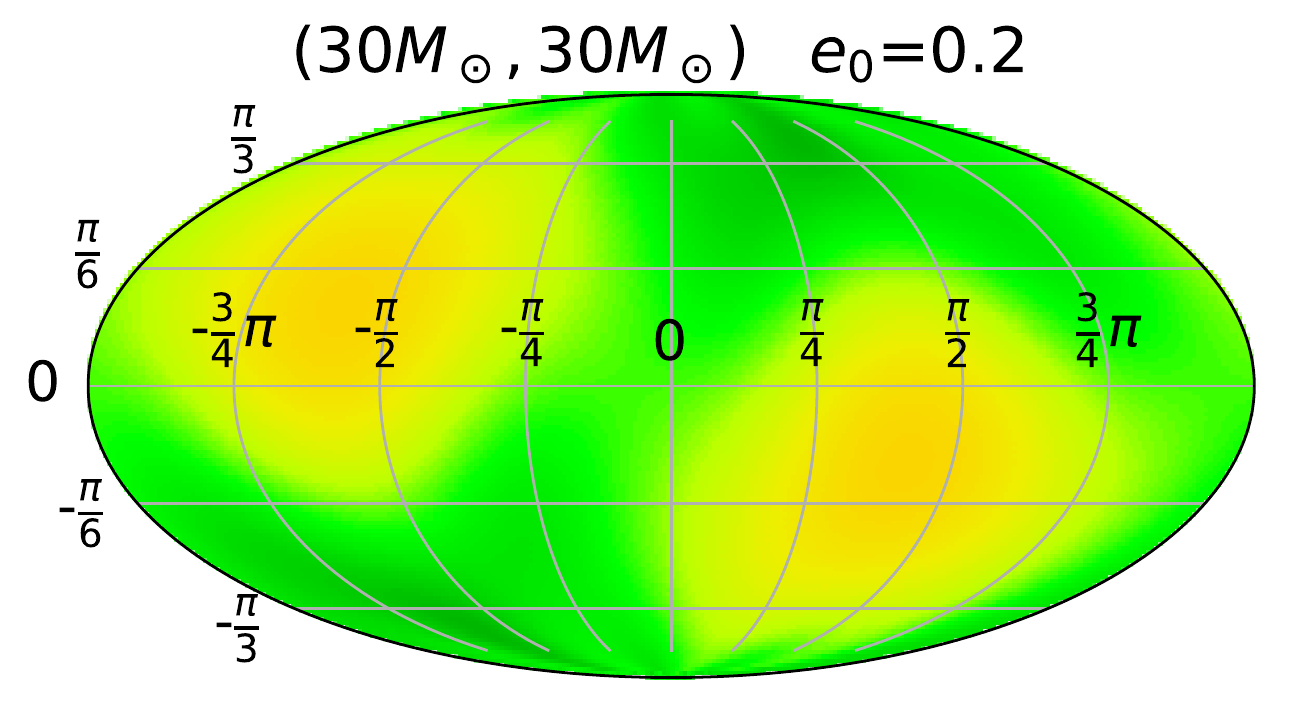}
	}
	\centerline{
		\includegraphics[width=\wid\textwidth]{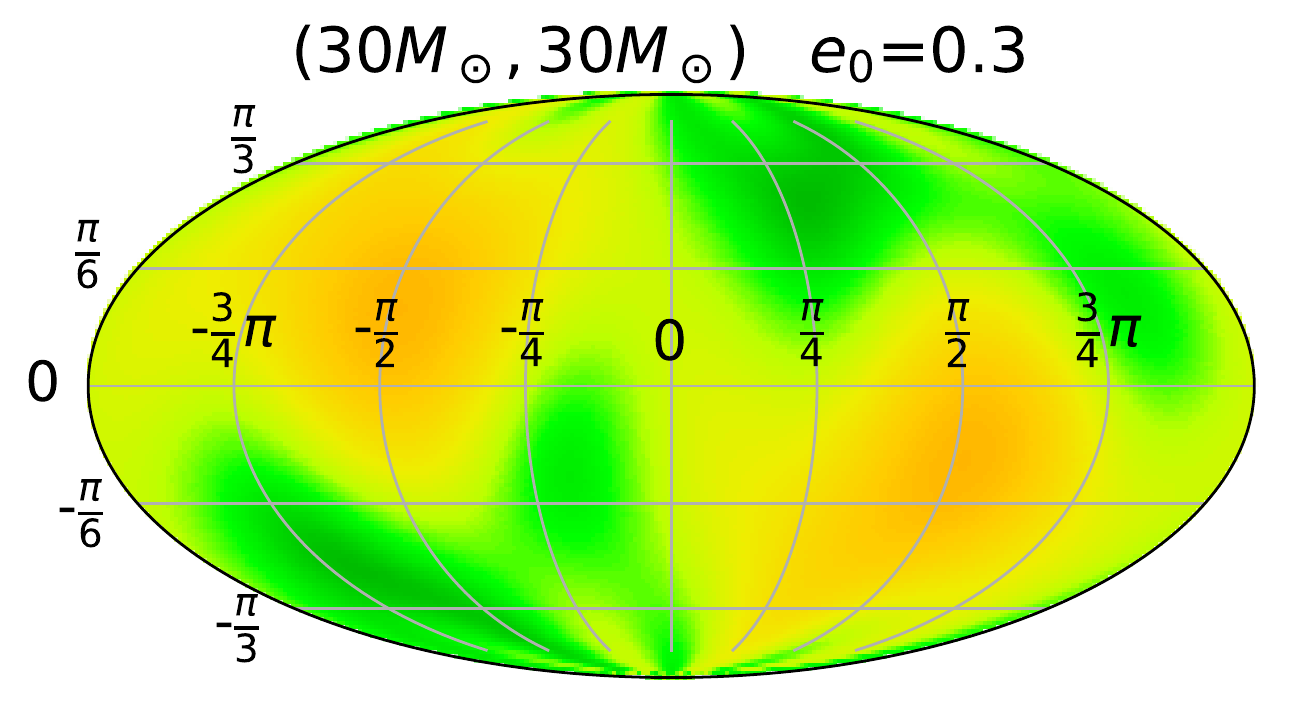}
		\includegraphics[width=\wid\textwidth]{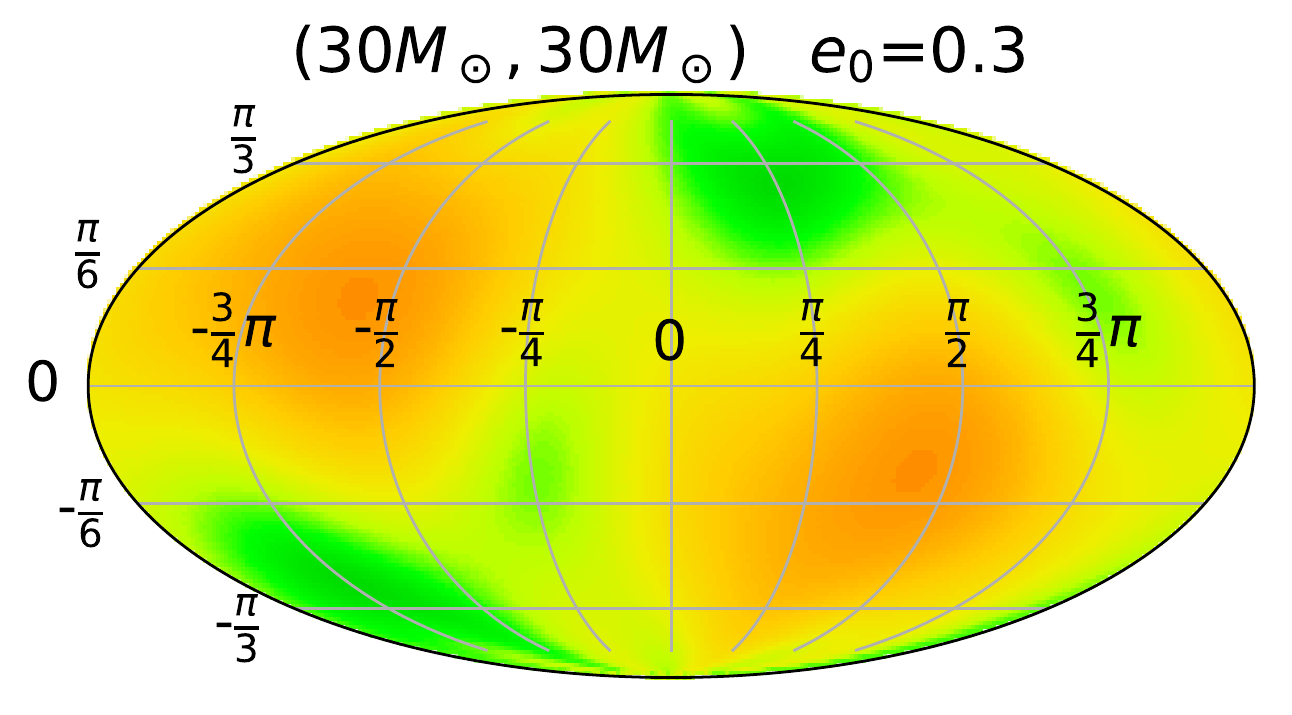}
		\includegraphics[width=\wid\textwidth]{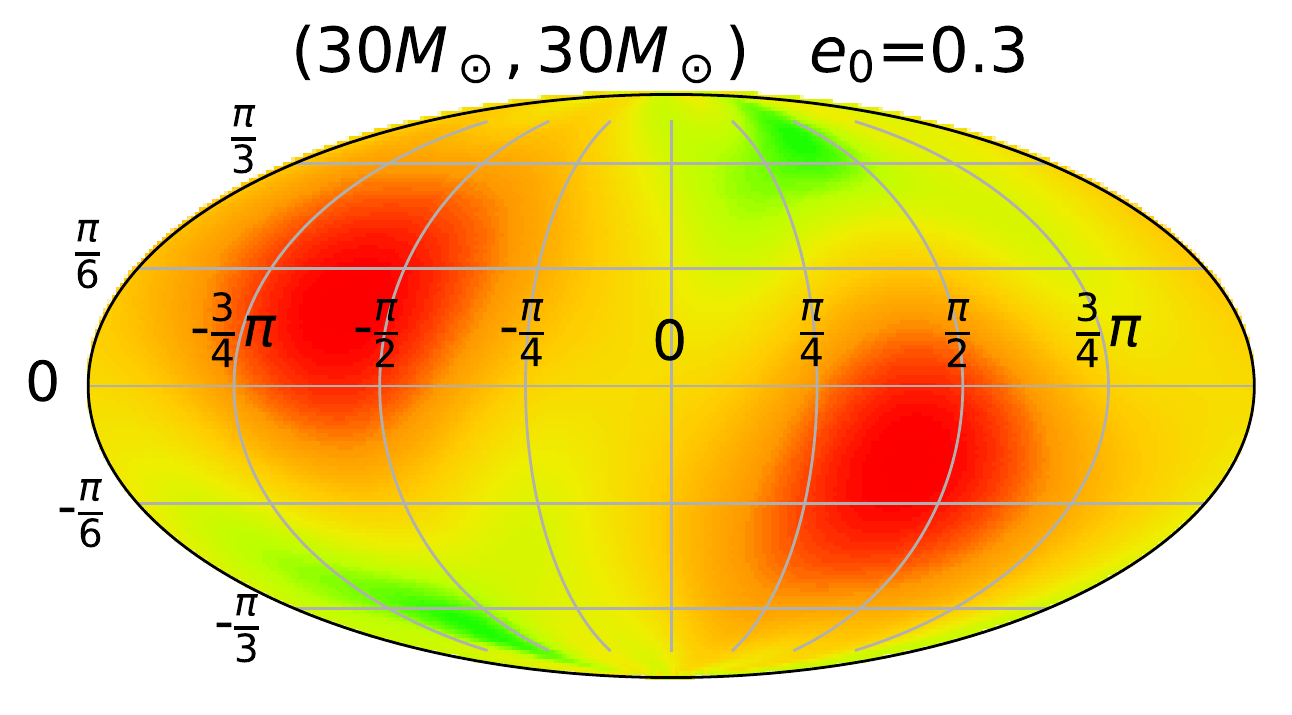}
	}
	\centerline{
		\includegraphics[width=\wid\textwidth]{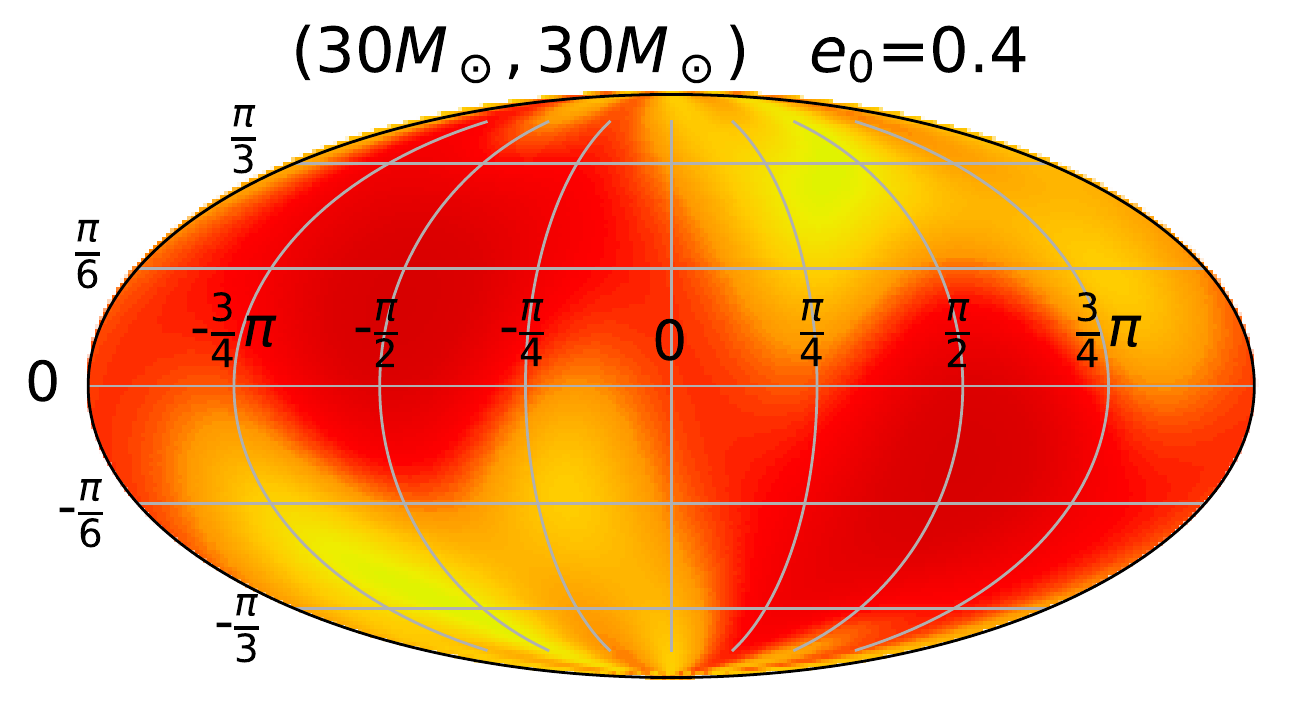}
		\includegraphics[width=\wid\textwidth]{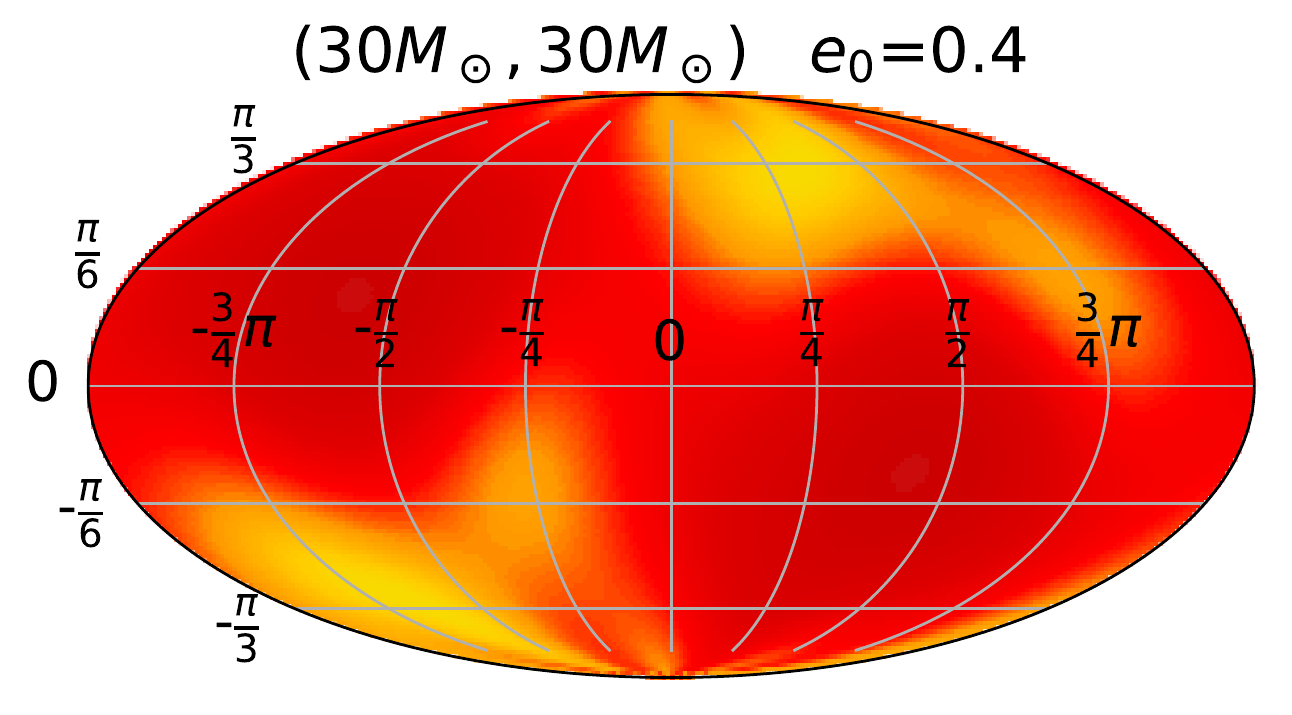}
		\includegraphics[width=\wid\textwidth]{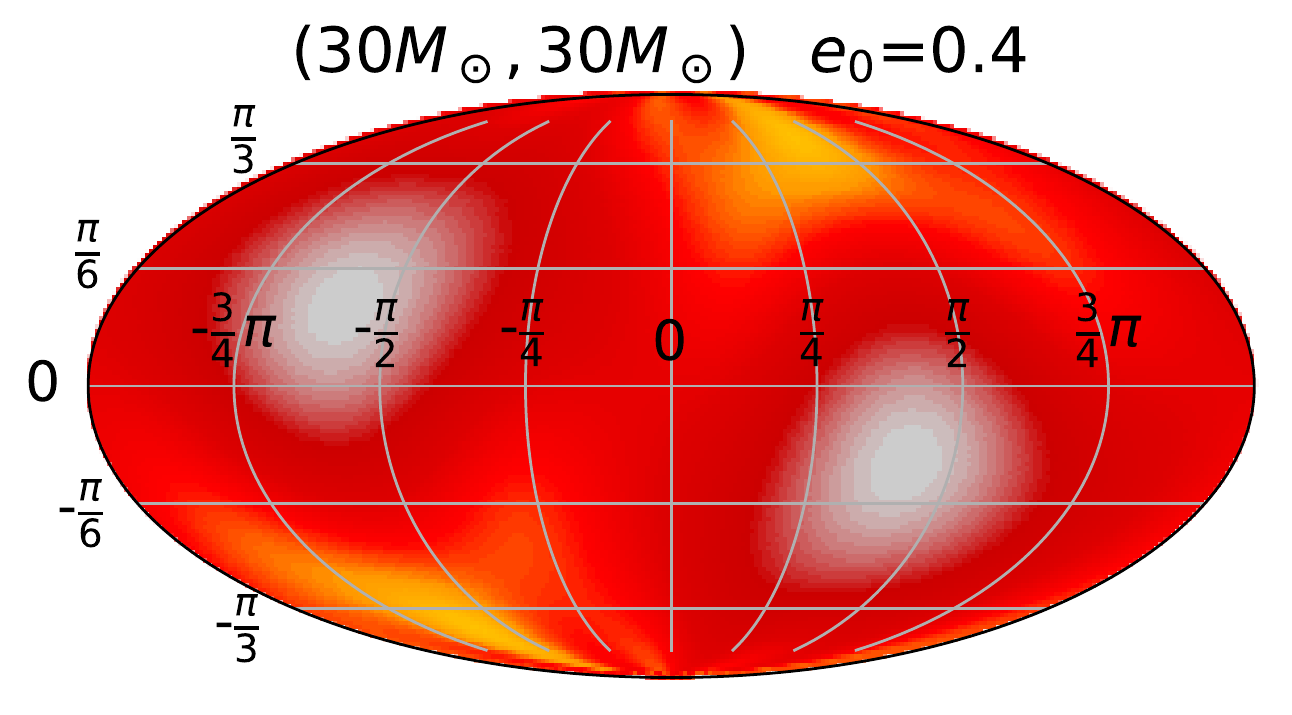}
	}
	\centerline{
		\includegraphics[width=\wid\textwidth]{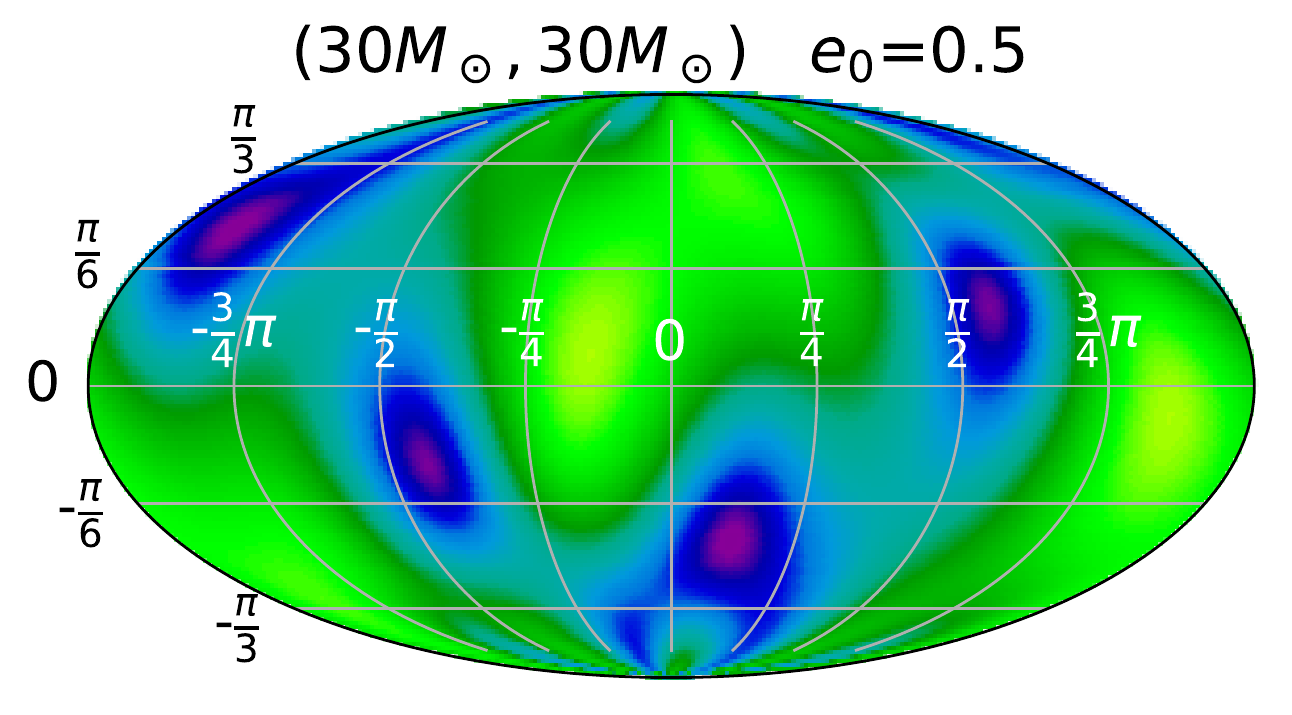}
		\includegraphics[width=\wid\textwidth]{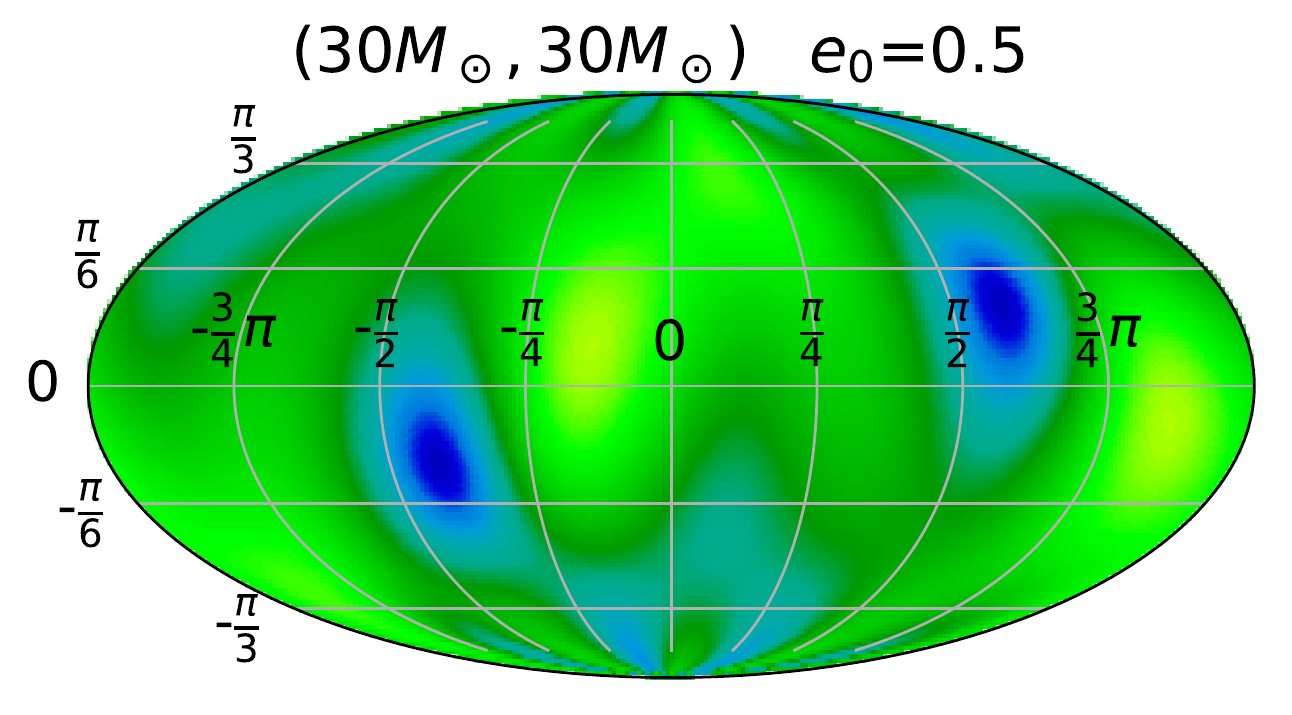}
		\includegraphics[width=\wid\textwidth]{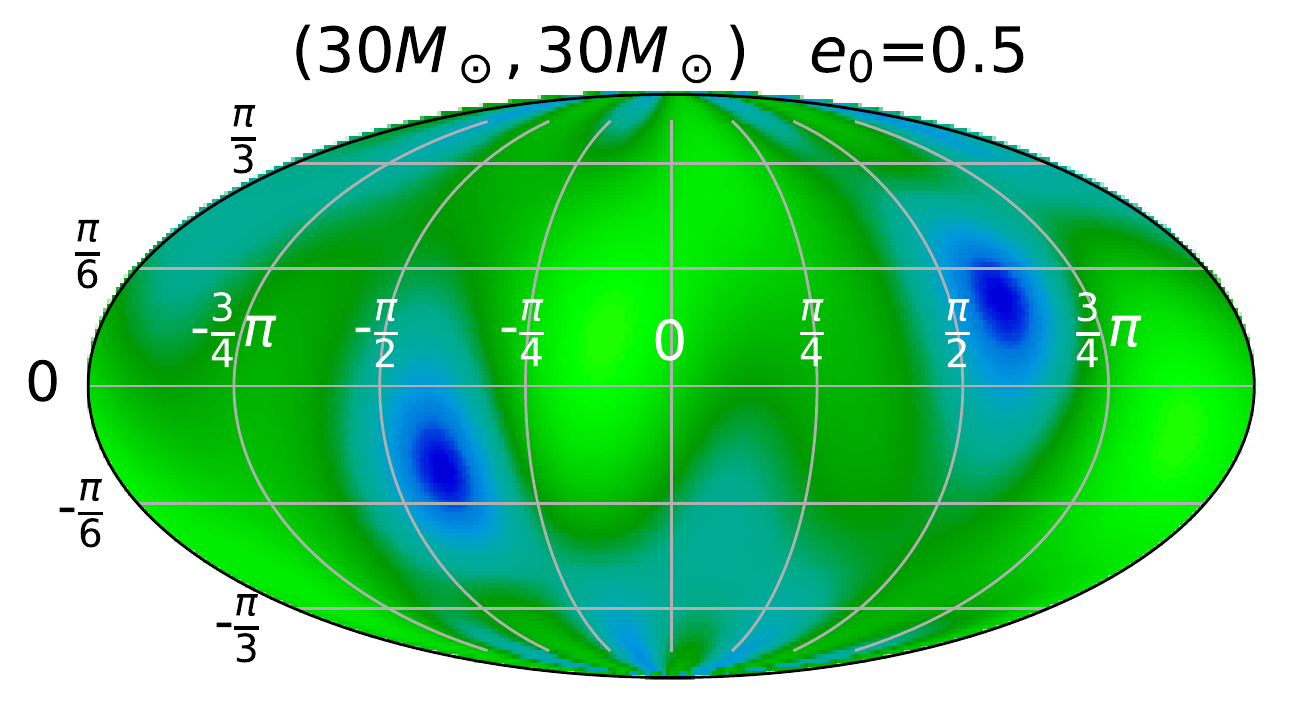}
	}
	\caption{As Figure~\ref{fig:equal_mass_LIGO_type} but now setting the binary inclination angle to 0.} 
	\label{fig:ligo_equal_opt}
\end{figure*}

\begin{figure*}
	\centerline{
		\includegraphics[width=\textwidth]{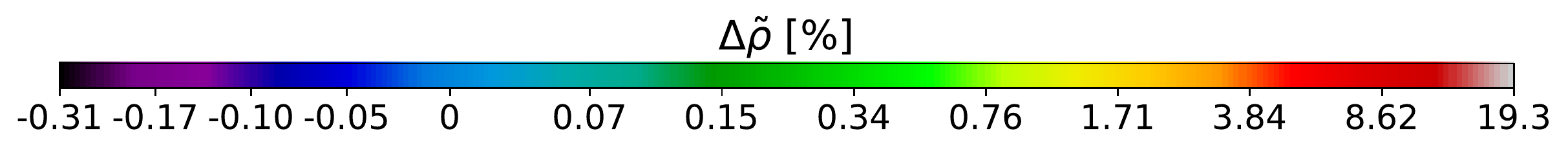}
	}
	\centerline{
		\includegraphics[width=\wid\textwidth]{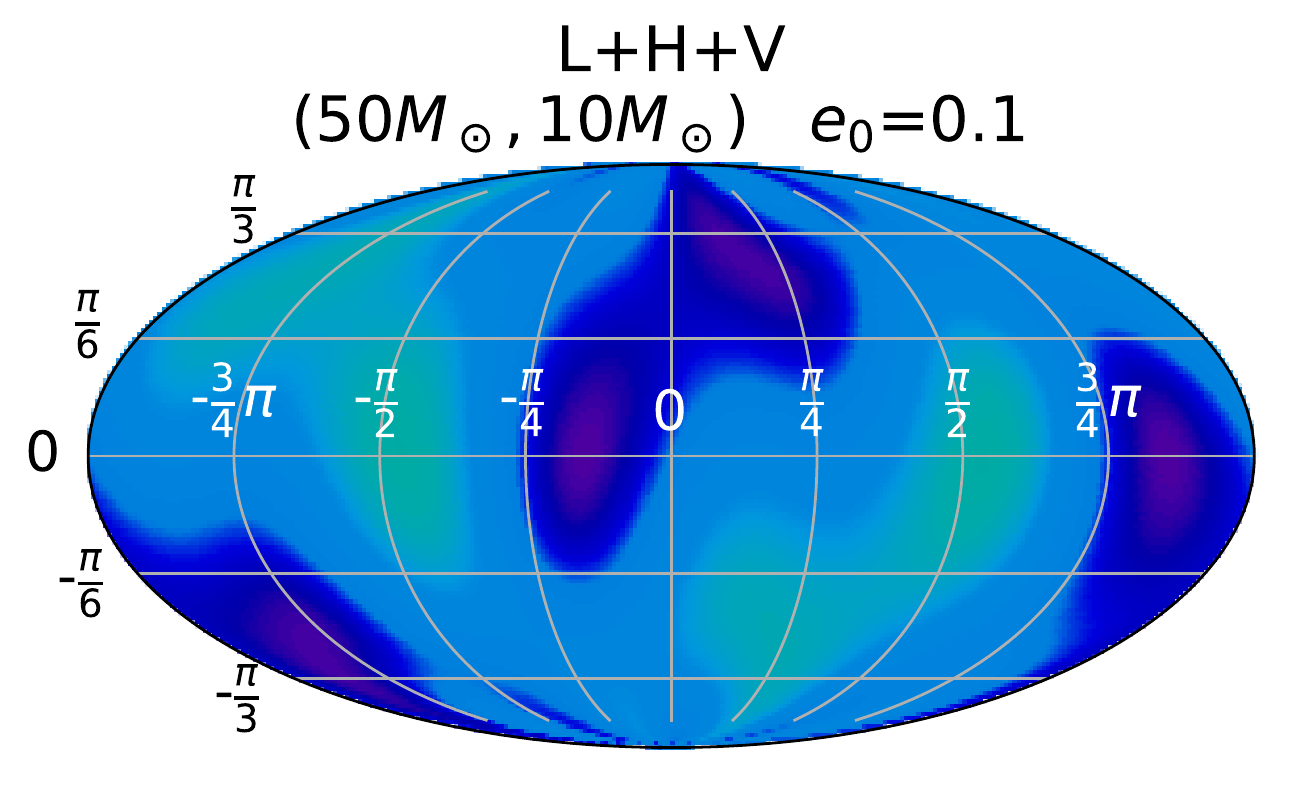}
		\includegraphics[width=\wid\textwidth]{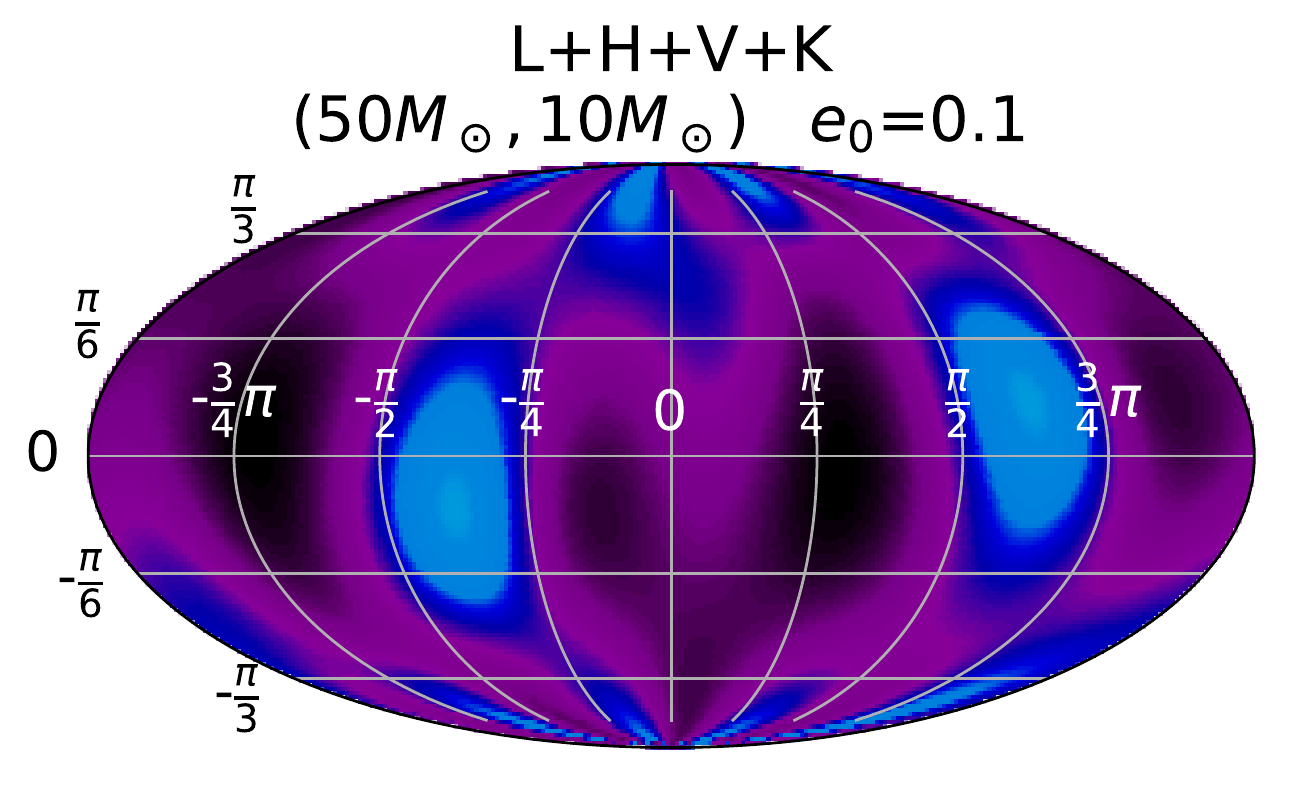}
		\includegraphics[width=\wid\textwidth]{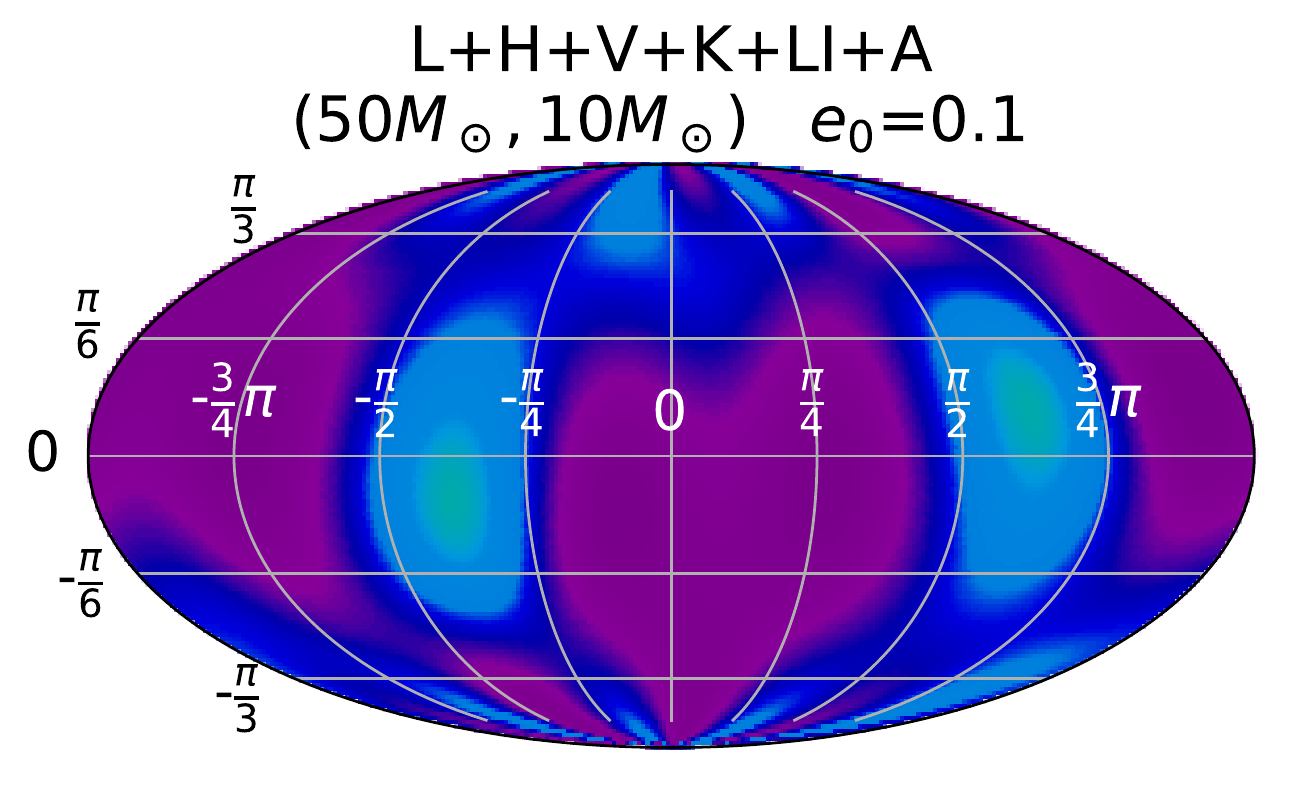}
	}
	\centerline{
		\includegraphics[width=\wid\textwidth]{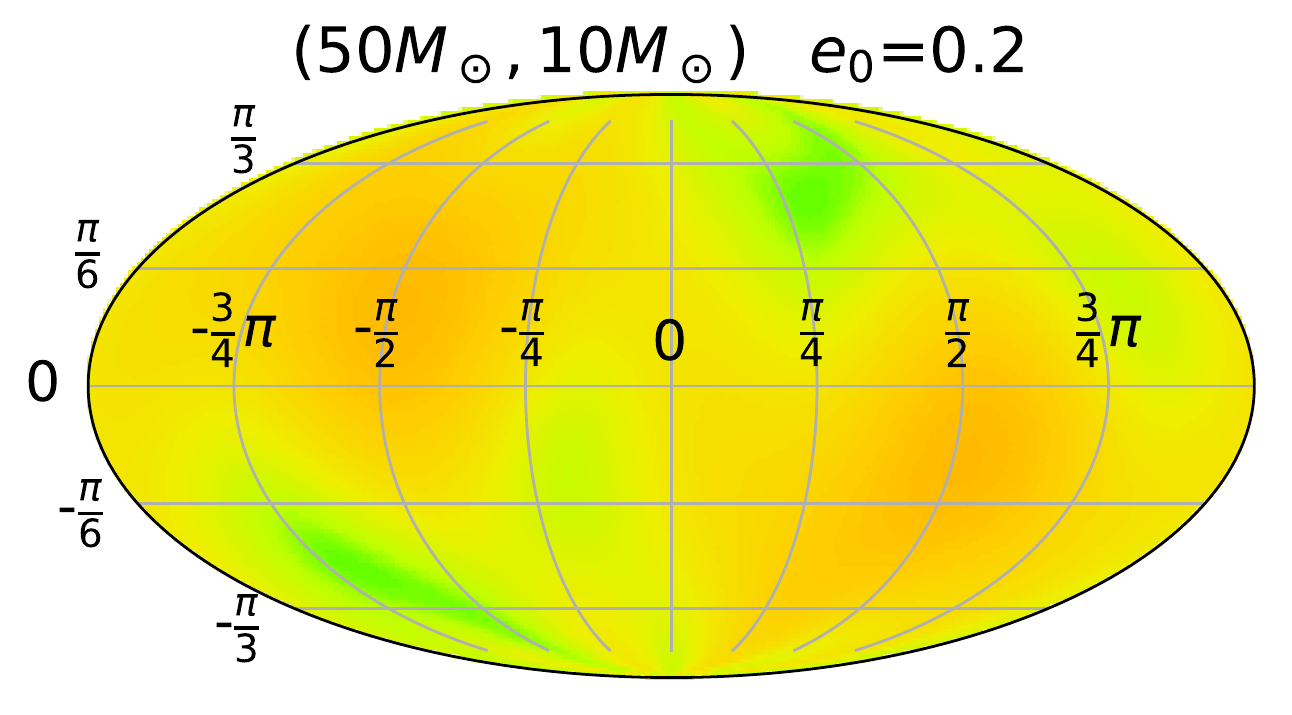}
		\includegraphics[width=\wid\textwidth]{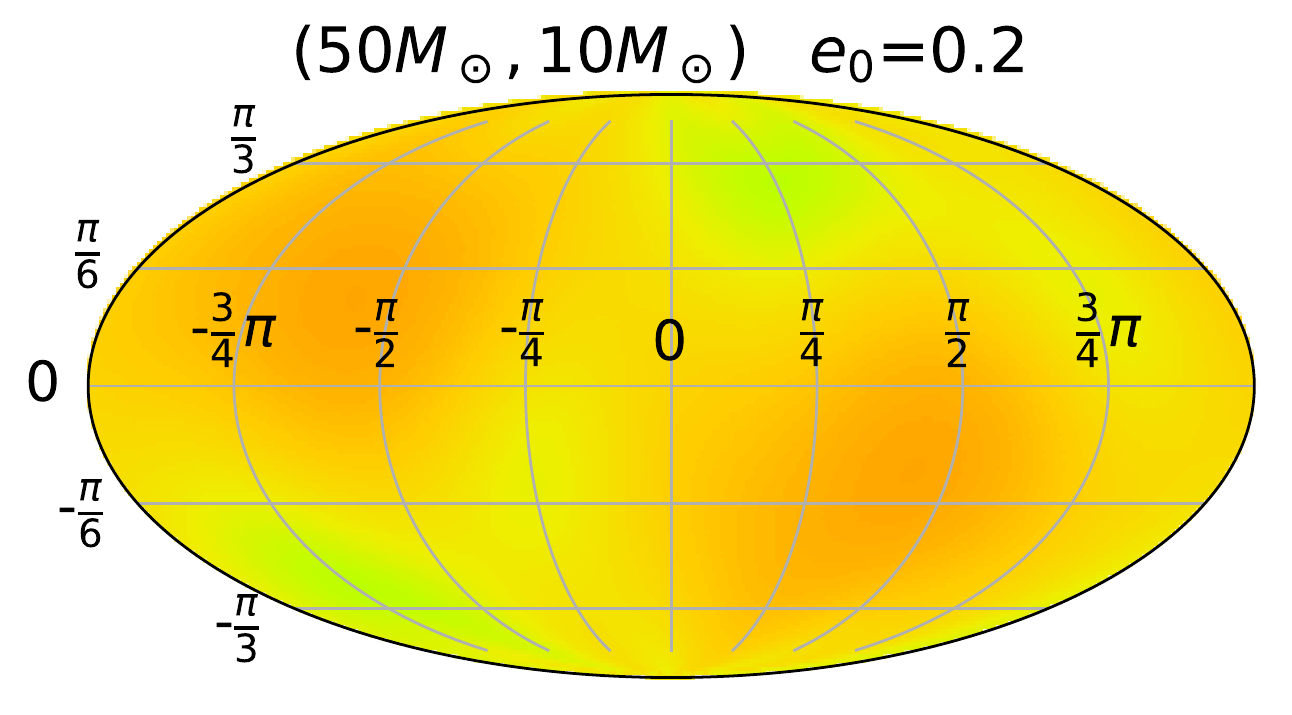}
		\includegraphics[width=\wid\textwidth]{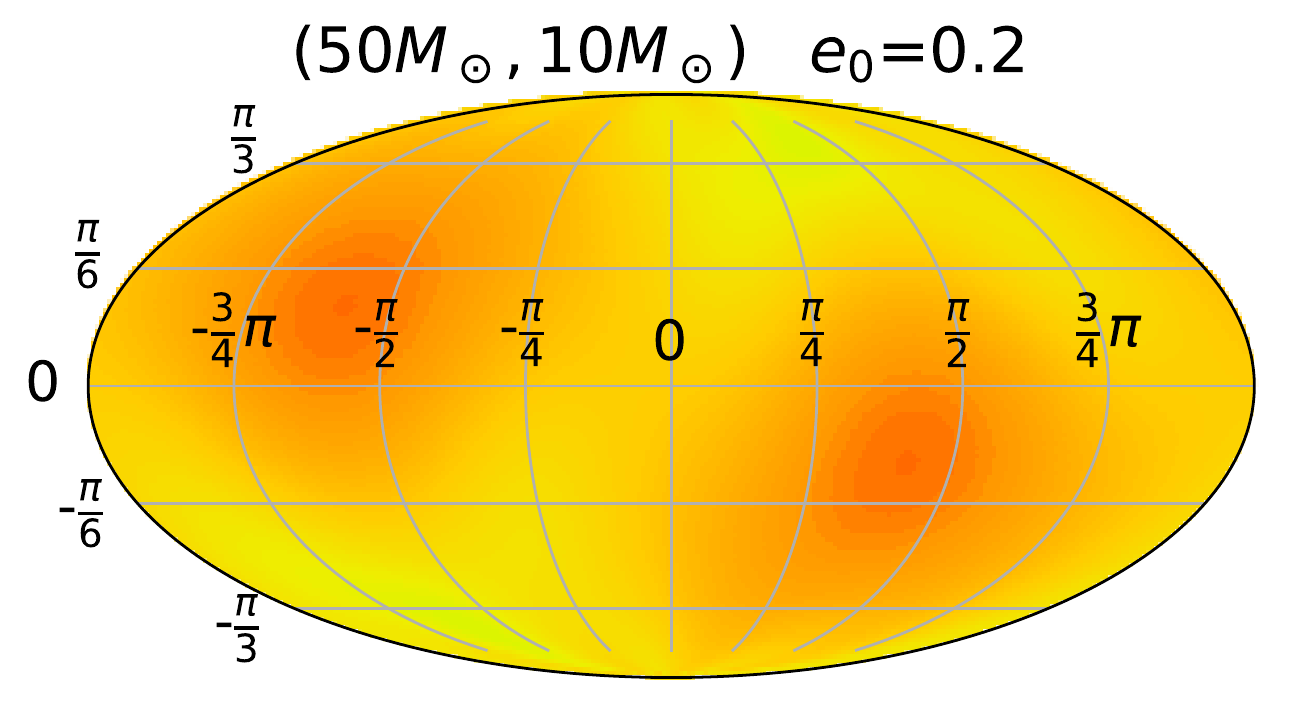}
	}
	\centerline{
		\includegraphics[width=\wid\textwidth]{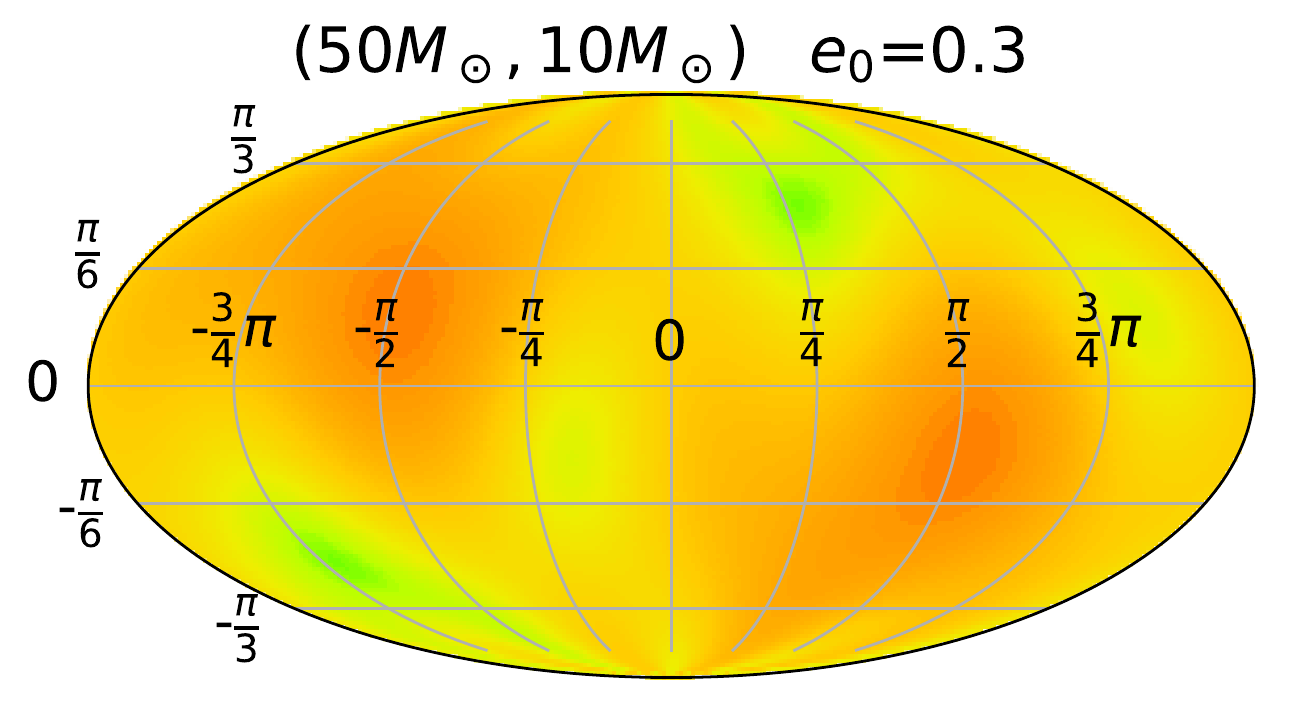}
		\includegraphics[width=\wid\textwidth]{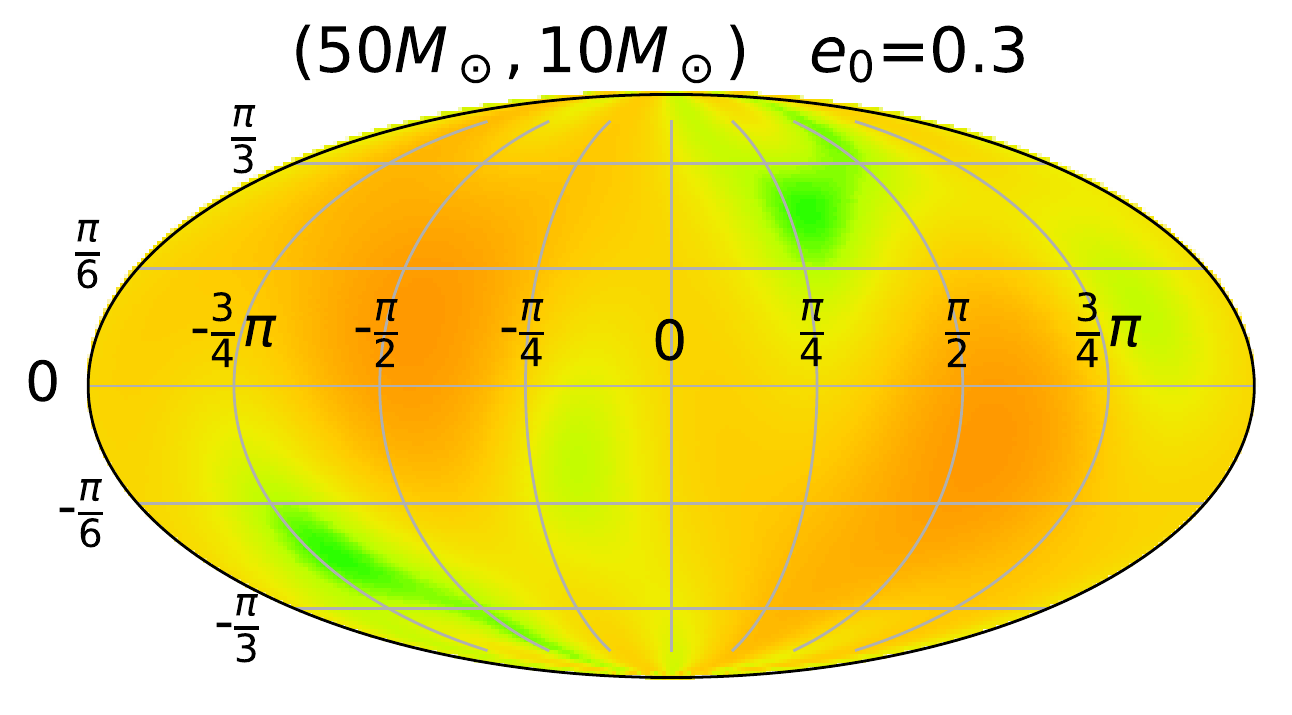}
		\includegraphics[width=\wid\textwidth]{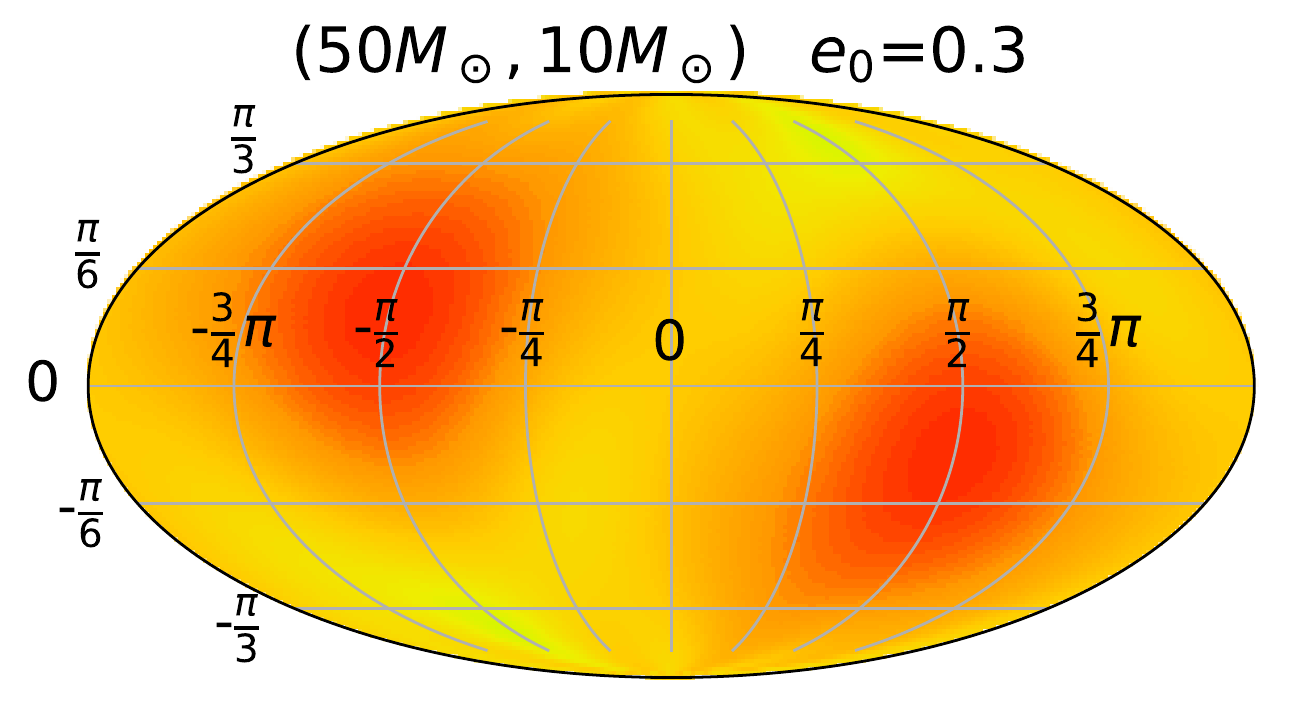}
	}
	\centerline{
		\includegraphics[width=\wid\textwidth]{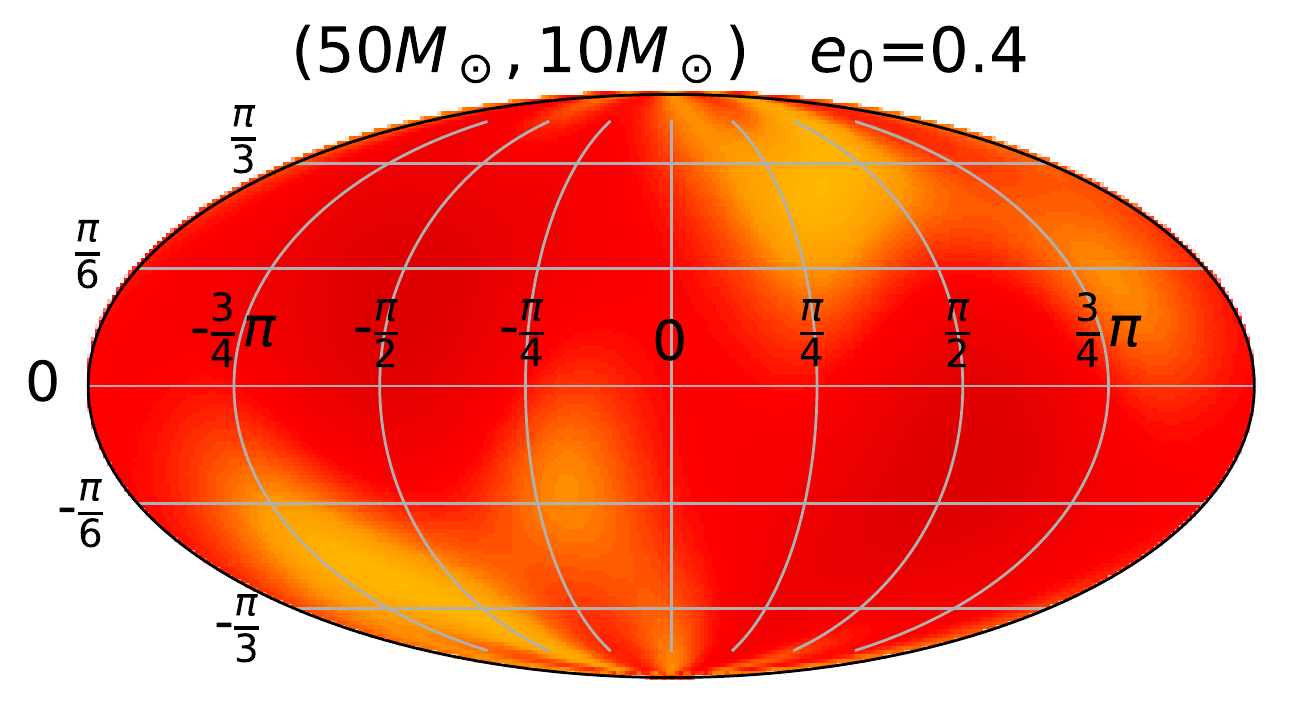}
		\includegraphics[width=\wid\textwidth]{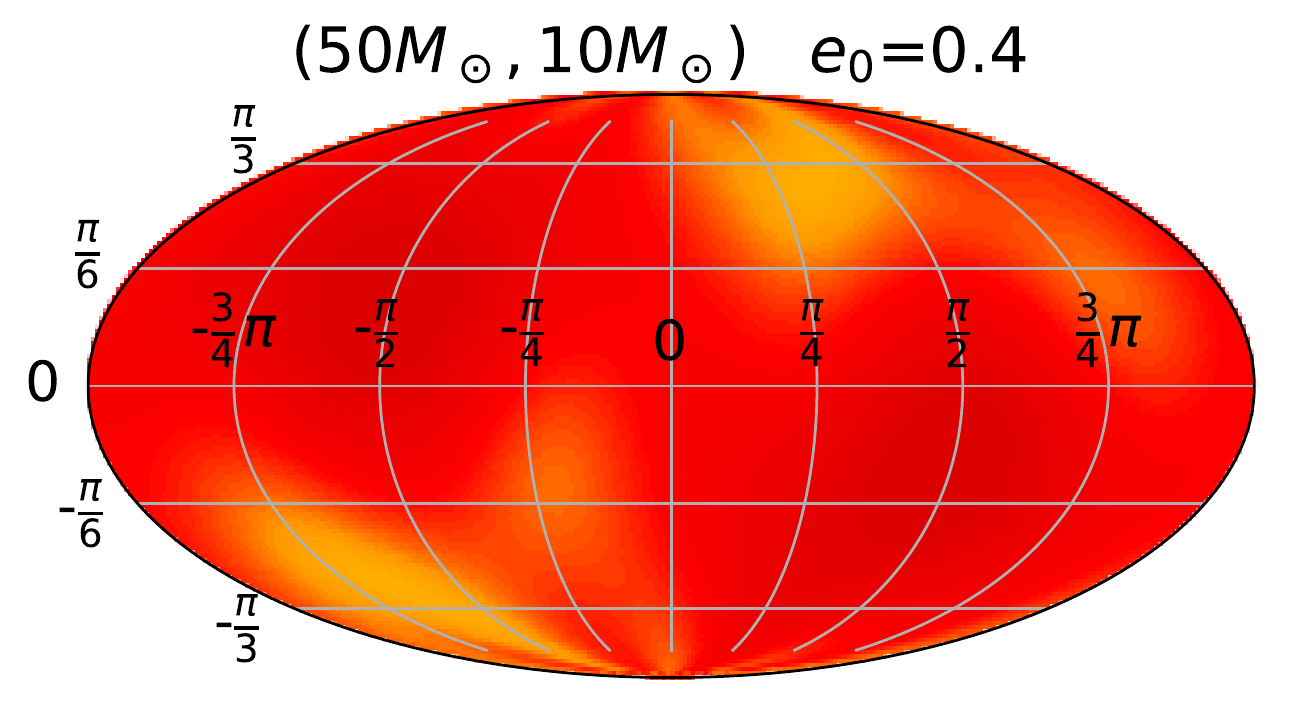}
		\includegraphics[width=\wid\textwidth]{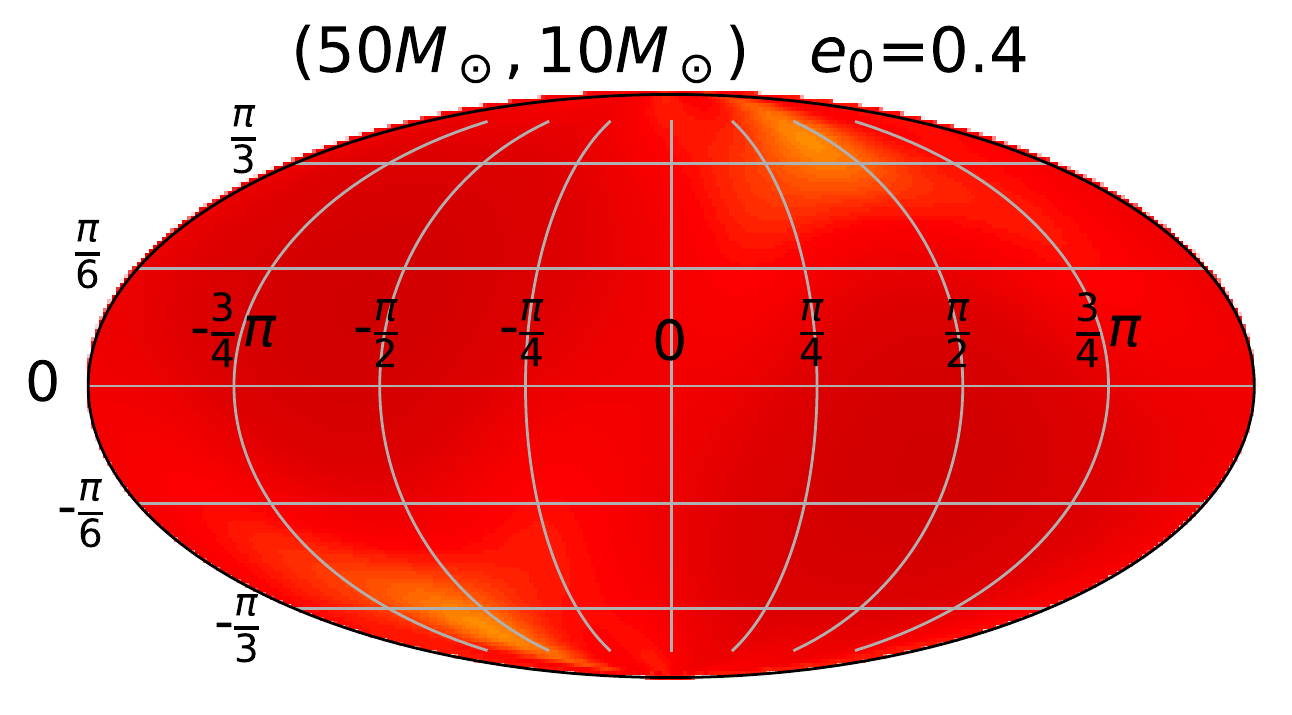}
	}
	\centerline{
		\includegraphics[width=\wid\textwidth]{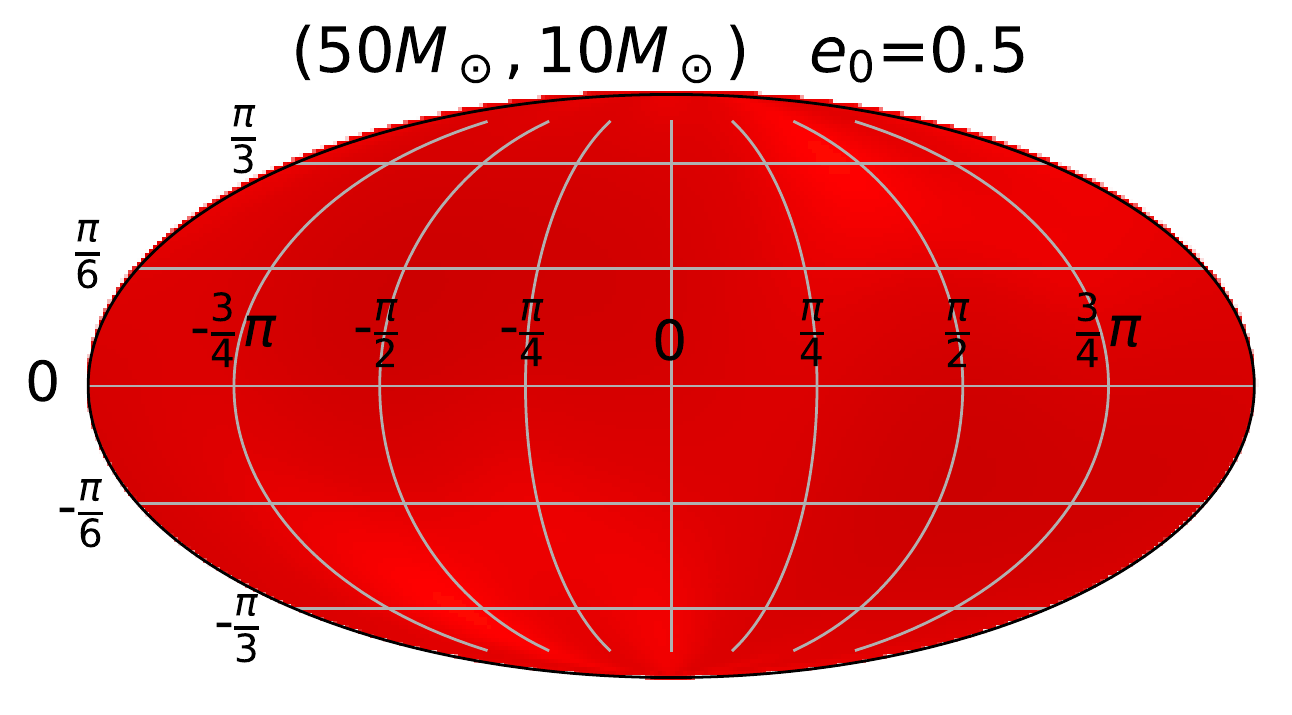}
		\includegraphics[width=\wid\textwidth]{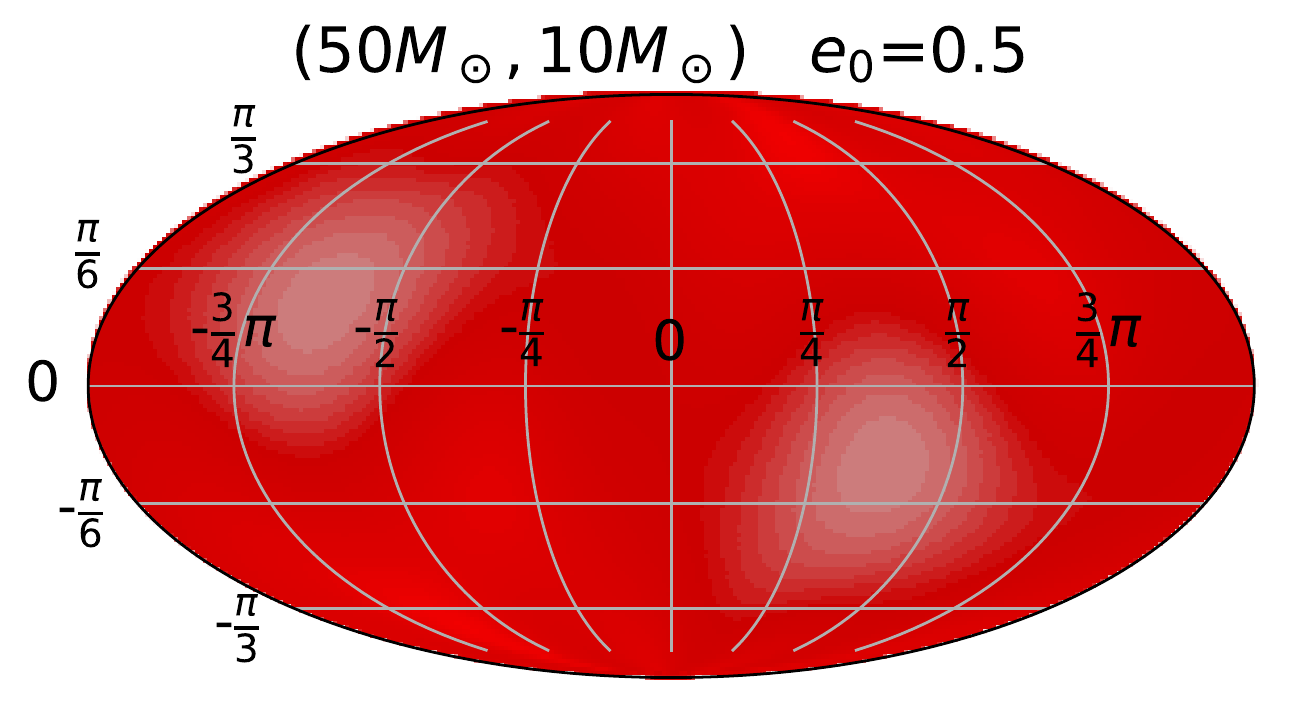}
		\includegraphics[width=\wid\textwidth]{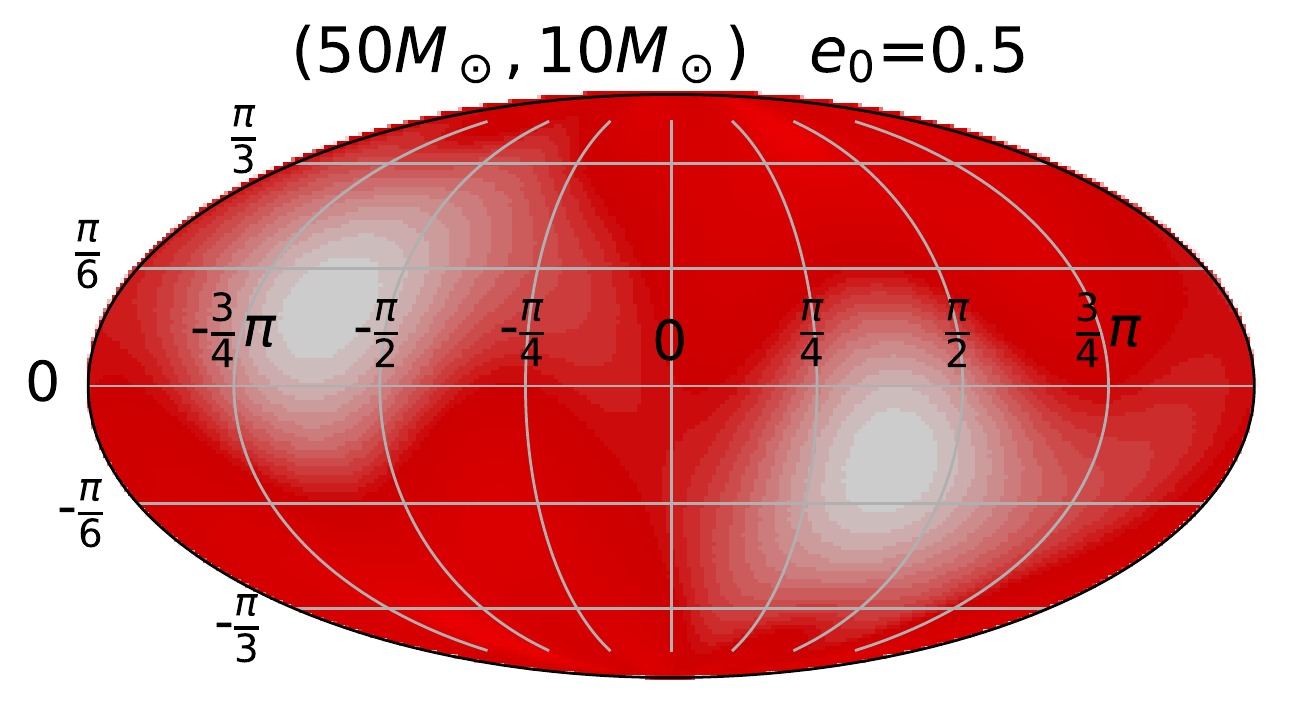}
	}
	\caption{As Figure~\ref{fig:ligo_equal_opt} but now for component masses \((50\msun,\,10\msun)\).} 
	\label{fig:ligo_five_opt}
\end{figure*}


\begin{figure*}
	\centerline{
		\includegraphics[width=\textwidth]{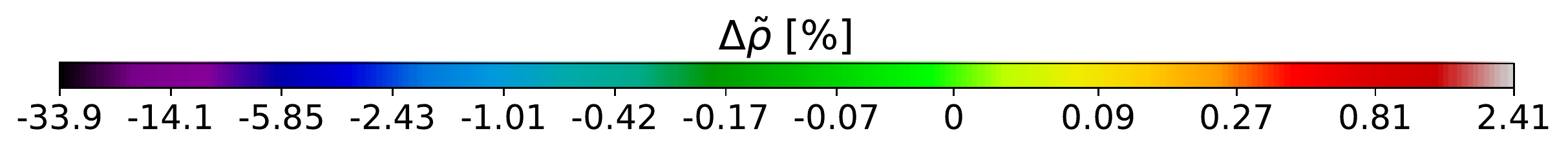}
	}
	\centerline{
		\includegraphics[width=\wid\textwidth]{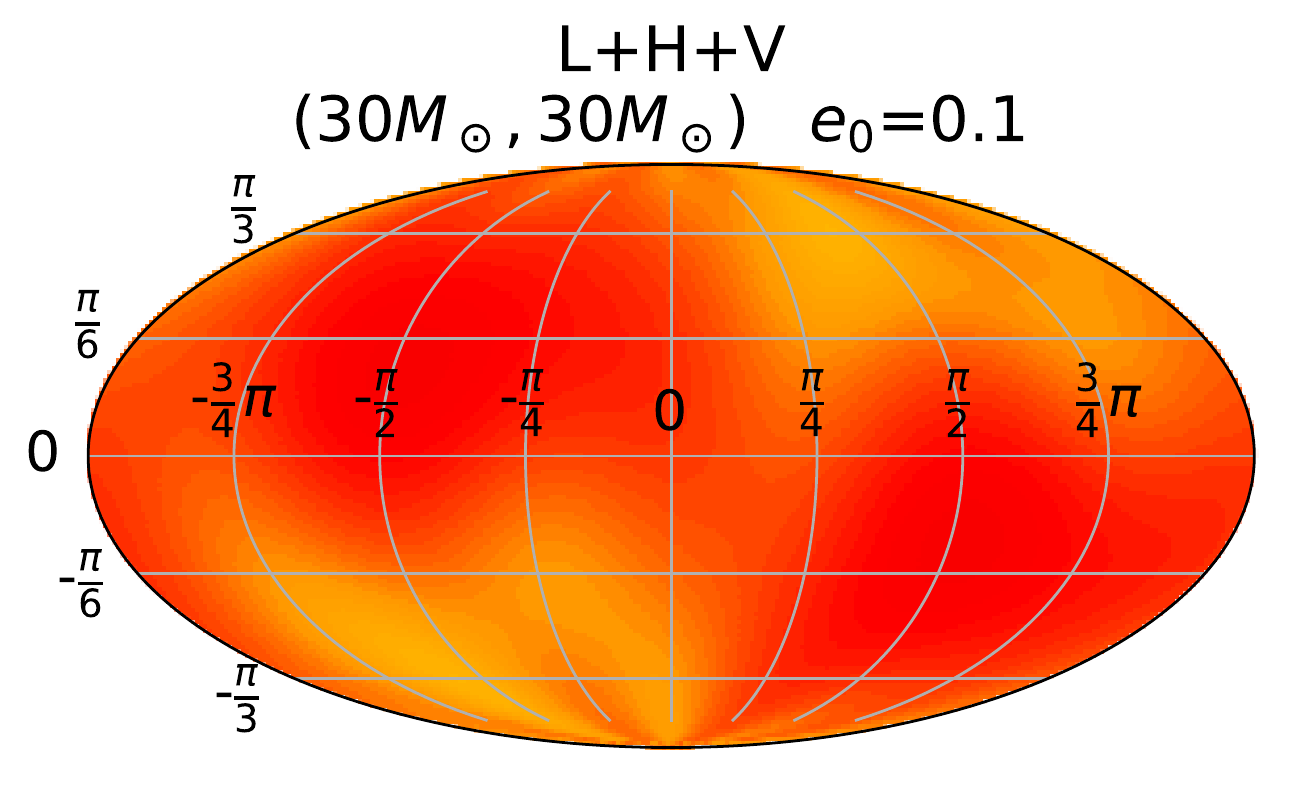}
		\includegraphics[width=\wid\textwidth]{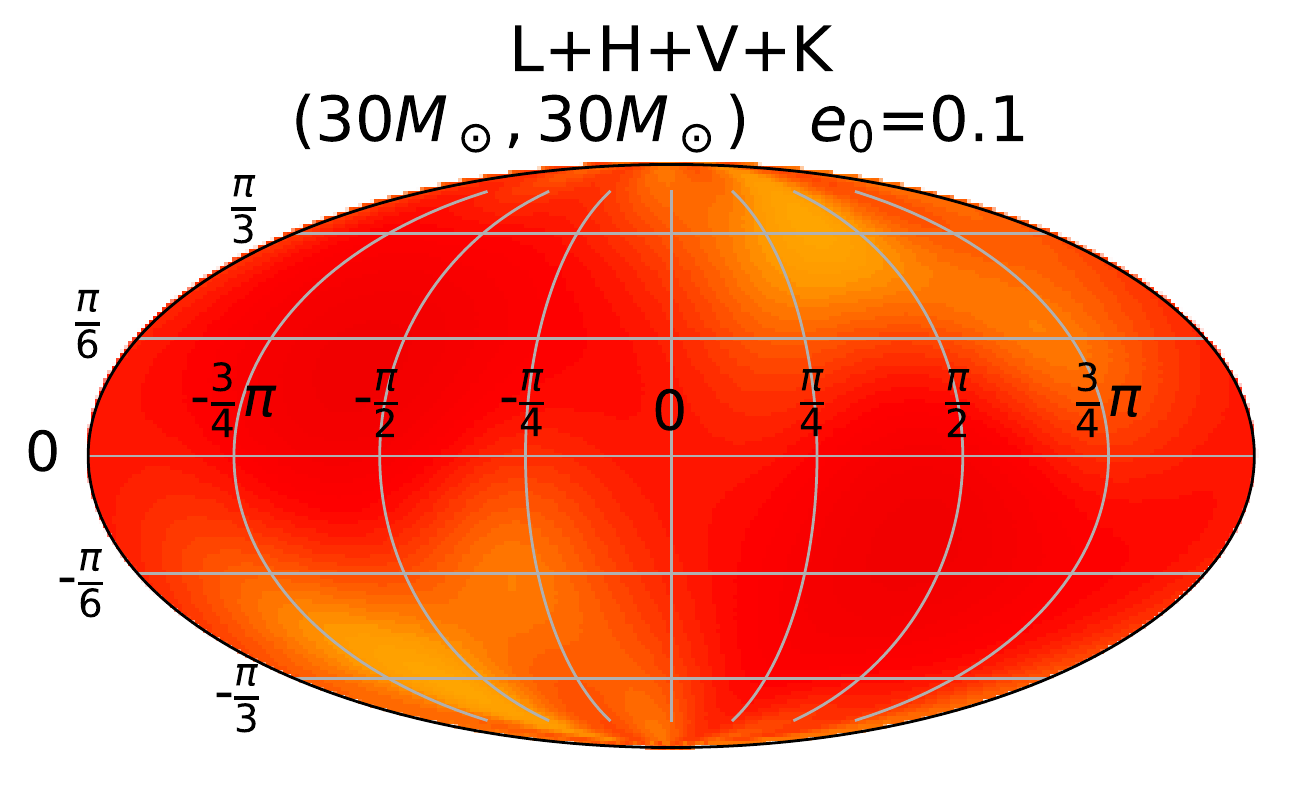}
		\includegraphics[width=\wid\textwidth]{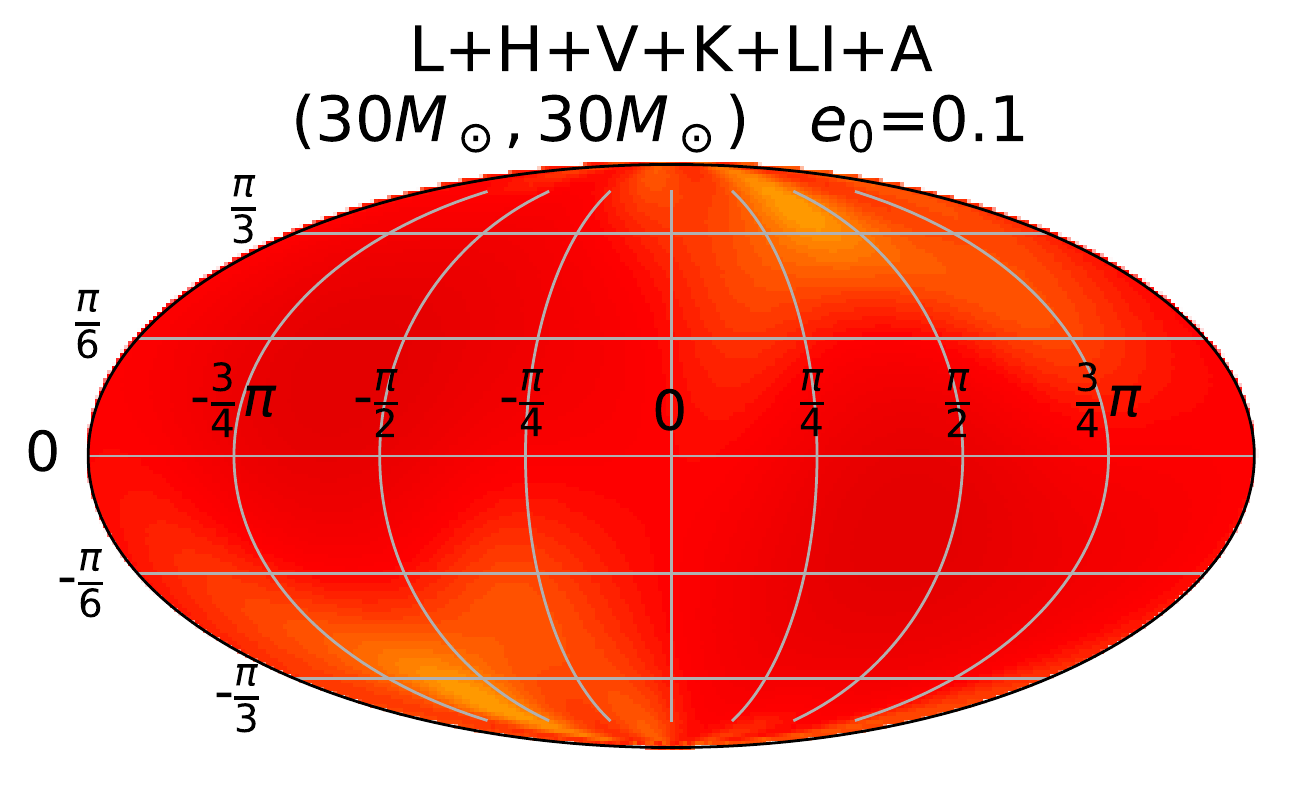}
	}
	\centerline{
		\includegraphics[width=\wid\textwidth]{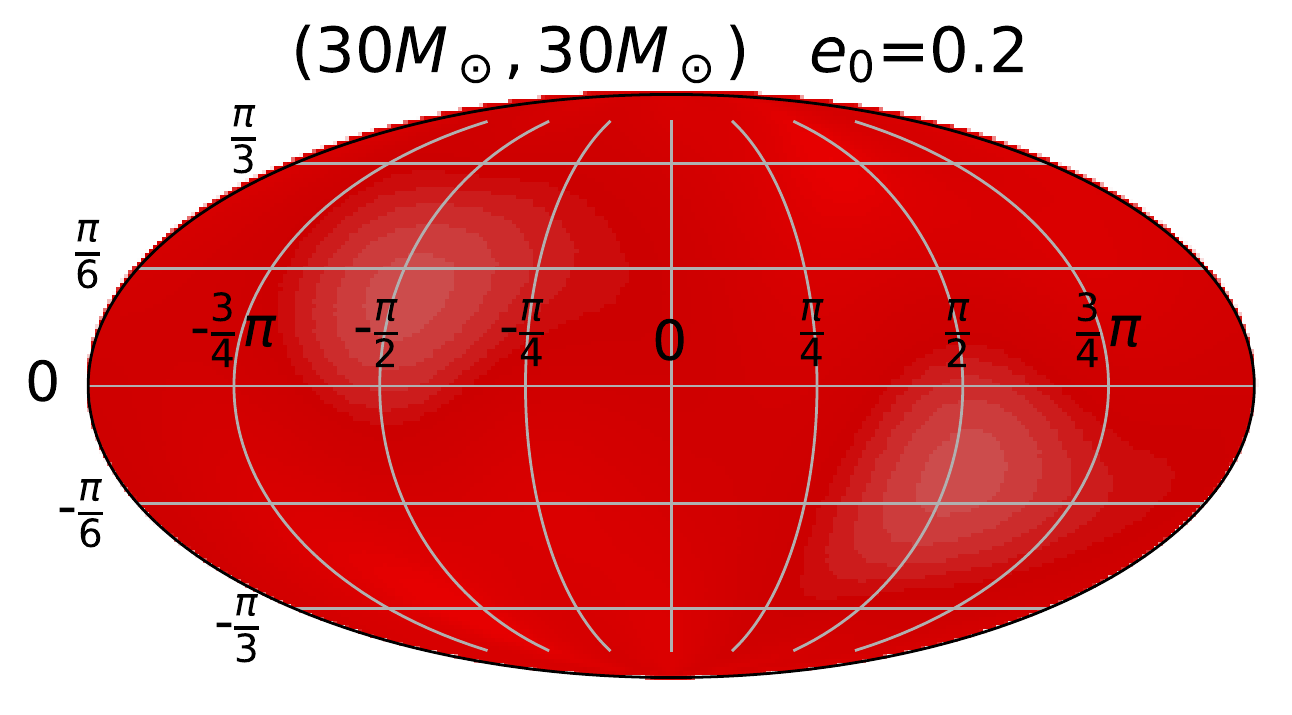}
		\includegraphics[width=\wid\textwidth]{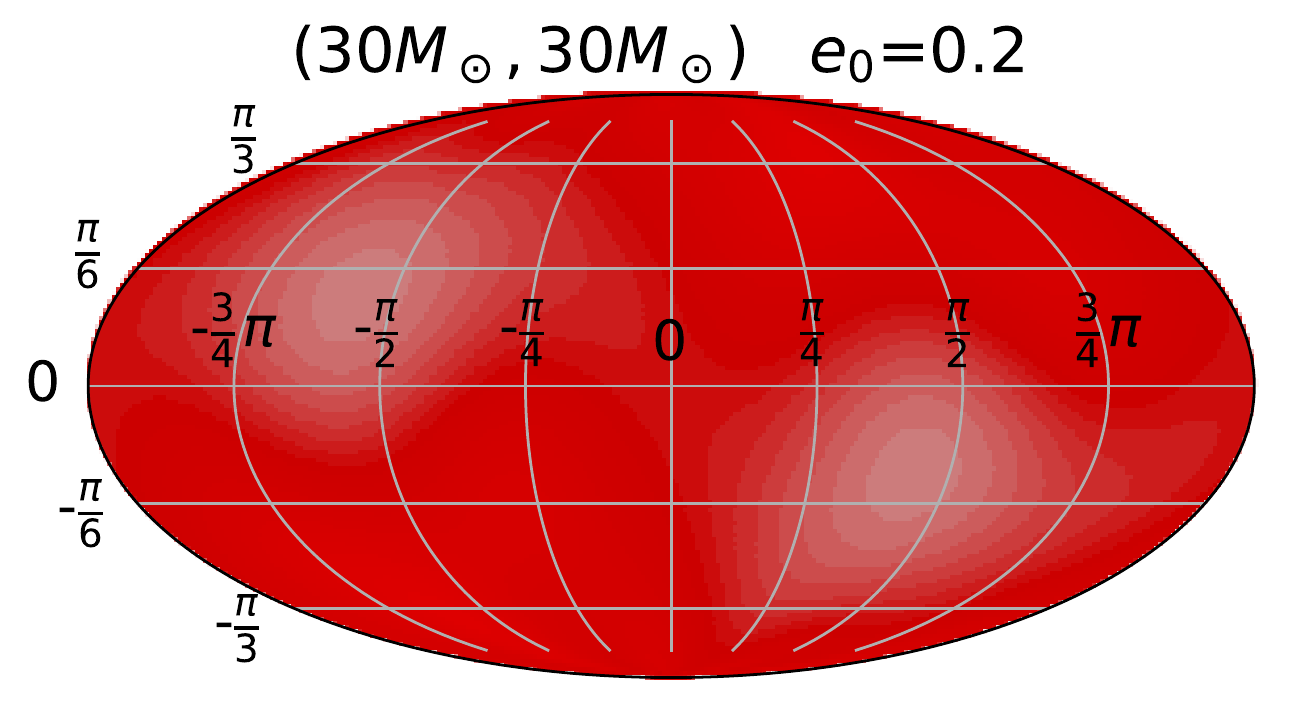}
		\includegraphics[width=\wid\textwidth]{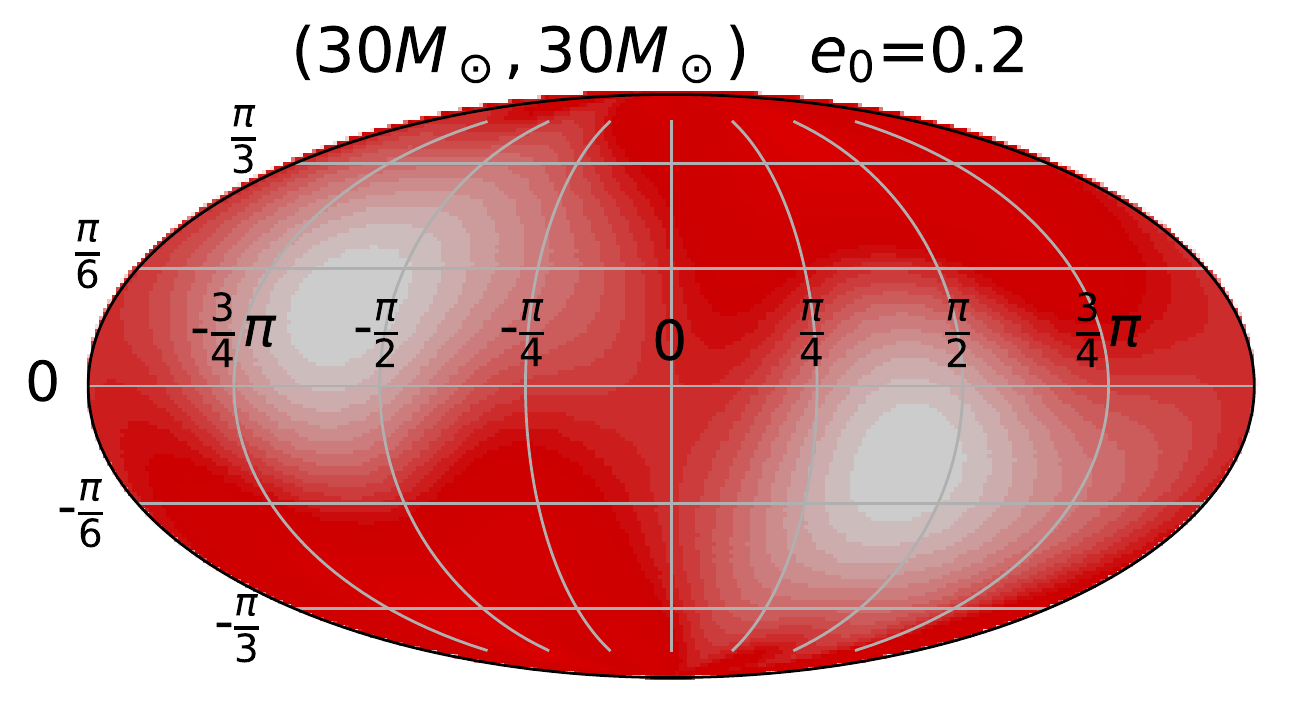}
	}
	\centerline{
		\includegraphics[width=\wid\textwidth]{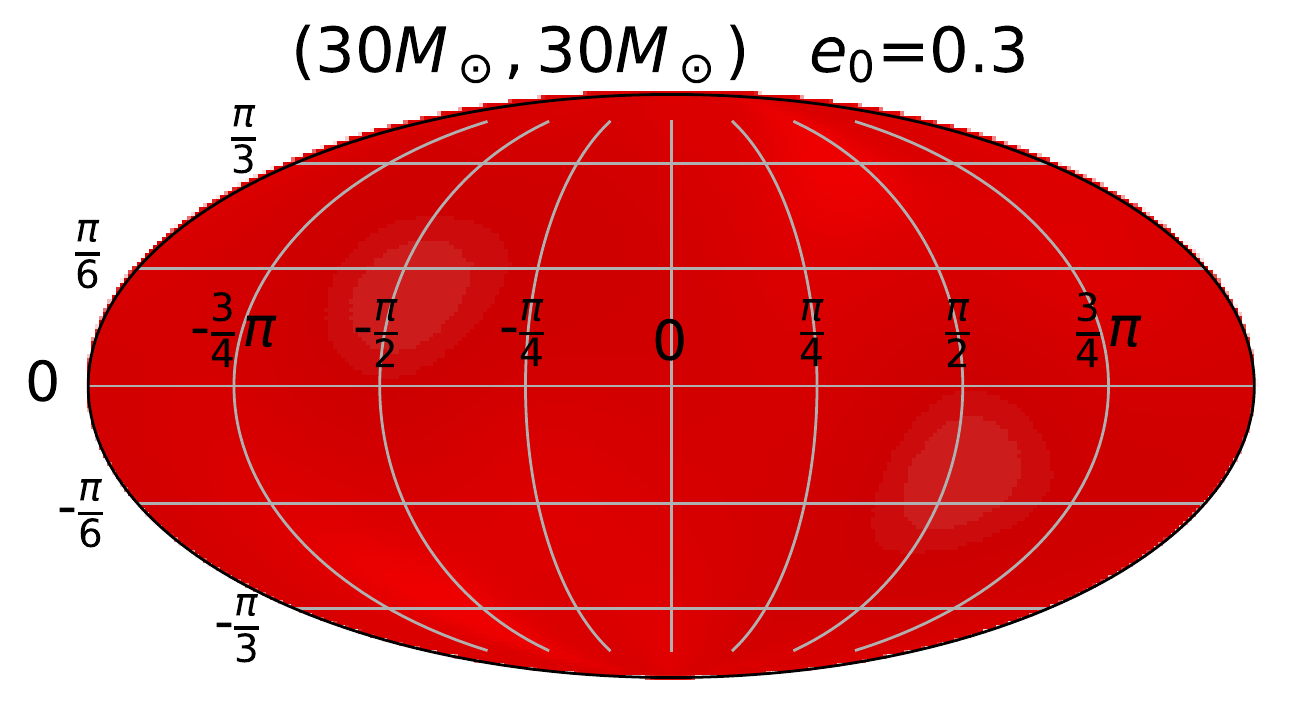}
		\includegraphics[width=\wid\textwidth]{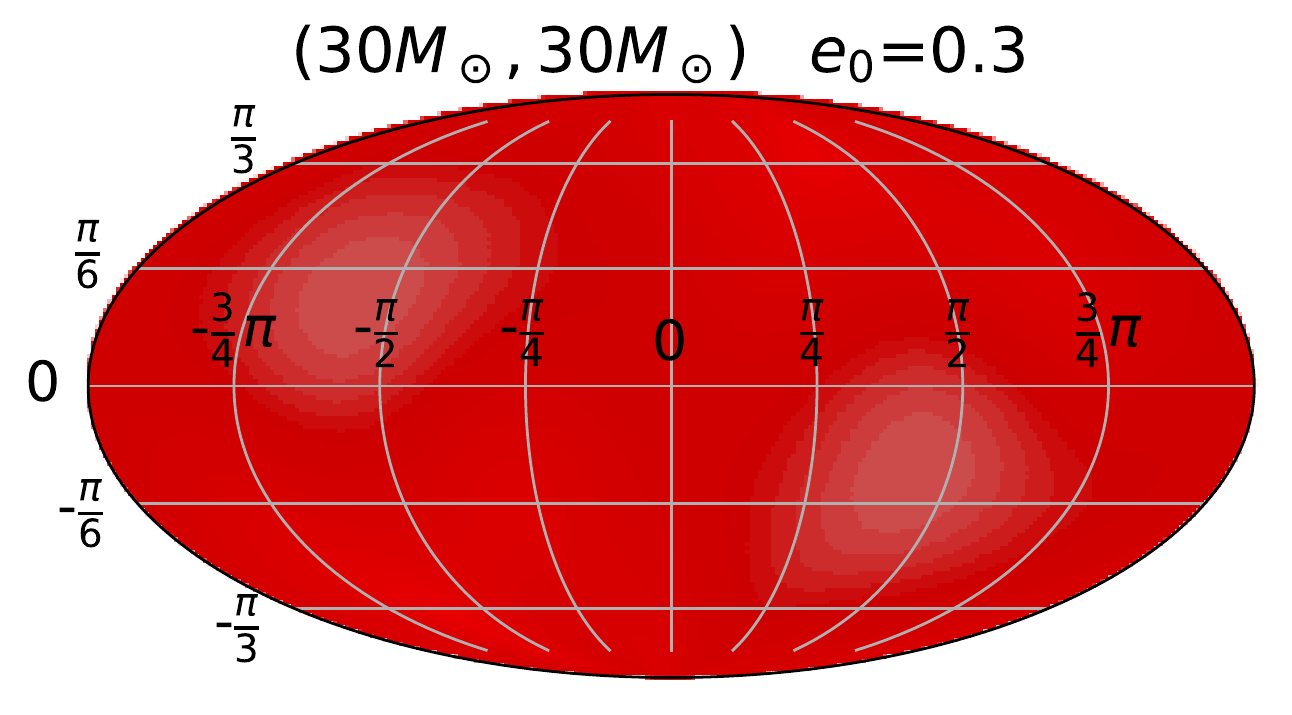}
		\includegraphics[width=\wid\textwidth]{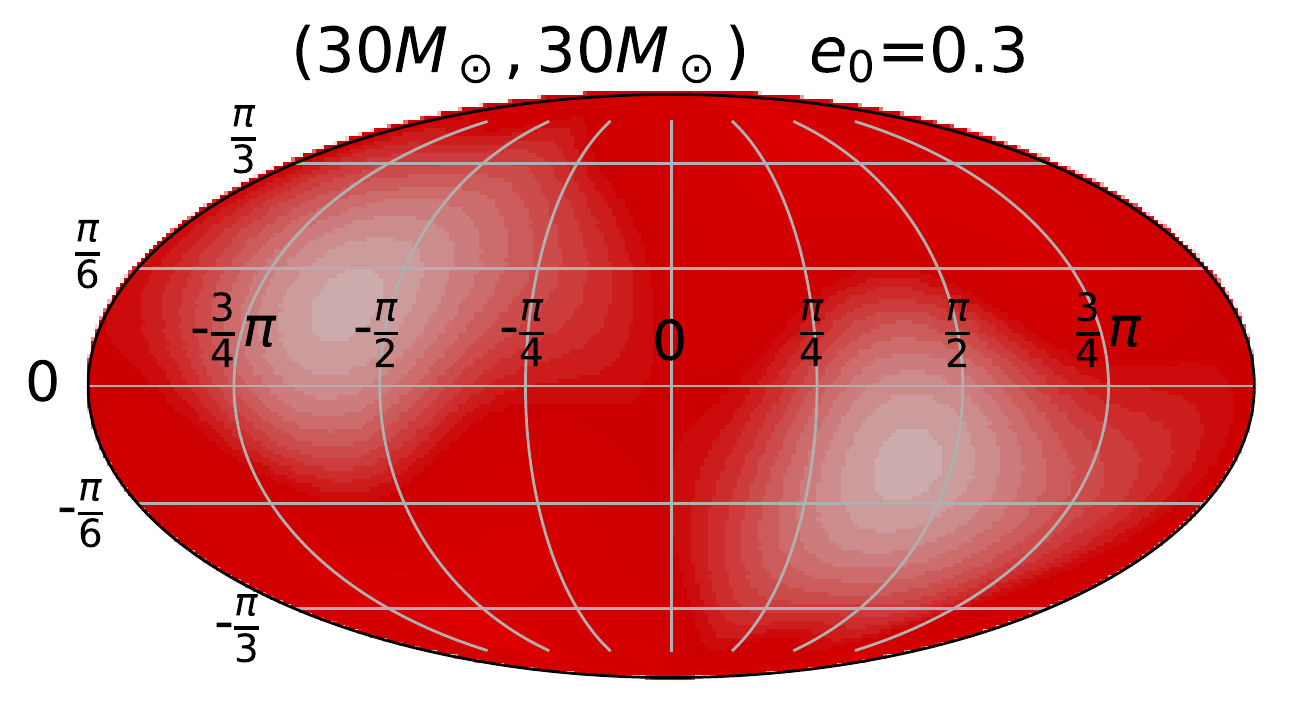}
	}
	\centerline{
		\includegraphics[width=\wid\textwidth]{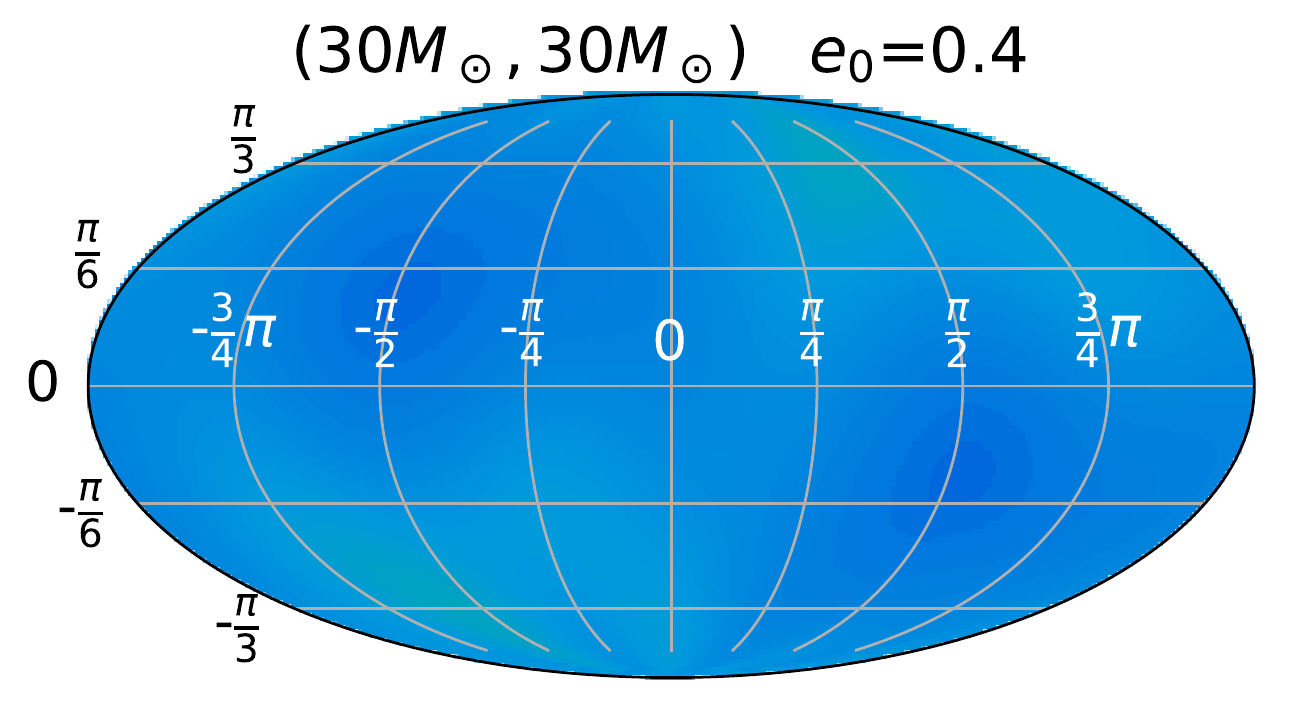}
		\includegraphics[width=\wid\textwidth]{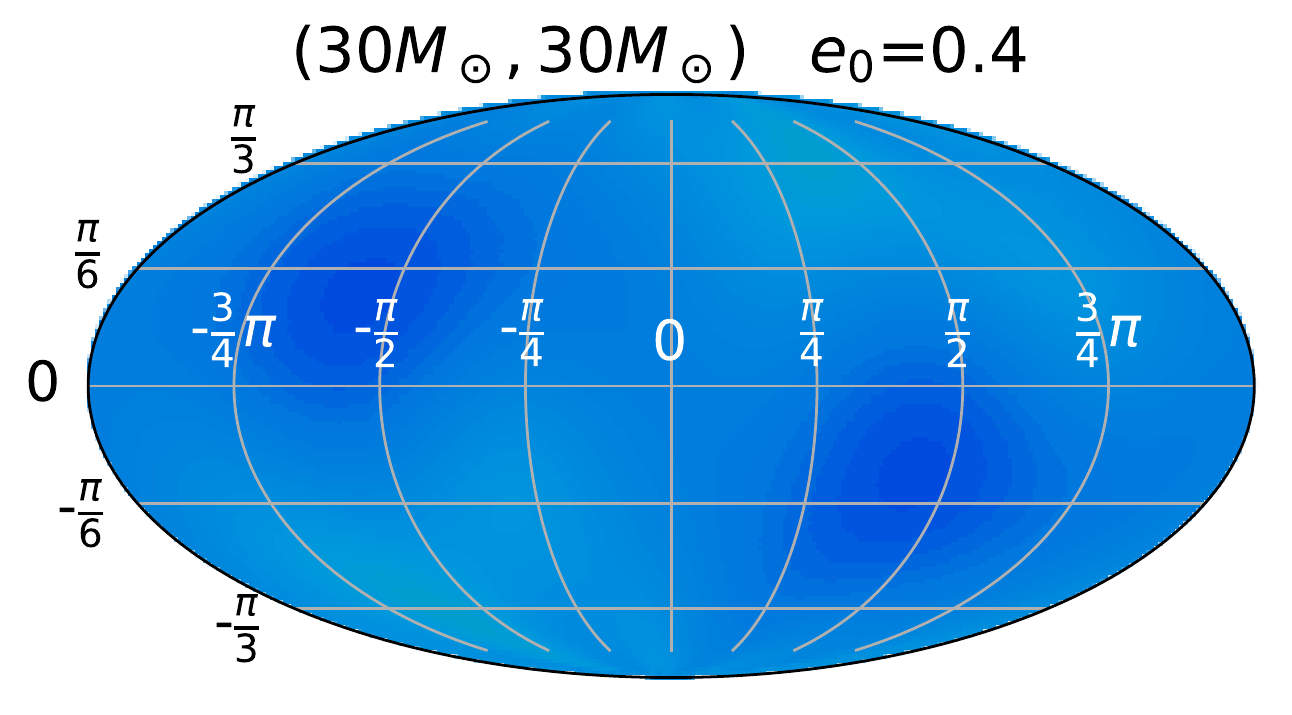}
		\includegraphics[width=\wid\textwidth]{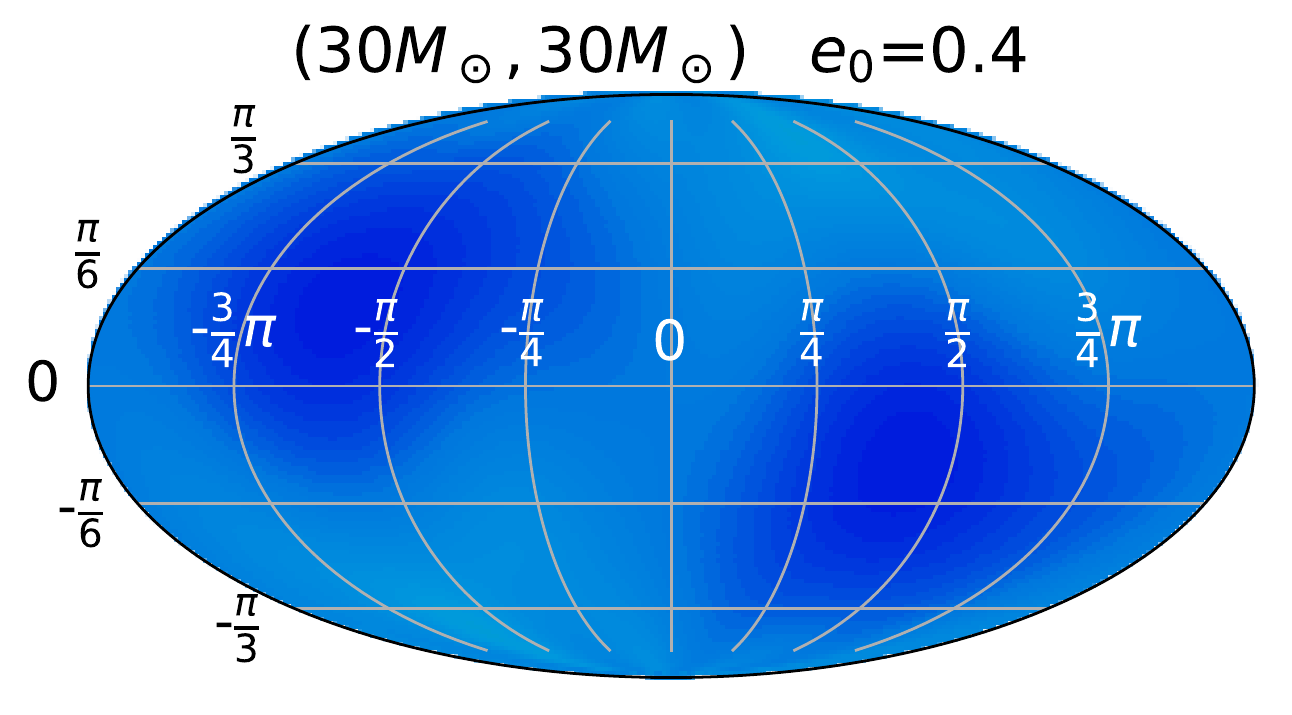}
	}
	\centerline{
		\includegraphics[width=\wid\textwidth]{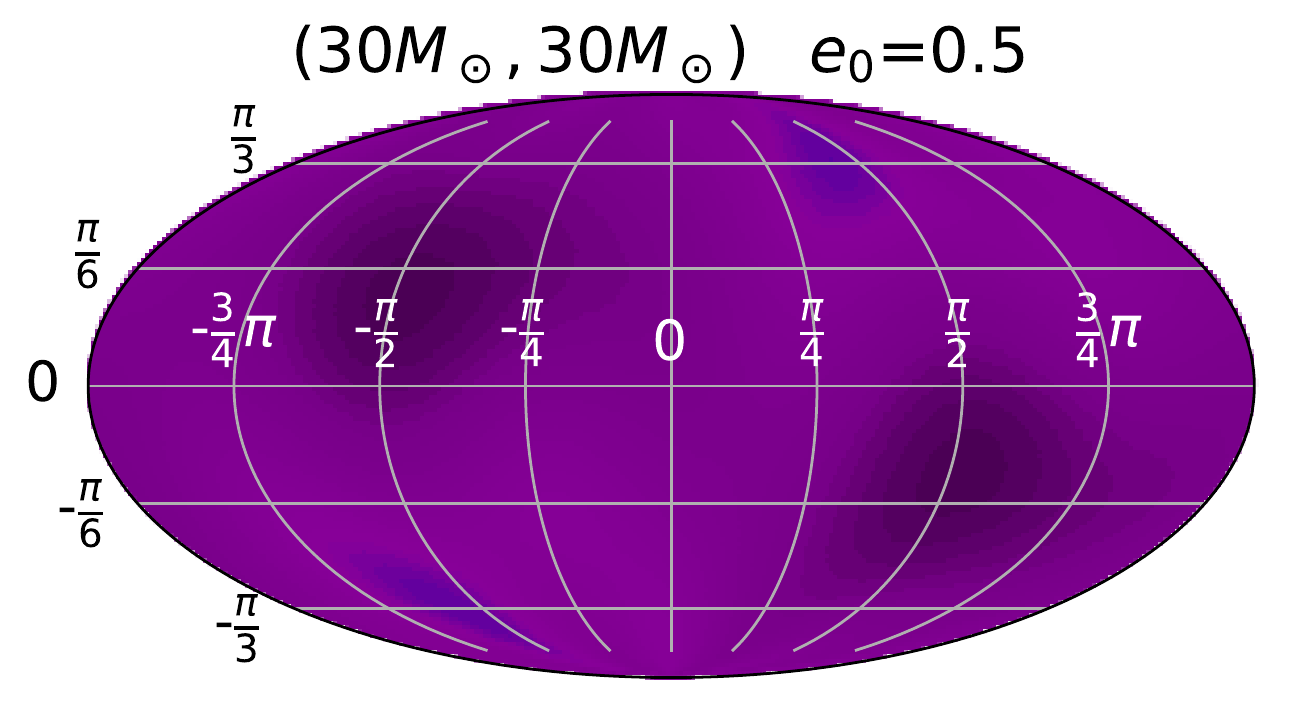}
		\includegraphics[width=\wid\textwidth]{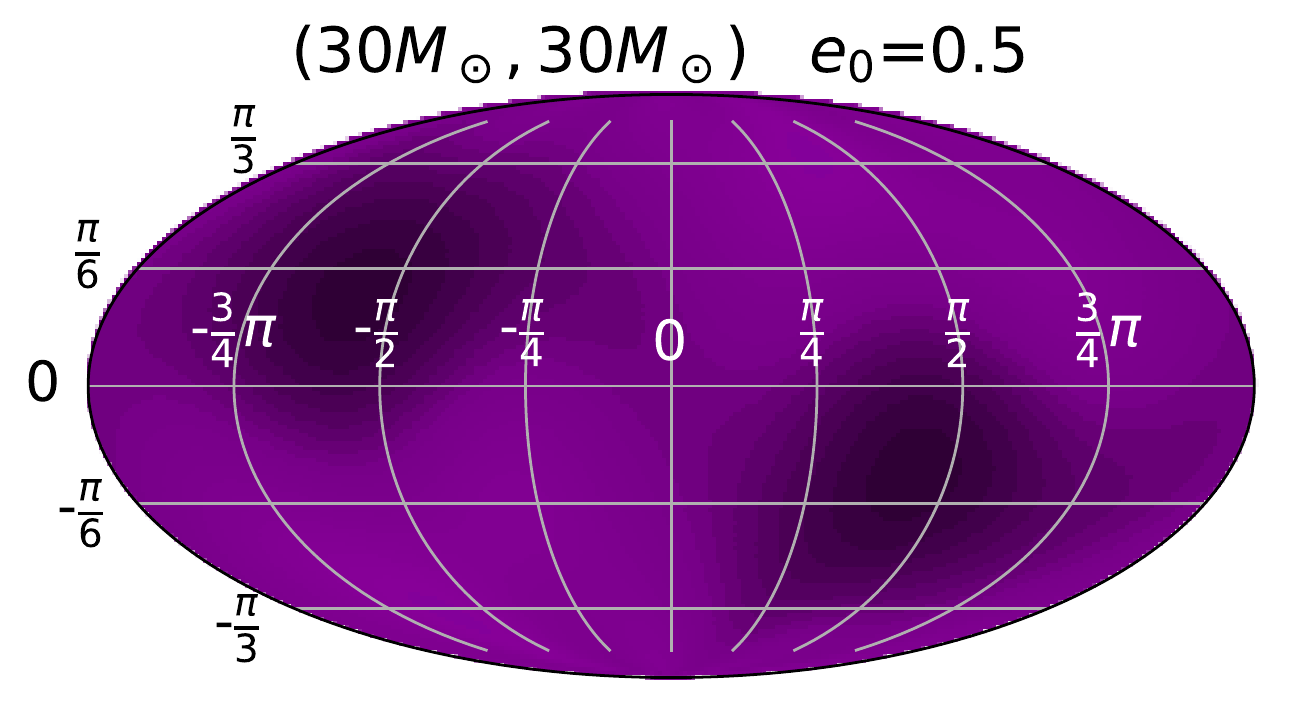}
		\includegraphics[width=\wid\textwidth]{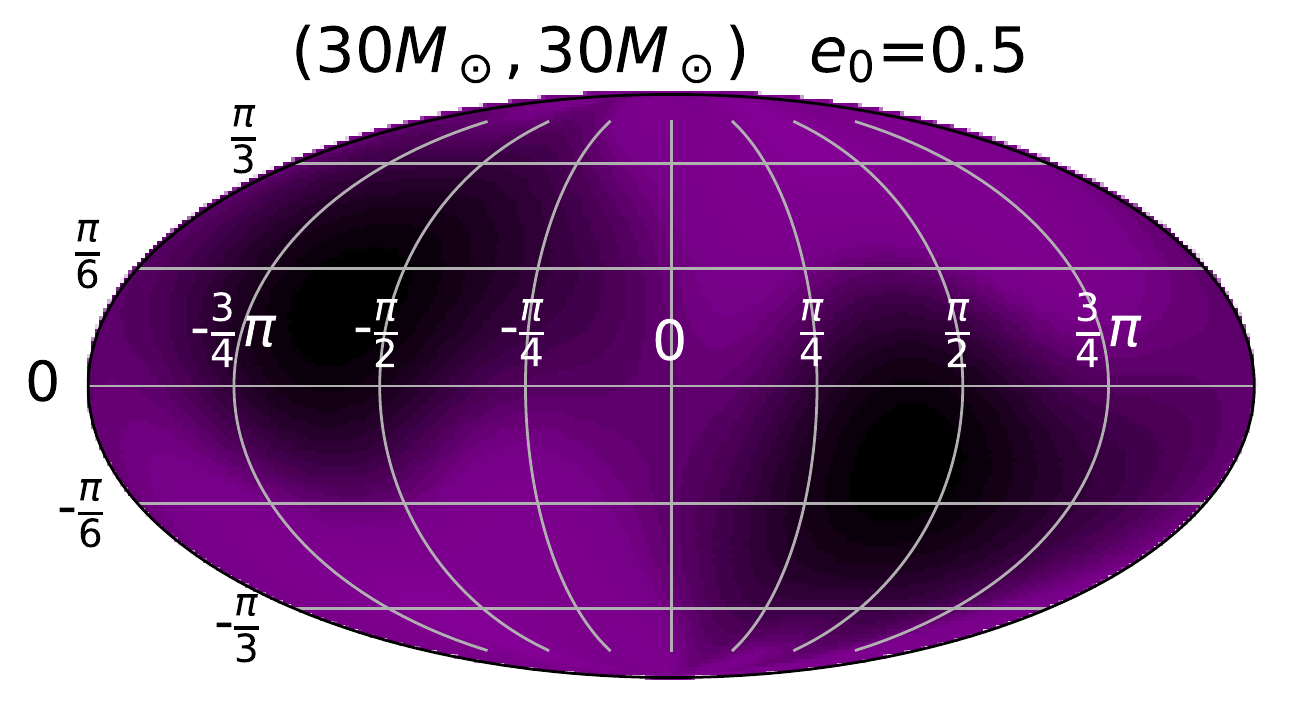}
	}
	\caption{As Figure~\ref{fig:ce_equal_mass} but now setting the binary inclination angle to 0.} 
	\label{fig:ce_equal_opt}
\end{figure*}

\begin{figure*}
	\centerline{
		\includegraphics[width=\textwidth]{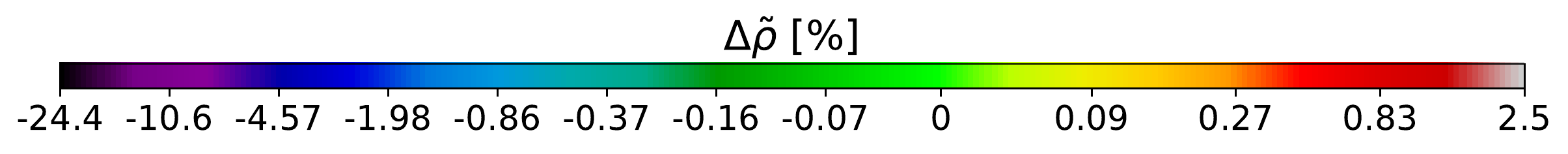}
	}
	\centerline{
		\includegraphics[width=\wid\textwidth]{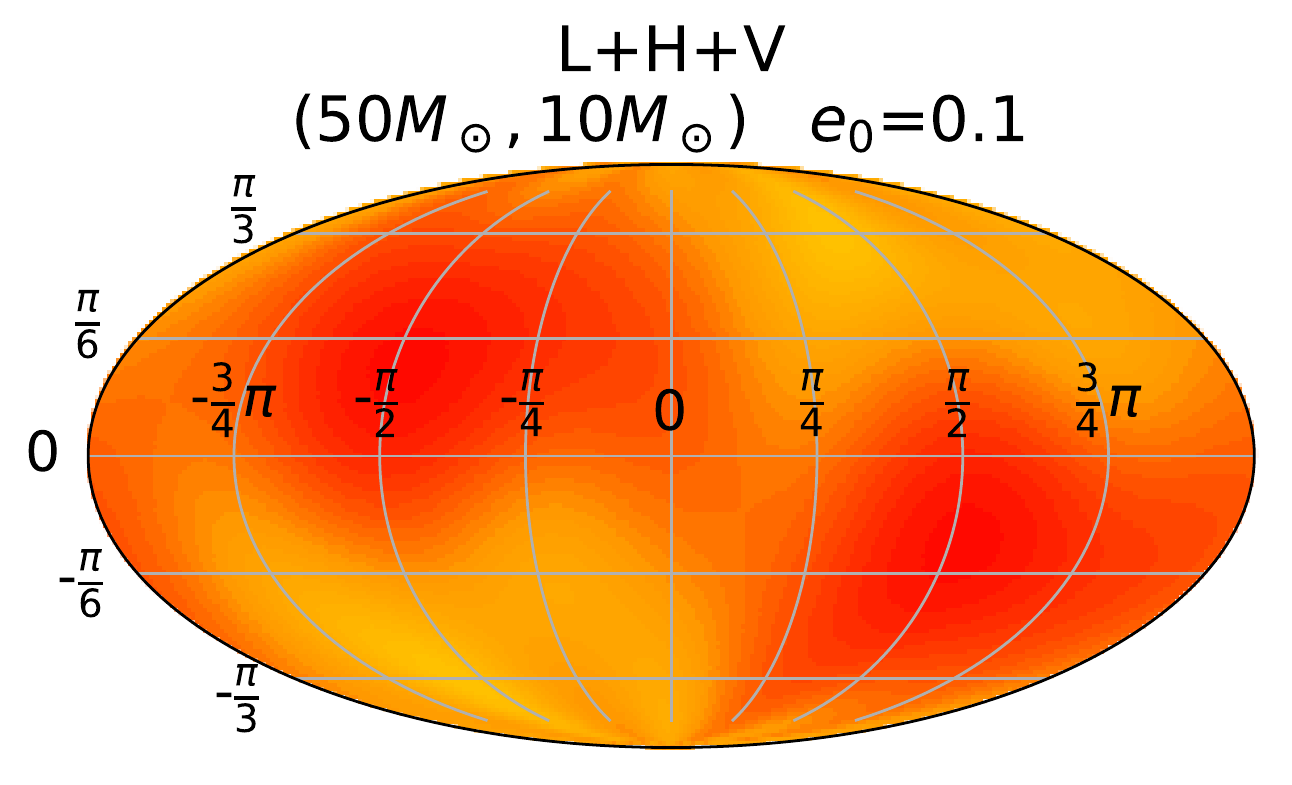}
		\includegraphics[width=\wid\textwidth]{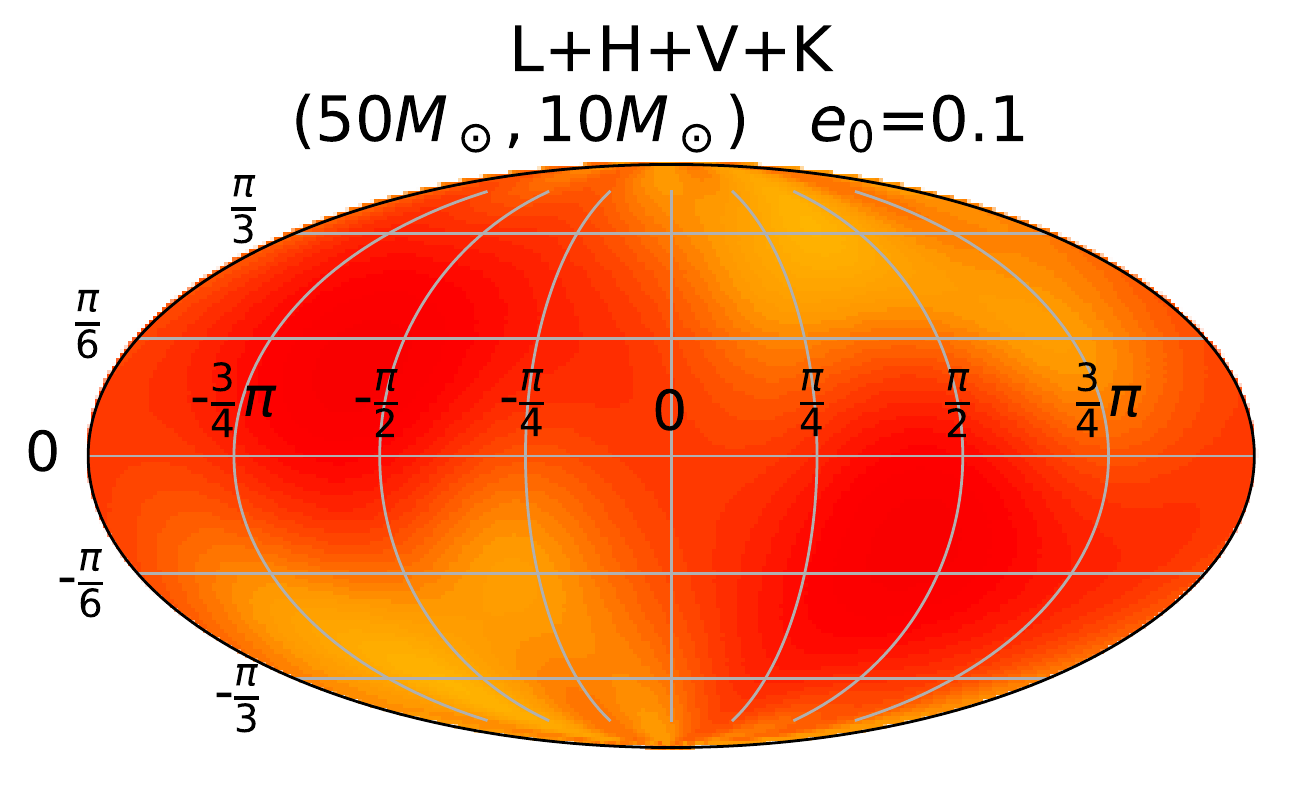}
		\includegraphics[width=\wid\textwidth]{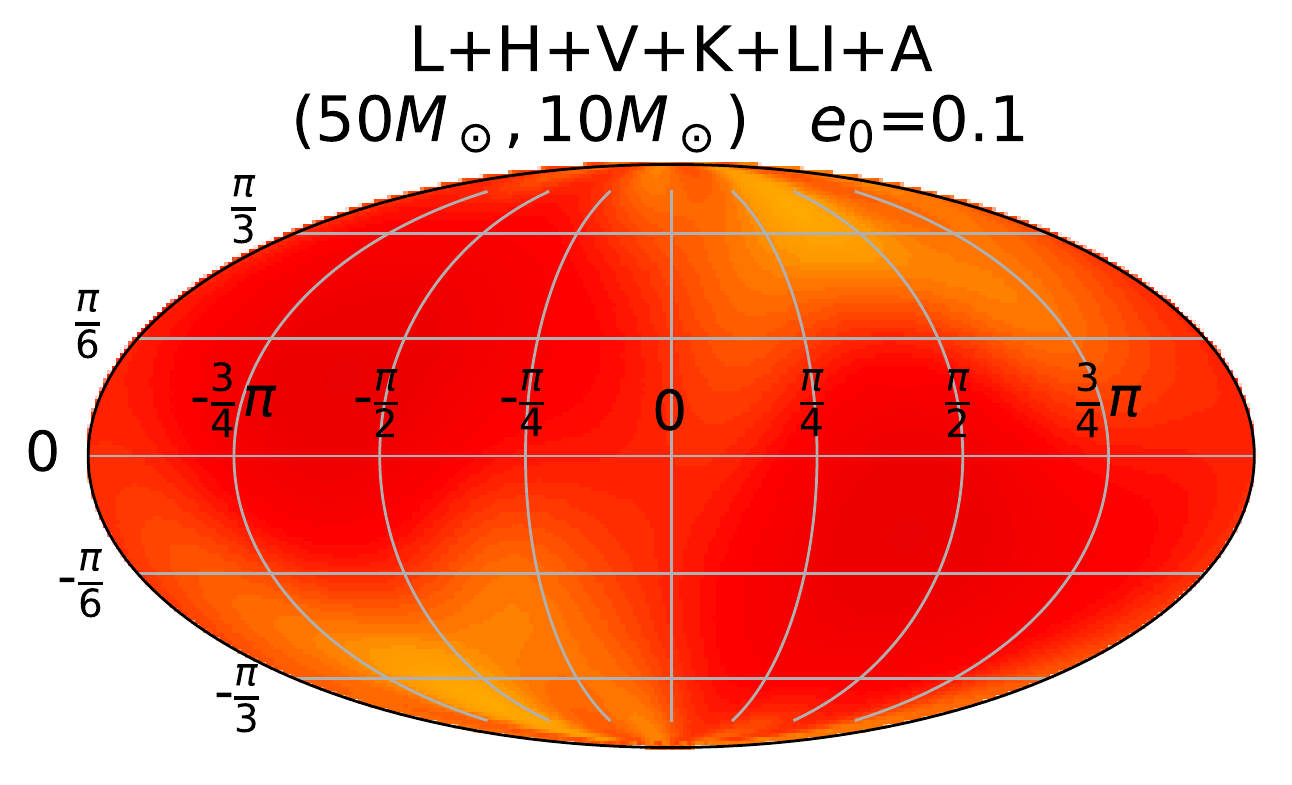}
	}
	\centerline{
		\includegraphics[width=\wid\textwidth]{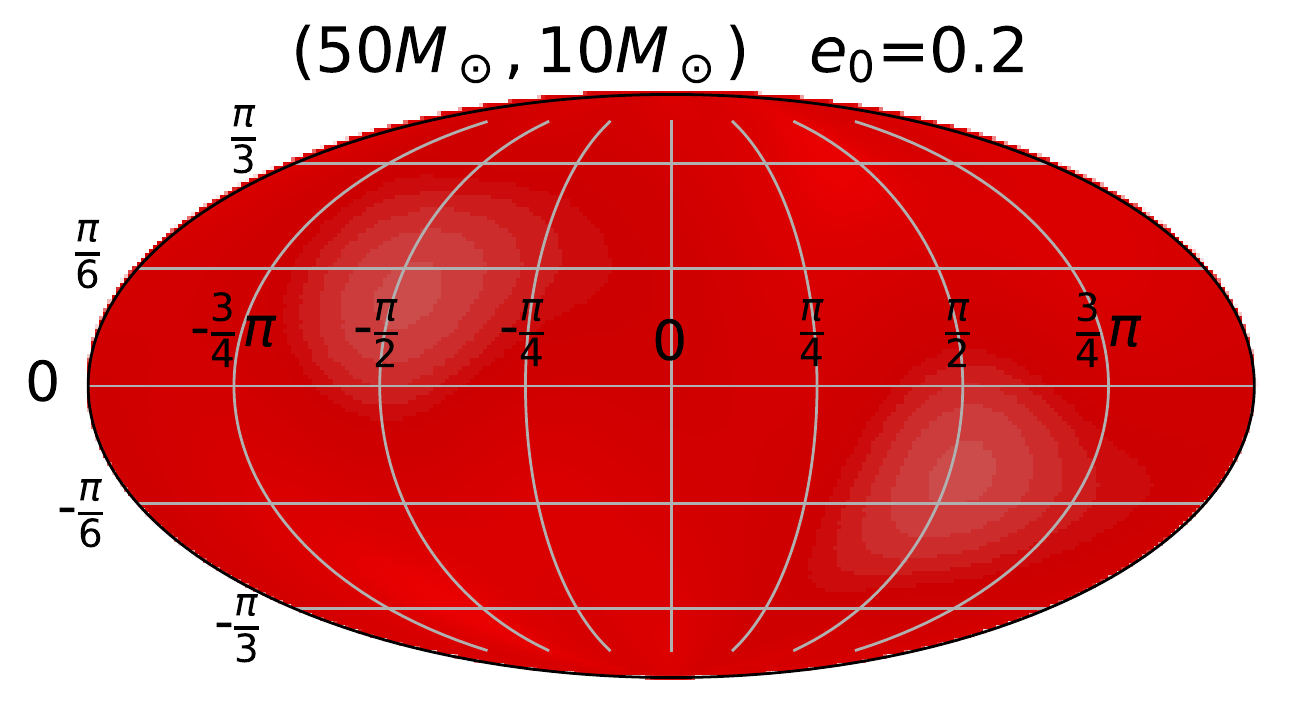}
		\includegraphics[width=\wid\textwidth]{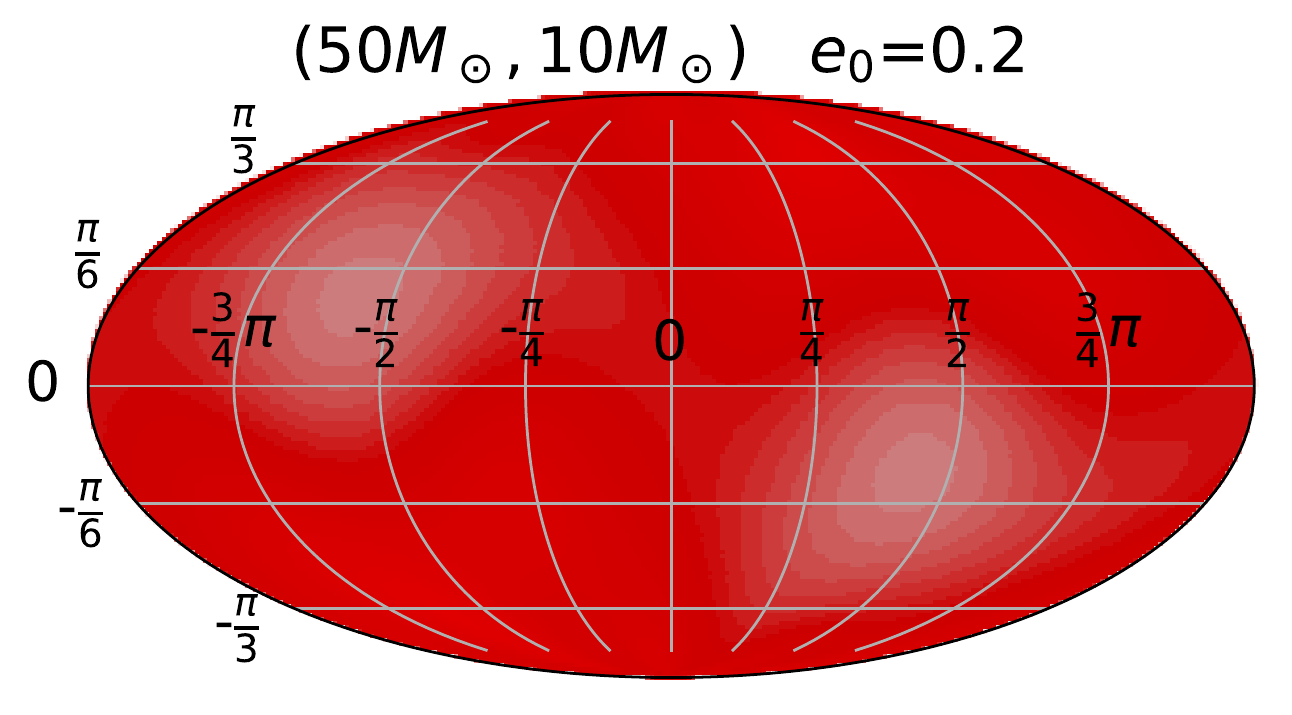}
		\includegraphics[width=\wid\textwidth]{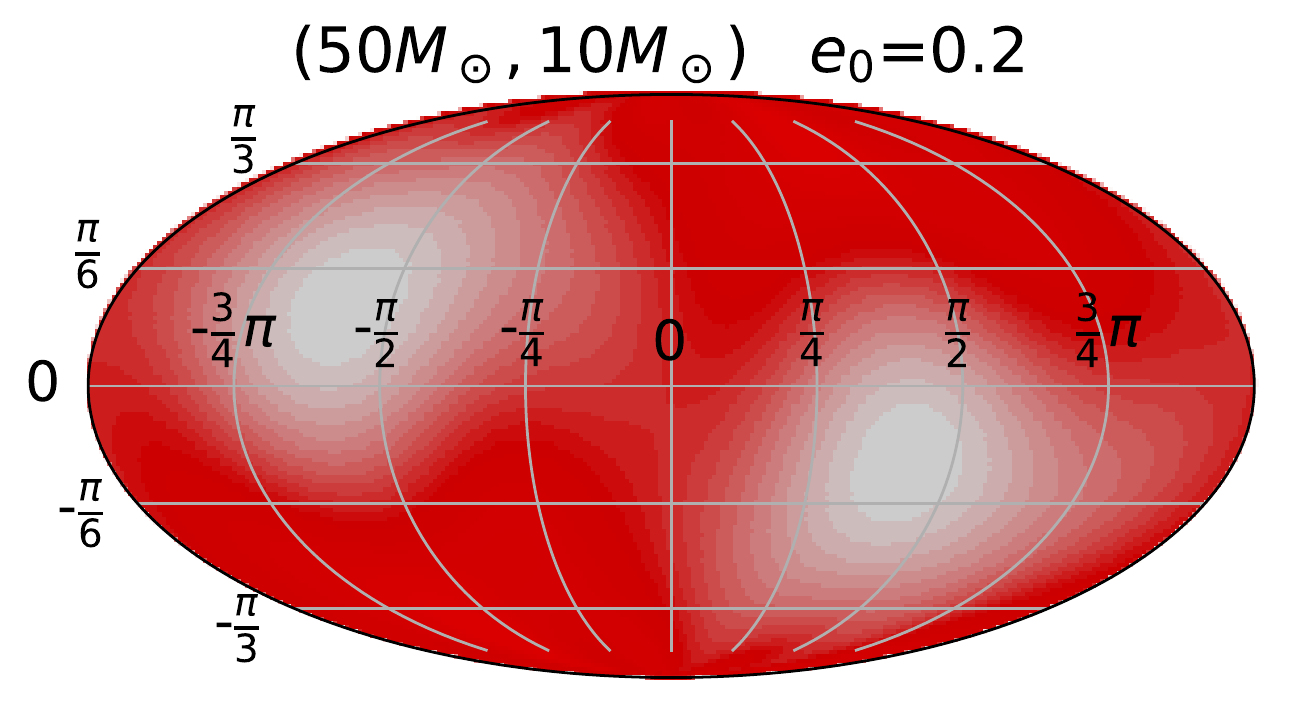}
	}
	\centerline{
		\includegraphics[width=\wid\textwidth]{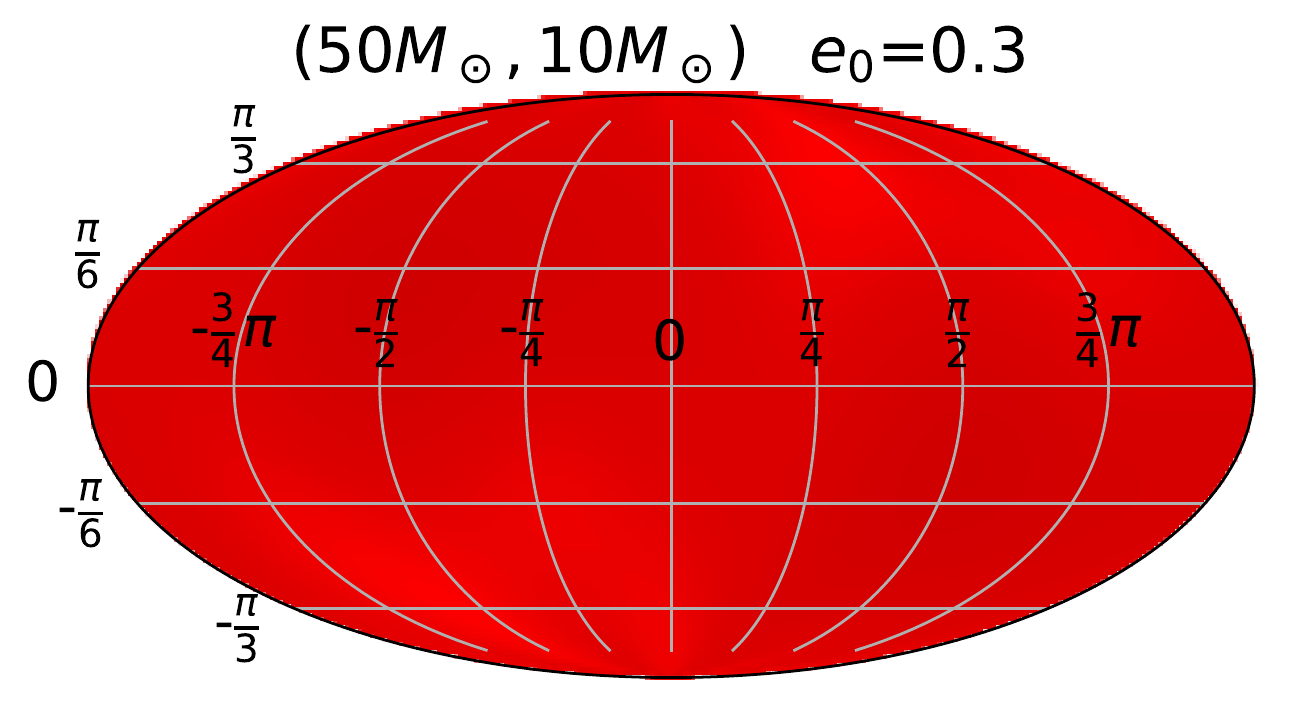}
		\includegraphics[width=\wid\textwidth]{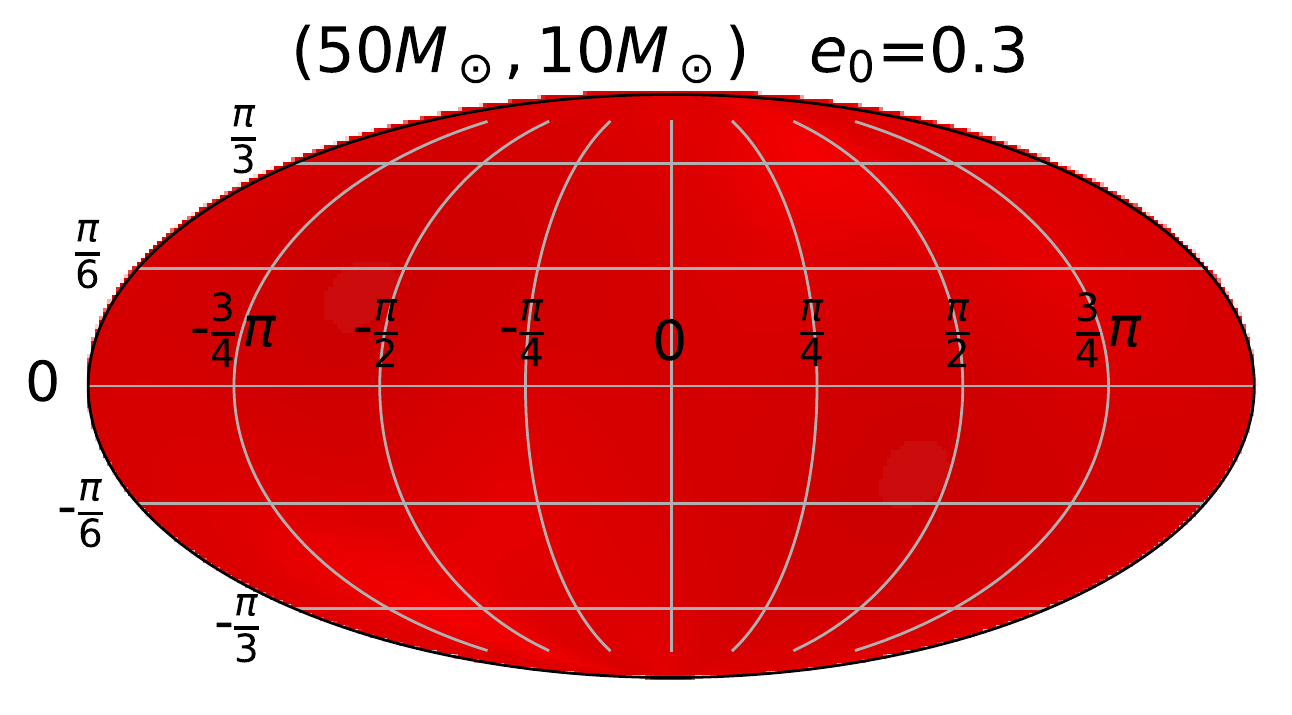}
		\includegraphics[width=\wid\textwidth]{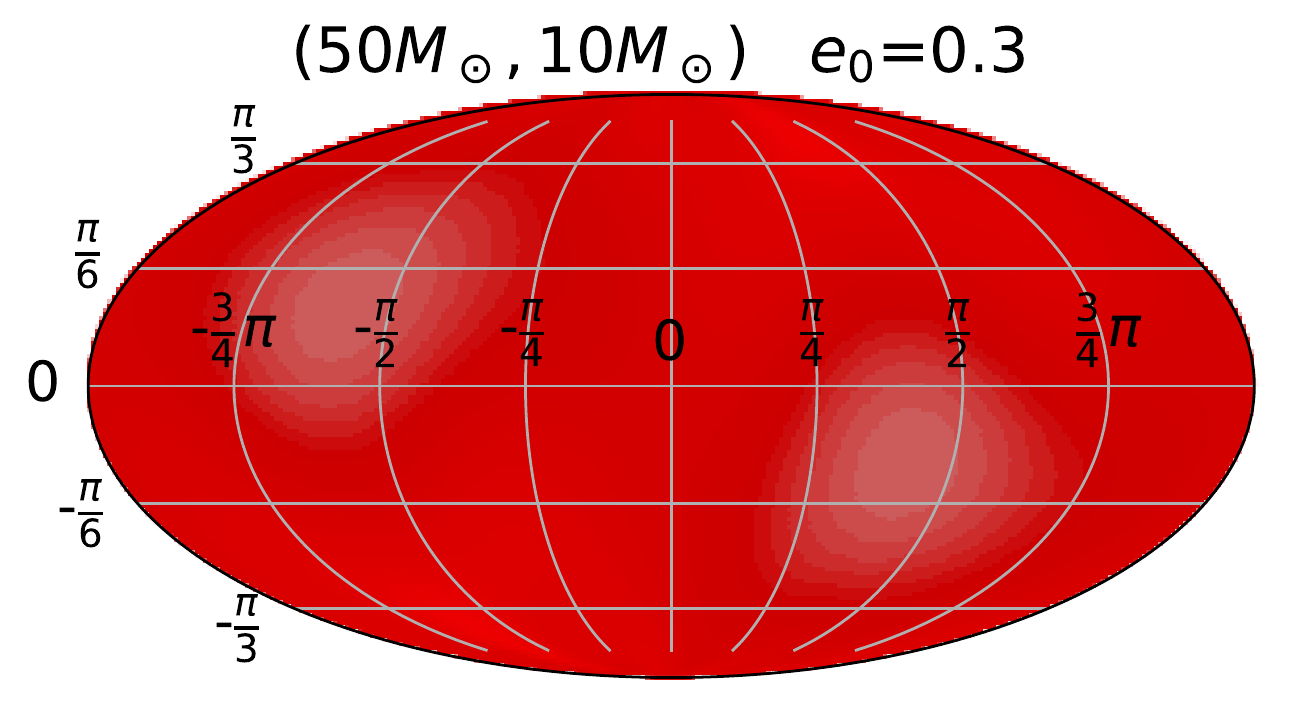}
	}
	\centerline{
		\includegraphics[width=\wid\textwidth]{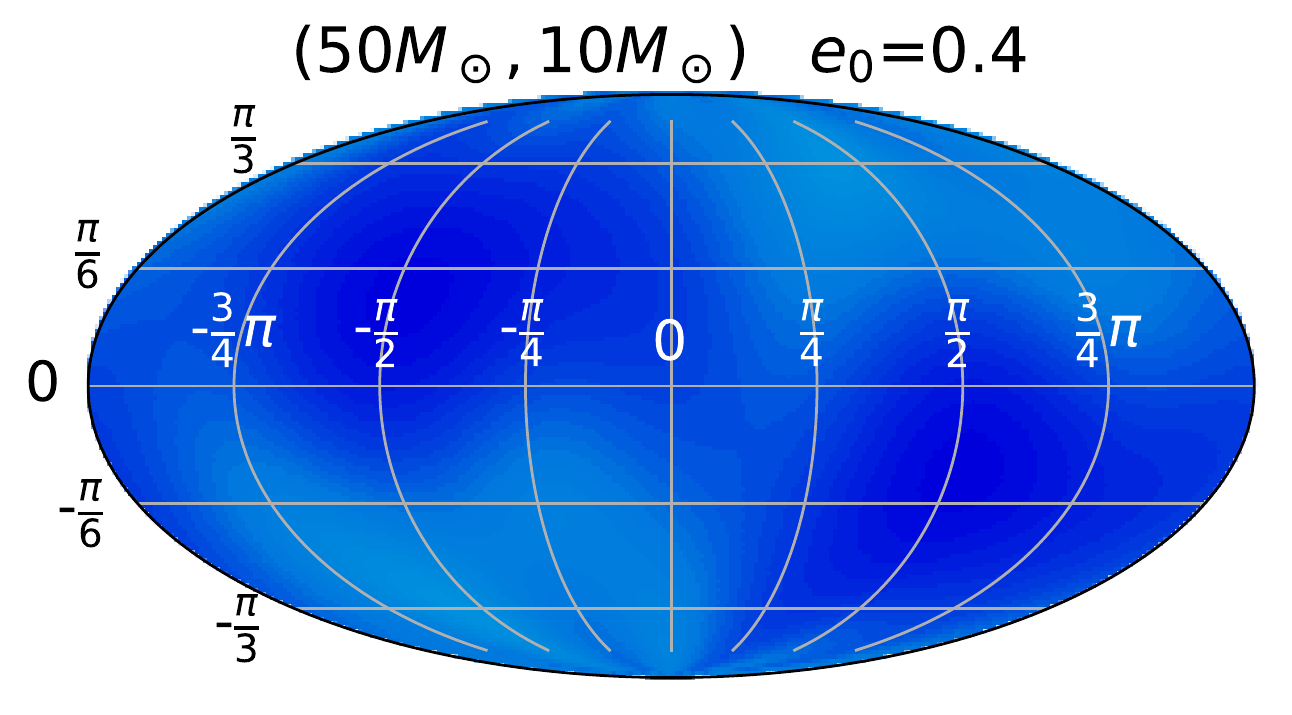}
		\includegraphics[width=\wid\textwidth]{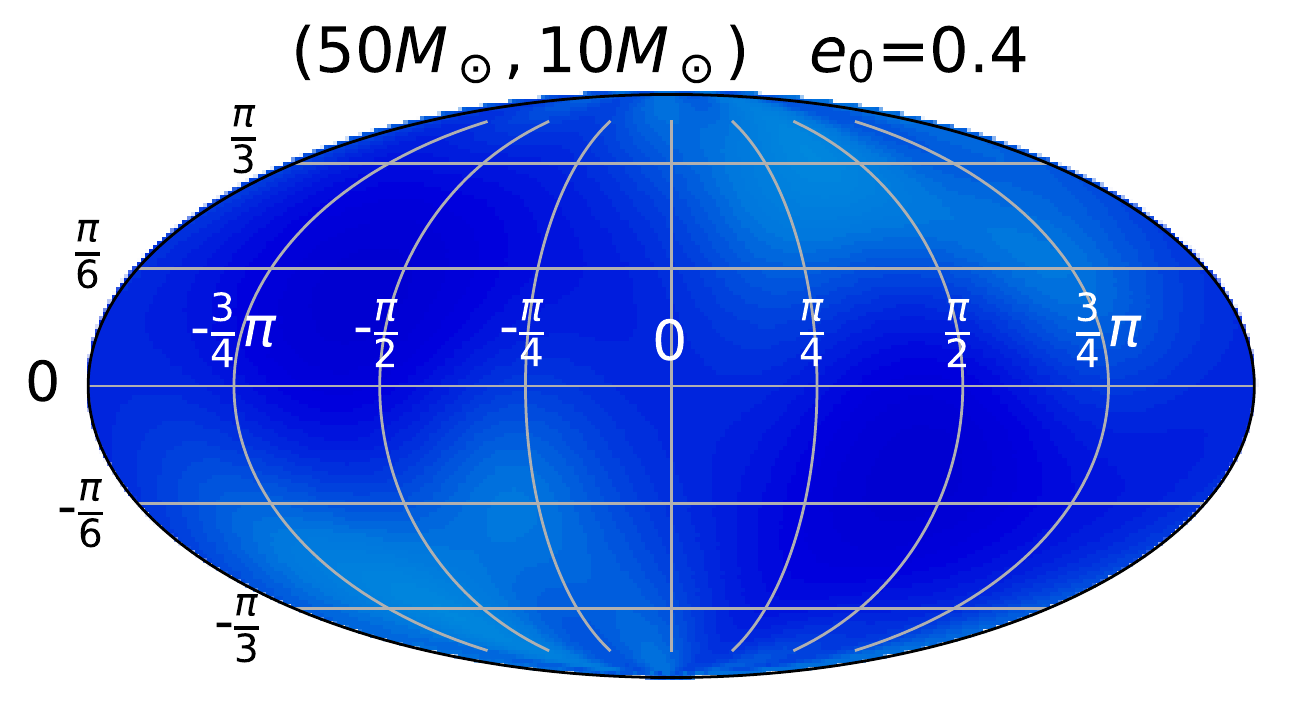}
		\includegraphics[width=\wid\textwidth]{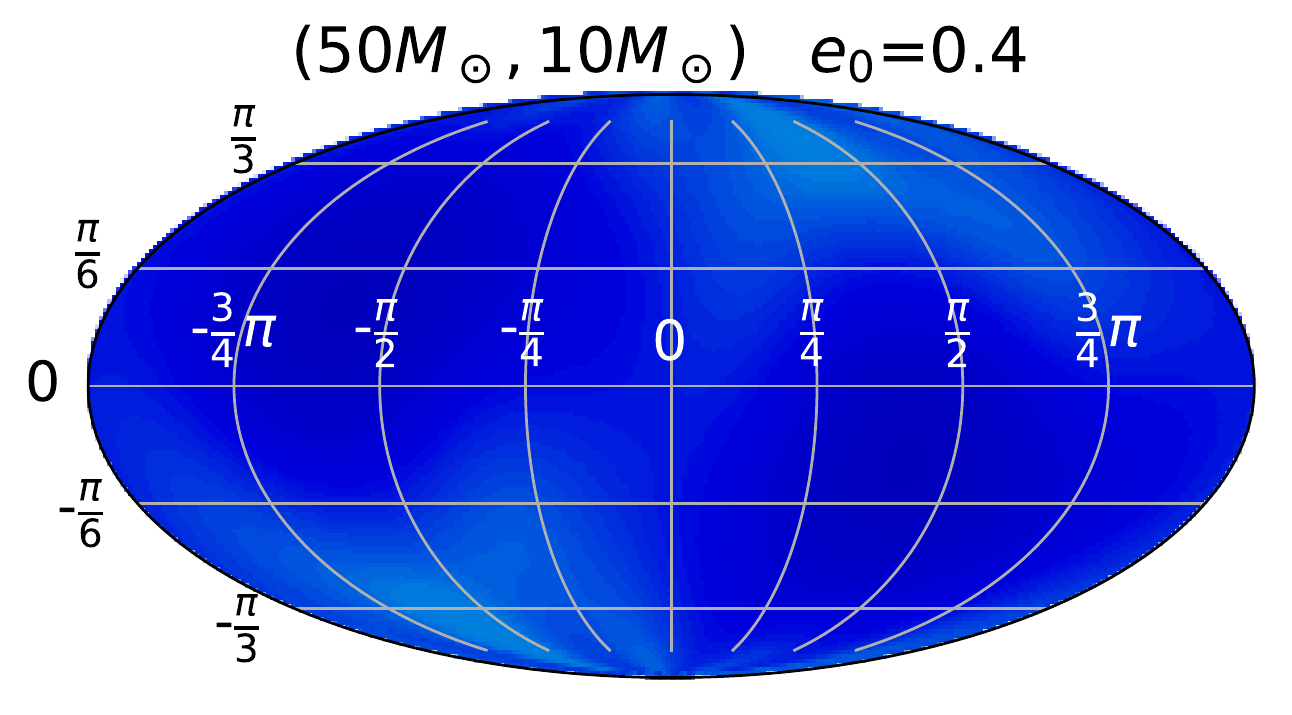}
	}
	\centerline{
		\includegraphics[width=\wid\textwidth]{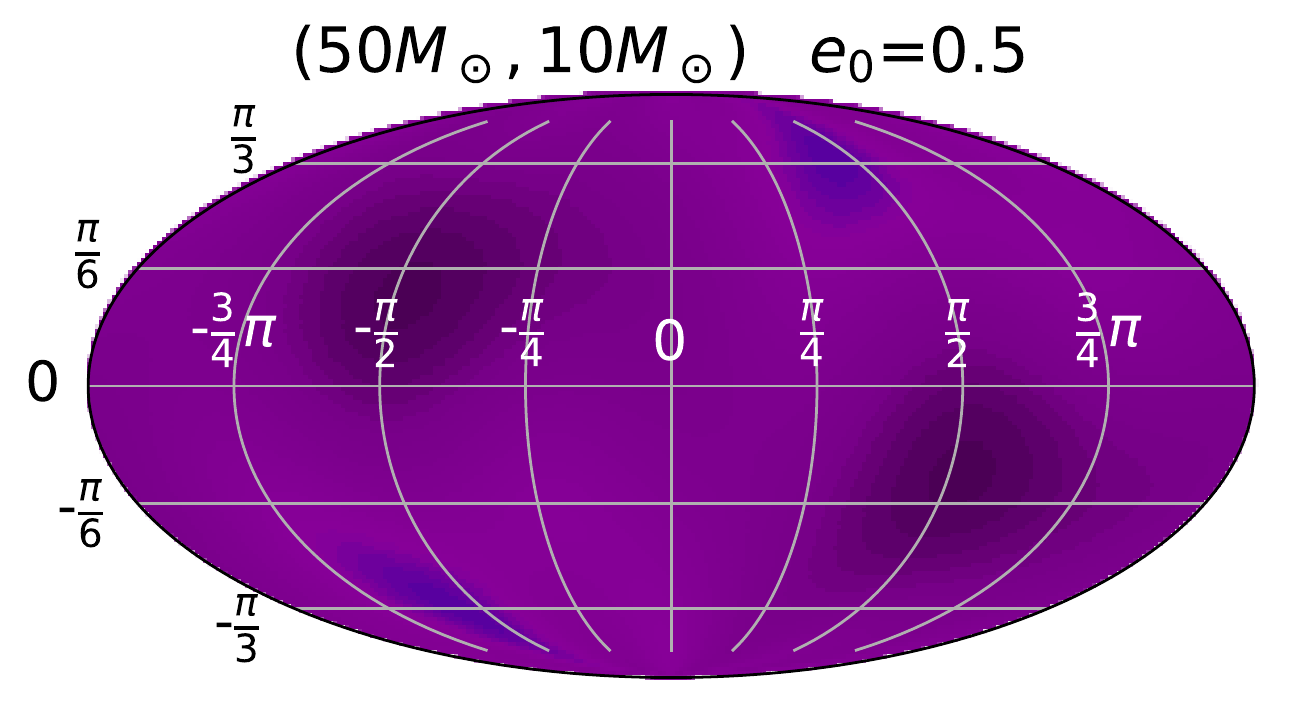}
		\includegraphics[width=\wid\textwidth]{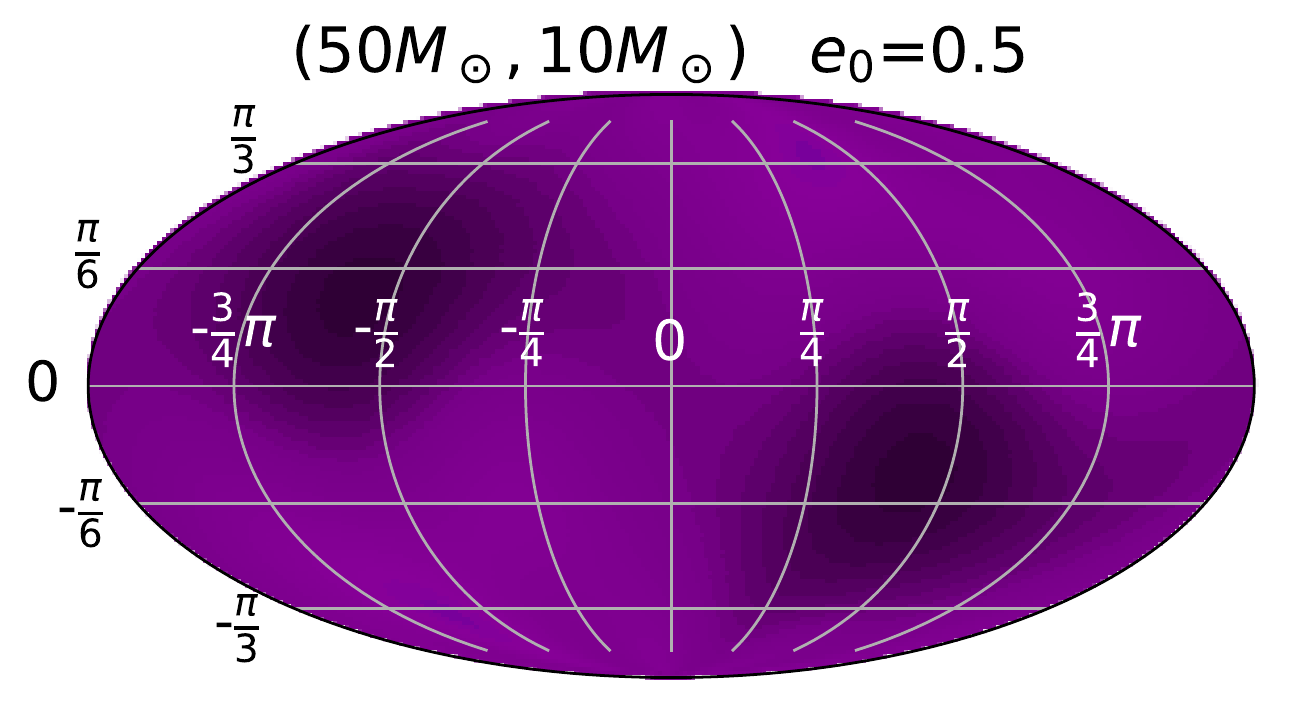}
		\includegraphics[width=\wid\textwidth]{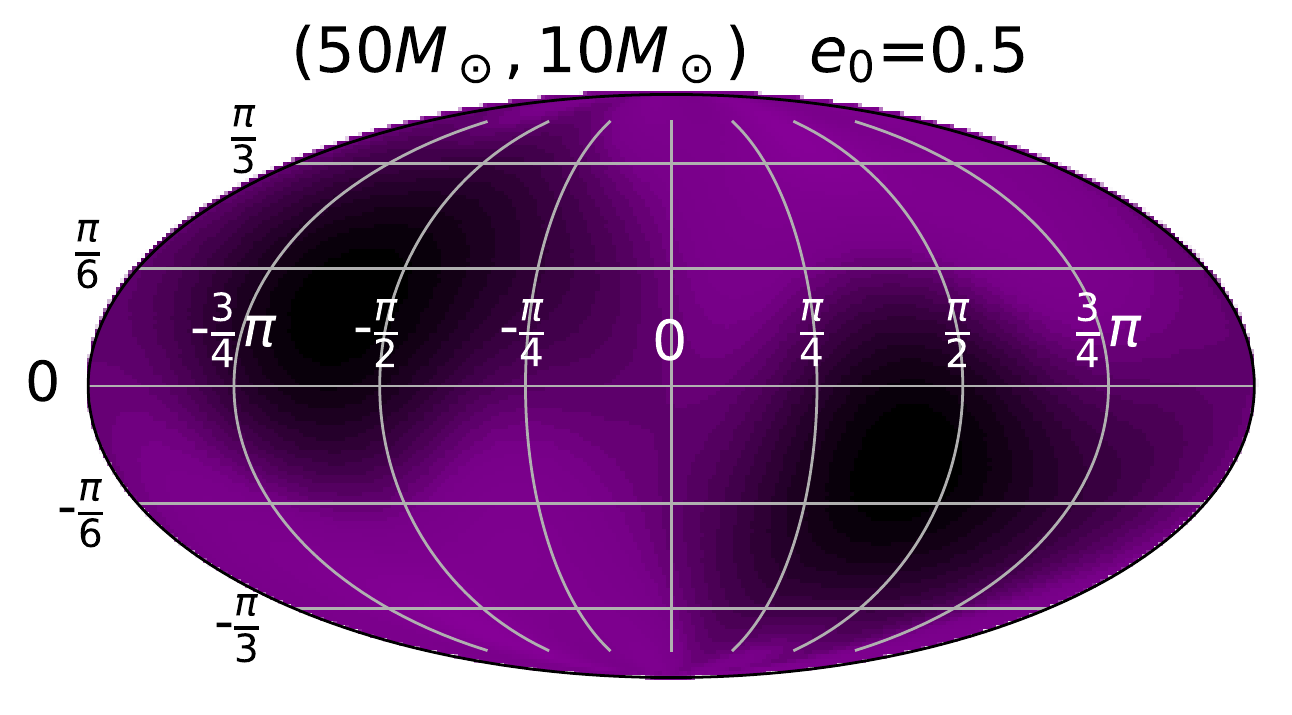}
	}
	\caption{As Figure~\ref{fig:ce_equal_opt} but now for component masses \((50\msun,\,10\msun)\).} 
	\label{fig:ce_five_opt}
\end{figure*}


\begin{figure*}
	\centerline{
		\includegraphics[width=\textwidth]{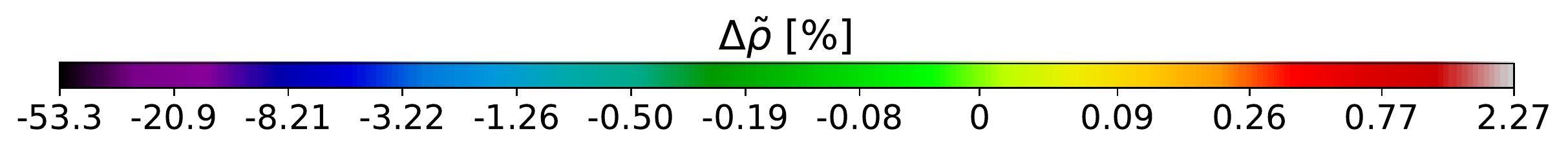}
	}
	\centerline{
		\includegraphics[width=\wid\textwidth]{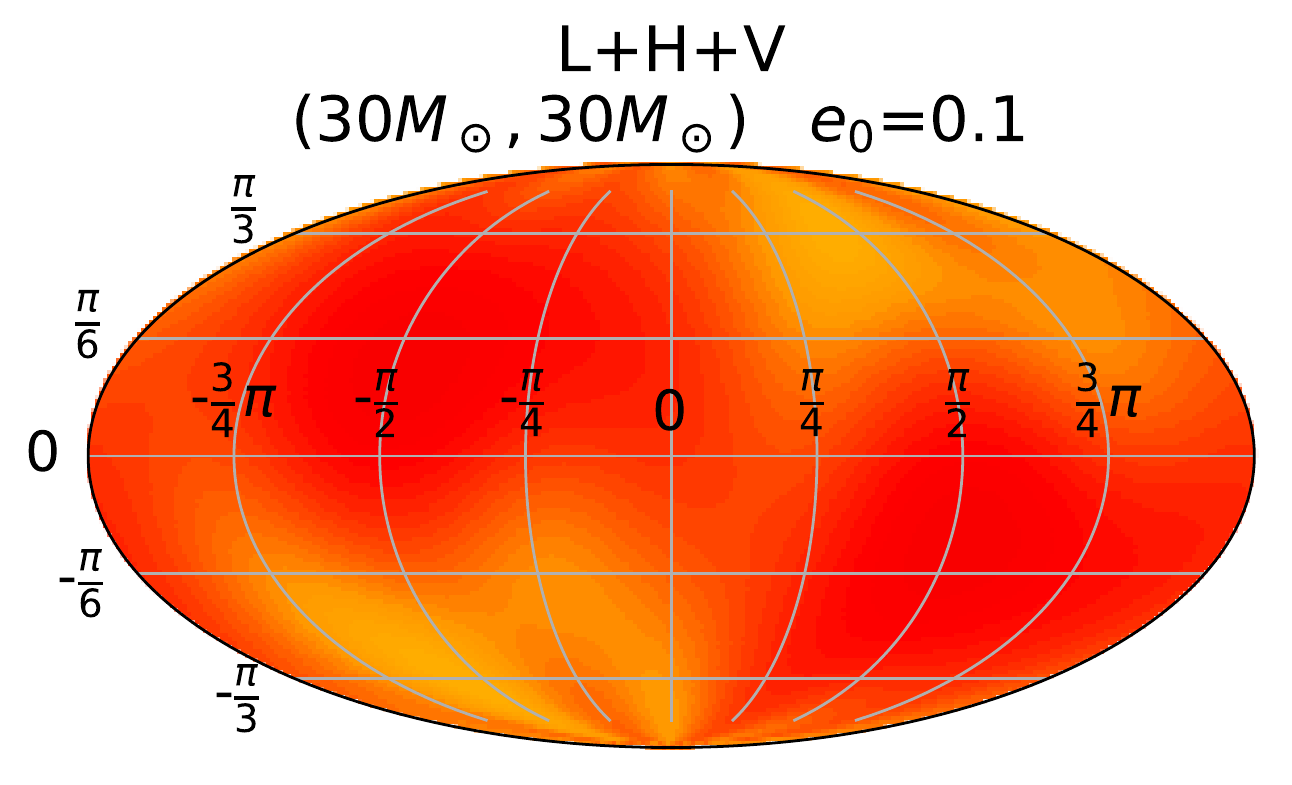}
		\includegraphics[width=\wid\textwidth]{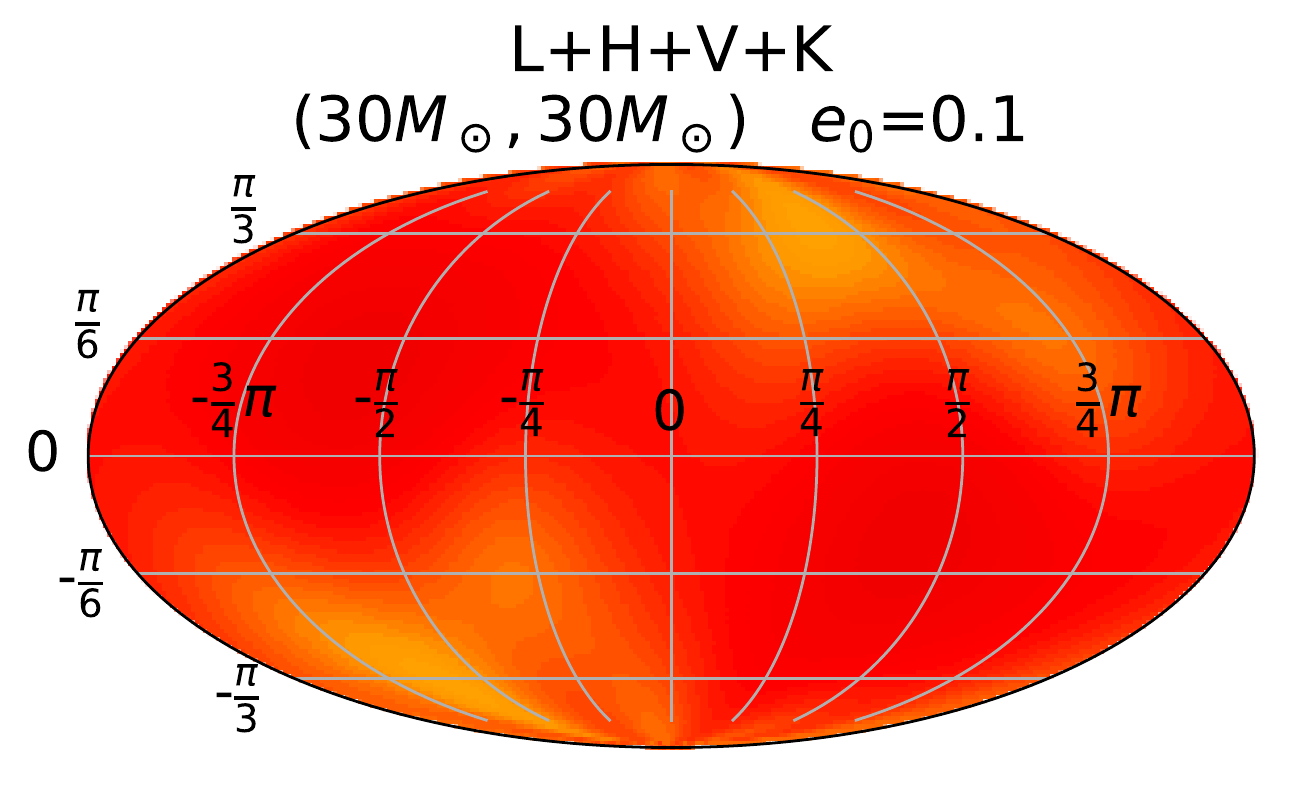}
		\includegraphics[width=\wid\textwidth]{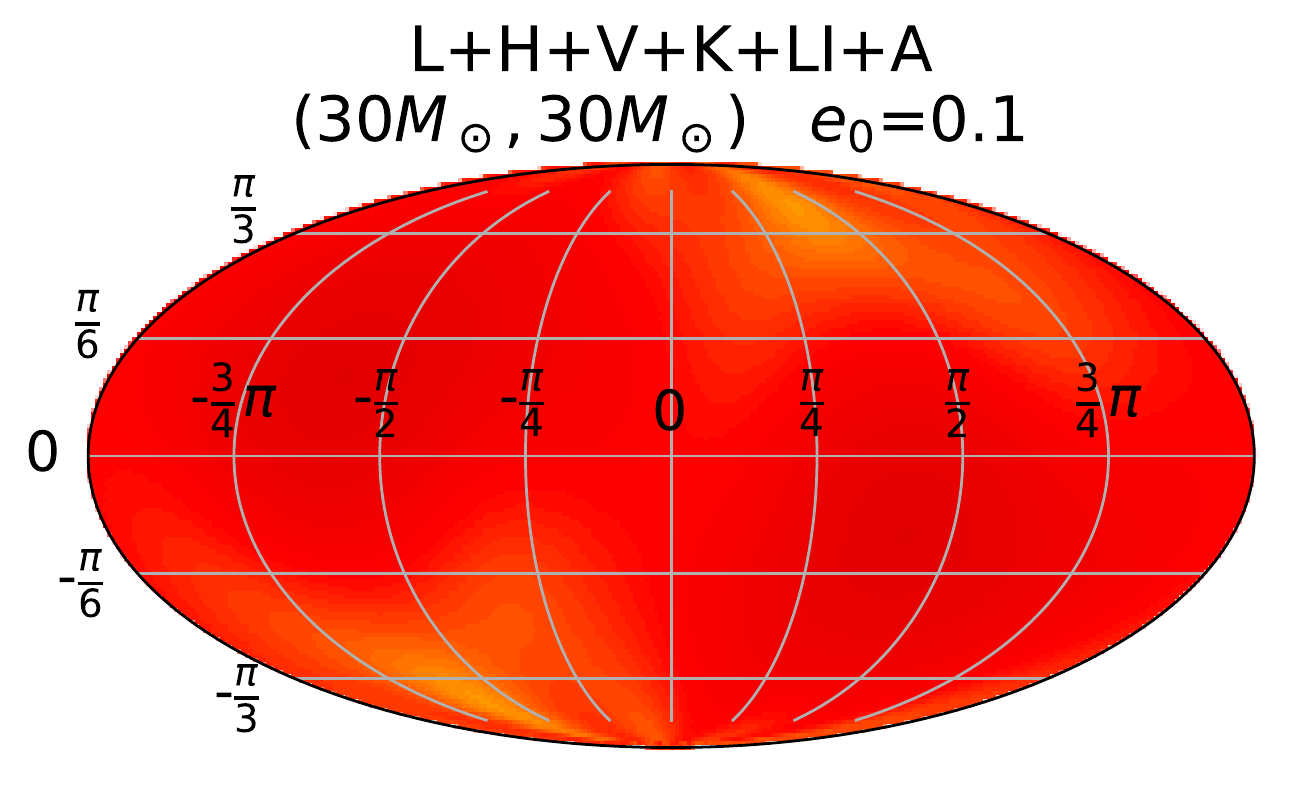}
	}
	\centerline{
		\includegraphics[width=\wid\textwidth]{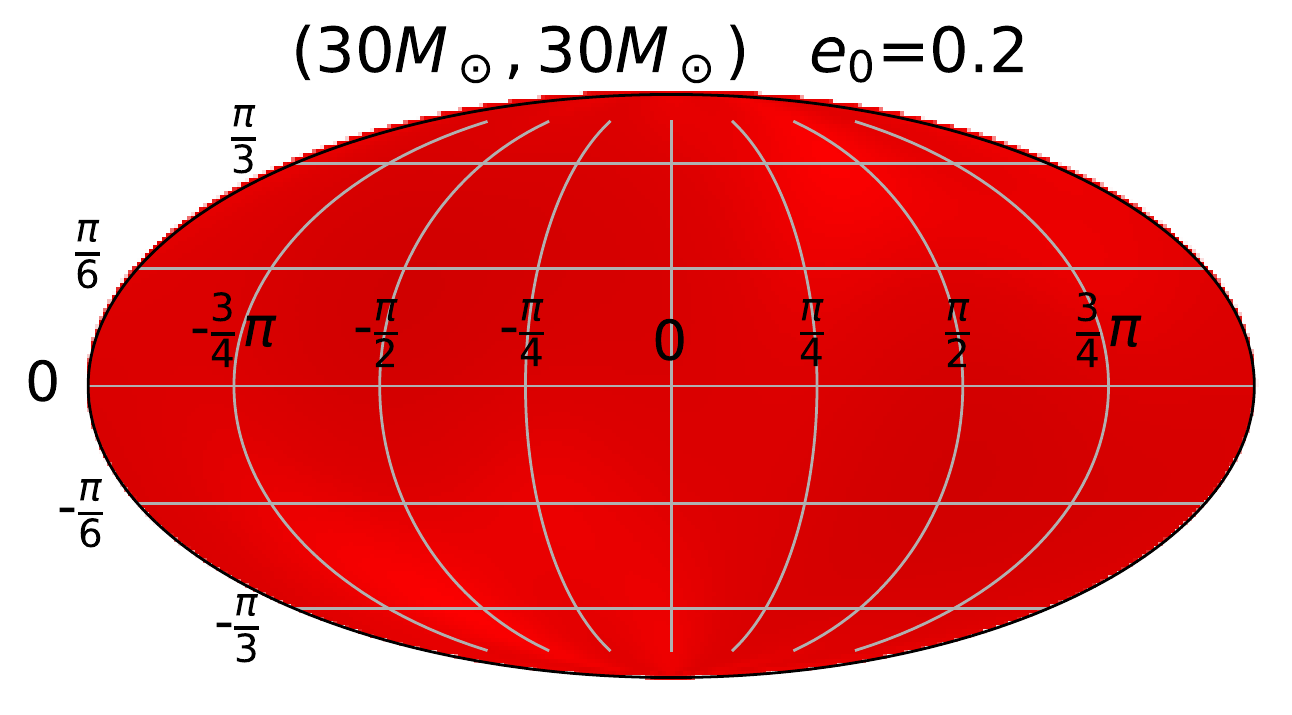}
		\includegraphics[width=\wid\textwidth]{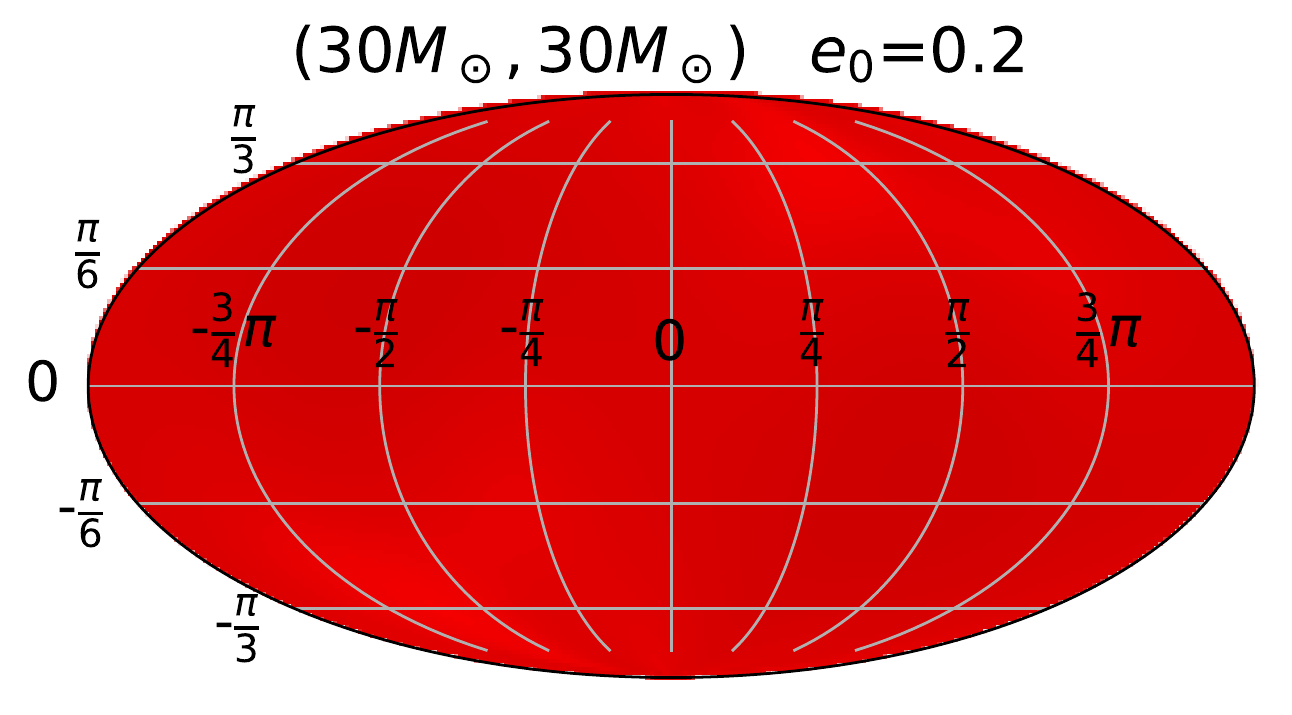}
		\includegraphics[width=\wid\textwidth]{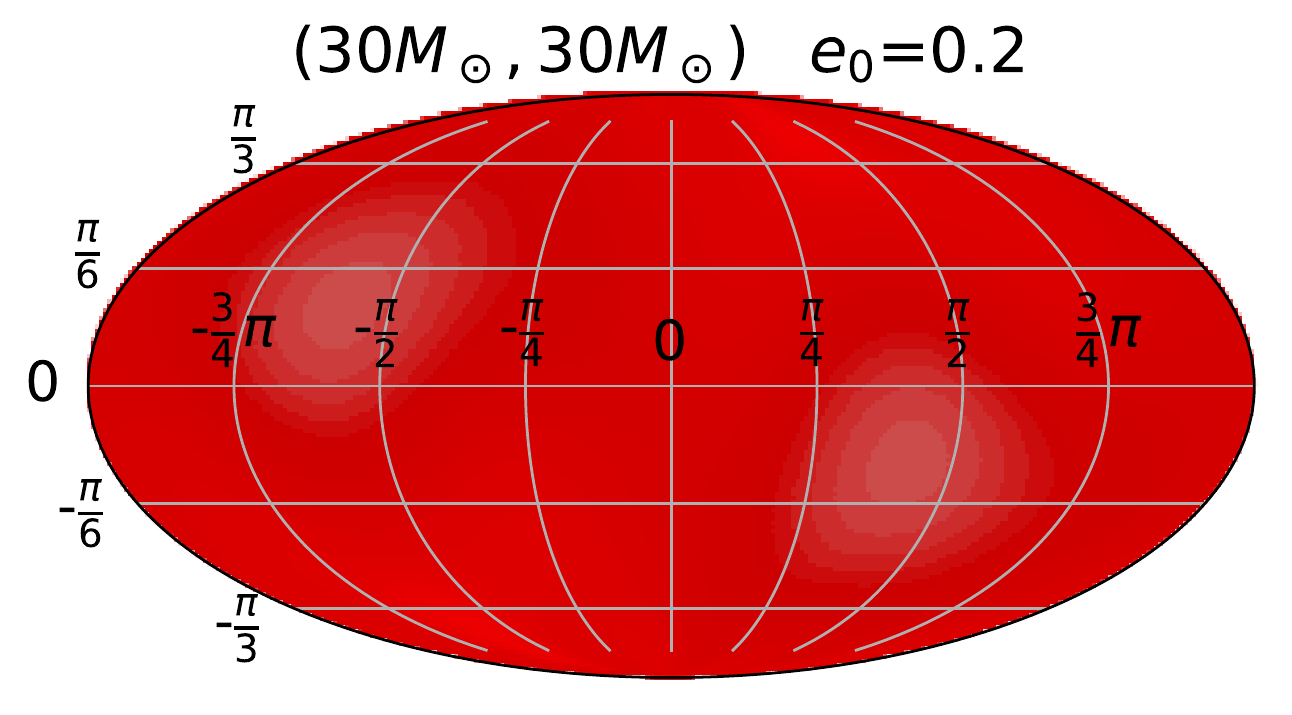}
	}
	\centerline{
		\includegraphics[width=\wid\textwidth]{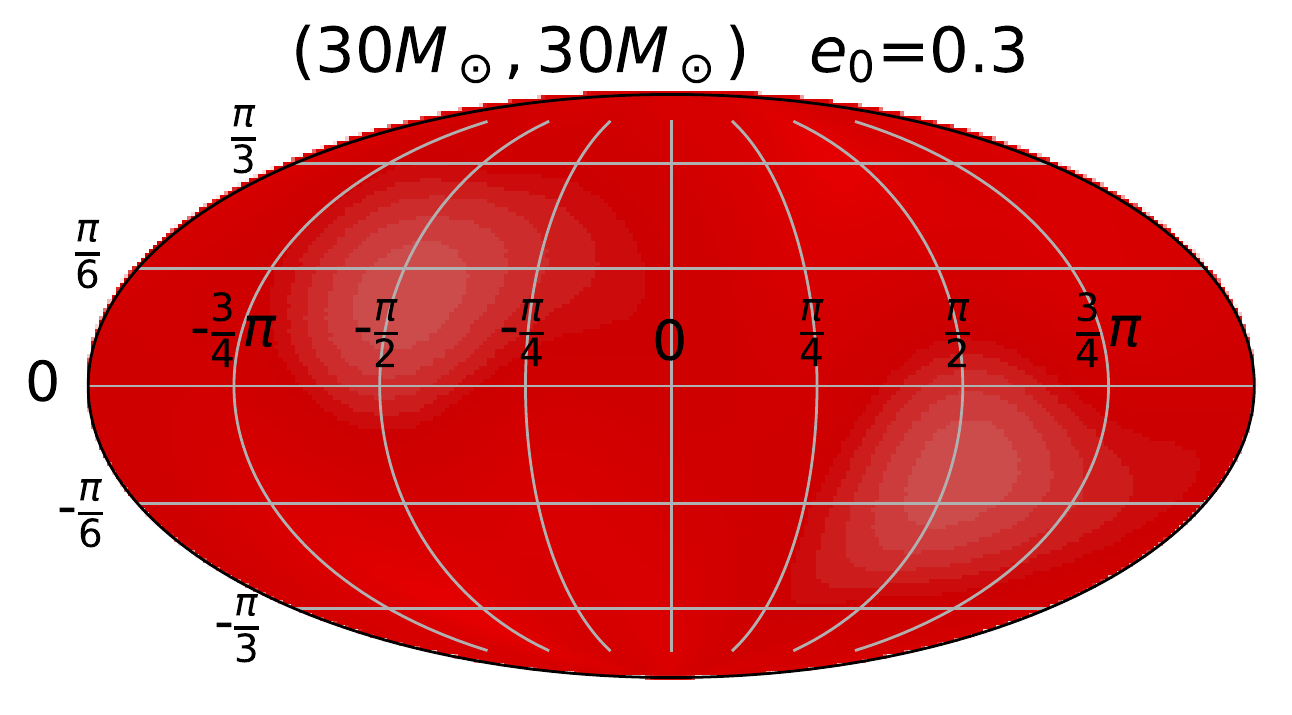}
		\includegraphics[width=\wid\textwidth]{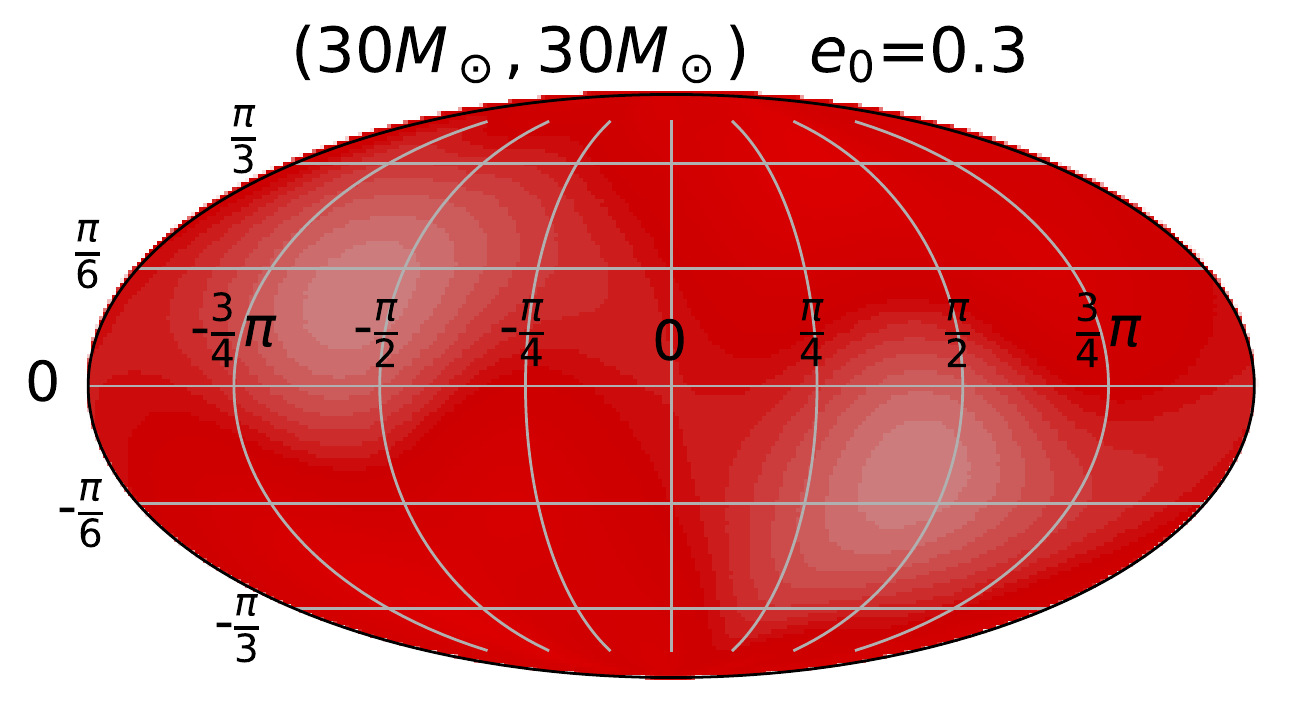}
		\includegraphics[width=\wid\textwidth]{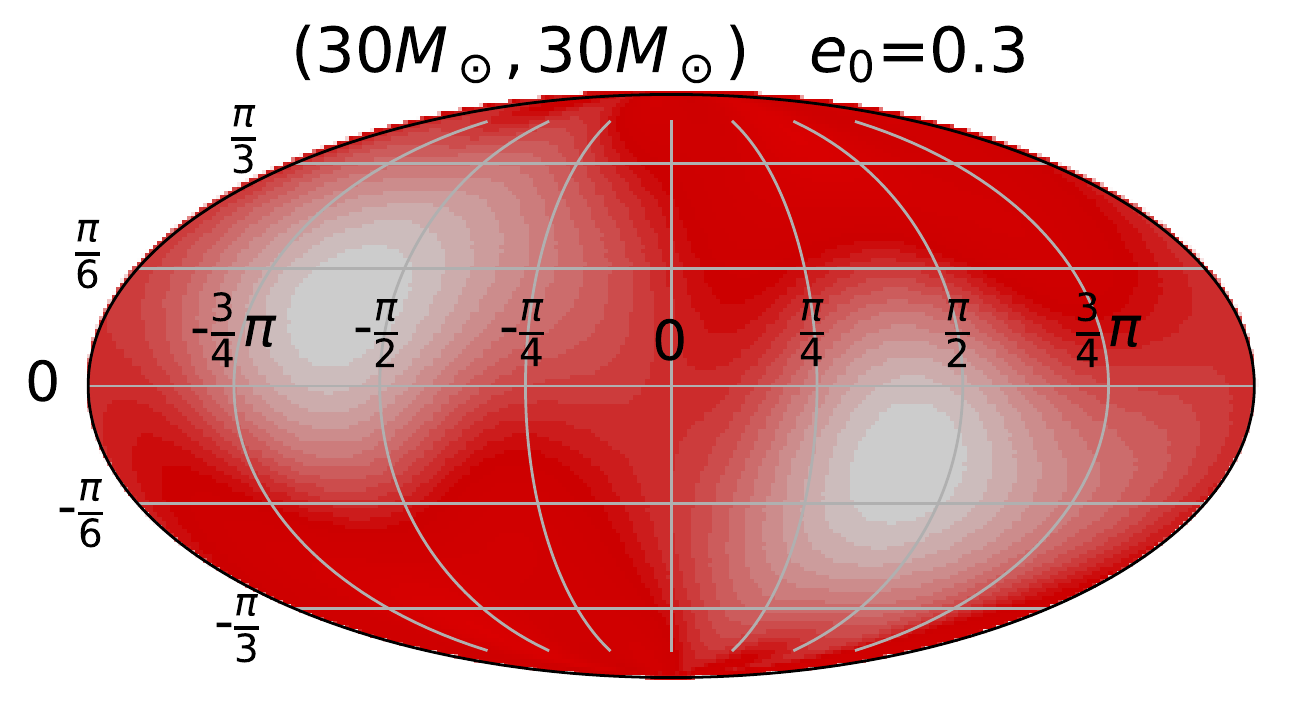}
	}
	\centerline{
		\includegraphics[width=\wid\textwidth]{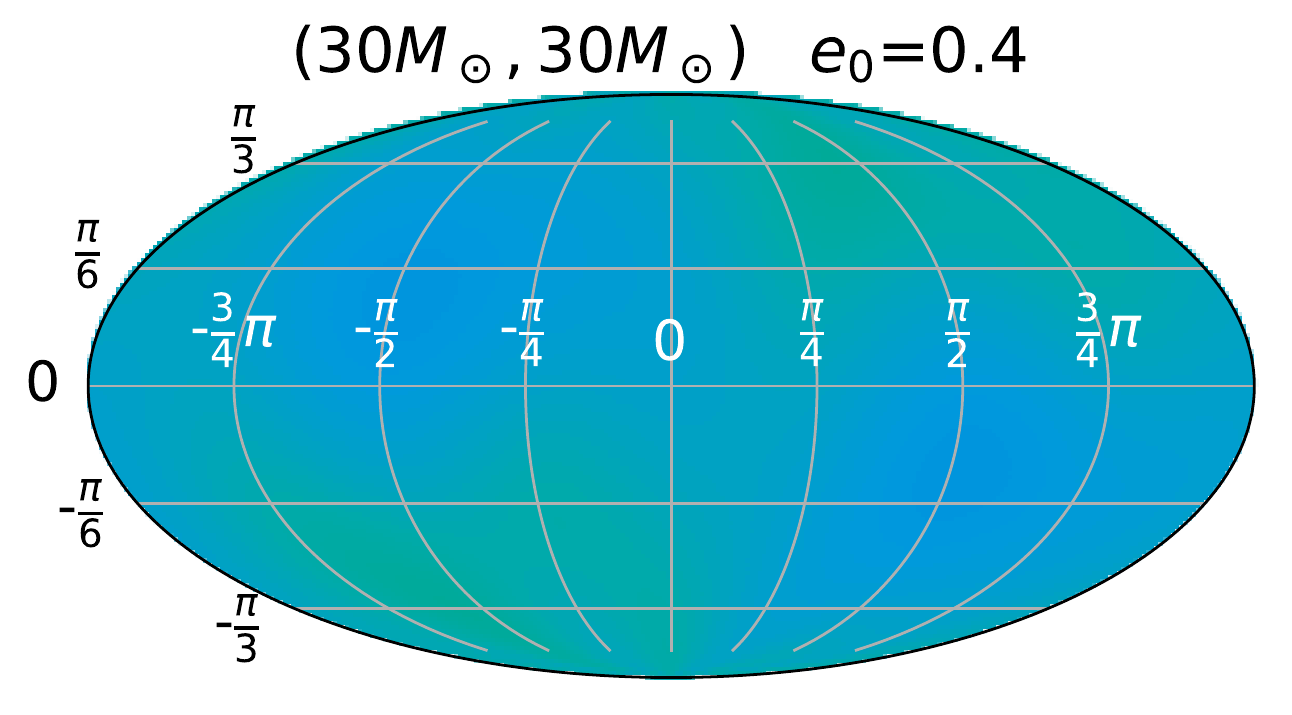}
		\includegraphics[width=\wid\textwidth]{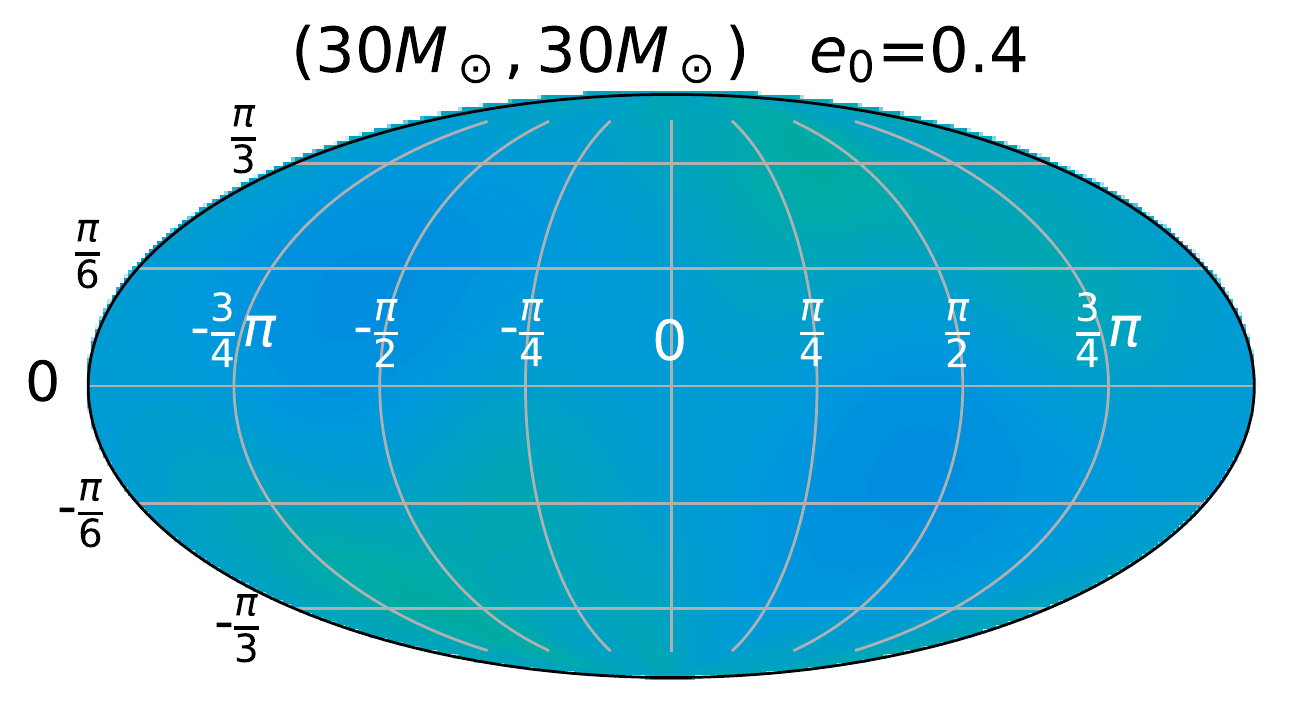}
		\includegraphics[width=\wid\textwidth]{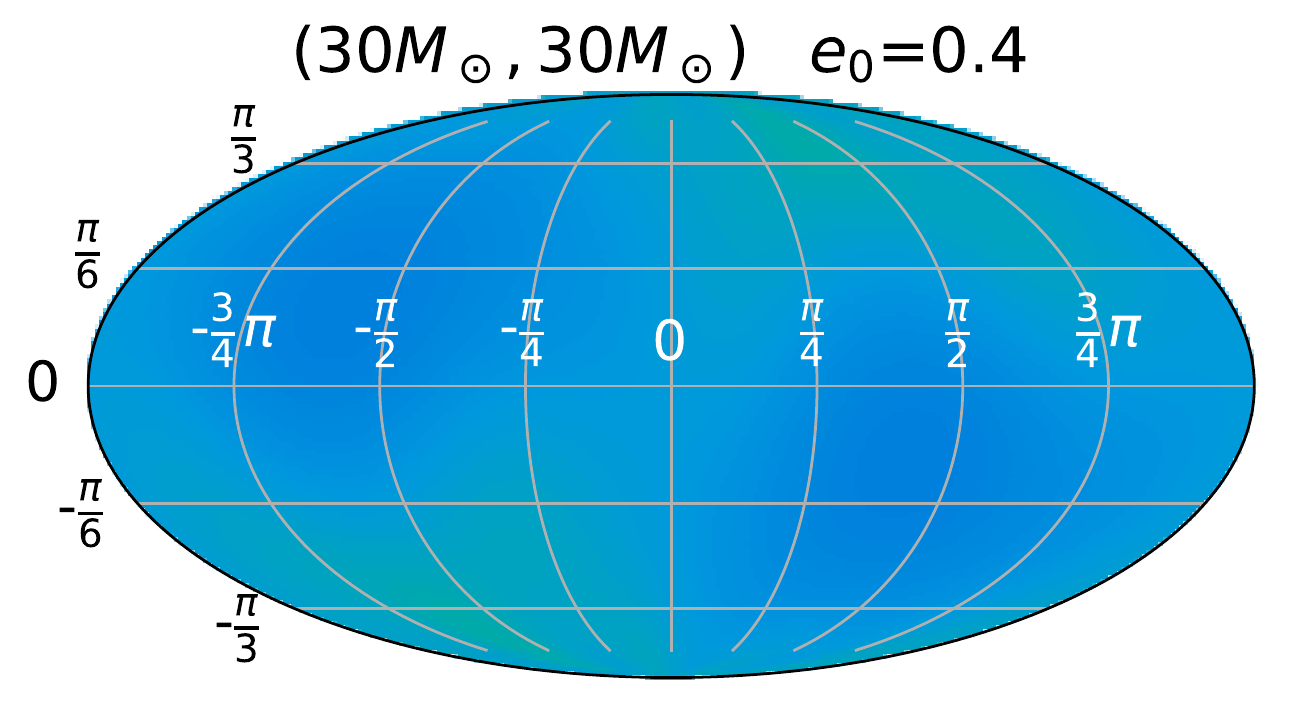}
	}
	\centerline{
		\includegraphics[width=\wid\textwidth]{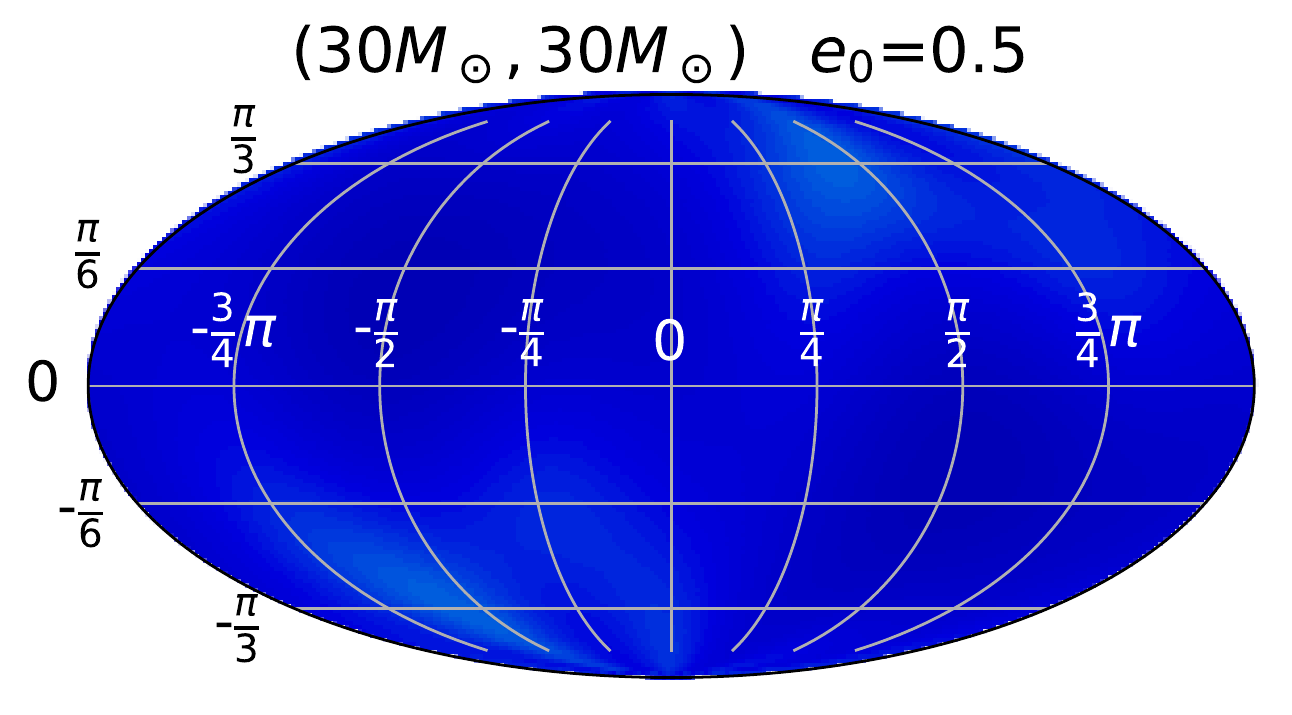}
		\includegraphics[width=\wid\textwidth]{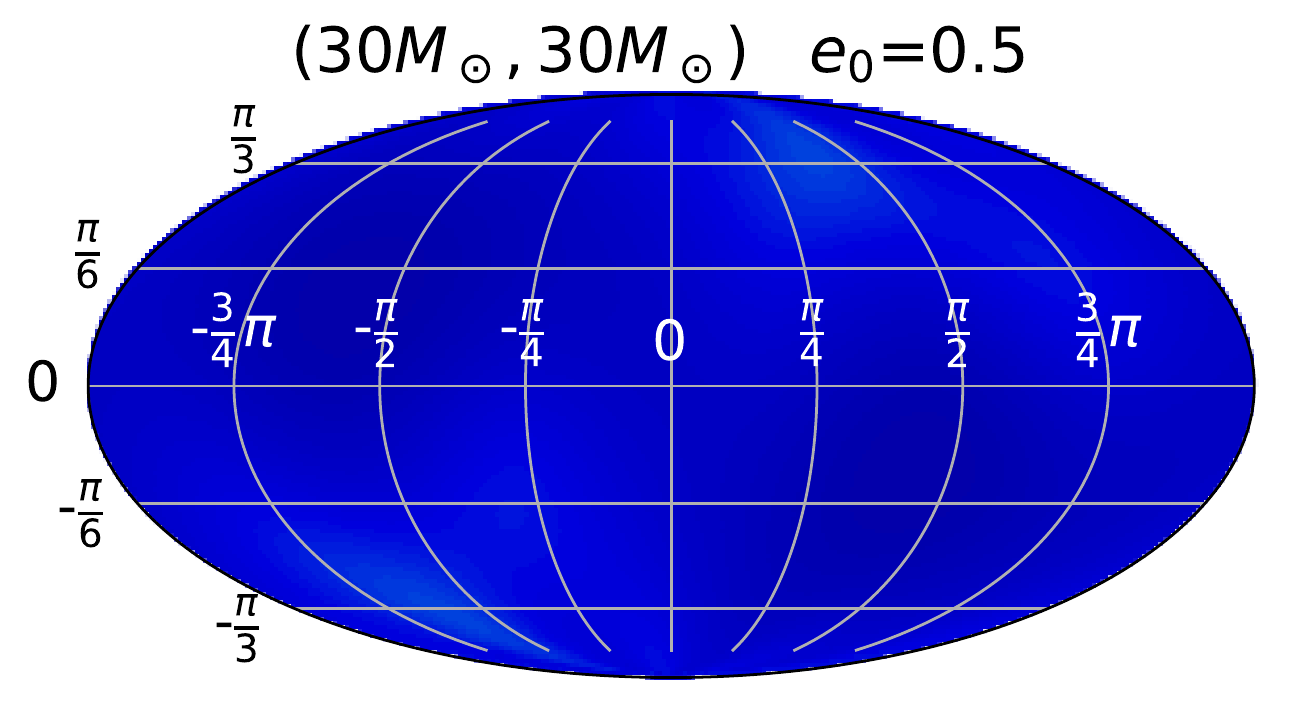}
		\includegraphics[width=\wid\textwidth]{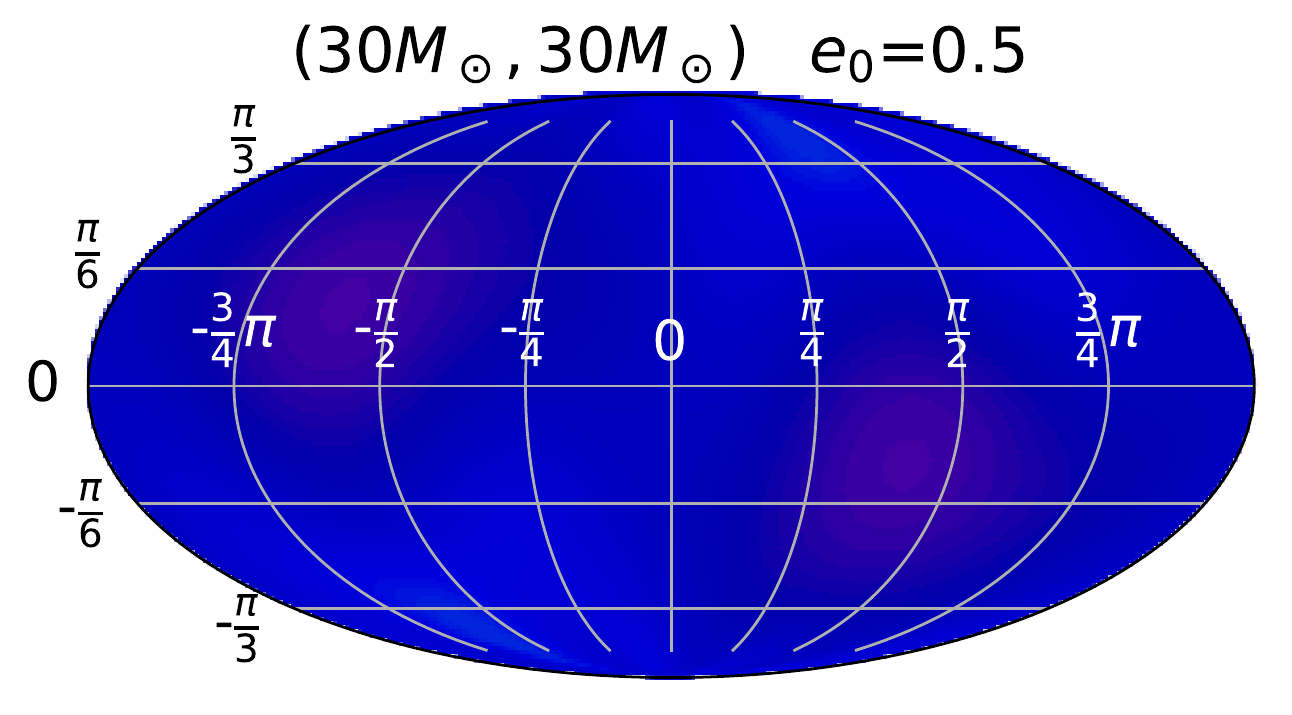}
	}
	\centerline{
		\includegraphics[width=\wid\textwidth]{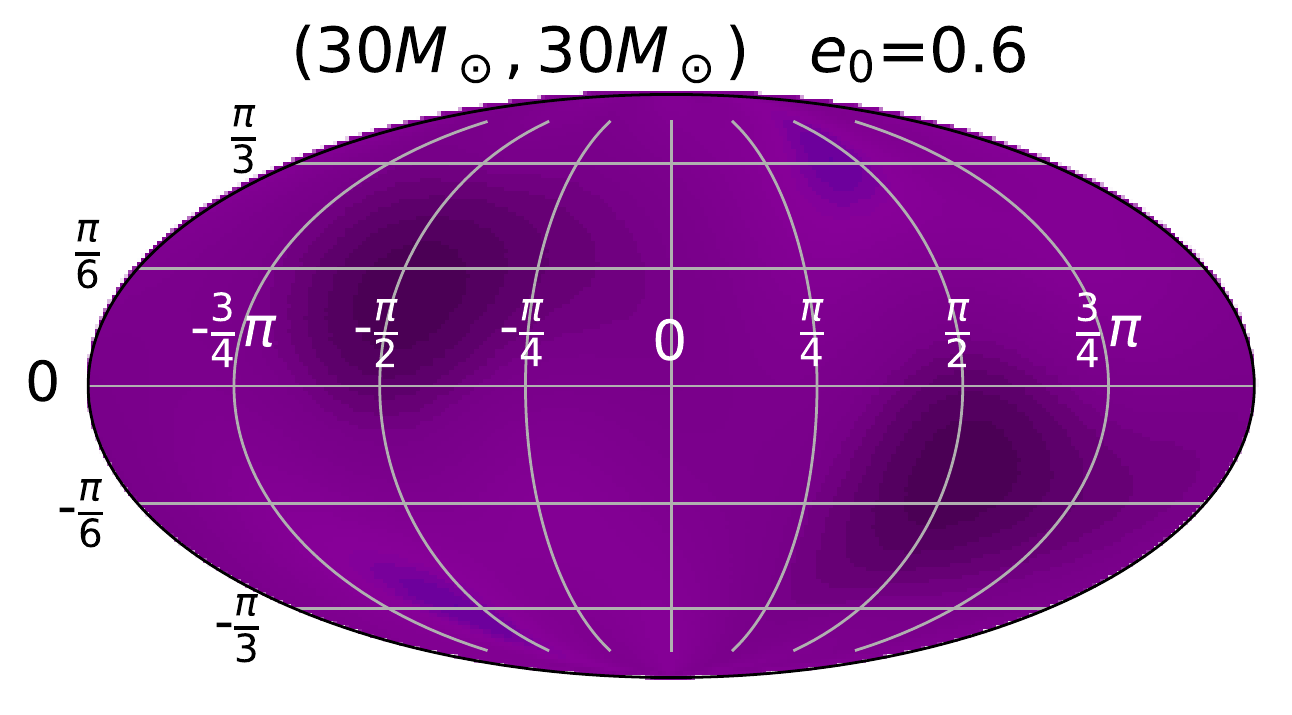}
		\includegraphics[width=\wid\textwidth]{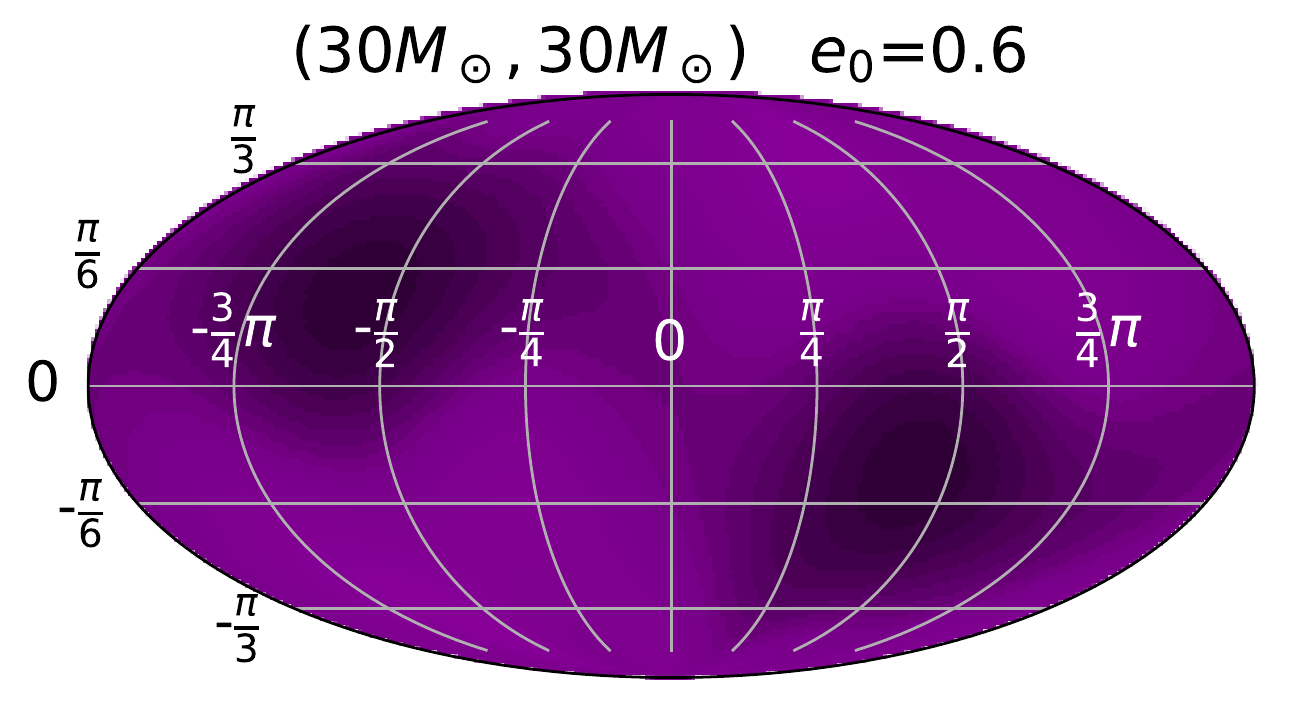}
		\includegraphics[width=\wid\textwidth]{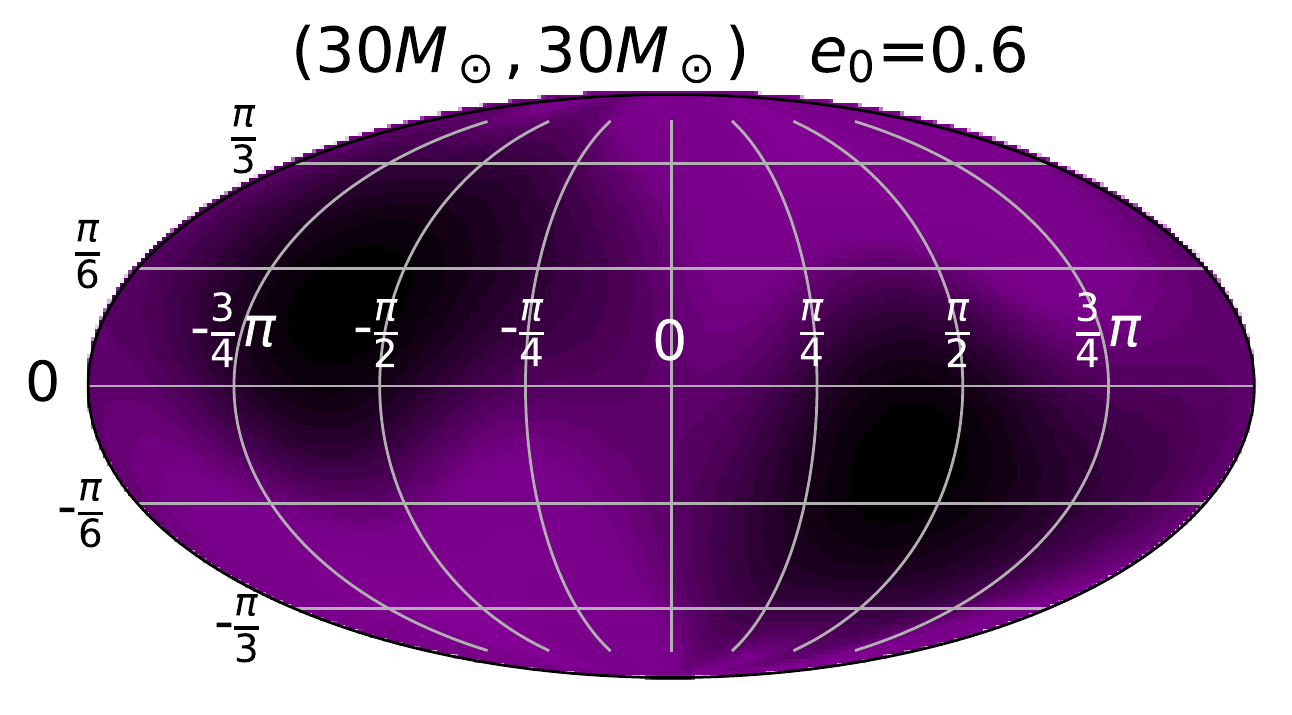}
	}
	\caption{As Figure~\ref{fig:etd_equal_mass} but now setting the binary inclination angle to 0.} 
	\label{fig:etd_equal_opt}
\end{figure*}

\begin{figure*}
	\centerline{
		\includegraphics[width=\textwidth]{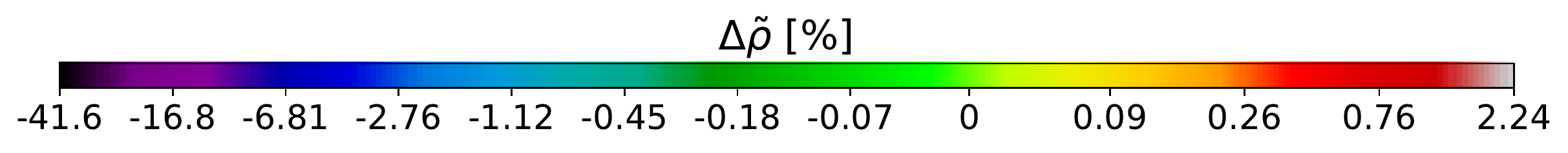}
	}
	\centerline{
		\includegraphics[width=\wid\textwidth]{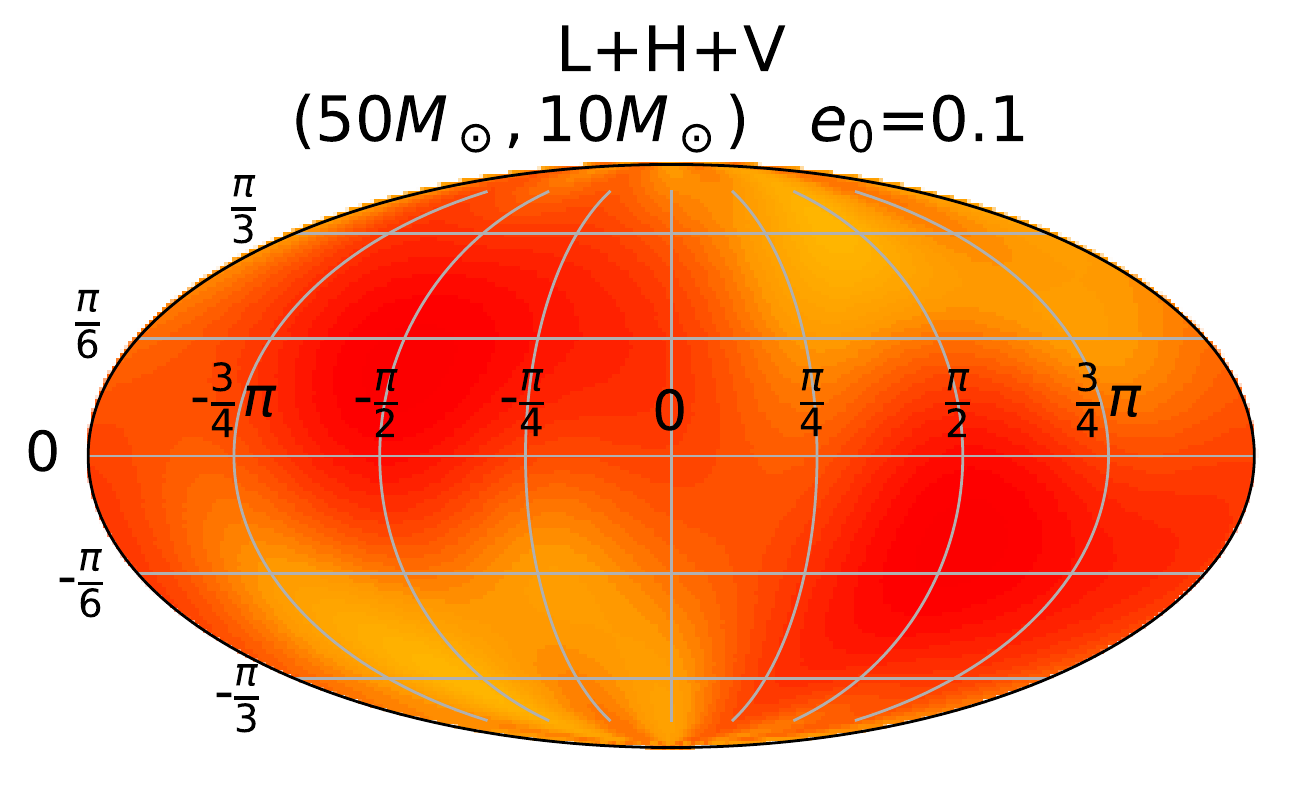}
		\includegraphics[width=\wid\textwidth]{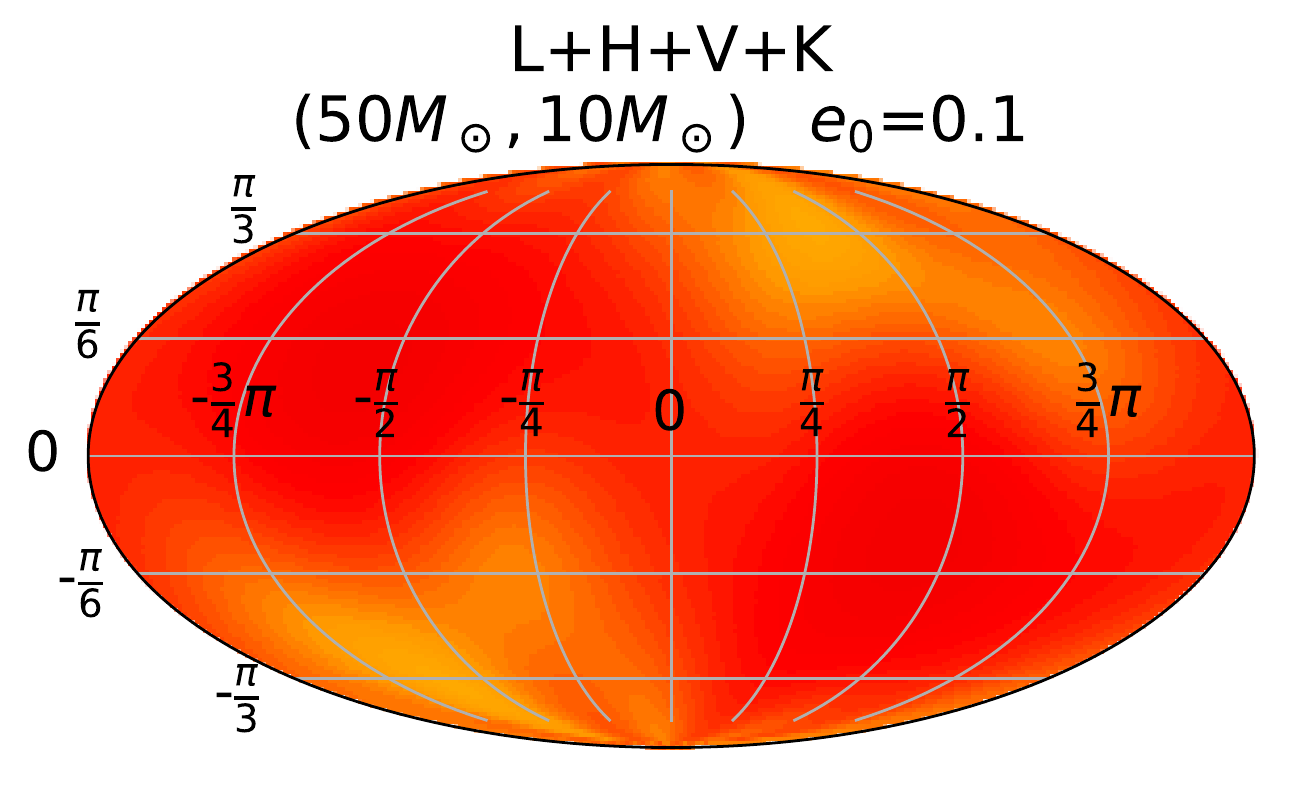}
		\includegraphics[width=\wid\textwidth]{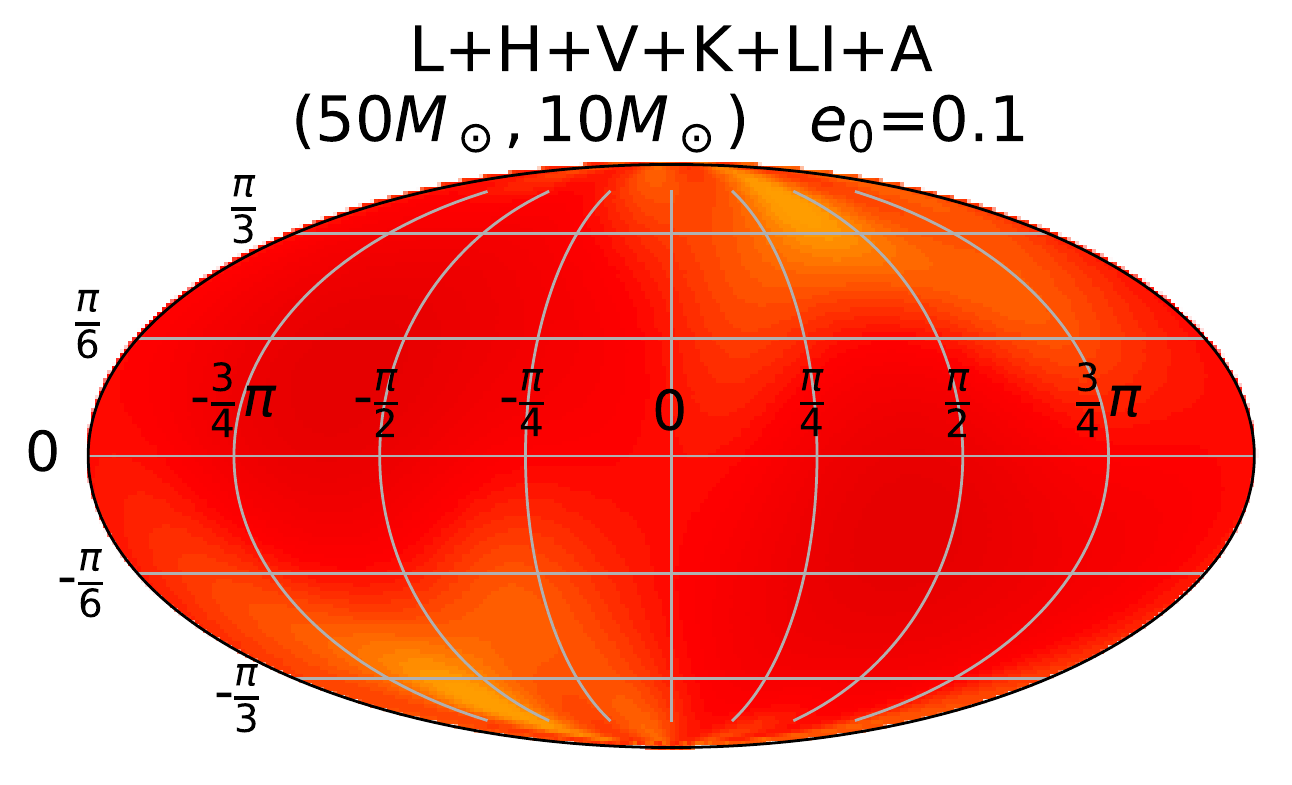}
	}
	\centerline{
		\includegraphics[width=\wid\textwidth]{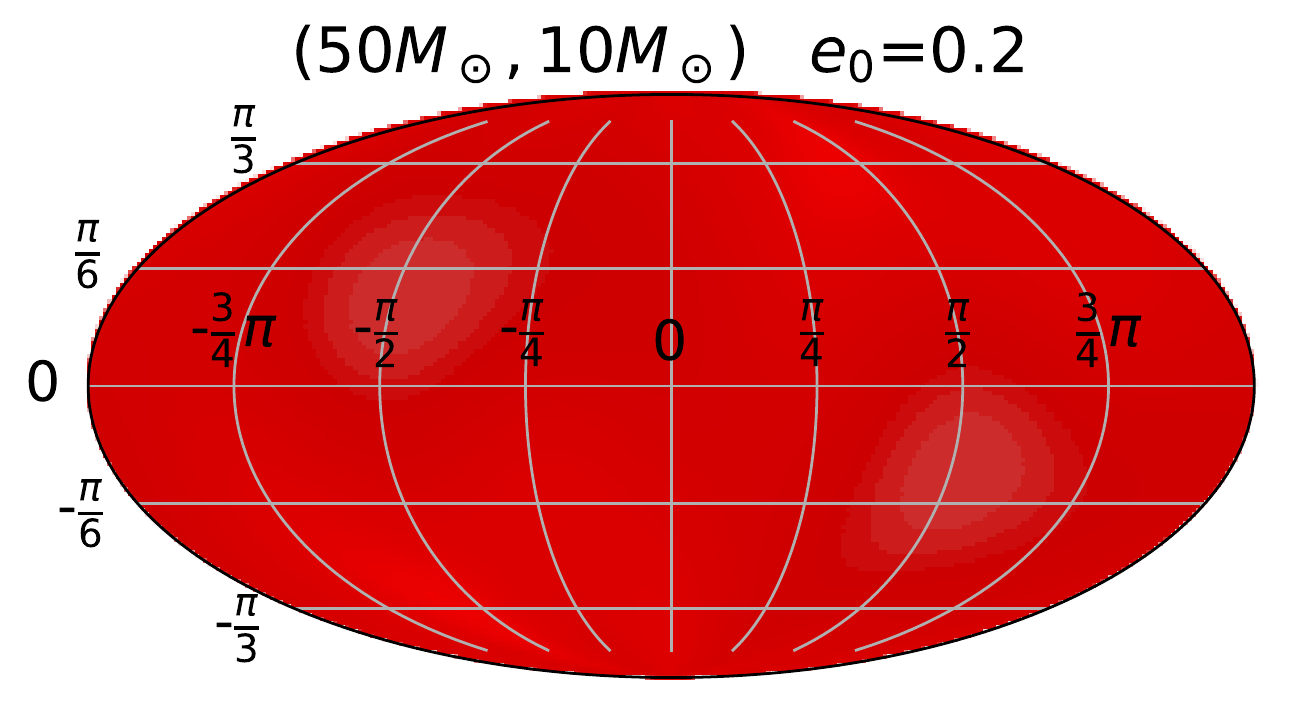}
		\includegraphics[width=\wid\textwidth]{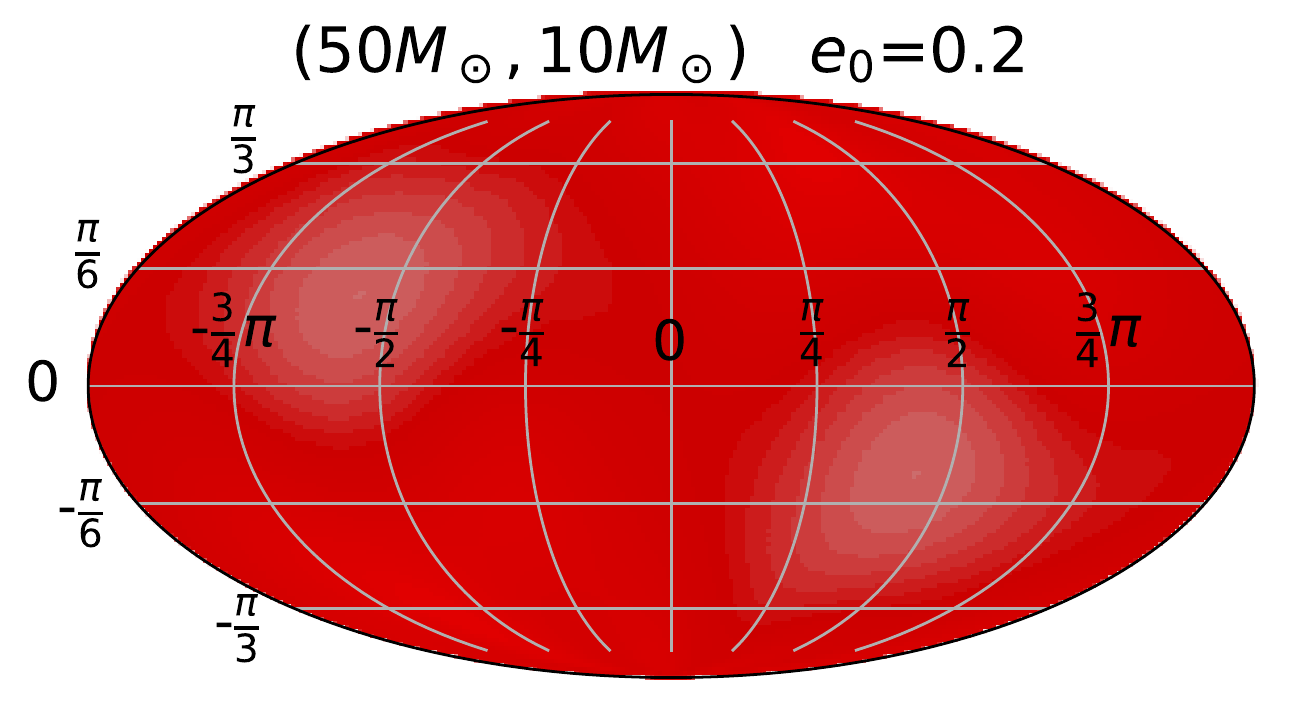}
		\includegraphics[width=\wid\textwidth]{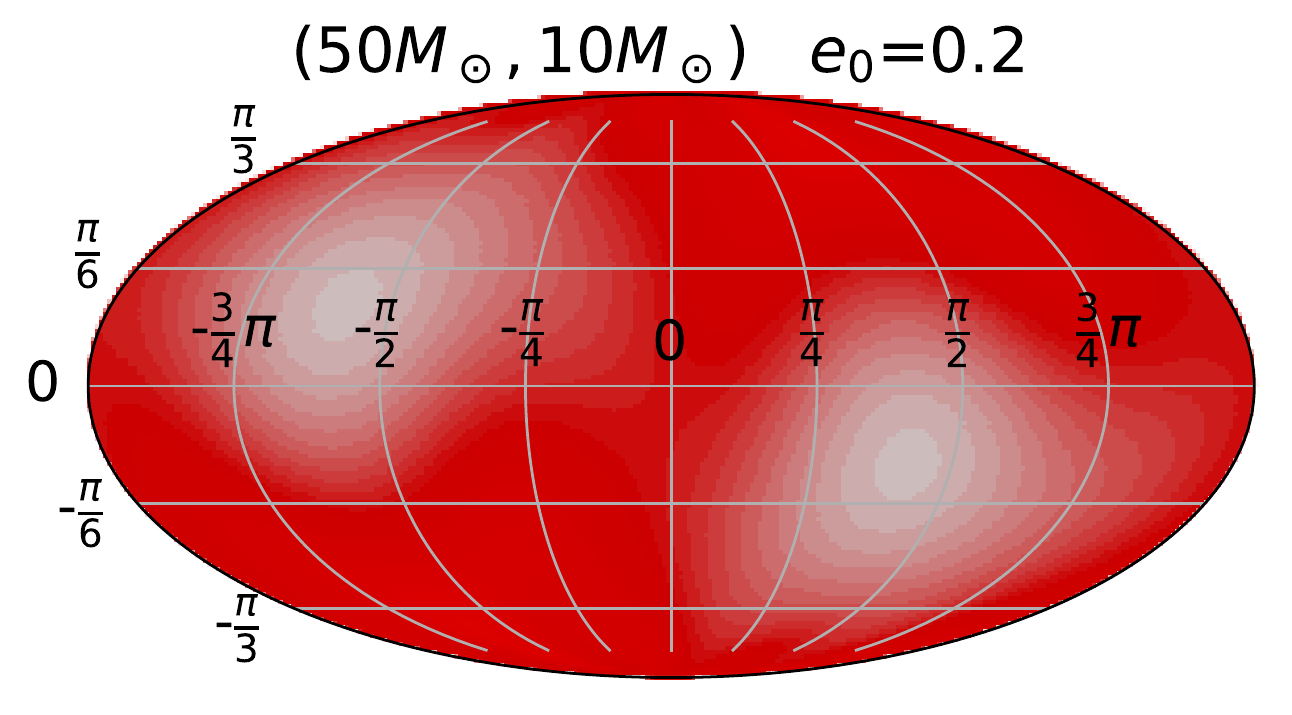}
	}
	\centerline{
		\includegraphics[width=\wid\textwidth]{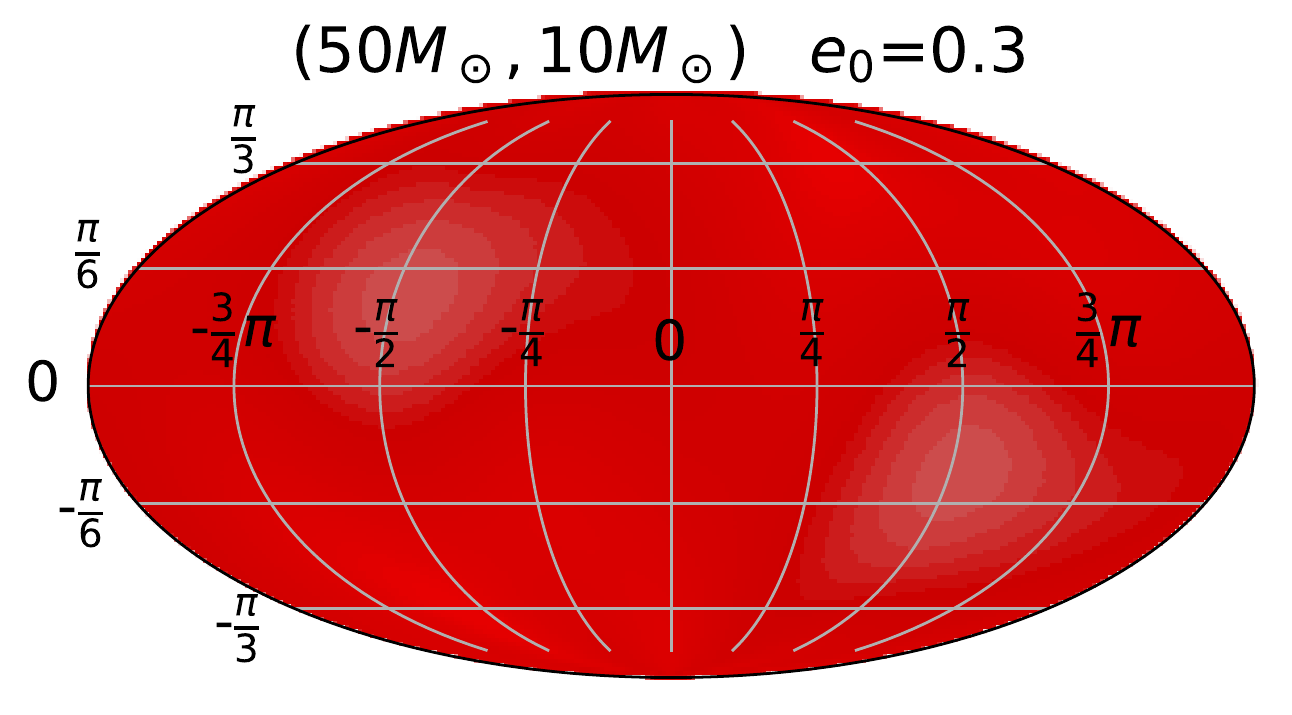}
		\includegraphics[width=\wid\textwidth]{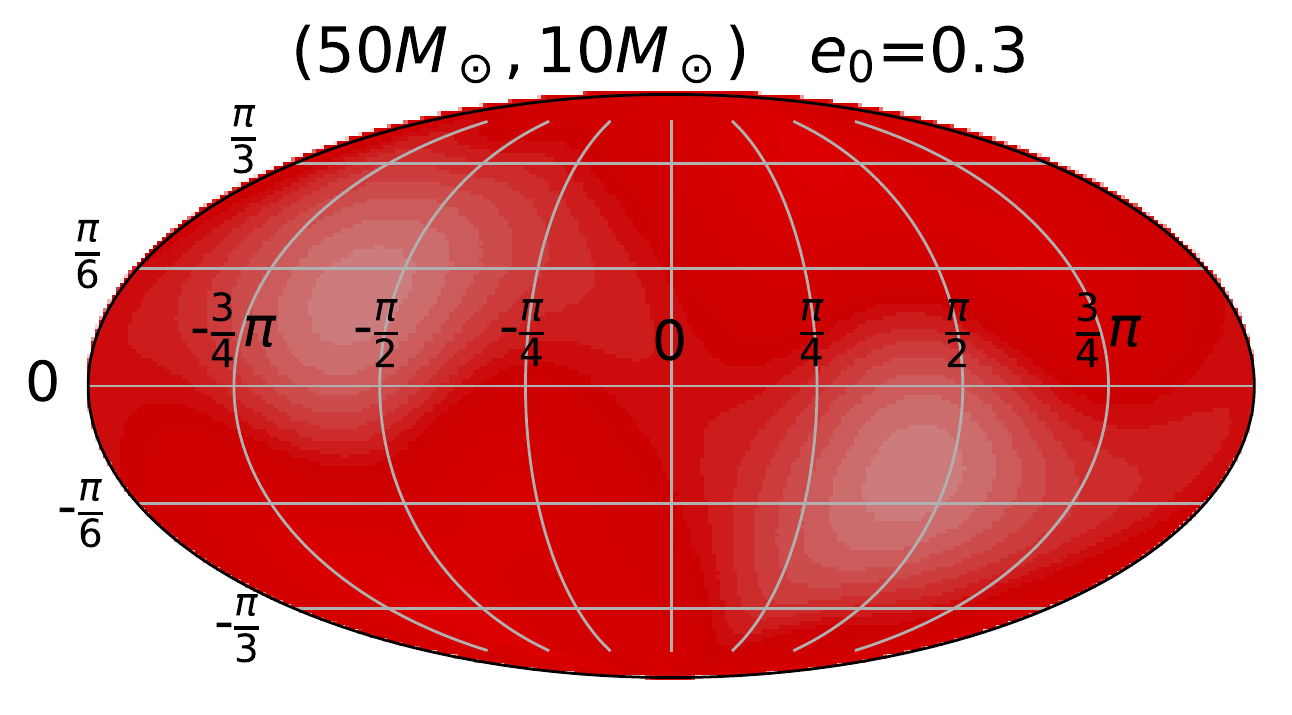}
		\includegraphics[width=\wid\textwidth]{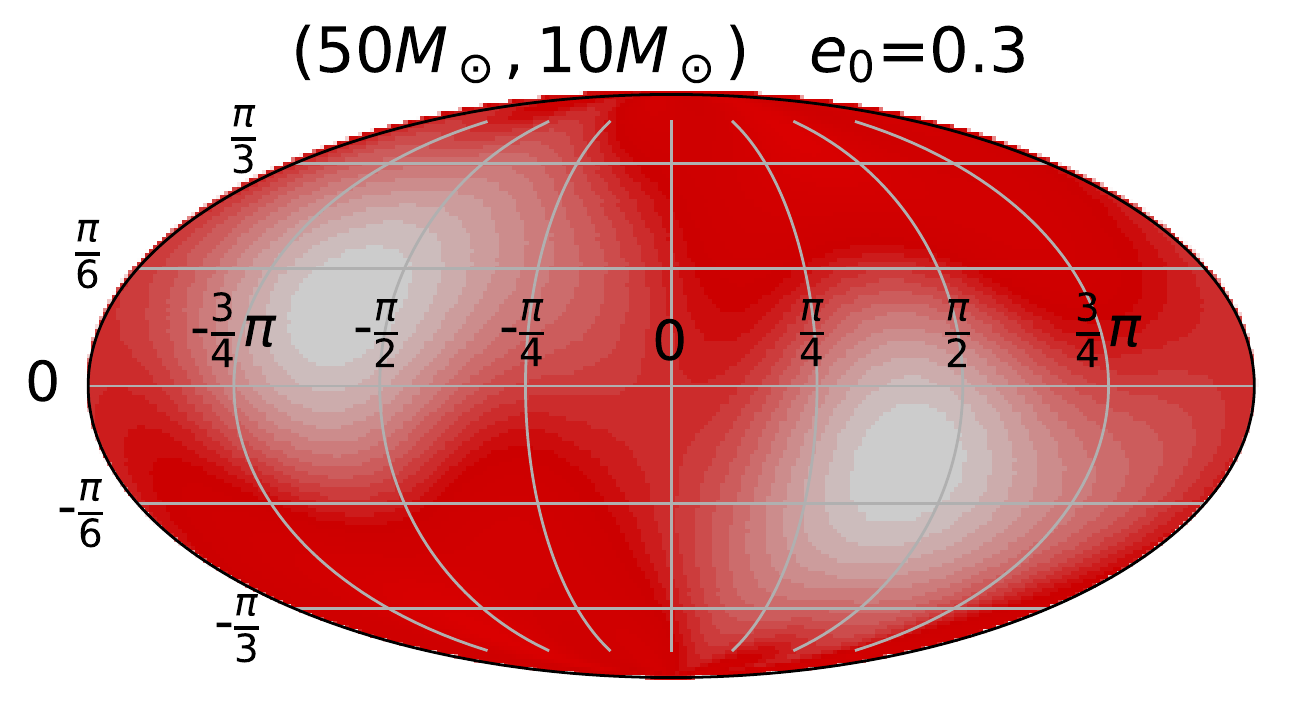}
	}
	\centerline{
		\includegraphics[width=\wid\textwidth]{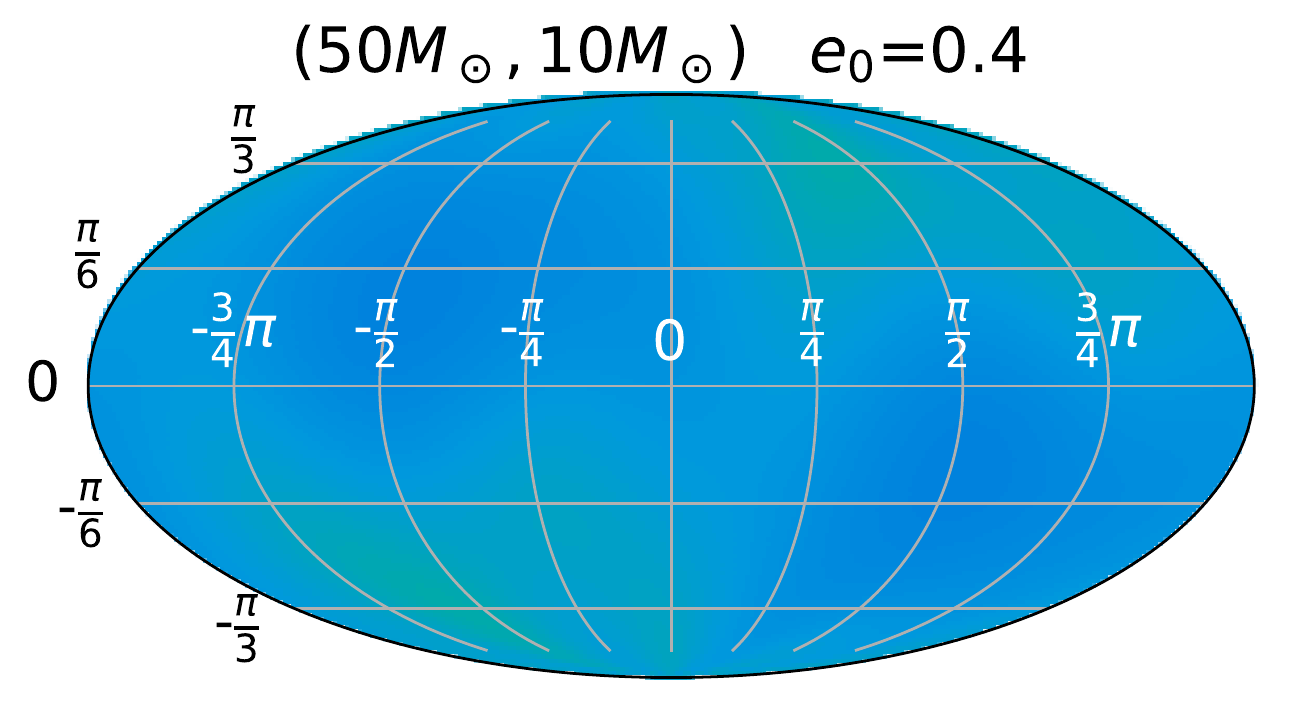}
		\includegraphics[width=\wid\textwidth]{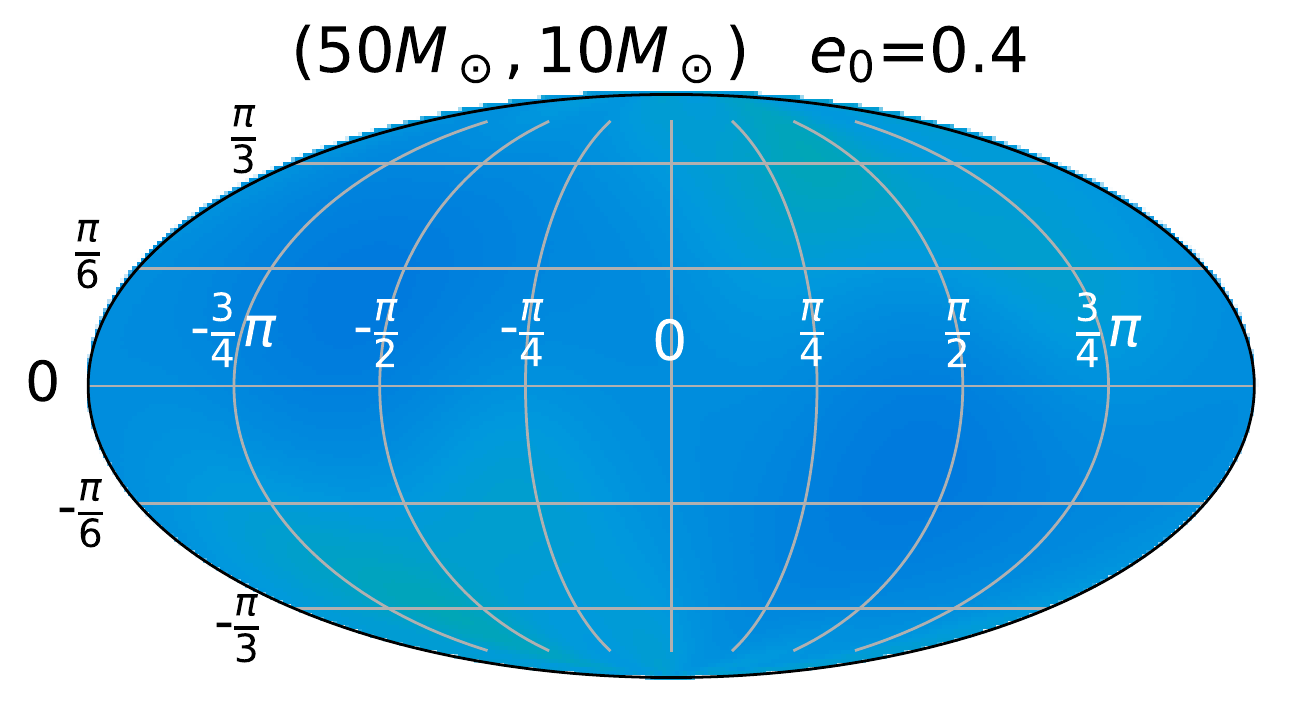}
		\includegraphics[width=\wid\textwidth]{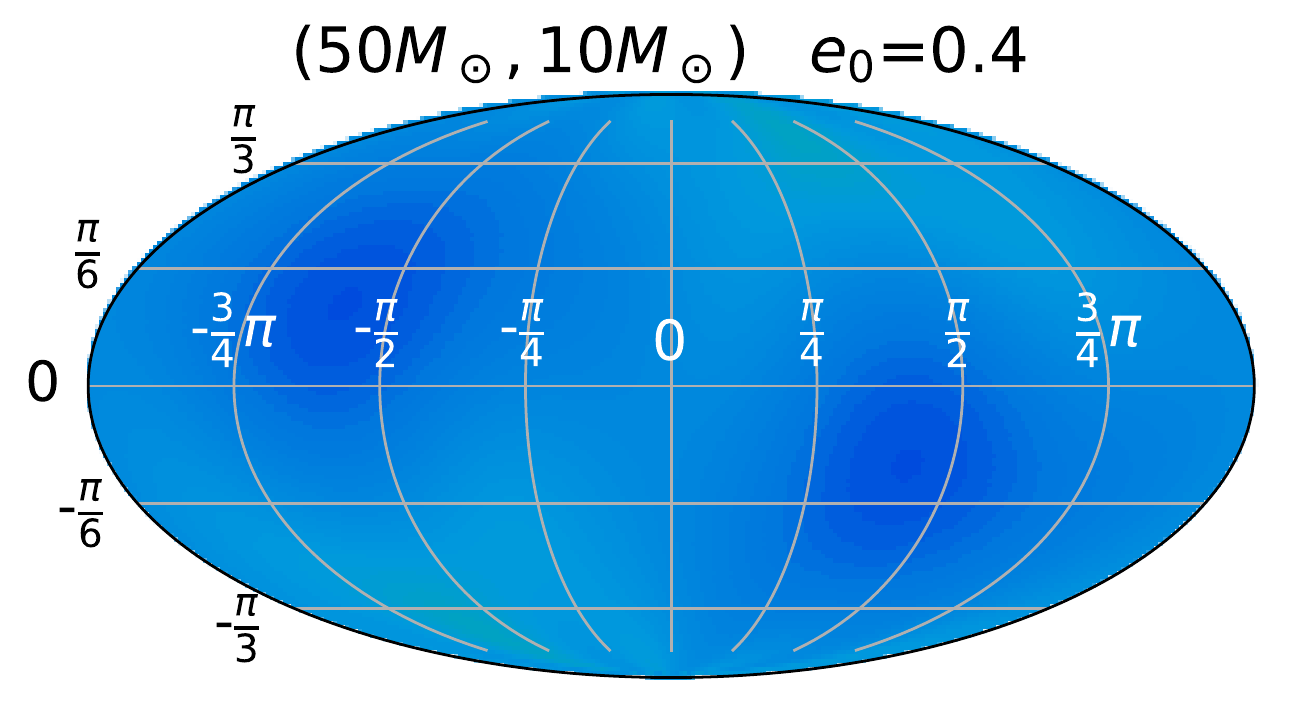}
	}
	\centerline{
		\includegraphics[width=\wid\textwidth]{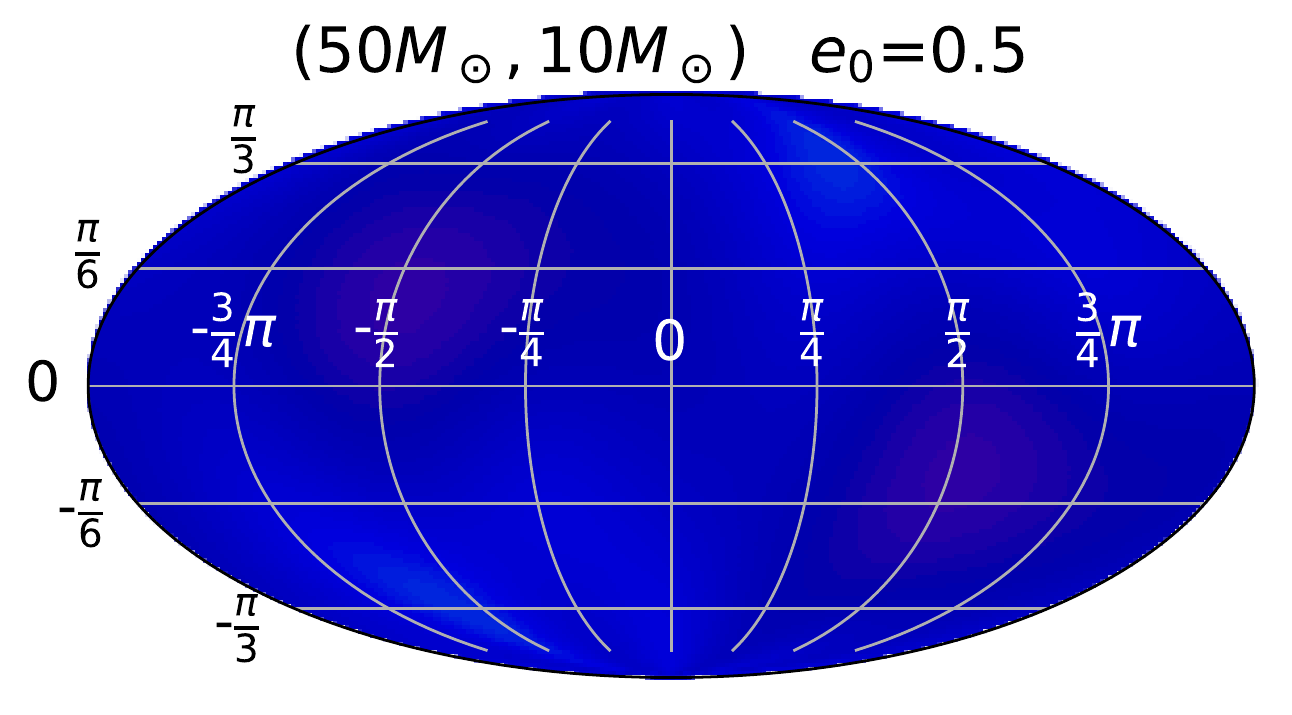}
		\includegraphics[width=\wid\textwidth]{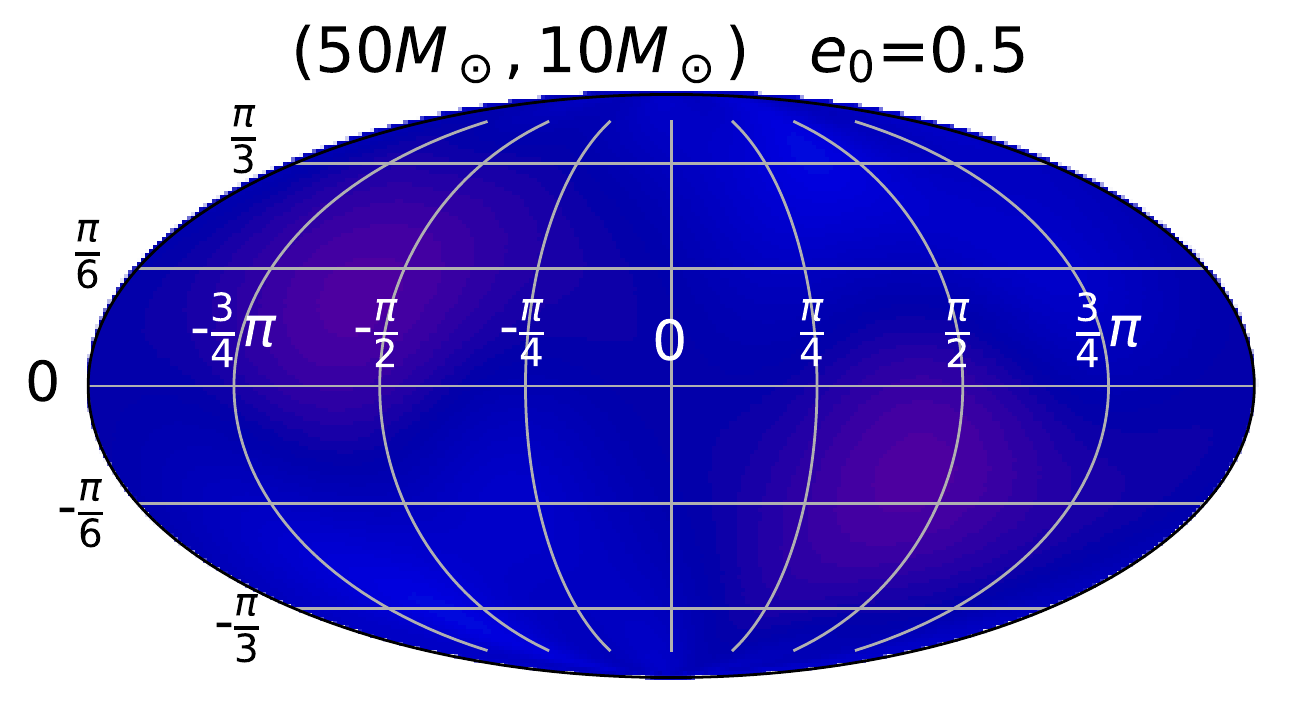}
		\includegraphics[width=\wid\textwidth]{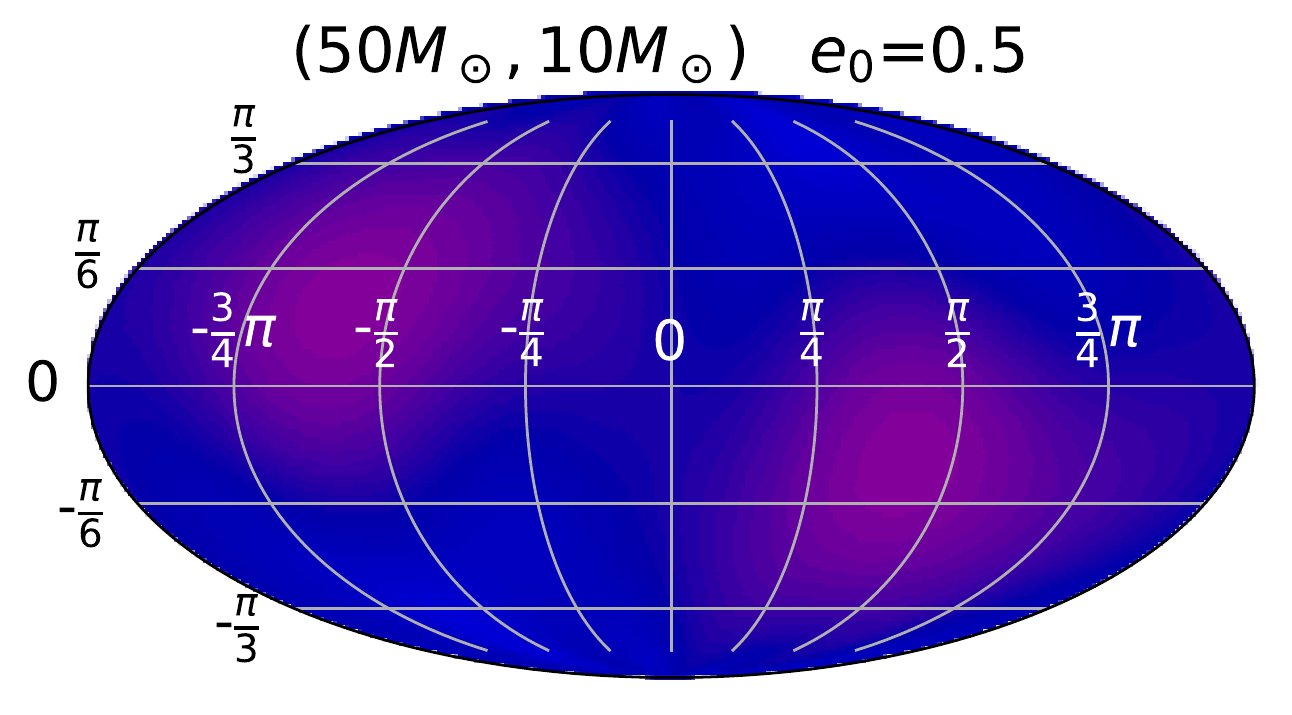}
	}
	\centerline{
		\includegraphics[width=\wid\textwidth]{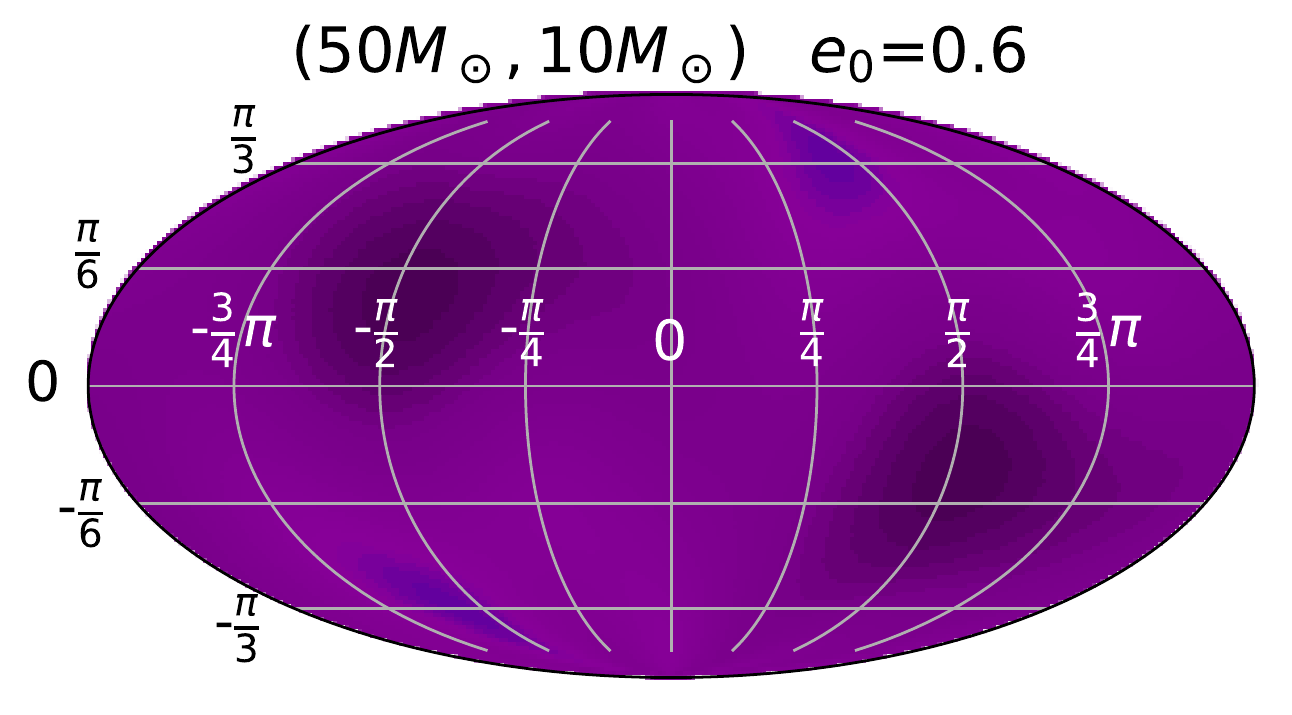}
		\includegraphics[width=\wid\textwidth]{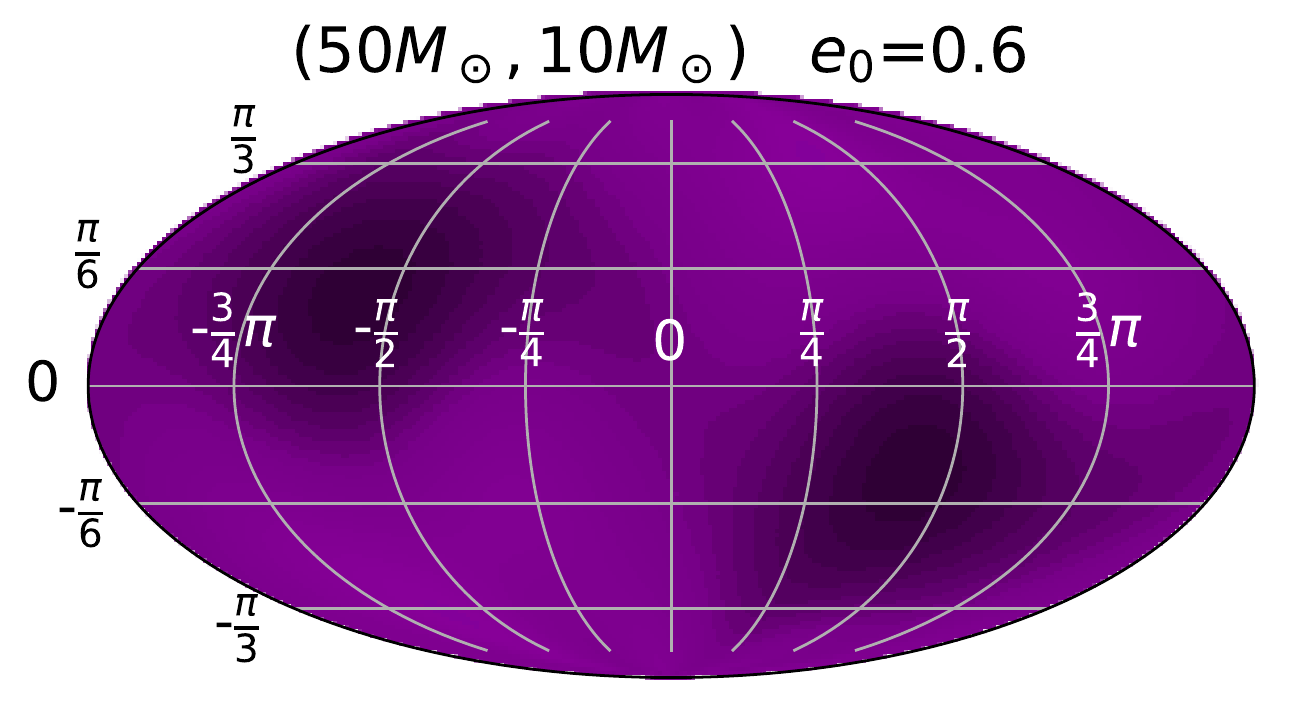}
		\includegraphics[width=\wid\textwidth]{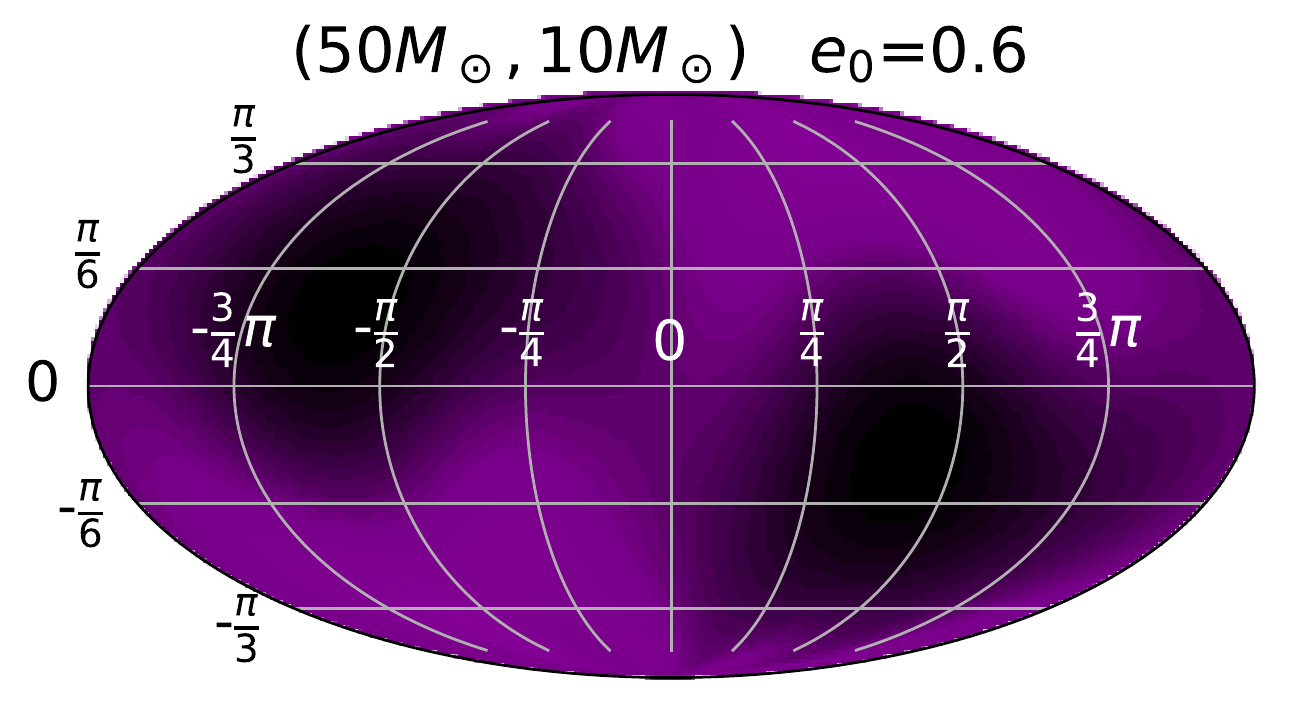}
	}
	\caption{As Figure~\ref{fig:etd_equal_opt} but now for component masses \((50\msun,\,10\msun)\).} 
	\label{fig:etd_five_opt}
\end{figure*}

\end{widetext}

\clearpage


\bibliography{references,dl_references,ecc_references}
\bibliographystyle{apsrev4-1}

\end{document}